\documentclass[11pt,a4paper]{article}
\pdfoutput=1
\usepackage{jheppub}

\usepackage[utf8]{inputenc}
\usepackage{graphicx}
\usepackage{amsmath,amsfonts,amssymb}
\usepackage{epsfig,color}
\usepackage[outdir=./]{epstopdf}
\usepackage[T1]{fontenc}
\usepackage{graphicx}% Include figure files
\usepackage{color} % use colored fonts
\usepackage{dcolumn}% Align table columns on decimal point
\usepackage{bm}% bold math
\usepackage{epsfig}
\usepackage[subfigure]{graphfig}
\usepackage{latexsym}
\usepackage{amsmath}
\usepackage{amsfonts}
\usepackage{amsxtra}
\usepackage{hyperref}
\usepackage{placeins}

\newcommand{\msbar}{\overline{\mbox{{\rm MS}}}}

\newcommand{\lwrsim}{\raise0.3ex\hbox{$<$\kern-0.75em\raise-1.1ex\hbox{$\sim$}}}
\def\krto{ {\,\,\lower .8ex\hbox {$\longrightarrow \atop k \rightarrow 0$}\,\,}}

\def\bea{\begin{eqnarray} }
\def\beq{\begin{eqnarray} }

\def\eea{\end{eqnarray}}
\def\eeq{\end{eqnarray}}

\preprint{\begin{tabular}{r}JLAB-THY-21-3409\end{tabular}}
\begin{document}

%%%%%%%%%%%%%%%%%%%%%%%%%%%%%%%%%%%%%%%%%%%%%%
%% title page 
%%%%%%%%%%%%%%%%%%%%%%%%%%%%%%%%%%%%%%%%%%%%%%

\title{The Continuum and Leading Twist Limits of Parton Distribution Functions in Lattice QCD}

\author[a]{Joseph Karpie}
\author[b,c]{, Kostas Orginos}
\author[d,c]{, Anatoly Radyushkin}
\author[e]{and Savvas Zafeiropoulos}
\author{\\(for the HadStruc Collaboration)}
\affiliation[a]{Physics Department, Columbia University, \\ New York City, New York 10027, USA}
\affiliation[b]{Department of Physics, The College of William \& Mary, \\ Williamsburg, VA 23187, USA}
\affiliation[c]{Thomas Jefferson National Accelerator Facility, \\ Newport News, VA 23606, USA}
\affiliation[d]{Old Dominion University, \\ Norfolk, VA 23529, USA}
\affiliation[e]{Aix Marseille Univ, Universit\'e de Toulon, CNRS, CPT, Marseille, France}
\emailAdd{jmk2289@columbia.edu}
\emailAdd{kostas@wm.edu}
\emailAdd{radyush@jlab.org}
\emailAdd{savvas.zafeiropoulos@cpt.univ-mrs.fr}

\abstract{
In this study, we present continuum limit results for the unpolarized parton distribution function of the nucleon computed in
lattice QCD. This study is the first continuum limit using the pseudo-PDF approach with Short Distance Factorization for factorizing lattice QCD calculable matrix elements. Our findings are also compared with the pertinent phenomenological determinations. Inter alia, we are employing the summation Generalized Eigenvalue Problem (sGEVP) technique in order to optimize our control over the excited state contamination which can be one of the most serious systematic errors in this type of calculations. A crucial novel ingredient of our analysis is the parameterization of systematic errors using Jacobi polynomials to characterize and remove both lattice spacing and higher twist contaminations, as well as the leading twist distribution. This method can be expanded in further studies to remove all other systematic errors.
}
  
\maketitle

\date{\today}

\flushbottom
%%%%%%%%%%%%%%%%%%%%%%%%%%%%%%%%%%%%%%%%%%%%%%%%%%%%%%%%%
%% body of the paper
%%%%%%%%%%%%%%%%%%%%%%%%%%%%%%%%%%%%%%%%%%%%%%%%%%%%%%%%%

\section{Introduction}
Ever since the pioneering deep inelastic lepton-proton scattering (DIS) experiments at SLAC in 1973 which yielded the first evidence for proton structure, the excitement towards the understanding of the fundamental constituents of the nucleon lead to a culmination of theoretical and experimental results. From the theoretical side the QCD factorization theorem allows for a separation of the hadronic cross sections into a perturbative, process dependent partonic cross section and nonperturbative, process independent parton distribution functions (PDFs). Thus, in order to decipher the information coming from the Large Hadron Collider (LHC) experiments, and in order to capitalize maximally the potential of the upcoming Electron-Ion Collider (EIC), it is of vital importance to accurately determine the PDFs. The calculation of the momentum distribution that bound quarks and gluons carry within the proton is a nonperturbative problem which due to the lightcone nature of the PDFs was elusive to lattice QCD calculations until very recently. Consequently, progress was being made through global fits of experimental data or modelling. PDFs have a huge phenomenological value since they constitute a fundamental limit for the Higgs boson characterization in terms of its couplings, they are the dominant systematics for precision Standard Model (SM) measurements, such as the $W$ boson mass, but also the biggest uncertainties for beyond the SM heavy particle production. Therefore a precise knowledge of the PDFs can help to rule out a broad class of BSM models. For a comprehensive review we refer the reader to~\cite{Gao:2017yyd}.

A series of lattice methodologies have been developed to circumvent the issues stemming from the light-cone nature of PDFs and in the recent years there has been a new dawn in the ab-initio determination of lightcone parton distributions via numerical lattice simulations. The first such approach~\cite{Liu:1993cv,Liu:1998um,Liu:1999ak} was to directly calculate the hadronic tensor and factorize it in a similar way that one adopts when interpreting DIS data. In this direct approach, one can study not only the DIS regime, but also the Resonance and Shallow Inelastic regimes, which makes it unique amongst the lattice approaches. A related approach~\cite{Aglietti:1998ur} was proposed to calculate Distribution Amplitudes where between the two currents a scalar quark will be inserted instead.
The most widely adopted approach, which is also the one most responsible for today's vast amount of activity in lattice calculations of PDFs, is the Large Momentum Effective Theory (LaMET) method~\cite{Ji:2013dva}. In this approach, a matrix element of an operator with a spacelike separation is Fourier transformed with respect to the separation length to get a so-called quasi-distribution. A factorization theorem is applied when the matrix element has sufficiently large external momentum, hence the name LaMET. 
The next approach~\cite{Detmold:2005gg,Chambers:2017dov,Karpie:2018zaz,Detmold:2021uru}, called OPE-without-OPE, is to calculate either the hadronic tensor or the matrix elements with spacelike separations used in later approaches. These matrix elements can be used to determine moments of the PDF through the OPE and from those moments determine the PDF. Attempts to reconstruct the PDF with the limited number of moments accessible from local matrix elements had already met with some success~\cite{Detmold:2003rq}. In the OPE-without-OPE approach, far more moments are accessible than through the calculation of local matrix elements and the PDF extraction can be systematically improved as more moments are obtains. 
Finally, the approach, which is adopted in this study, was proposed originally in~\cite{Braun:2007wv}, and now is usually called Short Distance Factorization (SDF), is to calculate a matrix element of operators with a spacelike separation, which are now typically named pseudo-distributions or lattice cross sections (LCS) when using this approach. The Operator Product Expansion (OPE) is used to relate these matrix elements to light cone matrix elements through a factorization theorem at small spacelike separation. Originally, SDF was proposed to determine meson distribution amplitudes~\cite{Braun:2007wv}, but the method was later independently reinvented by~\cite{Radyushkin:2017cyf,Ma:2017pxb}, who were motivated by LaMET, for calculations of the PDFs.
Since this approach would use the same type of matrix elements as in LaMET, LaMET and SDF are intimately related in their factorization theorems, but provide two distinct limits for approaching the light-cone distributions with their different power corrections. The power corrections for SDF are ordered by matrix elements of operators with distinct twist $t$ and are proportional to $(z^2)^{t-2}$. The power corrections of LaMET come from matrix elements with mixed twist and are proportional to $(p_3^{-2})^n$.

All these frameworks that target the ab-initio study of parton physics via lattice calculations in Euclidean space incited a feverish activity of the lattice community~\cite{Liu:1993cv,Liu:1998um,Liu:1999ak,Detmold:2005gg,Chambers:2017dov,Karpie:2018zaz,Detmold:2021uru,Detmold:2003rq,Braun:2007wv,Radyushkin:2017cyf,Ma:2017pxb,Ji:2013dva,Aglietti:1998ur,Musch:2010ka,Xiong:2013bka,Lin:2014yra,Lin:2014zya,Ji:2014gla,Ma:2014jla,Ji:2015jwa,Ji:2015qla,Monahan:2015lha,Alexandrou:2015rja,Li:2016amo,Chen:2016utp,Alexandrou:2016jqi,Monahan:2016bvm,Radyushkin:2016hsy,Zhang:2017bzy,Constantinou:2017sej,Alexandrou:2017huk,Chen:2017mzz,Orginos:2017kos,Ji:2017rah,Ji:2017oey,Stewart:2017tvs,Ishikawa:2017faj,Lin:2017ani,Hobbs:2017xtq,Jia:2017uul,Bali:2017gfr,Radyushkin:2017lvu,Ishikawa:2019flg,Radyushkin:2018cvn,Zhang:2018ggy,Izubuchi:2018srq,Xu:2018eii,Alexandrou:2018pbm,Chen:2018xof,Chen:2017mie,Chen:2018fwa,Jia:2018qee,Briceno:2018lfj,Alexandrou:2018eet,Liu:2018uuj,Bali:2018spj,Lin:2018qky,Radyushkin:2018nbf,Fan:2018dxu,Zhang:2018diq,Li:2018tpe,Braun:2018brg,Liu:2018hxv,Sufian:2019bol,Karpie:2019eiq,Cichy:2019ebf,Alexandrou:2019lfo,Bali:2019ecy,Hobbs:2019gob,Detmold:2019ghl,Chen:2019lcm,Izubuchi:2019lyk,Joo:2019jct,Joo:2019bzr,Ebert:2019tvc,Ji:2019sxk,Ji:2019ewn,Radyushkin:2019mye,Sufian:2020vzb,Green:2020xco,Chai:2020nxw,Shanahan:2020zxr,Lin:2020ssv,Braun:2020ymy,Joo:2020spy,Bhat:2020ktg,Ji:2020ect,Zhang:2020dkn,Fan:2020nzz,Chen:2020arf,Zhang:2020gaj,Zhang:2020dbb,Chen:2020iqi,Li:2020xml,Bhattacharya:2020jfj,Chen:2020ody,DelDebbio:2020cbz,DelDebbio:2020rgv,Gao:2020ito,Ji:2020byp,Alexandrou:2020tqq,Fan:2020cpa,Ji:2020brr,Alexandrou:2020zbe,Lin:2020rxa,Alexandrou:2020qtt,Lin:2020fsj,Sufian:2020wcv,Bringewatt:2020ixn,Zhang:2020rsx,Gao:2021hxl,Huo:2021rpe,Bhattacharya:2021boh}. 

In~\cite{Cichy:2018mum,Lin:2020rut,Constantinou:2020pek} one can find reviews of all the aforementioned lattice approaches to the extraction of light cone PDFs. 

The lattice regulator itself induces artifacts which contaminate any quantity that one wishes to compute. Ultimately, a continuum limit must be properly taken in order to remove this regulator dependence and to achieve the final result. The most widely used lattice actions and some observables are $\mathcal{O}(a)$ improved but this is not the case for the fermion bilocal matrix element that is the starting point of the quasi and pseudo-distributions. Also for the time being no Symanzik program has been developed for this quantity and consequently the cut-off effects are of $\mathcal{O}(a)$. This issue could in principle translate itself in sizeable cut-off effects and render the approach to the continuum limit quite tricky~\cite{Alexandrou:2020qtt,Lin:2020fsj}.
In this paper, we study the continuum limit of the nucleon unpolarized parton distributions employing the SDF method of Ioffe time distributions and employing three lattice ensembles with lattice spacings ranging from 0.0749 fm to 0.0483 fm. The finest lattice is finer than those used in the continuum limit of previous LaMET studies~\cite{Alexandrou:2020qtt,Lin:2020fsj}. These ensembles allow us to get a firm understanding of the size of lattice artifacts that one encounters in such studies and also allow us to study the extrapolation to the continuum limit. In general, there are different ways that one can take the continuum limit which is complicated due to the two physical scales in a lattice PDF calculation, the spacelike separation and the momentum of the hadron. In previous continuum limit studies~\cite{Alexandrou:2020qtt,Lin:2020fsj}, the ensembles are chosen such that one of the scales, the momentum, is all the same across the ensembles. In this manuscript, we advocate for a method which has not been studied beforehand in the literature. It allows for any ensemble to be used even if the separation and momenta scales cannot be exactly matched. Moreover it will allow usage of data from all available momenta and separations, within the limitations of the assumptions made for the LaMET or SDF factorization. 

The manuscript is organized as follows, in Sec.~\ref{sec:pseudo}, we review the SDF approach as applied to the matrix elements which define quark-PDFs and pseudo-PDFs. In Sec.~\ref{sec:nuisance}, we define the models which are used to describe the leading twist PDF and the nuisance terms which control the systematic errors. In Sec.~\ref{sec:latt}, we describe the lattice QCD methodologies used for extraction of matrix elements. In Sec.~\ref{sec:bayes}, we discuss how Bayes' theorem is used to determine the most probable model parameters and how the fits to the PDF and nuisance terms are employed. In Sec.~\ref{sec:pdf_results}, we present the results of fitting PDFs to a range of models and discuss how the modification of the nuisance terms as well as the Bayesian priors affect the final results. Finally in Sec.~\ref{sec:conc} we conclude our findings and discuss future studies.

\section{Ioffe time pseudo-distributions}\label{sec:pseudo}
As described in~\cite{Braun:1994jq}, parton distributions can be described in terms of a boost invariant matrix element called the Ioffe time distribution (ITD), whose Fourier transform gives the standard PDFs. The ITD, up to a factor of $2P^\alpha$, is a special case of the generic Lorentz covariant matrix element in Eq.~\eqref{eq:matelem}, which was first studied at length in~\cite{Musch:2010ka} prior to proposals to factorize it~\cite{Ji:2013dva,Radyushkin:2017cyf},
\beq\label{eq:matelem}
M^{\alpha}(p,z) =\langle p | \bar{\psi}(z)\frac{\lambda_3}{2} \gamma^\alpha W(z;0) \psi(0)|p\rangle \,,
\eeq
where $W(z;0)$ denotes a straight Wilson line of length $z$ and $\lambda_3$ is a flavor Pauli matrix to project onto the flavor non-singlet distribution, which is easier to handle in lattice QCD. This matrix element has the Lorentz decomposition
\beq
M^\alpha(p,z) = 2p^\alpha \mathcal{M}(\nu,z^2) + 2 z^\alpha \mathcal{N}(\nu,z^2)\, ,
\eeq
where $\nu=p\cdot z$ is here called the Ioffe time~\cite{Braun:1994jq}. For timelike separations, it is equal to the product of mass of the hadron and the time $t$ ($m\, t$) in the hadron's rest frame. This time $t$ is what is typically referred to as the Ioffe time in analyses of DIS~\cite{IOFFE1969123}. The ITD is given by the special case of lightlike separation $z = (0,z_-, 0_T)$ and $\alpha = +$, where in light cone coordinates, the $+$ and $-$ directions are defined by the direction of the hadron's momentum. The Fourier transform of the ITD with respect to $\nu$ gives the PDF where $x$ is the Fourier-conjugate variable to $\nu$. Due to this lightlike separation, the ITD cannot be directly calculated from lattice QCD, where the calculation takes place in Euclidean space which only allows spacelike separations.

Following the framework of Short Distance Factorization (SDF)~\cite{Braun:2007wv}, the Lorentz invariant function $\mathcal{M}$, which will be called the Ioffe time pseudo-distribution (pseudo-ITD)~\cite{Radyushkin:2017cyf}, can be related to the ITD through a factorization relationship. This term can be isolated from the purely higher twist distribution $\mathcal{N}$ by a choice of $z = (0,0,z_3,0)$, $p = (0,0,p_3,E)$, and $\alpha = 4$ using the Euclidean Cartesian notation. The OPE of the pseudo-ITD is given by
\beq\label{eq:ope}
\mathcal{M}(\nu,z^2) = \sum_{n=0}^\infty c_n(\mu^2z^2) a_n(\mu^2) \frac{(i \nu)^n}{n!}  + \mathcal{O}(z^2)\,,
\eeq
where $a_n$ are the Mellin moments of the PDF, $c_n$ are perturbatively calculable Wilson coefficients, and $\mathcal{O}(z^2)$ represents higher twist and target mass corrections. Here it is important to note that we use an unconventional definition of the moments 
\beq
a_n(\mu^2) = \int_{-1}^1 dx\, x^n f(x, \mu^2)\,.
\eeq
This sum can be rearranged into a more standard convolutional form
\beq\label{eq:matching}
\mathcal{M}(\nu,z^2) = \int_0^1 du \, C(u,\mu^2z^2) Q(u\nu,\mu^2) + \mathcal{O}(z^2)\,,
\eeq
where the kernel $C$ is the inverse Mellin transform of the Wilson coefficients $c_n$.
The dependence of this kernel, $C$, on $\mu^2z^2$ is precisely the Dokshitzer-Gribov-Lipatov-Altarelli-Parisi (DGLAP) scale evolution. There is also a contribution independent of the scales which depends on the choice of $\msbar$ to renormalize the ITD and whatever renormalization prescription is chosen for the pseudo-ITD. The kernel in this convolution has been calculated to $\mathcal{O}(\alpha_s)$ for multiple renormalization schemes~\cite{Radyushkin:2018cvn,Zhang:2018ggy,Izubuchi:2018srq}.
Recently, an $\mathcal{O}(\alpha_s^2)$ calculation has appeared in the literature~\cite{Li:2020xml}. For this study only the $\mathcal{O}(\alpha_s)$ kernel is used, since the precision of the data is less than the expected size of the $\mathcal{O}(\alpha_s^2)$ terms. In future work, the $\mathcal{O}(\alpha_s^2)$ kernel will be used to improve the theoretical accuracy of the factorization procedure. 

Instead of utilizing a typical lattice renormalization scheme, such as a regulator independent momentum subtraction (RI-MOM) scheme, it is useful to consider the renormalization group invariant (RGI) quantity
\beq\label{eq:ratio}
\mathfrak{M}(\nu,z^2) = \frac{M^0(p,z)M^0(0,0)}{M^0(0,z)M^0(p,0)}\,,
\eeq
which is called the reduced pseudo-ITD~\cite{Radyushkin:2017cyf}. The matrix elements involving the local vector current, $M^0(0,0)$ and $M^0(p,0)$,  and the pseudo-ITD, $M^0(p,z)$ and $M^0(0,z)$, are all multiplicatively renormalizable~\cite{Ishikawa:2017faj}. The latter depends only on the separation $z$ through the  divergences related to   the Wilson line~\cite{Dotsenko:1979wb,Polyakov:1980ca,Ishikawa:2017faj}. Within this double ratio the renormalization constants all cancel explicitly and nonperturbatively, in a way independent of the renormalization scheme, making the reduced pseudo-ITD a RGI quantity~\cite{Karpie:2018zaz}. 
The matching of this object to the $\msbar$ ITD lacks the scheme dependent systematic errors which have been observed in calculations of the related quasi-PDF quantities~\cite{Stewart:2017tvs,Alexandrou:2019lfo}, which so far have always used different variants of RI-MOM schemes. 
Additionally, this object has dramatically reduced higher twist errors compared to the RI-MOM renormalized matrix elements as originally suggested in~\cite{Radyushkin:2017cyf} and directly observed by~\cite{Gao:2020ito}. Besides the aforementioned unnecessary complications, the ratio is free of the pathologies of fixed gauge renormalization as well as of the undesirable systematic effects that plague any RI-MOM type of calculation. 
The higher twist, as well as lattice spacing, finite volume, and unphysical pion mass, systematic errors are all being reduced, and this fact has been observed in~\cite{Joo:2019bzr,Joo:2019jct,Joo:2020spy,Gao:2020ito}. 
Finally this particular choice of ratio cancels correlated fluctuations between the terms in the numerator and denominator for small momenta and for small separation data, leading to a measurable improvement of the signal-to-noise ratio of the pertinent matrix element. This may prove crucial in studying statistically noisier cases such as the pion quark PDF, gluon PDF, and quark disconnected matrix elements that we wish to address in the future. A related ratio has been described in~\cite{Gao:2020ito}, which utilizes matrix elements with non-zero momenta only. This ratio leads to a more complicated matching relationship connecting the RGI ratio and the PDF. 

Since three of the four matrix elements in Eq.~\eqref{eq:ratio} defining the reduced pseudo-ITD, after renormalization in a RGI scheme, are equal to unity up to lattice spacing and higher twist errors, they do not modify the matching relationship in Eq.~\eqref{eq:matching}. We separate the scale dependent DGLAP contribution and the scale independent scheme dependent contribution to the kernel $C$ up to $\mathcal{O}(\alpha_s)$ as
\beq
C(u,\mu^2z^2) = \delta(1-u) + \frac{\alpha_s C_F}{2\pi}\left[ \log \left (\mu^2z^2 \frac{e^{2\gamma_E+1}}{4} \right ) B(u) + L(u)\right]\,,
\eeq
where
\beq
B(u) = \left[\frac{1+u^2}{1-u}\right]_+\,,
\qquad
L(u) = \left[4\frac{\log(1-u)}{1-u} -2 (1-u) \right]_+
\eeq
and $C_F$ is the fundamental Casimir of SU(3) and $\gamma_E$ is the Euler-Mascheroni constant~\cite{Radyushkin:2018cvn,Zhang:2018ggy,Izubuchi:2018srq}.

As can be seen in Eq.~\eqref{eq:ope}, the (reduced) pseudo-ITD can be directly related to the moments of the PDF without going through the ITD itself. A Taylor expansion of the ITD with respect to the Ioffe time can be written as 
\beq
\mathfrak{M}(\nu,z^2) = \sum_n m_n(z^2) \frac{(i\nu)^n}{n!} + \mathcal{O}(z^2)\,,
\eeq
where $m_n(z^2)$, called the pseudo-moments, are the Mellin moments of the pseudo-PDF, which is the Fourier transform with respect to $\nu$ of the (reduced) pseudo-ITD. The pseudo-moments have a multiplicative matching relationship to the $\msbar$ PDF moments given by
\beq
m_n(z^2) = c_n(\mu^2 z^2) a_n(\mu^2) + \mathcal{O}(z^2) \,
\eeq
where $c_n$ are the Mellin moments of the matching kernel $C$. These matching coefficients have been calculated to $\mathcal{O}(\alpha_s)$ for the reduced pseudo-PDF
\beq
c_n(\mu^2z^2) = 1 + \frac{\alpha_s C_F}{2\pi}\left[ \log \left (\mu^2z^2 \frac{e^{2\gamma_E+1}}{4} \right ) \gamma_n + l_n\right]
\eeq
where 
\beq 
\gamma_n = \int_0^1 B(u) u^n = \frac{1}{(n+1)(n+2)} - \frac 12 -2 \sum_{k=2}^{n+1} \frac 1k
\eeq 
are the well known anomalous dimensions of the moments of the PDF and the scheme dependent term is given by 
\beq 
l_n = \int_0^1 L(u) u^n = 2\left[ \left(\sum_{k=1}^n \frac 1k \right)^2 + \sum_{k=1}^n \frac1{k^2} + \frac 12 - \frac{1}{(n+1)(n+2)}\right]\,.
\eeq 

One could easily fit the Ioffe-time dependence of the reduced pseudo-ITD for each separation $z^2$ independently to obtain the pseudo-moments. Studying the $z^2$ dependence of the pseudo-moments, as well as the resulting $\msbar$ PDF moments, one can try to estimate the size of systematic errors. Deviations of $\msbar$ PDF moments originating from the low $z^2$ data could signal large lattice spacing errors and deviations from large $z^2$ data could signal large higher twist errors. A wide region of $z^2$ where the resulting $\msbar$ PDF moments are independent of $z^2$ would indicate a window of opportunity where the data are free of these systematic effects. It is only within such a window that an ITD derived from the reduced pseudo-ITD can be trustworthy without other methods of removing systematic errors.

Instead of simultaneously handling the real and imaginary components of the complex $\mathfrak{M}$, it is helpful to separate the $CP$ even and odd contributions which are related to $q_-(x) = f(x) + f(-x) = q(x) - \bar{q}(x)$ and $q_+(x)= f(x) - f(-x) =q(x) + \bar{q}(x)$ respectively, where $f$ is defined in the window [-1,1] while $q$, $\bar{q}$, $q_-$, and $q_+$ are defined in the window [0,1]. These PDFs can be individually extracted from the real and imaginary components separately. The components are factorized as
\bea
{\rm Re}\, \mathfrak{M}(\nu,z^2) = \int_0^1 dx \, \mathcal{K}_{R}(x\nu,\mu^2 z^2) q_-(x,\mu^2)+ \mathcal{O}(z^2)\nonumber\\
{\rm Im}\, \mathfrak{M}(\nu,z^2) = \int_0^1 dx \,\mathcal{K}_{I}(x\nu,\mu^2 z^2) q_+(x,\mu^2)+ \mathcal{O}(z^2)\,,
\eea
where
\bea
\mathcal{K}_{R}(x\nu,\mu^2 z^2) = \int_0^1 du \, C(u,\mu^2z^2) \cos(u\nu x)\nonumber\\
\mathcal{K}_{I}(x\nu,\mu^2 z^2) = \int_0^1 du \, C(u,\mu^2z^2) \sin(u\nu x)\,.
\eea
Use of these matching kernels which factorize directly to the PDF removes the need for the intermediate determination of the $\msbar$ ITD. Unfortunately, they prove to be complicated functions whose direct numerical evaluation is inefficient when incorporated into the analysis of the matrix elements computed from lattice QCD. In Sec.~\ref{sec:parameterization}, we adopt a power series approximation to the convolution integrals that the above kernel functions participate in which allows for efficient computations within the available range of the Ioffe time. With sufficient number of terms, this power series approximates the convolution integrals to numerical precision.

\section{Determination of the continuum limit PDF and nuisance parameters}\label{sec:nuisance}
The continuum limit is a critical step in any precision lattice calculation. In this study, we take advantage of the symmetries of the reduced pseudo-ITD to parameterize the lattice spacing correction to the continuum limit, as well as the higher twist effects. 
The continuum PDF is also parameterized and a simultaneous analysis of all three ensembles obtains the continuum limit PDF with higher twist contamination removed. This method of adding ``nuisance parameters'' to parameterize the systematic errors of experimental cross sections is also used in the phenomenological extractions of PDFs. Such a combined analysis approach can also be used with results obtained with different pion masses, lattice spacings, matrix elements, and even lattice actions given appropriate parameterizations of those effects. Ultimately, one can imagine taking all published lattice matrix elements and analyzing them within this approach, given sufficiently novel nuisance parameterizations, just as a global phenomenological fit is performed using experimental data with vastly different systematic errors. In order to minimize the dependence of the effect of nuisance parameters, in this study only higher twist and lattice spacing errors are considered for data with the same physical quark mass and lattice action. Future work will study the extension of this method to include other effects.

It is important to note that the coefficients of the lattice spacing errors can be functions of the Ioffe time. Previous parameterizations of lattice spacing errors for parton observables have only used simple dependences on the Ioffe time, which all diverge as $\nu\to \infty$. In~\cite{Sufian:2020vzb,Gao:2020ito}, the Ioffe time dependence of lattice spacing errors was equivalent to $\frac{a^2}{z^2} \nu^2$ which is the simplest possible dependence. In~\cite{Joo:2019bzr}, correction terms were modeled at fixed lattice spacing and given a higher order polynomial dependence. In this study, the Ioffe-time dependence is studied further by employing functions whose large Ioffe time behavior is finite as it physically should be. 

In the recent work~\cite{Alexandrou:2020qtt,Zhang:2020gaj}, the Ioffe-time dependence is taken into account by fitting the lattice spacing dependence at fixed LaMET scale $p_3$, or at least requiring $p_3$'s to be sufficiently close in physical units and extrapolating their data, or an interpolation of it, in $z$. In pseudo-PDF studies, as in this work, matrix elements originating from many different SDF scales $z^2$ are simultaneously utilized. Studying a wide range of scales allows us to access systematic errors arising from lattice spacing effects or higher twist contributions. In particular, at short distances where higher twist effects are suppressed, lattice spacing errors arise, while at large distances higher twist effects may dominate.  
Therefore, a window in $z^2$, where both systematic errors are suppressed can be identified in order to extract the universal leading twist matrix element. In order to perform the simple extrapolations of~\cite{Alexandrou:2020qtt,Zhang:2020gaj}, the gauge ensembles must have specifically tuned lattice spacings  and volumes to allow for the same scales, $p_3$ or $z^2$, to occur in each ensemble. In most sets of ensembles, only few momenta or values of $z^2$ coincide in physical units since tuning lattice spacings to be integer multiples of each other is difficult. An analysis of 8 values of $z^2$ per ensemble, as is done in this study, would not be feasible without an overwhelming computational cost in gauge ensemble generation.

Besides the continuum limit, there is a significant complication due to the integral relation between the matrix element and the PDF. The inversion of this integral relation is a numerically ill-defined problem when there is a limited range of Ioffe times as there is in a lattice QCD calculation. An infinite number of solutions that fit the matrix elements exist, so procedures must be chosen to select the best class of solutions. There exist many classes of these procedures which have been proposed for lattice QCD calculations of PDFs~\cite{Orginos:2017kos,Lin:2017ani,Karpie:2019eiq,Cichy:2019ebf,Alexandrou:2020tqq}. The most popular amongst phenomenological global analyses of PDFs are parametric solutions. In this procedure a functional form for the PDF is written down based on a set of parameters whose values are tuned to represent the data. This style of solution is used in this study. Many methods also exist which do not rely on parameterizing the unknown functions.  Though they lack functional forms, non-parametric solutions also have their own uncontrolled systematic errors and are not fundamentally better than parametric solutions. To arrive at a proper systematic error analysis of the resulting PDF, the systematic errors of any of these procedures must be tested, typically by comparing results from several approaches. In this study, the parameterizations of the data are varied in order to study the parametrization dependence and access the associated systematic errors. In future work, more diverse parameterizations can be used to obtain a better estimate of the  variance due to model choices. 

\subsection{Separating continuum PDFs from systematic errors}
The $CP$ symmetry implies that the reduced pseudo-ITD has the property
\begin{equation}\label{eq:pitd_cp_sym}
    \mathfrak{M}(p,z,a) = \mathfrak{M}^\ast(-p,z,a) = \mathfrak{M}^\ast(p,-z,a) = \mathfrak{M}(-p,-z,a)\,,
\end{equation}
which we used when constructing the summed three-point correlation functions to increase the statistical precision by averaging, after appropriate complex conjugations, the correlation functions with positive and negative momenta and separations. The relation $\mathfrak{M}(p,z,a) = \mathfrak{M}(-p,-z,a)$ restricts lattice spacing errors with odd powers of $a$ to be functions of $a|p|$ and $a/|z|$. A Taylor expansion in lattice spacing gives the continuum reduced pseudo-ITD $\mathfrak{M}_{\rm cont}$ and lattice spacing corrections
\begin{equation}
\mathfrak{M}(p,z,a) = \mathfrak{M}_{\rm cont} (\nu,z^2) + \sum_{n=1} \left(\frac{a}{|z|} \right)^n P_n(\nu) + (a\Lambda_{\rm QCD})^n R_n(\nu)     \,.
\end{equation}
 With an $O(a)$ improved lattice action, the lattice spacing errors related to the momentum $p$, must come in from the momentum transfer. This feature is known in the improvement of the local vector current~\cite{Martinelli:1997zc}, the case of $z=0$, where the local vector current mixes with the divergence of the tensor current. The operators discussed in~\cite{Chen:2017mie} also demonstrate these features when considering the hadronic matrix elements in question. These momentum transfer effects are necessary for the studies of Generalized Parton Distributions, but not for the PDF. There is also potential $z^2$ dependence on the lattice spacing coefficient functions, $P_n$ and $R_n$. Those effects which can come from logarithmic perturbative corrections, higher twist contributions, or target mass corrections are additionally suppressed either by $\alpha_s$, $\Lambda_{\rm QCD}^2 z^2$, or $m^2 z^2$ respectively on top of the suppression by ${a}/{|z|}$ and $a\Lambda_{\rm QCD}$. These $z^2$ dependencies are neglected here.

 The relationship between the reduced pseudo-ITD and the ITD is through a convolution with Wilson coefficient function. Ultimately, the ITD is not the goal of this study, but instead its Fourier transform, the PDF. We adopt an approach analogous to~\cite{Izubuchi:2019lyk,Fan:2020nzz,Gao:2020ito} where the intermediate ITD is not required, but a parameterization of the PDF is directly related to the reduced pseudo-ITD. Unlike~\cite{Izubuchi:2019lyk,Fan:2020nzz,Gao:2020ito}, the PDF is related to the leading twist reduced pseudo-ITD through its moments. The higher twist power corrections are added as nuisance terms similar to the lattice spacing terms. The functional form is given by
\begin{equation}
    \mathfrak{M}_{\rm cont}(\nu,z^2) = \mathfrak{M}_{\rm lt}(\nu,z^2) + \sum_{n=1} (z^2\Lambda_{\rm QCD}^2)^n B_n(\nu)\,.
\end{equation}
 in terms of the leading twist continuum limit reduced pseudo-ITD, $\mathfrak{M}_{\rm lt}$, and the higher twist distributions $B_n$. In principle, the higher twist distributions could have non-trivial $z^2$ dependence. Similarly to the lattice spacing terms, these effects which come from perturbative corrections and target mass effects are additionally suppressed by powers of $\alpha_s$ or $m^2 z^2$ respectively and are neglected in the remainder of this study.

In principle, there exist higher twist power corrections and lattice spacing errors of all orders. With these errors sufficiently under control, only the leading contributions are significant. We therefore make the approximation that $P_n  = R_n = B_n=0$ for $n>1$.  

\subsection{Parameterization of unknown functions}\label{sec:parameterization}
Extracting PDFs from matrix elements using a functional form to parametrize them may induce unwanted model dependence. Therefore, a careful study of such parametrization-dependent systematic error is required.  
For that purpose, the functional forms used should be varied in order to understand how certain choices affect the final result. In previous lattice PDF studies~\cite{Orginos:2017kos,Sufian:2019bol,Joo:2019jct,Joo:2019bzr,Sufian:2020vzb,Joo:2020spy,Bhat:2020ktg,Bringewatt:2020ixn}, the chosen functional forms are similar to those used in phenomenological analyses of PDFs~\cite{Sato:2019yez,Hou:2019qau,Bailey:2020ooq,Moffat:2021dji}. Progress has also been made on the application of neural networks to parameterize the PDF~\cite{Ball:2017nwa,Cichy:2019ebf,DelDebbio:2020rgv}. In this work, all of the unknown functions, $q_-(x)$, $q_+(x)$, $P_1(\nu)$, $R_1(\nu)$, and $B_1(\nu)$, are parameterized using Jacobi polynomials. 

The Jacobi polynomials, $j_n^{(\alpha,\beta)}(z)$, are  defined in the interval $[-1,1]$ and they satisfy the orthogonality relation
\begin{align}
\int_{-1}^{1} dz (1-z)^\alpha (1+z)^\beta  j_n^{(\alpha,\beta)}(z) j_m^{(\alpha,\beta)}(z) = \tilde{N}_n^{(\alpha,\beta)} \delta_{n,m}\,,
\end{align}
for $\alpha,\beta>-1 $. For the purposes of this study, it is useful to change variables to $x = \frac{1-z}{2}$ or $z = 1- 2 x $. This transformation maps the interval $[-1,1]$ to the interval $[0,1]$
and the orthogonality weight becomes  $ (1-z)^\alpha (1+z)^\beta  = 2^{\alpha+\beta} x^\alpha(1-x)^\beta$. We therefore introduce the transformed Jacobi polynomials $J^{(\alpha,\beta)}_n(x)$, which are referred to as Jacobi polynomials from now on, as
\begin{align}
J^{(\alpha,\beta)}_n(x) = \sum_{j=0}^n \omega_{n,j}^{(\alpha,\beta)} x^j\,,
\end{align}
with
\beq
\omega_{n,j}^{(\alpha,\beta)} = \binom{n}{j}  \frac{(-1)^j}{n!}   \frac{\Gamma(\alpha+n+1)\Gamma(\alpha+\beta+n+j+1)}{ \Gamma(\alpha+\beta+n+1) \Gamma(\alpha+j+1)}\,.
\eeq
The orthogonality relation becomes
\begin{align}
\int_{0}^{1} dx\, x^\alpha (1-x)^\beta  J_n^{(\alpha,\beta)}(x) J_m^{(\alpha,\beta)}(x)   = N_n^{(\alpha,\beta)} \delta_{n,m}\,,
\end{align}
where
\beq
N_n^{(\alpha,\beta)} = \frac{1}{2n + \alpha + \beta + 1} 
\frac{\Gamma(\alpha+n+1)\Gamma(\beta+n+1)}{n!\,\Gamma(\alpha+\beta+n+1)}\,.
\eeq
One thing to note is that there exists a formula that relates Jacobi polynomials for different values of the weight parameters, $\alpha$ and $\beta$. This formula reads as following
\begin{align}
 J_n^{(\alpha,\beta)}(x) = \sum_{m=0}^n  \hat{c}^n_m(\alpha, \alpha'; \beta,\beta')J_m^{(\alpha',\beta')}(x) \,,
 \label{eq:JacobiCONV}
\end{align} 
where the coefficients $\hat{c}^n_m(\alpha, \alpha'; \beta,\beta')$ are analytically known. 
Finally, it can be shown that the coefficients of the Jacobi polynomials satisfy the orthogonality relationship
\beq
\sum_{i,j=0}^{\infty}   \omega^{(\alpha,\beta)}_{n,i} B(\alpha+i+j+1,\beta+1) \omega^{(\alpha,\beta)}_{m,j}= N^{(\alpha,\beta)}_n \delta_{n,m}\,,
\eeq
where $B(a,b)$ is the beta function. Since the Jacobi polynomials form a complete basis of functions in the interval of [0,1], the PDFs can be written as 
\beq
q_\pm(x) = x^\alpha (1-x)^\beta \sum_{n=0}^\infty \,_\pm d_n^{(\alpha,\beta)} J_n^{(\alpha,\beta)}(x)\,
\eeq
for any $\alpha$ and $\beta$. The choice of those parameters does affect the convergence of the coefficients $_\pm d_n^{(\alpha,\beta)}$. In practice, one needs to truncate the series introducing in this way some model dependence which can be easily controlled. The control of the truncation can be improved if one fits for the optimal  values of $\alpha$ and $\beta$ for that given order of truncation. In  other words, the rate of convergence of the series can be optimized by tuning the values of $\alpha$ and $\beta$.  One way to understand why tuning of $\alpha$ and $\beta$ can result in improved convergence of the series is to realize that phenomenological considerations tell us that the Jacobi weight is a good approximation to the shape of the PDF, therefore if 
$\alpha,\beta$ are tuned to roughly match the shape of the PDF, the Jacobi polynomials need only to approximate a smooth, slowly varying function with small coefficients. Using Eq.~\ref{eq:JacobiCONV}, we can easily convert an expansion of the PDF in terms of $(\alpha,\beta)$ Jacobi polynomials to one with  $(\alpha',\beta')$ Jacobi polynomials. The transformation of the expansion coefficients is linear and if a truncation of the series up to order $N$ is used the linear transformation involves only coefficients up to that order. Finally, there also exists a linear transformation which connects these coefficients and the Mellin moments of the PDF given by
\beq\label{eq:jacobi_moments}
_\pm d_n^{(\alpha,\beta)} = \frac{1}{ N^{(\alpha,\beta)}_n} \sum_{j=0}^n \omega^{(\alpha,\beta)}_{n,j} a_{j}^\pm
\eeq
where $a_{n}^\pm = \int_0^1 dx\, x^{n}q_\pm(x)$, so this parameterization can be thought as another way to parameterize the PDF by a set of its moments.

To determine the relationship between the reduced pseudo-ITD and the parameters of the PDF, the matching kernels $\mathcal{K}_{R,I}$ are expanded in terms of Jacobi polynomials. It can be shown that the kernels can be written as
\bea
\mathcal{K}_R(x\nu, \mu^2z^2) = \sum_{n=0}^\infty \frac{\sigma_n^{(\alpha,\beta)}(\nu,\mu^2z^2)}{N^{(\alpha,\beta)}_n} J^{(\alpha,\beta)}_n(x) \nonumber \\
\mathcal{K}_I(x\nu, \mu^2z^2) = \sum_{n=0}^\infty \frac{\eta_n^{(\alpha,\beta)}(\nu,\mu^2z^2)}{N^{(\alpha,\beta)}_n} J^{(\alpha,\beta)}_n(x) \,,
\eea
with
\bea
\sigma^{(\alpha,\beta)}_{n}(\nu,z^2\mu^2) = \sum_{j=0}^{n}\sum_{k=0}^\infty \frac{(-1)^k}{(2k)!}c_{2k} (z^2\mu^2) \omega^{(\alpha,\beta)}_{n,j}\, B(\alpha+2k+j+1,\beta+1)\,  \nu^{2k}  \nonumber \\
\eta^{(\alpha,\beta)}_{n}(\nu,z^2\mu^2) =  \sum_{j=0}^n  \sum_{k=0}^\infty \frac{(-1)^k}{(2k+1)!}c_{2k+1} (z^2\mu^2)  \omega^{(\alpha,\beta)}_{n,j} B(\alpha+2k+j+2,\beta+1) \nu^{2k+1} \,.
\eea
Numerically, the sum over $k$ can be performed to a sufficiently high order ($k\sim30$) to achieve  convergence to double precision accuracy in the relevant range of Ioffe time. Given this expansion, the leading twist reduced-pseudo ITD can be written as the truncated sums
\bea\label{eq:jacobi_pdf}
{\rm Re}\, \mathfrak{M}_{\rm lt}(\nu,z^2) = 1 + \sum_{n=1}^{N_-} \sigma^{(\alpha,\beta)}_{n}(\nu,z^2\mu^2) _-d_n^{(\alpha,\beta)} \nonumber \\
{\rm Im}\, \mathfrak{M}_{\rm lt}(\nu,z^2) = \sum_{n=0}^{N_+-1} \eta^{(\alpha,\beta)}_{n}(\nu,z^2\mu^2) _+d_n^{(\alpha,\beta)}\,.
\eea
Similarly, the nuisance parameters can be introduced in $x$ space and the unknown functions $B_1$, $P_1$, $Q_1$, and $R_1$ are constructed from similar sums. 
\bea
{\rm Re} \, B_1(\nu) = \sum_{n=1}^{N_{R,b}} \sigma^{(\alpha,\beta)}_{0,n}(\nu) b_{R,n}^{(\alpha,\beta)}\quad, \qquad {\rm Im} \, B_1(\nu) = \sum_{n=1}^{N_{I,b}} \eta^{(\alpha,\beta)}_{0,n}(\nu) b_{I,n}^{(\alpha,\beta)}  \nonumber \\
{\rm Re} \, P_1(\nu) = \sum_{n=1}^{N_{R,p}} \sigma^{(\alpha,\beta)}_{0,n}(\nu) p_{R,n}^{(\alpha,\beta)} \quad, \qquad {\rm Im} \, P_1(\nu) = \sum_{n=1}^{N_{I,p}} \eta^{(\alpha,\beta)}_{0,n}(\nu) p_{I,n}^{(\alpha,\beta)}  \nonumber \\
{\rm Re} \, R_1(\nu) = \sum_{n=1}^{N_{R,r}} \sigma^{(\alpha,\beta)}_{0,n}(\nu) r_{R,n}^{(\alpha,\beta)} \quad , \qquad {\rm Im} \, R_1(\nu) = \sum_{n=1}^{N_{I,r}} \eta^{(\alpha,\beta)}_{0,n}(\nu) r_{I,n}^{(\alpha,\beta)} \,,
\eea
where
\bea
\sigma^{(\alpha,\beta)}_{0,n}(\nu) = \int_0^1 dx\, \cos(\nu x)  x^\alpha (1-x)^\beta  J_n^{(\alpha,\beta)}(x) \nonumber \\
\eta^{(\alpha,\beta)}_{0,n}(\nu) = \int_0^1 dx\, \sin(\nu x)  x^\alpha (1-x)^\beta  J_n^{(\alpha,\beta)}(x) \,,
\eea
which are the leading $O(\alpha_s^0)$ order of $\sigma_n$ and $\eta_n$.

Unlike parameterizing with a polynomial form in Ioffe time, these functional forms are better behaved in the large Ioffe time regime. Unlike a polynomial in $\nu$, one does not expect these nuisance terms to grow indefinitely with Ioffe time, but instead eventually falling to zero as the ITD does. In~\cite{Braun:2018brg}, a calculation using renormalon methods showed the ratio of the ITD and the leading power correction plateaus as $\nu$ grows indicating that the higher twist contribution eventually decays to zero at large Ioffe time. Similarly, the size of the lattice spacing error is not expected to grow infinitely with $\nu$. For a fixed $z^2$, it is expected to ultimately go to zero as $\nu$ increases.

\begin{figure}[!htp]
\centering
\includegraphics[width=0.48\textwidth]{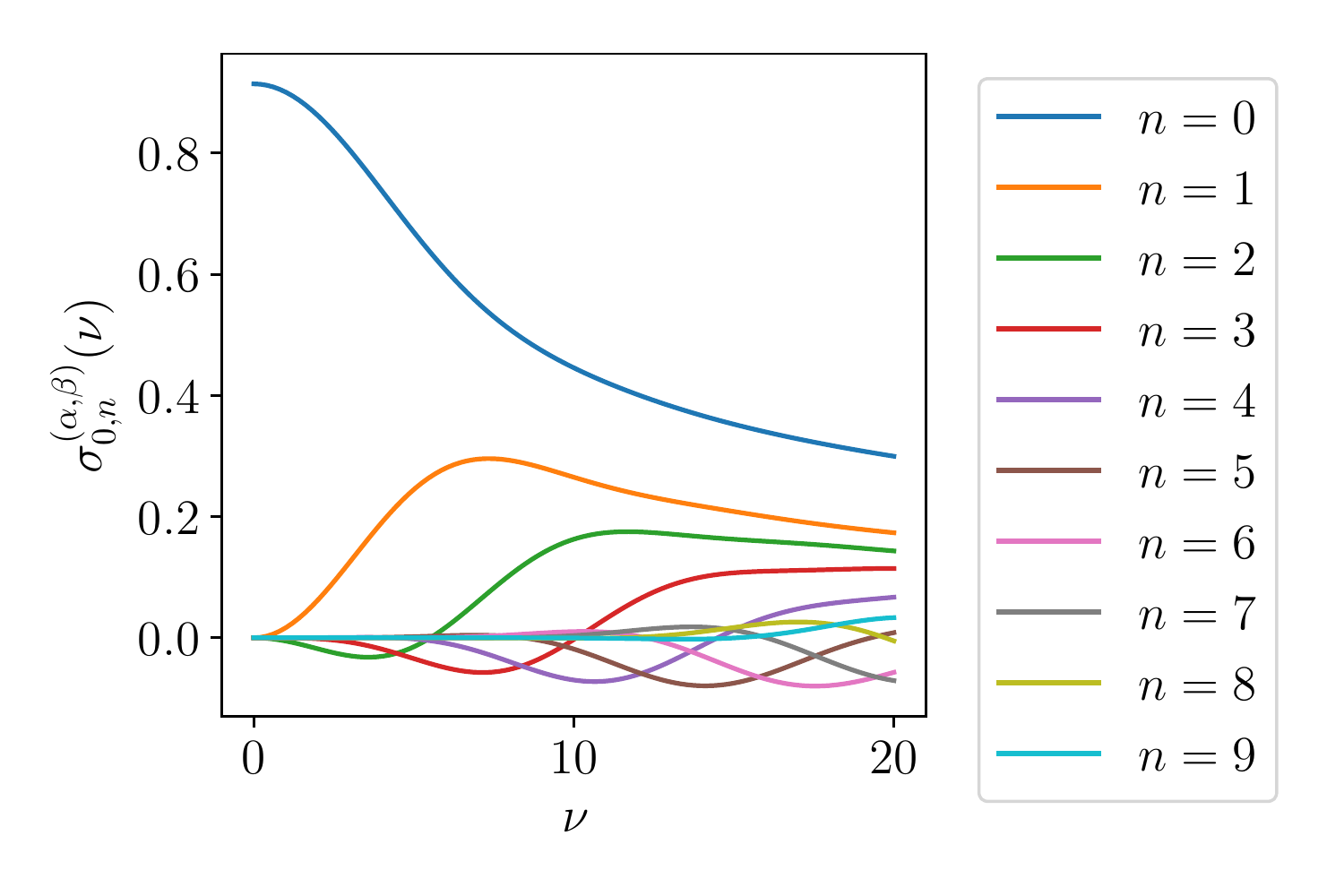}
\includegraphics[width=0.48\textwidth]{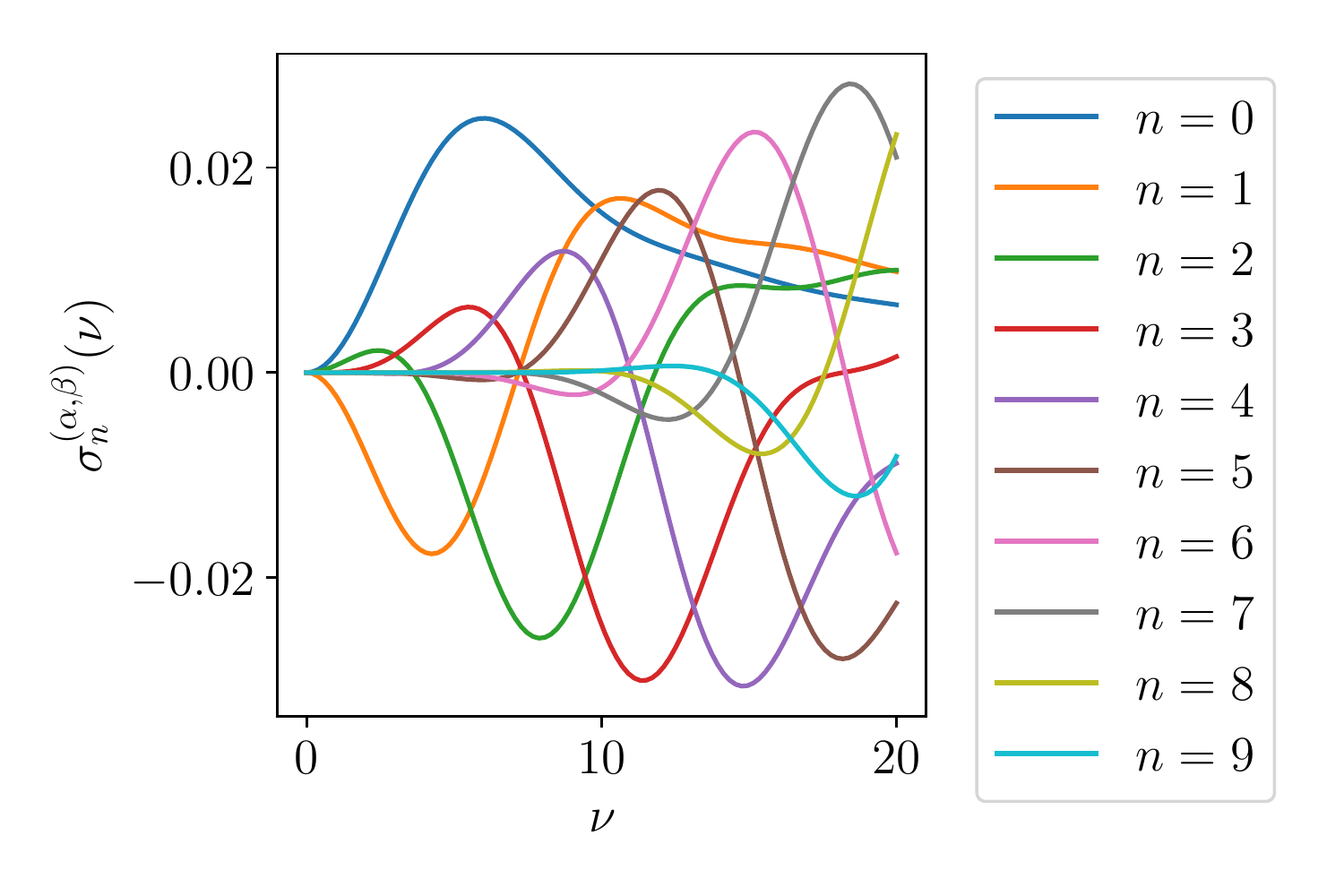}
\caption{\label{fig:sigma} The functions $\sigma^{(\alpha,\beta)}_{0,n}(\nu)$ (Left) and $\sigma_n^{({\rm NLO})}(\nu,z^2\mu^2) = \sigma^{(\alpha,\beta)}_{n}(\nu,z^2\mu^2) - \sigma^{(\alpha,\beta)}_{0,n}(\nu)$ (Right) for $\alpha=-0.5$ and $\beta=3$ over a range of $n$. For the NLO contribution, the value of $z^2 = 4 * 0.065$ fm, $\mu = 2$ GeV, and $\alpha_s = 0.3$ were chosen as a typical example which will be used in this study. The peaked structures of $\sigma^{(\alpha,\beta)}_{0,n}$ and $\sigma_n^{({\rm NLO})}$ mean that only certain $n$ significantly contribute in the limited range of Ioffe time. The size of $\sigma_n^{({\rm NLO})}$, relative to $\sigma^{(\alpha,\beta)}_{0,n}$, leads to a small perturbative contribution as desired.  }
\end{figure}

\begin{figure}[!htp]
\centering
\includegraphics[width=0.48\textwidth]{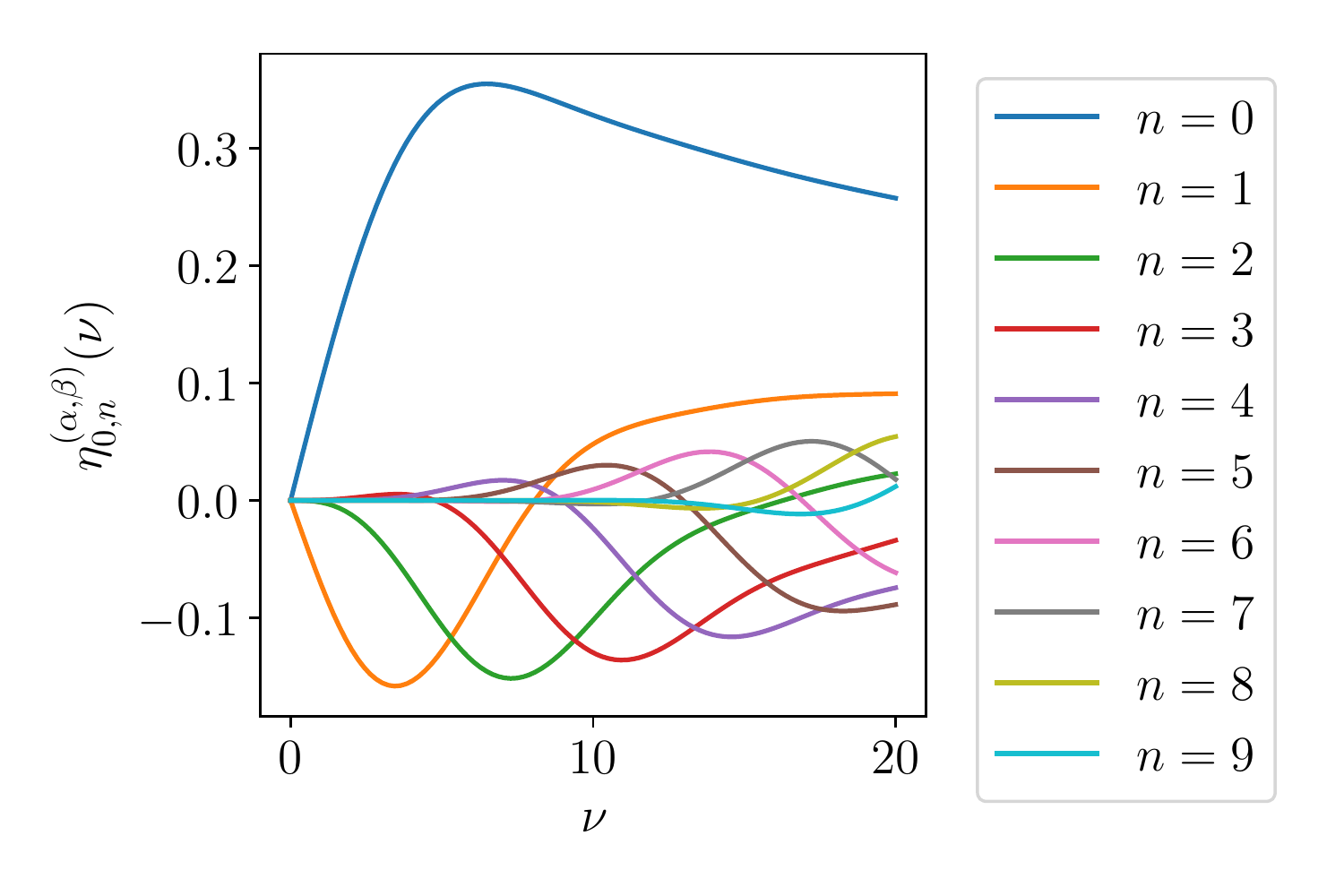}
\includegraphics[width=0.48\textwidth]{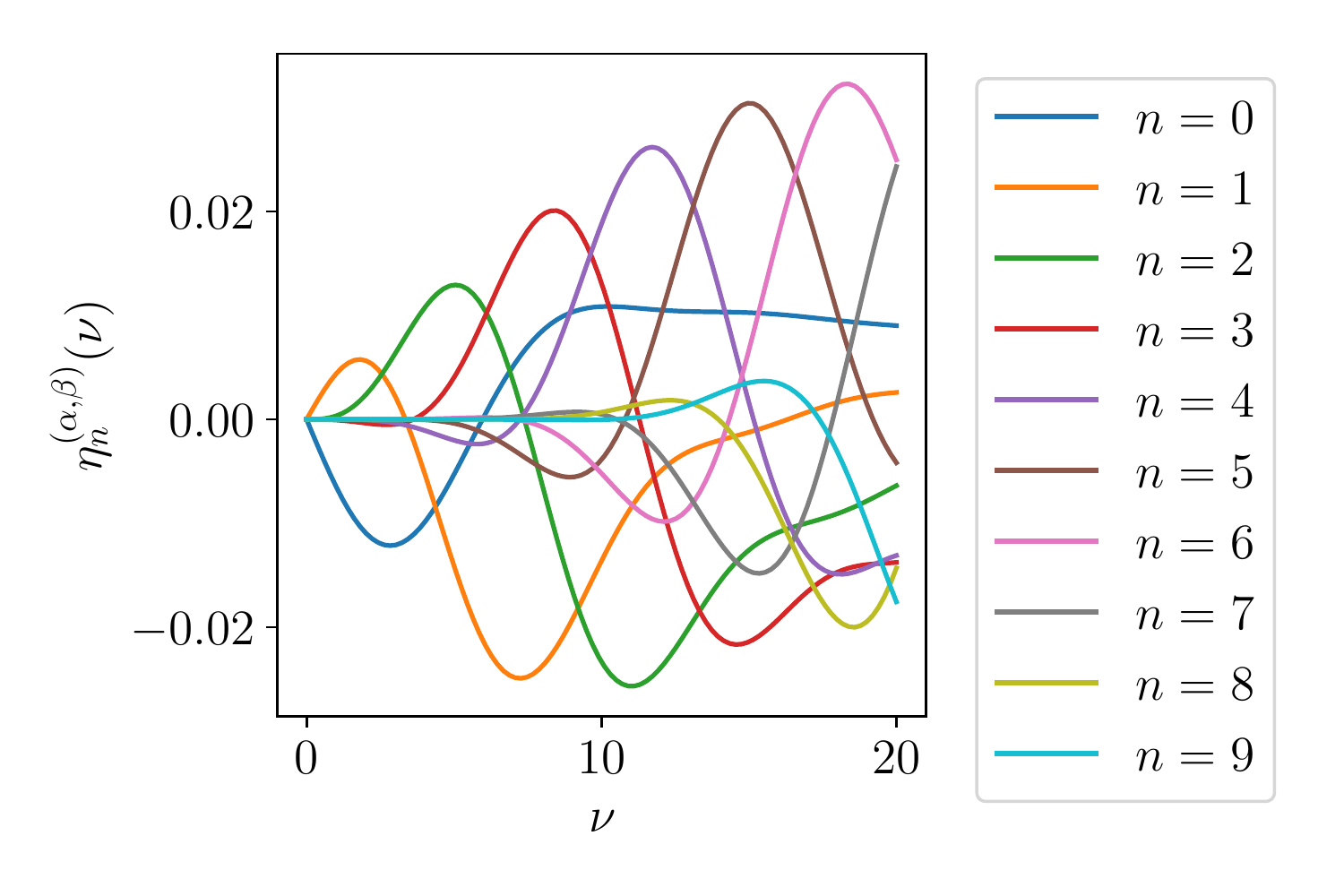}
\caption{\label{fig:eta} The functions $\eta^{(\alpha,\beta)}_{0,n}(\nu)$ (Left) and $\eta_n^{({\rm NLO})}(\nu,z^2\mu^2) = \eta^{(\alpha,\beta)}_{n}(\nu,z^2\mu^2) - \eta^{(\alpha,\beta)}_{0,n}(\nu)$ (Right) for $\alpha=-0.5$ and $\beta=3$ over a range of $n$. For the NLO contribution, the value of $z^2 = 4 * 0.065$ fm, $\mu = 2$ GeV, and $\alpha_s = 0.3$ were chosen as a typical example which will be used in this study. The peaked structures of $\eta^{(\alpha,\beta)}_{0,n}$ and $\eta_n^{({\rm NLO})}$ mean that only certain $n$ significantly contribute in the limited range of Ioffe time. The size of $\eta_n^{({\rm NLO})}$, relative to $\eta^{(\alpha,\beta)}_{0,n}$, leads to a small perturbative contribution as desired. }
\end{figure}

Not only do the $\sigma$ and $\eta$ functions have better large Ioffe time behavior, but they also appear to dominate only in a given region of Ioffe time ordered by $n$. Figs~\ref{fig:sigma} and~\ref{fig:eta} show the functions over a range of $n$. As can be seen, these functions have a peak region and fall to zero as Ioffe time increases, albeit slowly, and the peaks are ordered by $n$. Since our data exist within a limited range of Ioffe time, the terms whose peaks are beyond this region do not contribute significantly. More so, since the (pseudo-)ITD is believed to decay towards zero without any large values at larger Ioffe times, the values of the parameters with larger $n$ will be small as well. This expected convergence of the series and the known shape of the the $\sigma$ and $\eta$ functions in the available range of Ioffe time can be used as natural guides for when to truncate the series without significant chance of losing vital information on the PDF's structure.

This final functional form is capable of removing lattice spacing and higher twist dependencies which spoil the leading twist reduced pseudo-ITD. By testing with various functional forms, the model dependent systematic error can be studied. Using the Akaike information criterion (AIC), a weighted average of these models produces a final continuum limit PDF with the model dependence smoothed out in a statistically meaningful way, especially when sufficiently many distinct models are used. In future studies more adventurous parameterizations of the PDF and the nuisance parameters, such as a neural network, can be included alongside these fits into the weighted average.

\section{Lattice QCD calculation}\label{sec:latt}
This study utilizes three ensembles of configurations with decreasing lattice spacing. These ensembles have two flavors of dynamical  Wilson clover fermions and pion mass around 440 MeV. The specific parameters of these ensembles are given in Table~\ref{tab:lat}. The lattice spacings of the configurations are 0.0749, 0.0652, and 0.0483 fm. The finer two ensembles were generated by the CLS effort~\cite{Fritzsch:2012wq} while the coarsest was generated by the authors for this study. These ensembles allow for a controlled continuum limit extrapolation which is a necessary step for precision calculations of PDFs. 
Apart from that, the finest lattice spacing employed in this study is half compared to our previous studies allowing us to reach much higher momenta and smaller separations.

%------ensembles
\begin{table*}[t] 
\centering
\begin{tabular}{ l | c c | c c c | c | c c }
ID & ~$a$(fm)~ & ~$M_\pi$(MeV) & ~$\beta~$ & ~$c_{\rm SW}$~ & ~$\kappa$~ &  $L^3 \times T$ & $N_{\rm cfg}$\\\hline\hline
$\widetilde{A}$5 & 0.0749(8) & 446(1) & 5.2 & 2.01715 & 0.13585 & $32^3 \times 64$ & 1904 \\\hline
E5 & 0.0652(6) & 440(5) & 5.3 & 1.90952 & 0.13625 & $32^3 \times 64$ & 999 \\\hline
N5 & 0.0483(4) & 443(4) & 5.5 & 1.75150 & 0.13660&  $48^3 \times 96$ & 477 \\\hline \hline\end{tabular}
\caption{\label{tab:lat}\footnotesize Parameters for the lattices generated by the CLS collaboration using two flavors of $\mathcal{O}(a)$ improved Wilson fermions. More details about these ensembles can be found in~\cite{Fritzsch:2012wq}. }
\end{table*}

 The nucleon interpolating fields are constructed with Gaussian smearing~\cite{Allton:1993wc} and momentum smearing~\cite{Bali:2016lva}. The source field is always be smeared, and an unsmeared and a smeared sink field is used. These scenarios are referred to as ``SP'' (standing for smeared-point) and ``SS'' (standing for smeared-smeared) respectively. For both of these scenarios, three values of the momentum smearing parameter $\zeta$ are used. To implement the momentum smearing, prior to the Gaussian smearing step, the gauge links are modified by 
\beq
U_\mu(x) \to e^{i\frac{2\pi}L \zeta x_3} U_\mu(x) \,,
\eeq
in order to smear only the direction parallel to the momentum. The smearing parameters are chosen to increase the overlap to the ground state, and thereby the signal-to-noise ratio, for correlation functions over a range of momenta.

The matrix elements are calculated using the summation Generalized Eigenvalue Problem (sGEVP) technique~\cite{Bulava:2011yz} to have optimal control over the excited state contamination, as described in Sec.~\ref{sec:sgevp}. Summation techniques have proven to be extremely powerful in controlling excited state errors~\cite{Bouchard:2016heu} and have been used in a number of lattice calculations of PDFs~\cite{Orginos:2017kos,Karpie:2017bzm,Fan:2018dxu,Alexandrou:2019lfo,Izubuchi:2019lyk,Joo:2019jct,Joo:2019bzr}. These methods have dramatically reduced excited state contamination $O(e^{-\Delta T})$ compared to typical ratio methods $O(e^{-\Delta T/2})$. These methods are necessary for efficient calculations especially for future work with physical pion masses where $\Delta$ is smaller making excited states persistent for larger $T/a$ which consequently increases the computational cost needed to achieve equivalent statistical precision. To obtain comparable statistical precision of a summation method calculation with $N$ measurements, a ratio method calculation can be estimated to require $N^2$ measurements. 

\subsection{sGEVP matrix element extraction}\label{sec:sgevp}
Excited state contamination is a problem which can interfere with the lattice calculation of any matrix element. Large Euclidean times are required for isolating the ground state from this exponentially decaying contamination. This necessity is plagued by an exponentially decaying signal-to-noise ratio of the correlators as the Euclidean time is increased. Particularly for physical pion mass calculations with high momentum hadrons, a very large number of samples is necessary for precision calculations with large Euclidean times when using the typical ratio method for calculating matrix elements. The sGEVP method is a combination of two techniques which dramatically improves the scenario. Summation methods drastically increase the rate of decay of the excited state contributions. The GEVP method can be used to create an optimal operator which overlaps with the ground state and is orthogonal to the lowest lying excited states. The combination of these two methods allow for significant control over excited state contamination in a computationally efficient manner. 

In a summation technique, one extracts the matrix element from the large Euclidean time behavior of a ratio
\beq
R(T) = \sum_t \frac{C_3(T,t)}{C_2(T)}\,,
\eeq
where $C_3$ and $C_2$ are typical two and three point correlators, with source and sink interpolating field separation $T$, and operator time $t$. At large Euclidean times, the difference of this ratio at two times is proportional to the matrix element $M$
\beq\label{eq:sum_meff}
 M_{\rm {eff}}(T) = \frac{1}{\tau} (R(T) - R(T+\tau)) = M +  Ae^{-\Delta T} + B T e^{-\Delta T}+ \dots \,,
 \eeq
 where the ellipses indicate terms originating from higher excited states.
Summation techniques reduce the contributions from excited states of a correlator at time $T$ from $O(e^{-\Delta T/2})$, where $\Delta$ is the excited state energy gap, to $O(e^{-\Delta T})$. Since the signal-to-noise ratio of correlators decays as $O(e^{-E T})$, where $E$ is the energy of the state, this feature is critical for efficient high momenta calculations. When considering the exponentially growing signal to noise ratio of the correlators, this improvement in excited state contamination means that if the summation method requires $N$ measurements to obtain a desired precision, more traditional methods would require $N^2$ measurements for a point with equivalent excited state contamination. This advantage may also be critical for efficient calculations of pion quark PDFs, gluon PDFs, and quark disconnected contributions, which are notoriously more noisy than the connected quark operators used here.

Attempts to increase the overlap of the interpolating field with the ground state, and ideally also lower the overlap with low excited states, has generated a number of smearing procedures including the Gaussian and momentum smearing techniques used in this study. An approach, orthogonal and complimentary to these methods, is the GEVP technique. One considers a matrix of correlators, $C(T)$, with a basis of interpolating fields which overlap with the desired state. Then one solves the GEVP equation
\beq
C(T) v_n(T,t_0) = \lambda_n(T,t_0) C(t_0) v_n(T,t_0)\,,
\eeq
where $\lambda_n$ and $v_n$ are the $n^{\rm th}$ generalized eigenvalues and eigenvectors. 

With a sufficiently well chosen basis, the generalized eigenvectors of this matrix will overlap with the individual states and be largely orthogonal to the others. This allows one to choose the linear combination of interpolating fields which overlaps with the desired state be it ground state or excited state.  These optimized operators can then be used to form the three point correlation function. This improved overlap allows for a decreased minimum Euclidean time for the matrix element fit improving the efficiency of the calculation. This approach has been used very successfully in identifying multiple energy levels for hadron spectroscopy~\cite{Hansen:2020otl,Woss:2020ayi,Alexandru:2020xqf,Khan:2020ahz,Johnson:2020ilc} and in the determination of matrix elements~\cite{Egerer:2018xgu,Egerer:2020hnc}.

The combination of the summation and GEVP methods~\cite{Bulava:2011yz} is a powerful technique for improving the excited state contamination in a matrix element calculation. The effective matrix element is given by the difference of the ratio
\beq
R(T,t_0) =  \frac{ v_n^\dagger(T,t_0) \left[K(t)\lambda(T,t_0)^{-1} - K(t_0) \right] v_n(T,t_0)}{ v_n^\dagger(T,t_0) C(t_0) v_n(T,t_0)}\,, \nonumber
\eeq
\beq
M^{\rm{eff}}(T,t_0) =\frac{1}{\tau} ( R(T,t_0) - R(T+\tau,t_0))
\eeq
where $K(T)$ is the sum over operator insertion time of the three point correlation matrix. This method has the combined advantages of the increased exponential decay and the reduced overlap of excited state contamination.

\begin{figure}[!htp]
\centering
\includegraphics[width=0.48\textwidth]{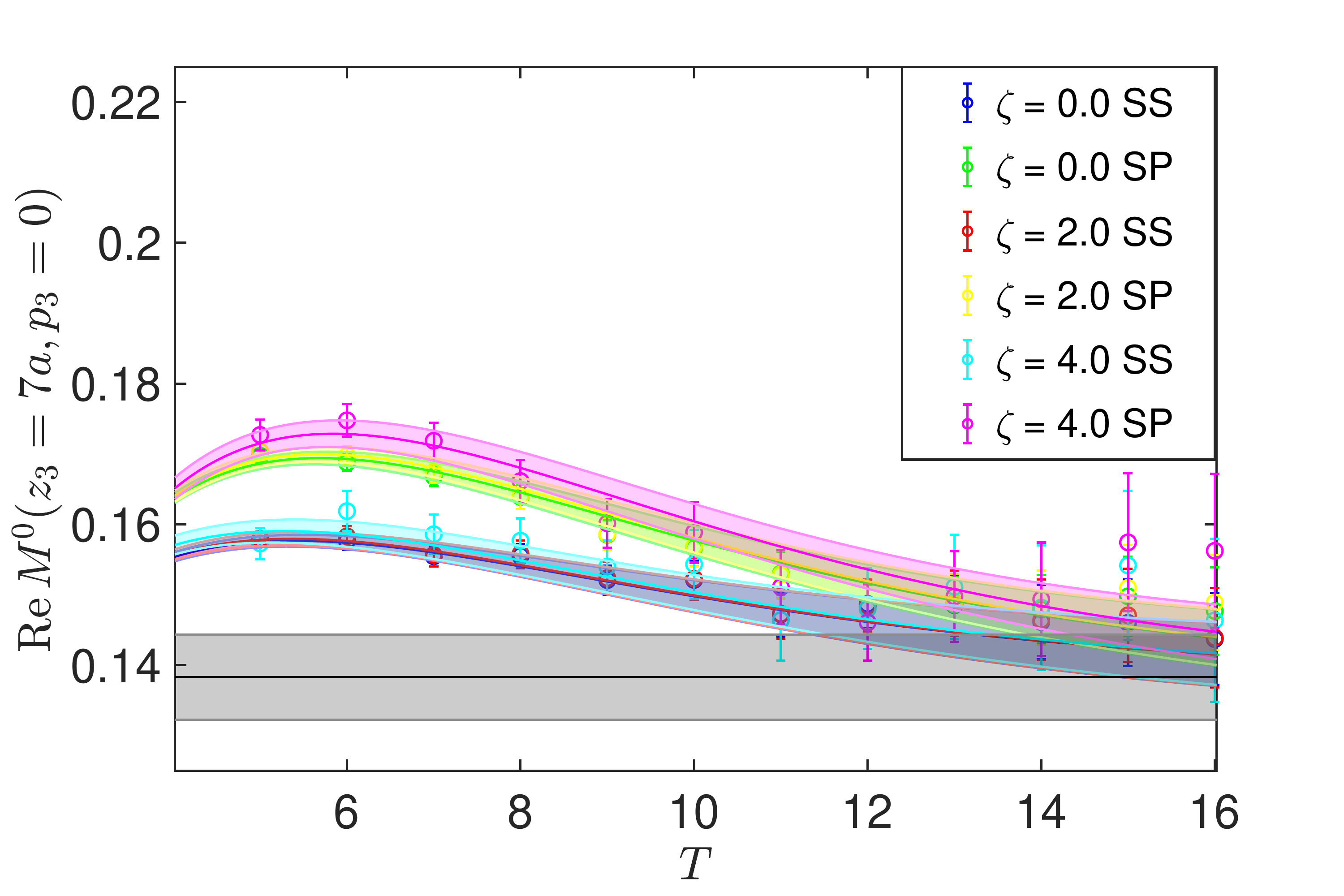}
\includegraphics[width=0.48\textwidth]{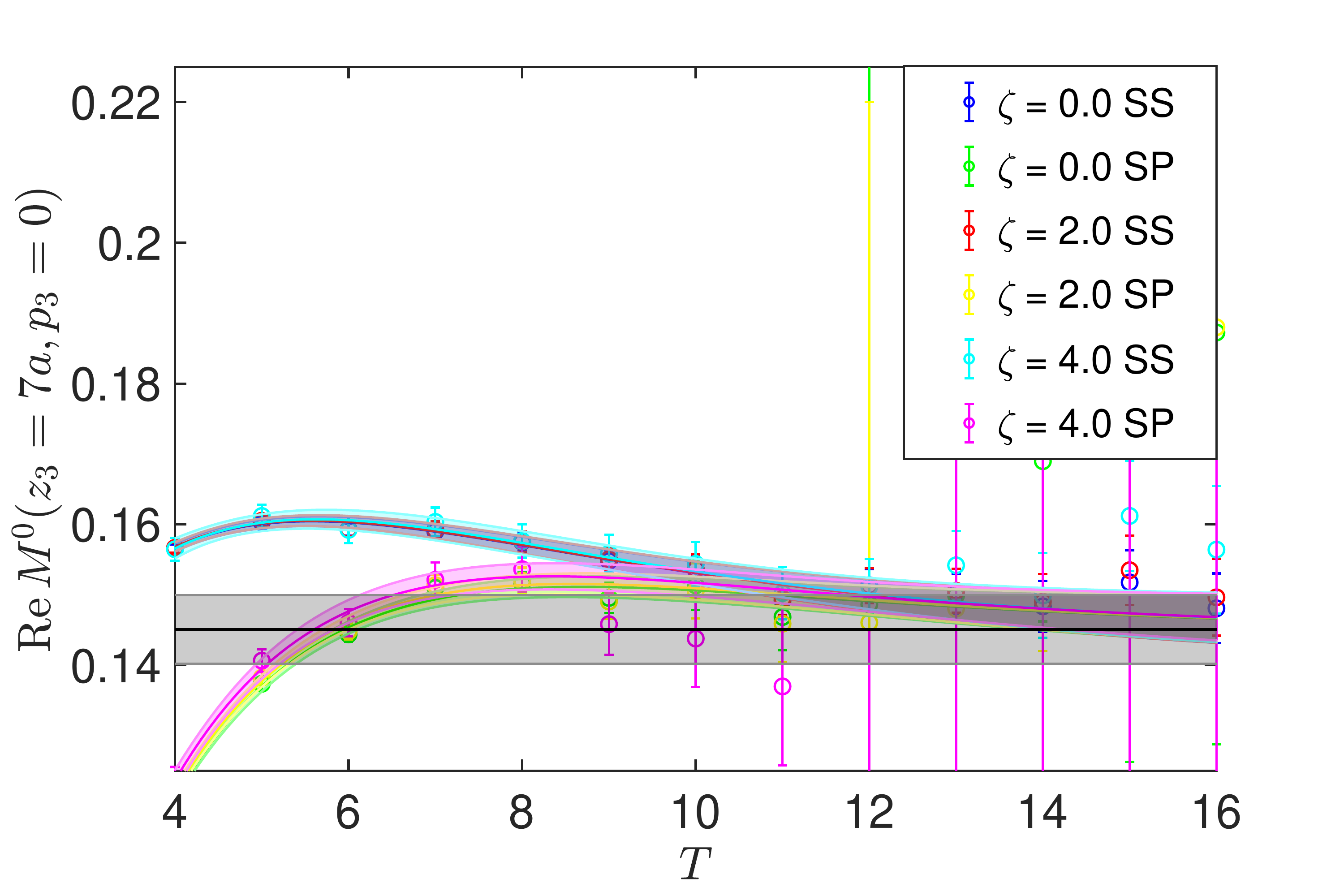}
\caption{\label{fig:gevp_comparison} The correlators using the summation technique (right) and the sGEVP technique (left). The correlators with smeared source and sink fields (SS) show little difference between the two techniques. On the other hand, the correlators with smeared sources and point sinks (SP) show a dramatic improvement in excited states. }
\end{figure}

In Fig~\ref{fig:gevp_comparison}, the summation technique with a single operator is compared to the sGEVP. The correlations with smeared source and sink interpolating fields already had relatively small excited state contamination in the summation technique. The sGEVP results do not significantly change those data. On the other hand, the correlators with smeared source and point sink interpolating fields had a large reduction in excited state contamination. The plateau region is reached significantly earlier when using the sGEVP technique. 

This example demonstrates how even a minimal application of the sGEVP using only three local nucleon operators can create some control over the excited state contamination. Other applications of the GEVP method~\cite{Khan:2020ahz,Egerer:2018xgu,Egerer:2020hnc} have used many more operators which are specifically selected to overlap with more excited states than these local operators. Our future applications to PDFs will utilize a larger basis of operators to have an even more substantial effect. The sGEVP method will be crucial for calculations at physical pion mass at high momenta where the correlation function may only be precise in a limited range of Euclidean time.

\begin{figure}[!htp]
\centering
\includegraphics[width=0.48\textwidth]{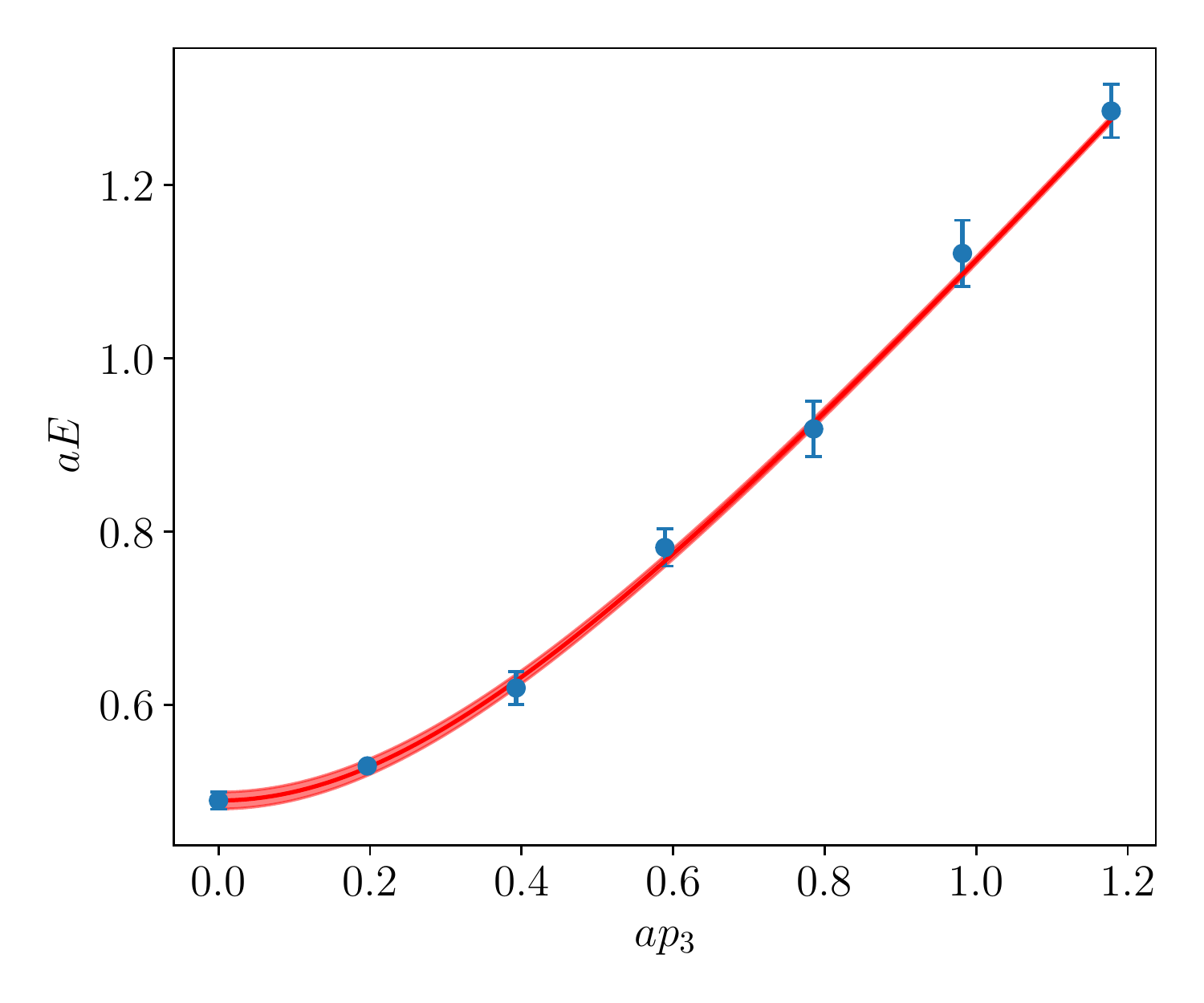}
\includegraphics[width=0.48\textwidth]{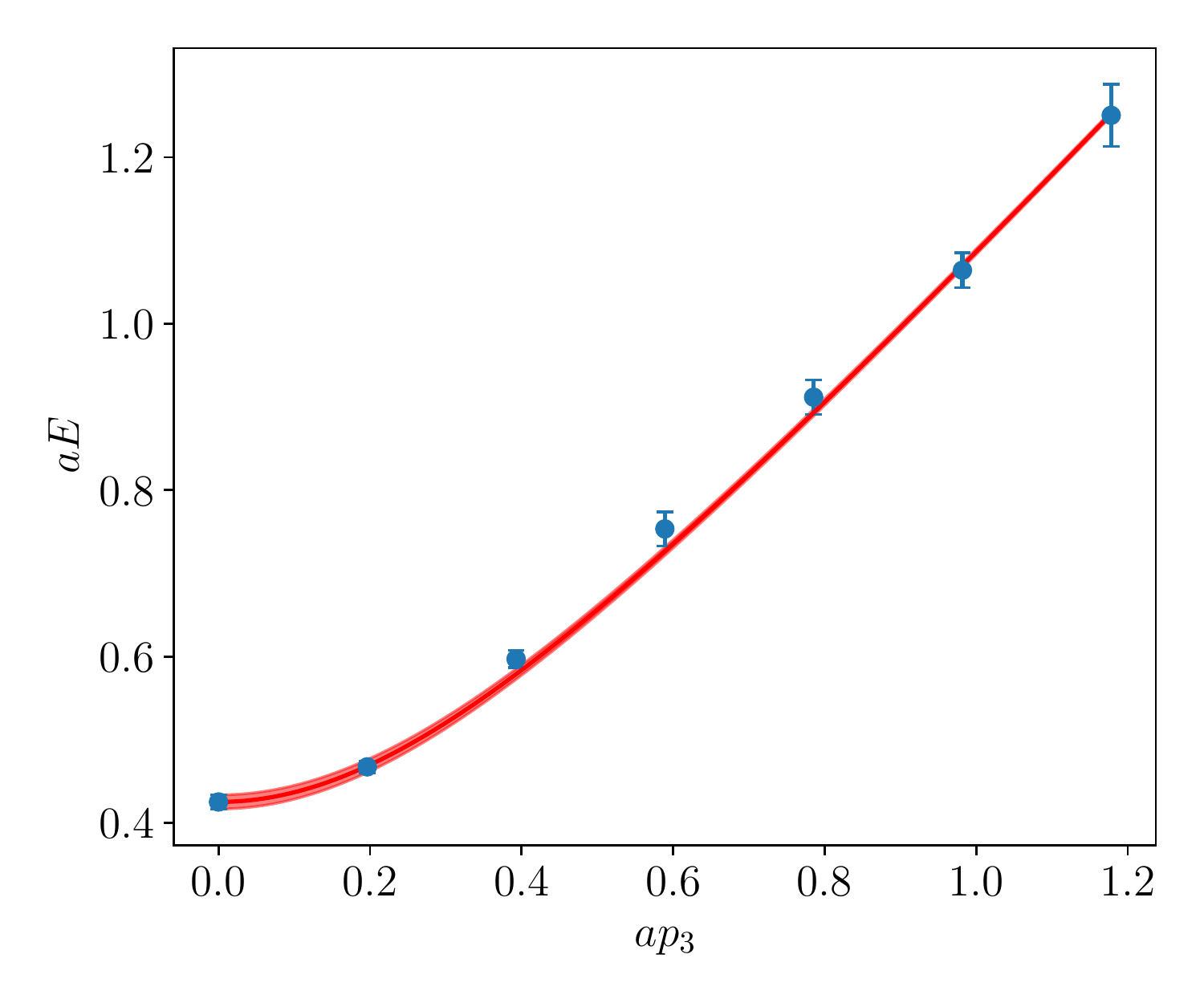}
\includegraphics[width=0.48\textwidth]{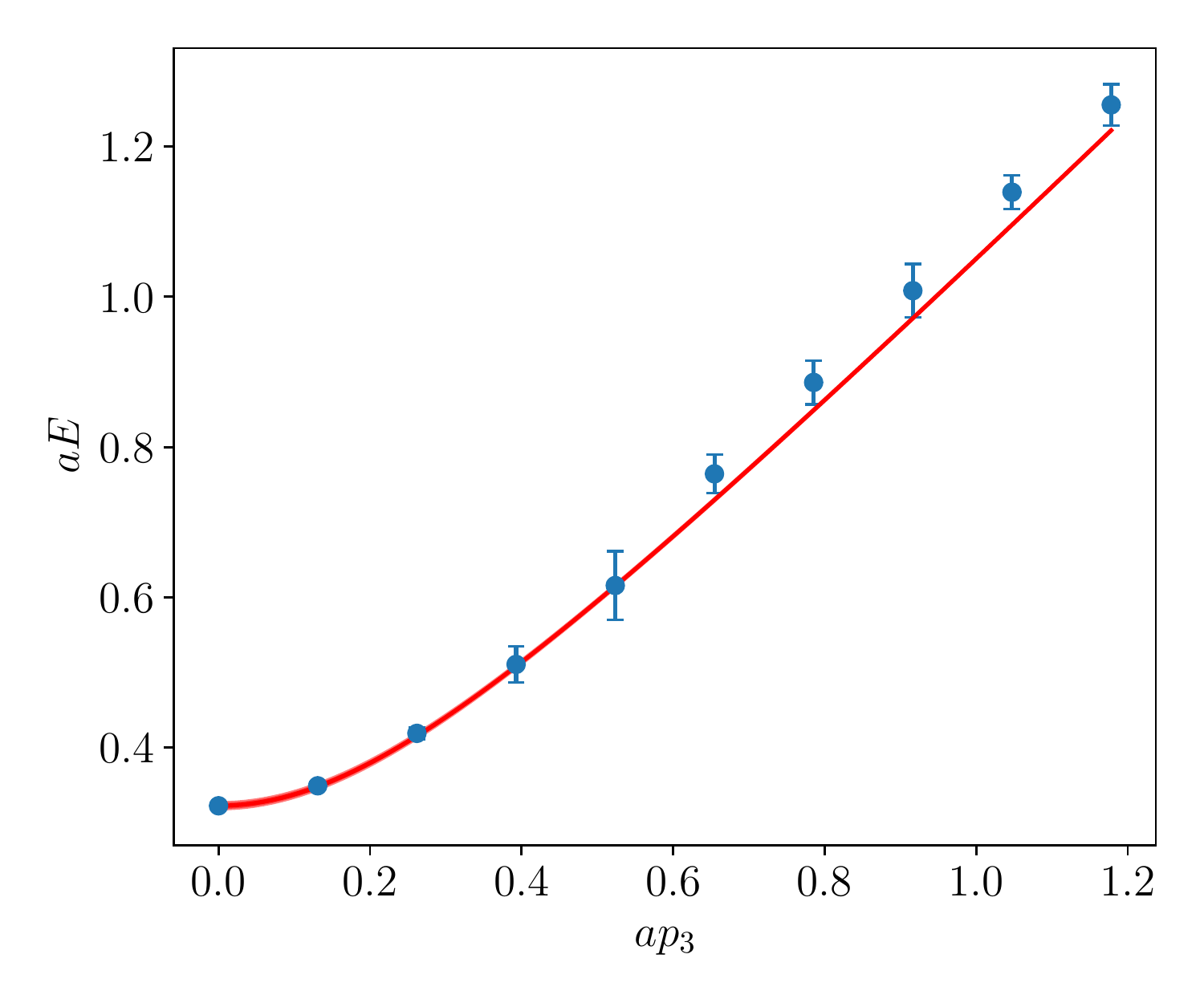}
\caption{\label{fig:meff} The results of fitting the effective mass of the principle correlator $\lambda_0(T,t_0)$ from the moving states. The ensemble $\widetilde{\rm A}_5$, ${\rm E}_5$, and ${\rm N}_5$, are in the upper right, upper left, and bottom respectively. The data points represent the fits to the effective mass and the curve is the continuum dispersion relationship from the rest frame. The energy levels are in agreement with the continuum relationship until the largest momenta where slight deviations occur.}
\end{figure}

Fig.~\ref{fig:meff} displays the results of the nucleon energy as function of the nucleon momentum. These energies are extracted by fitting the time dependence of the principle correlator $\lambda_0(T,t_0)$. As it is evident that these energy levels are  consistent with the continuum dispersion relationship within the range of available momenta. Since the momentum smearing parameter was not tuned specifically for each momentum state, the errors do not monotonically increase as they would if the same smearing was used or if the momentum smearing parameter was optimized for each momentum state. 

\subsection{Fitting matrix elements}
The sGEVP is applied to each scenario of smearing parameters individually. It is likely that modifying the operators by only changing smearing parameters will not drastically change its overlap with the ground and excited states. This means combining them within the sGEVP will have little effect. This feature can be seen in Fig.~\ref{fig:gevp_comparison}, where the effective matrix elements with different $\zeta$ are largely consistent within errors. With the same overlap they cannot significantly improve the cancellation of higher state effects. Instead, combinations of these six smearing scenarios are simultaneously fit to obtain a common matrix element and an excited state mass. When the signal-to-noise ratio for some of smearing scenarios is poor, they are excluded from the fit, for example large $\zeta$ at small $p$ or vice versa. 

There exists a systematic error from the particular choices of the maximum and minimum values of $T$ used within the fits for the matrix elements. The maximum value was chosen based upon the statistical noise of the correlation functions at those times. When the noise was sufficiently large that the fit result was not significantly affected, the maximum value was set. The minimum value was chosen to minimize the $\chi^2/{\rm d.o.f.}$ of the fit. The change of the central values when fitting with a minimum time decreased by a single time slice is used, in order to estimate the systematic error from the choice of minimum time. The square of this systematic error is added to the diagonal of the covariance matrix for the remainder of the analysis. The majority of the data points do not see a dramatic increase in error, but some do highlighting the importance of this analysis. 
\begin{figure}[!htp]
\centering
\includegraphics[width=0.48\textwidth]{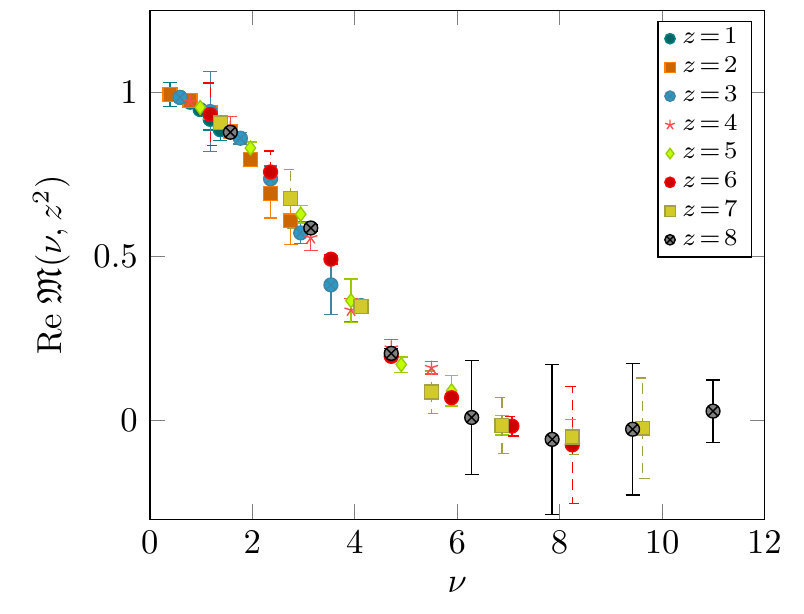}
\includegraphics[width=0.48\textwidth]{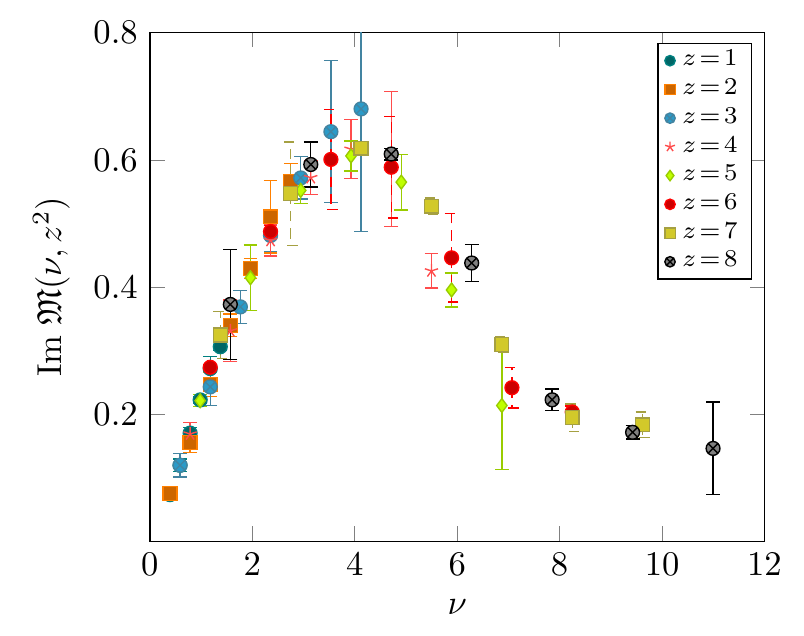}
\includegraphics[width=0.48\textwidth]{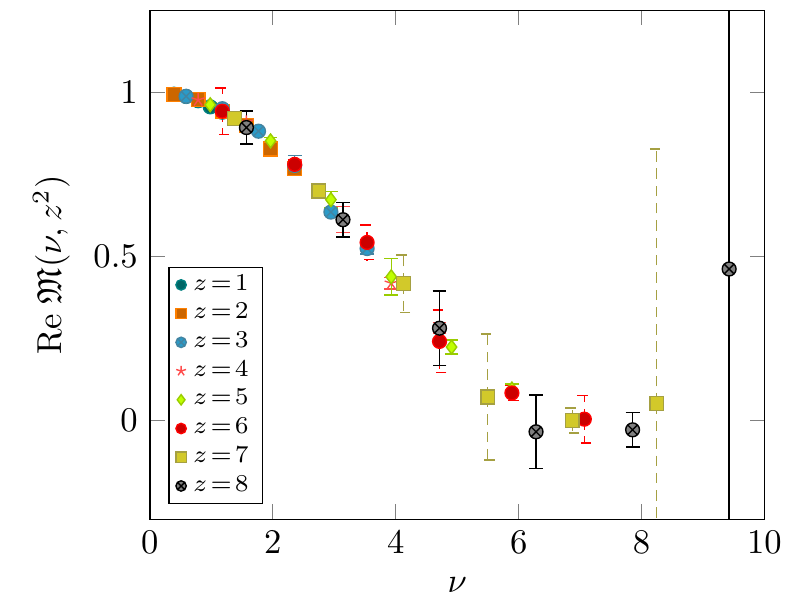}
\includegraphics[width=0.48\textwidth]{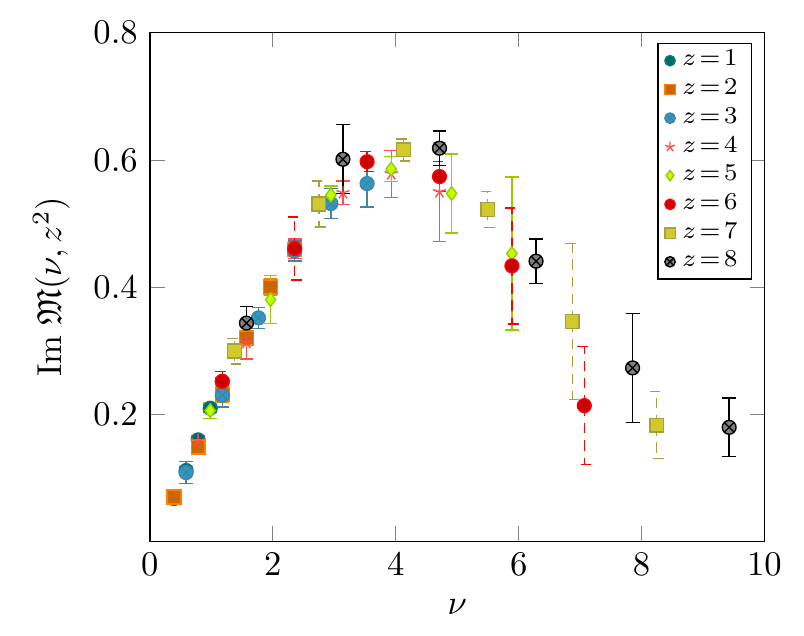}
\includegraphics[width=0.48\textwidth]{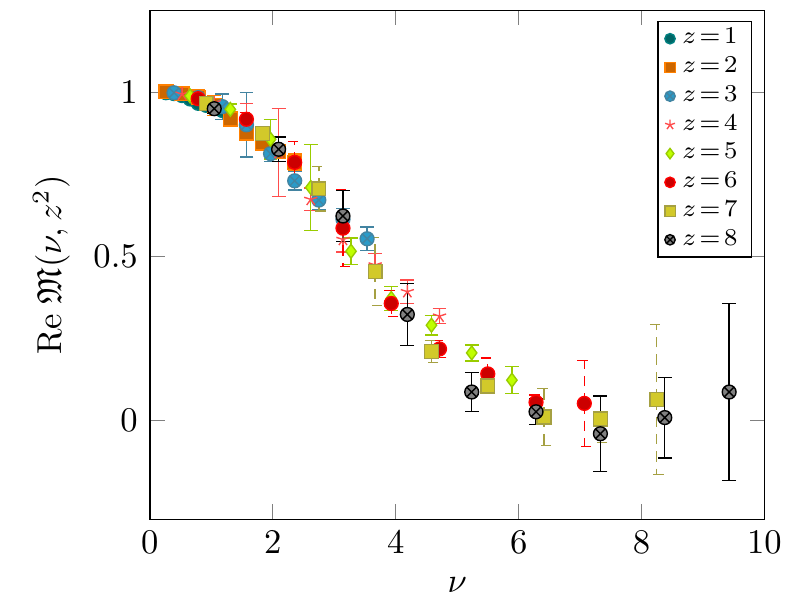}
\includegraphics[width=0.48\textwidth]{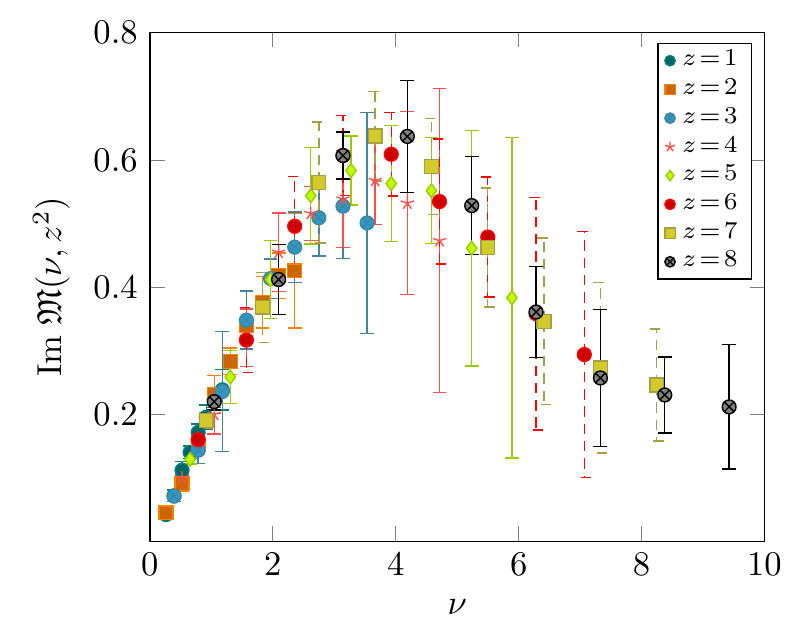}

\caption{\label{fig:reduced_ITDs}The real (LHS) and the imaginary (RHS) part of the reduced ITDs. The first row contains the results of the coarsest ensemble $\widetilde{\rm A}_5$, the second row the results of the ensemble ${\rm E}_5$ with the intermediate lattice spacing and the third row depicts the results for the ensemble ${\rm N}_5$ with the finest lattice spacing. The statistical and systematic errors are added in quadrature.}
\end{figure}

\begin{figure}[!htp]
\centering
\includegraphics[width=0.48\textwidth]{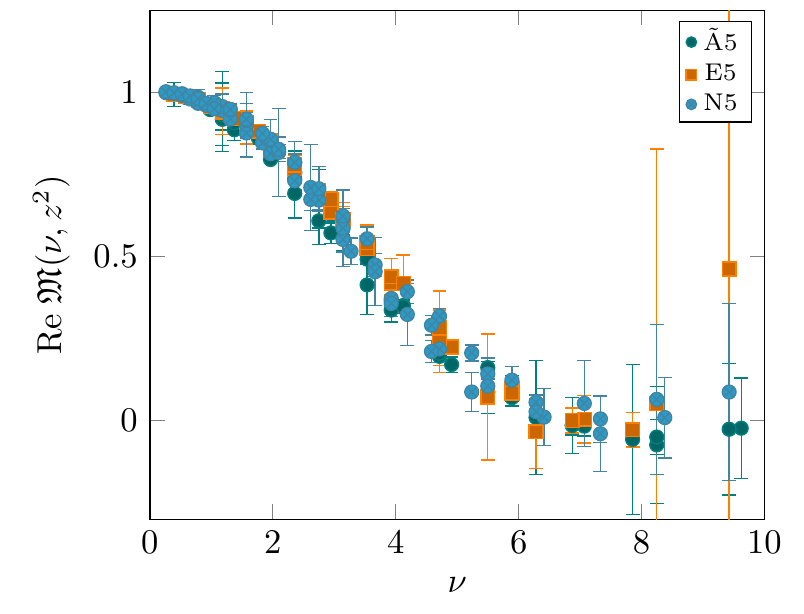}
\includegraphics[width=0.48\textwidth]{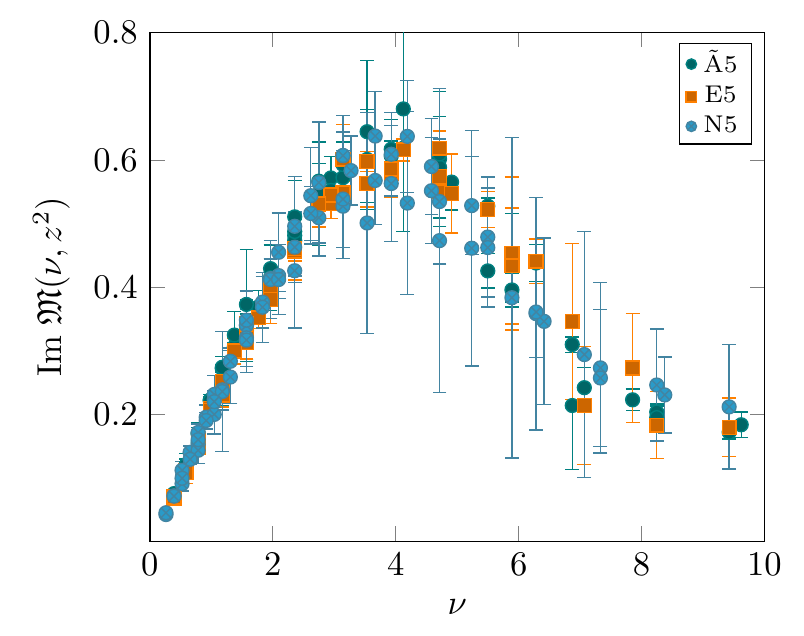}
\caption{\label{fig:all_reduced_ITDs}The real (LHS) and the imaginary (RHS) part of the reduced ITDs of the three lattice ensembles used in this study. We see that for the range of Ioffe times that is covered by our data the three ensembles have a pretty good overlap. The statistical and systematic errors are added in quadrature.}
\end{figure}
\section{Fits with Bayesian Priors}\label{sec:bayes}
In order to determine the PDF from our lattice matrix elements, we create a model to describe our data in terms of the PDF and various systematic errors as described in Sec.~\ref{sec:nuisance}. Let $\mathfrak{M}^L(\nu,z^2)$ be the lattice matrix elements while $\mathfrak{M}(\nu,z^2,\theta)$ be the matrix element from our model which depends on a set of parameters $\theta$. These parameters are the exponents $\alpha$, $\beta$, and the linear coefficients of the Jacobi series for the PDF and the nuisance terms. 

We attempt to determine the most likely values of the unknown parameters $\theta$ given our lattice matrix elements, $\mathfrak{M}^L$ and some prior information, $I$ by using Bayes' theorem, which states
\beq
P\left[\theta | \mathfrak{M}^L, I \right] = \frac{P\left[\mathfrak{M}^L |\theta \right]P\left[\theta |  I\right]}{P\left[\mathfrak{M}^L | I\right]}\,.
\eeq
Here $P\left[\theta | \mathfrak{M}^L, I \right]$ is the posterior distribution, which describes the probability distribution that a given set of parameters are the true parameters given a set of data and prior information. $P\left[\mathfrak{M}^L |\theta \right]$ is the probability distribution of the data given a set of model parameters. $P\left[\theta |  I\right]$ is the prior distribution  which describes the probability distribution of a set of parameters given some previously held information about it. Finally, $P\left[\mathfrak{M}^L | I\right]$ is the marginal likelihood or evidence which describes the probability that the data are correct given the previously held information. Ultimately, since the evidence does not depend on the model parameters it will be an unnecessary normalization for finding the most likely parameters.

The probability distribution of the lattice matrix elements, due to the central limit theorem,  may be written as the quadratic distance functional $P[\mathfrak{M}^L|\theta] \propto {\rm exp}[-\frac{\chi^2}2]$,
\begin{align}
\chi^2=\sum_{k,l} ( \mathfrak{M}^L_k - \mathfrak{M}_k ) C_{kl}^{-1} ( \mathfrak{M}^L_l - \mathfrak{M}_l ),
\end{align}
where the indices $k,l$ run over all our matrix elements and 
\begin{align}
C_{kl}=\frac{1}{N(N-1)}\sum_i \big(  \mathfrak{M}^{L,i}_k - \mathfrak{M}^L_k \big)\big( \mathfrak{M}^{L,i}_l - \mathfrak{M}^L_l \big),
\end{align}
is the covariance matrix of the $N$ samples (denoted as $\mathfrak{M}^{L,i}_k  $ ) of the matrix elements $\mathfrak{M}^L_k$. In the absence of any prior information, finding the most probable set of model parameters is done by minimizing $\chi^2$. 

The prior distributions are chosen to encode some expectations or requirements on the fit parameters. A simple example of how this could be done is by setting bounds on a fit. If one desires a model parameter $\theta_i$ to be limited to the range $[a,b]$, then the prior distribution is given by $P\left[\theta_i |  I \right] = (b-a)^{-1} \theta(x-a)\theta(b-x)$ where $\theta(x)$ is the Heaviside step function. The PDF is known to be dominated by the leading behavior $x^\alpha(1-x)^\beta$ and the other terms should be small corrections to this. Therefore we give the PDF model parameters $_\pm d_n^{(\alpha,\beta)}$ priors which are normal distributions, with a mean and width of $d_0$ and $\sigma_d$ respectively. In Sec.~\ref{sec:pdf_results}, we use normal distributions centered about 0, but change the widths in order to study its effects. Similarly for the nuisance terms, we expect their parameters to be small corrections to the dominant PDF and use a normal distribution for them, whose widths are smaller than those of the PDF parameters. The mean and width of these are given by $c_0$ and $\sigma_c$.

The prior distributions for $\alpha$ and $\beta$ could also be normal distributions, but they have other restrictions. First, $\alpha$ and $\beta$ must be greater than -1. This is an explicit restriction from the definition of the Jacobi polynomials, but also has a physical interpretation. If $\alpha$ or $\beta$ were equal or less than -1, then the integral of the PDF, which is related to the total number of quarks in the proton, would diverge.  Furthermore, we do not expect $\beta < 0$, since we expect that the parton distribution function vanishes at $x=1$.
In order to enforce the restrictions of $\alpha > -1$ and $\beta > 0$, their prior distributions are log-normal distributions,
\beq
P(x,\mu,\sigma,x_0) = \frac{1}{(x-x_0)\sigma \sqrt{2\pi}} e^{-\frac{[\log(x-x_0) - \mu]^2}{2\sigma^2}}
\eeq
where $\mu$ is the mean and $\sigma^2$ the variance of the distribution of $\log(x-x_0)$, and $x_0$ is the lower bound of the log-normal distributions. 
The mean $\mu_x$ and variance $\sigma_x^2$  of the variable $x$ are related to the log-normal distribution parameters by the following formulae,
\begin{align}
    \mu =& \log\Big( \frac{\mu_x-x_0}{\sqrt{(\mu_x-x_0)^2 + \sigma_x^2}}\Big)\\
    \sigma =& \log\Big( 1 + \frac{\sigma_x^2}{(\mu_x-x_0)^2}\Big).
\end{align}
  
The most likely parameters of the model are found by maximizing the posterior distribution.  This is performed by minimizing the negative log of the posterior distribution $L^2 = -2 \log(P\left[\theta | \mathfrak{M}^L, I \right]) + C$, where $C$ is the normalization of the posterior which is independent of the model parameters.
This is a relatively simple task because apart from $\alpha$ and $\beta$ all other parameters of the model enter linearly and therefore the minimization with respect to any of these parameters can be done analytically at fixed $\alpha$ and $\beta$. Subsequently, a non-linear minimization of $L^2$, which is now a function only of $\alpha$ and $\beta$, can be done with a non-linear minimizer.
As a consequence, in principle one can easily minimize $L^2$ with a large number of Jacobi polynomial terms as the non-linear minimization is always two dimensional. This is a well known technique called Variable Projection (VarPro)~\cite{varpro}. 
 
\section{PDF results}\label{sec:pdf_results}
As discussed before the PDF is related to the lattice matrix elements through a convolution integral relation. Extracting the PDF from the lattice matrix elements involves the solution of an inverse problem and therefore the resulting PDF depends on the method used to solve it introducing a new systematic error that requires careful study.  The statistical error of a single choice of solution, such as the discrete Fourier transform, the Backus-Gilbert method, or a fit to a particular model PDF, may significantly underestimate the true uncertainty on the PDF. This feature can clearly be seen in the few studies which have compared alternative methods or varied models~\cite{Sufian:2020vzb,Bhat:2020ktg,Alexandrou:2020qtt}. In this theme, we want to study  many, though rather interrelated, models which vary both the number of parameters as well as the prior distributions. However, the prior distributions have to be chosen carefully to reflect accurately our prior knowledge.

In the following analysis, the PDF scale is taken to be $\mu=2$ GeV, which results in the two flavor  $\overline{MS}$ $\alpha_s(\mu=2 $ GeV$)  = 0.245$, with $\Lambda_{\rm QCD} = 268$ MeV.
 
\subsection{Inclusion of nuisance terms}\label{sec:which_terms}
There are no {\em ab initio} estimates of the sizes of the lattice spacing nuisance terms $P_1$ and $R_1$. In principle, one could calculate matrix elements of the operators discussed in~\cite{Chen:2017mie} and develop a Symanzik improvement style program. In~\cite{Braun:2018brg}, an estimate of the size of the higher twist effects was made using a method based upon the fact that renormalon effects must cancel with the higher twist term. They demonstrated that the reduced pseudo-ITD had strikingly smaller higher twist effects than the pseudo-ITD, as expected. In the range of Ioffe time for this calculation, the improvement is at least a factor of 5. For the middle of this Ioffe time range it is closer to a factor of 10. We anticipate that the lattice spacing nuisance terms for the reduced pseudo-ITD will also be smaller than the pseudo-ITD due to the same cancellation within the ratio.

As discussed above, in this work, we use a parameterization of these unknown functions and study their effect on the fits of the PDF. First, it is important to understand which nuisance terms are more necessary than others. A common way approaching this is to iteratively add the terms and see the effect on the $L^2/$d.o.f. In order to study this effect, every combination of the leading twist PDF and the nuisance terms is fit to the data.

For simplicity, in this test the continuum leading twist term has two Jacobi polynomials for the PDF and one Jacobi polynomial for the possible nuisance terms. As shown in Figs.~\ref{fig:sigma} and~\ref{fig:eta}, the contributions of very high order Jacobi polynomials are small all the way up to Ioffe times at the upper end of the available range. The effect of varying the numbers of terms are studied in Sec.~\ref{sec:vary_numbers}. The widths and means of the prior distributions are the same as the model ``default'' in Tab.~\ref{tab:single_model}. 
Tab.~\ref{tab:which_terms} shows the $L^2$/d.o.f. and $\chi^2/$d.o.f. of the models with all possible combinations of nuisance terms. There is a clear decrease in $L^2$/d.o.f. when $P_1$ is included into the fit for both the real and the imaginary component. This effect is anticipated, since the small $z$ data, which are most sensitive to $P_1$ because they are affected more by lattice spacing errors, are generally more precise than the large $z$ data for any given momentum. The precision in combination with the expected lattice spacing errors  give a larger impact on the $\chi^2$ and therefore $L^2$. 
This feature of statistics, along with the ability to use small momentum data which are exponentially more precise than large momentum data, shows an advantage of the SDF approach over LaMET. The limitations of SDF require that the more precise low $z$ data are used, where the limitations of LaMET require the noisier larger $p$ data to be used. To reach the same precision for those points orders of magnitude of greater computational resources are required.

\begin{figure}[!htp]
\centering
\includegraphics[width=0.48\textwidth]{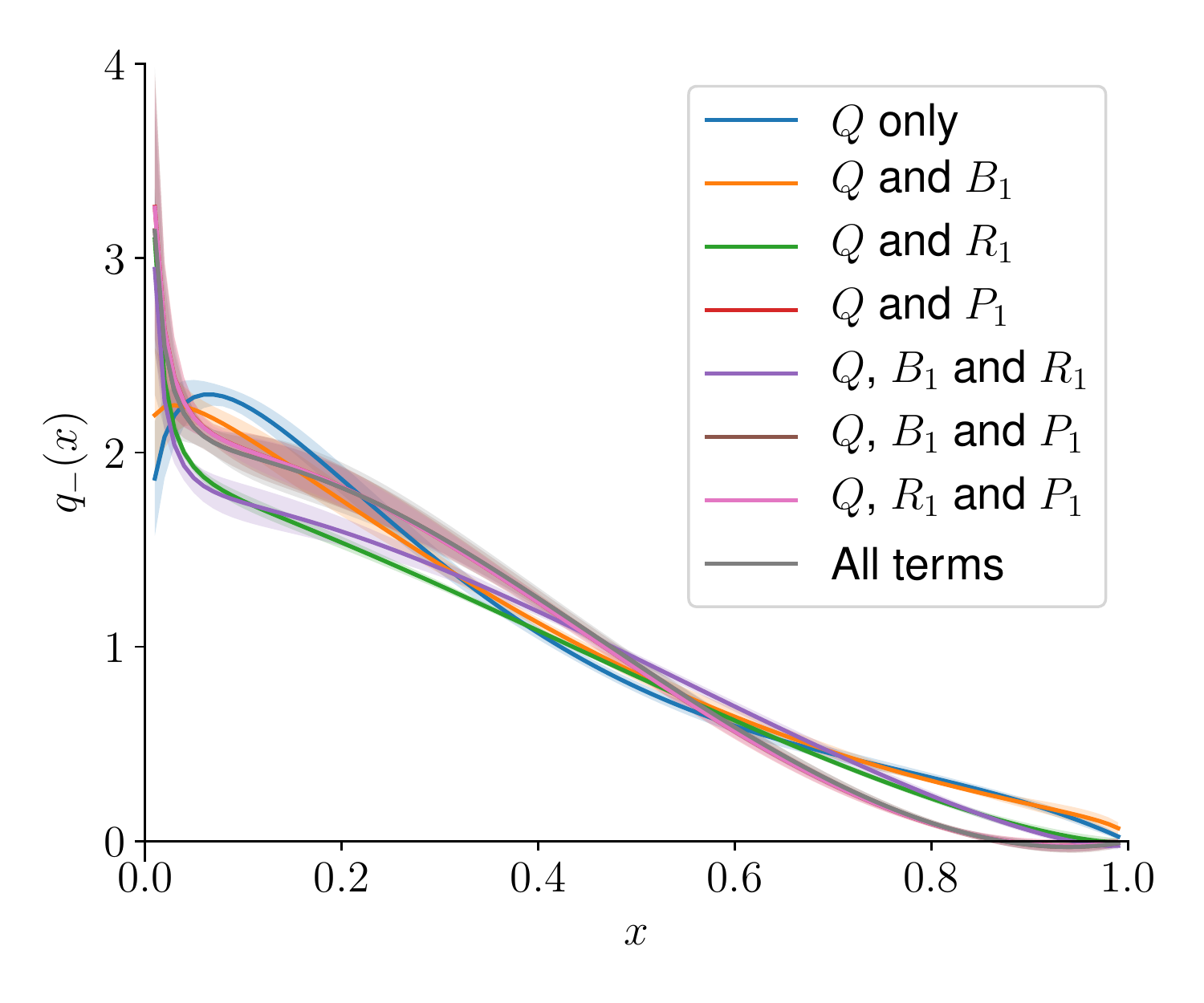}
\includegraphics[width=0.48\textwidth]{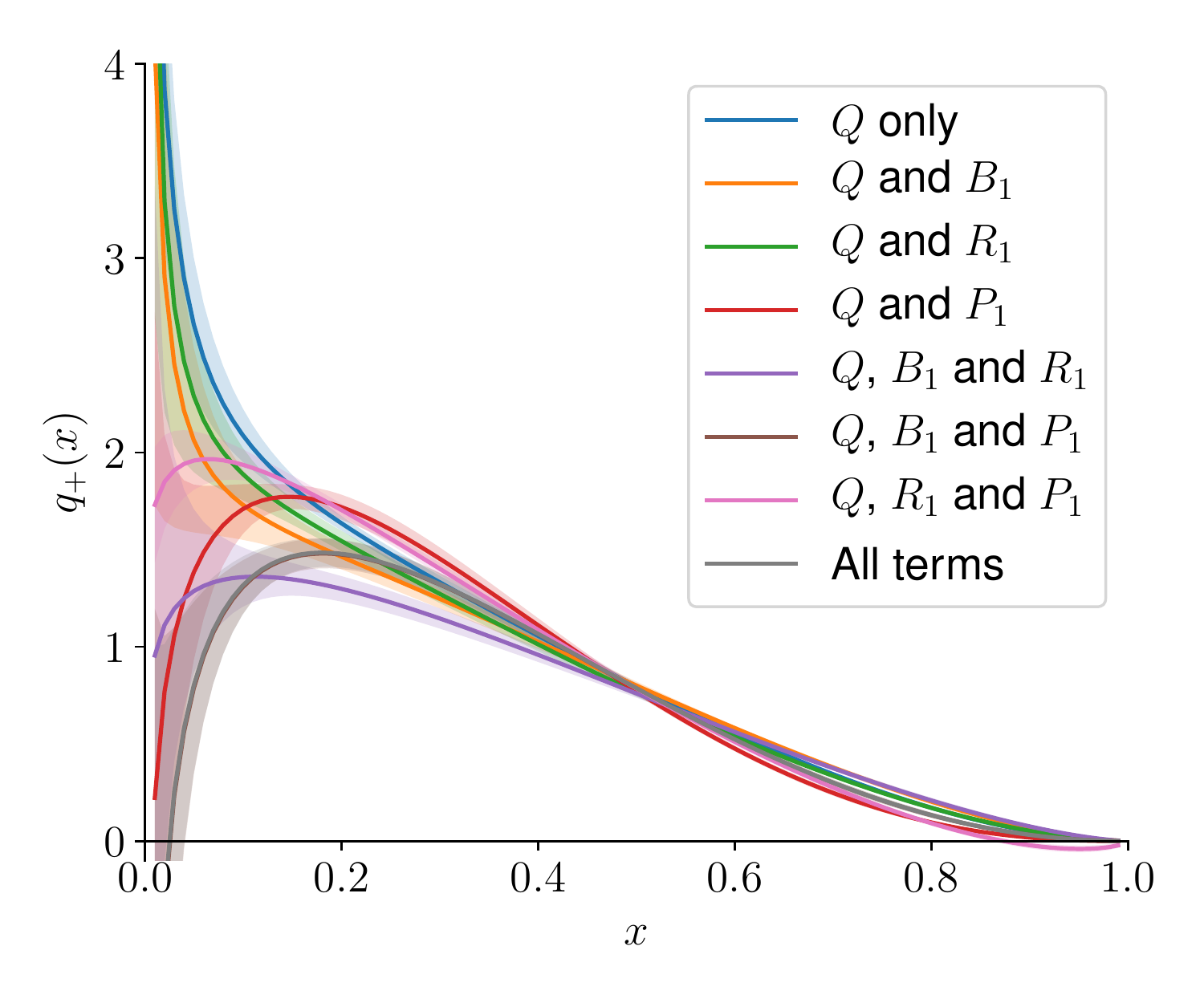}
\caption{\label{fig:which_terms} The results of fitting with various nuisance terms included.}
\end{figure}

The effects of $B_1$ and $R_1$ are less clear. The improvement of $L^2$/d.o.f. is only modest. The higher twist contribution is most sensitive to the largest $z$ data, which are statistically noisier and therefore affect the $\chi^2$ less. Deviations caused by neglecting $B_1$ may not be expected to generate larger contributions to the $\chi^2$. Unfortunately, no such argument can be made for $R_1$ terms which are agnostic to $z$. As  we show in Sec.~\ref{sec:study_priors}, the data are not sensitive to this term at all. A final thing to note is that the close values of both $L^2$ and $\chi^2$ imply that the data provide a significant part of the contribution to $L^2$, not the prior distributions. As seen in~\ref{sec:study_priors}, when the priors are restrictive, the difference between $\chi^2$ and $L^2$ is noticeably larger. Fig.~\ref{fig:which_terms} shows the PDFs which result from these fits. 
The shape of $q_\pm$ changes substantially when the $P_1$ term is added in the large $x$ region, consistent with the fact that the $P_1$ term affects mostly the small $\nu$ range which in turn controls the large $x$ region of the PDF. 

%------models
\begin{table*}[t] 
\centering
\begin{tabular}{ l | c c | c c  }
model & Real $L^2$/d.o.f. & Real $\chi^2$/d.o.f. & Imag $L^2$/d.o.f. & Imag $\chi^2$/d.o.f. \\\hline\hline
$Q$ only & 3.173 & 3.094 & 3.146 & 3.095\\\hline
$Q$ and $B_1$ & 2.721 & 2.479 & 3.054 & 2.969\\\hline
$Q$ and $R_1$ & 3.028 & 2.748 & 3.068 & 2.871\\\hline
$Q$ and $P_1$ & 0.876 & 0.809 & 1.186 & 1.088\\\hline
$Q$, $B_1$, and $R_1$ & 2.610 & 2.057 & 2.917 & 2.619\\\hline
$Q$, $B_1$, and $P_1$ & 0.852 & 0.723 & 1.020 & 0.888\\\hline
$Q$, $R_1$, and $P_1$ & 0.881 & 0.763 & 1.289 & 1.063\\\hline
All terms & 0.857 & 0.727 & 1.026 & 0.893\\\hline
\hline\end{tabular}
\caption{\label{tab:which_terms}\footnotesize The $L^2$/d.o.f. and $\chi^2/$d.o.f. of models using 2 Jacobi polynomials for the PDF and 1 Jacobi polynomial for the various nuisance terms from fits to the real and imaginary components of the reduced pseudo-ITD. The change in the $L^2$/d.o.f. is a metric to judge the necessity of various nuisance terms. The most dramatic decreases occur when $O(\frac az)$ nuisance terms are included.}
\end{table*}

\subsection{Effects of prior distributions}\label{sec:study_priors}
In this section, we consider a set of prior distributions which can be studied in detail while fixing the number of parameters. The effects of the prior distributions in this model are modified in order to study the stability of the final results. The correlations between the resulting parameters, as well as comparison of their fluctuations to the prior distribution, can be used to identify which terms are being controlled by the data and which by the priors. These terms can then be modified or removed in order to test their relevance. The models being used in this study are described by Tab.~\ref{tab:single_model}.

%------models
\begin{table*}[t] 
\centering
\begin{tabular}{ l | c c c c | c c | c c | c c | c c }
name& $N_\pm$ & $N_{R/I, b}$ & $N_{R/I,r}$ & $N_{R/I,p}$ & $\alpha_0$ & $\sigma_{\alpha}$ & $\beta_0$ & $\sigma_{\beta}$ & $d_0$ & $\sigma_{d}$ & $c_0$ & $\sigma_{c}$\\\hline\hline
default & 2 & 1 & 1 & 1 & 0 & 0.4 & 3 & 1 & 0 & 0.5 & 0 & 0.1 \\\hline
wide & 2 & 1 & 1 & 1 & 0 & 0.8 & 3 & 2 & 0 & 1 & 0 & 0.5 \\\hline
thin & 2 & 1 & 1 & 1 & 0 & 0.2 & 3 & 0.5 & 0 & 0.25 & 0 & 0.05 \\\hline
limited & 2 & 0 & 0 & 1 & 0 & 0.4 & 3 & 1 & 0 & 0.5 & 0 & 0.1 \\\hline
\hline\end{tabular}
\caption{\label{tab:single_model}\footnotesize The configurations of the models used to determine the continuum PDF. The model is modified to test the stability of the method.}
\end{table*}

\begin{table*}[t] 
\centering
\begin{tabular}{ l | c c | c c}
name & Real $L^2/$d.o.f. & Real $\chi^2/$d.o.f. & Imag $L^2/$d.o.f. & Imag $\chi^2/$d.o.f. \\\hline\hline
default & 0.857 & 0.750 & 1.027 & 0.944\\\hline
wide & 0.726 & 0.708 & 0.899 & 0.893 \\\hline
thin & 1.281 & 0.966 & 1.415 & 1.168 \\\hline
limited & 0.876 & 0.809 & 1.187 & 1.148 \\\hline
\hline\end{tabular}
\caption{\label{tab:single_model_chi2}\footnotesize The $L^2$ and $\chi^2$ of the models used to determine the continuum PDF given in Tab.~\ref{tab:single_model}. }
\end{table*}

\begin{figure}[!htp]
\centering
\includegraphics[width=0.48\textwidth]{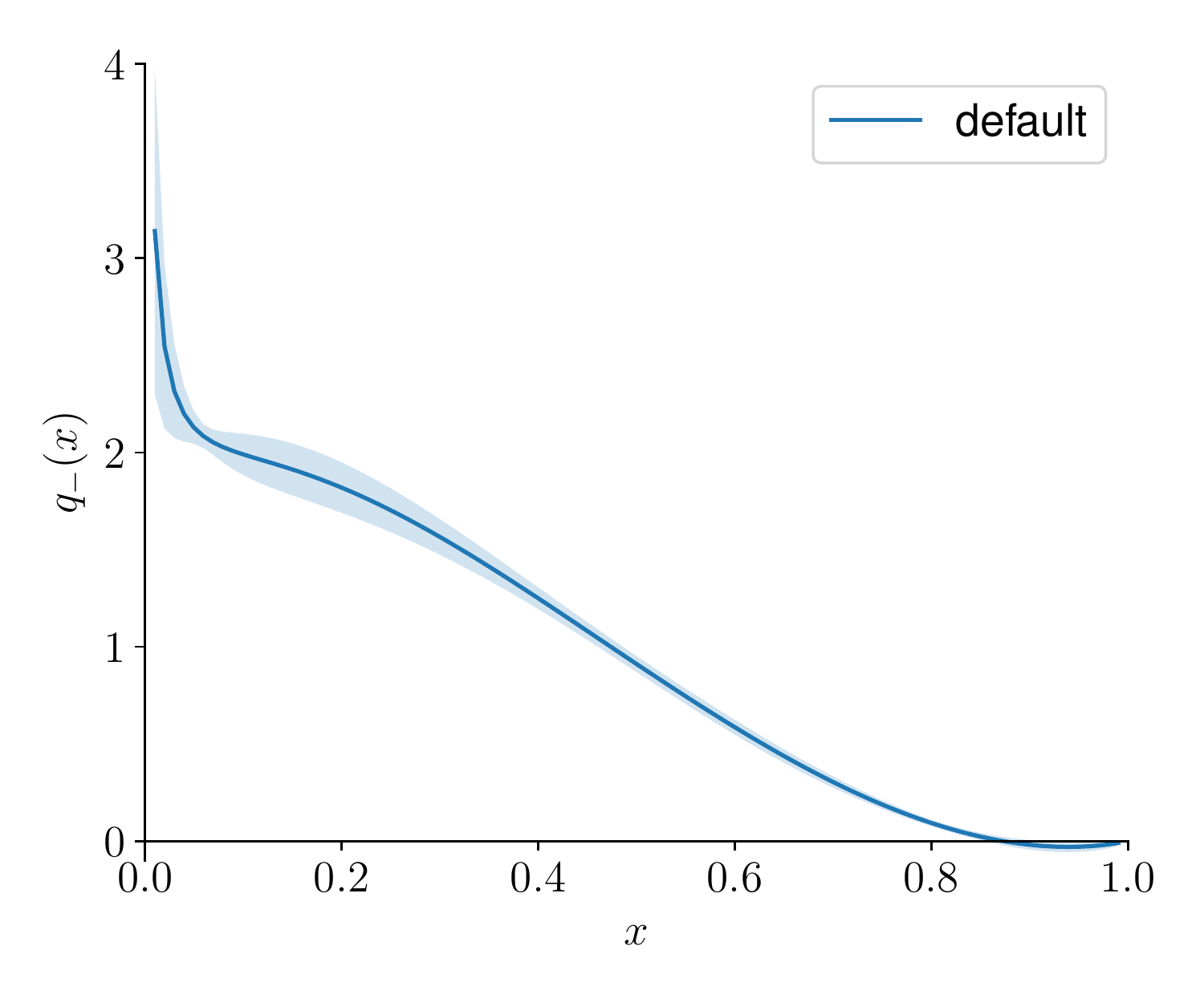}
\includegraphics[width=0.48\textwidth]{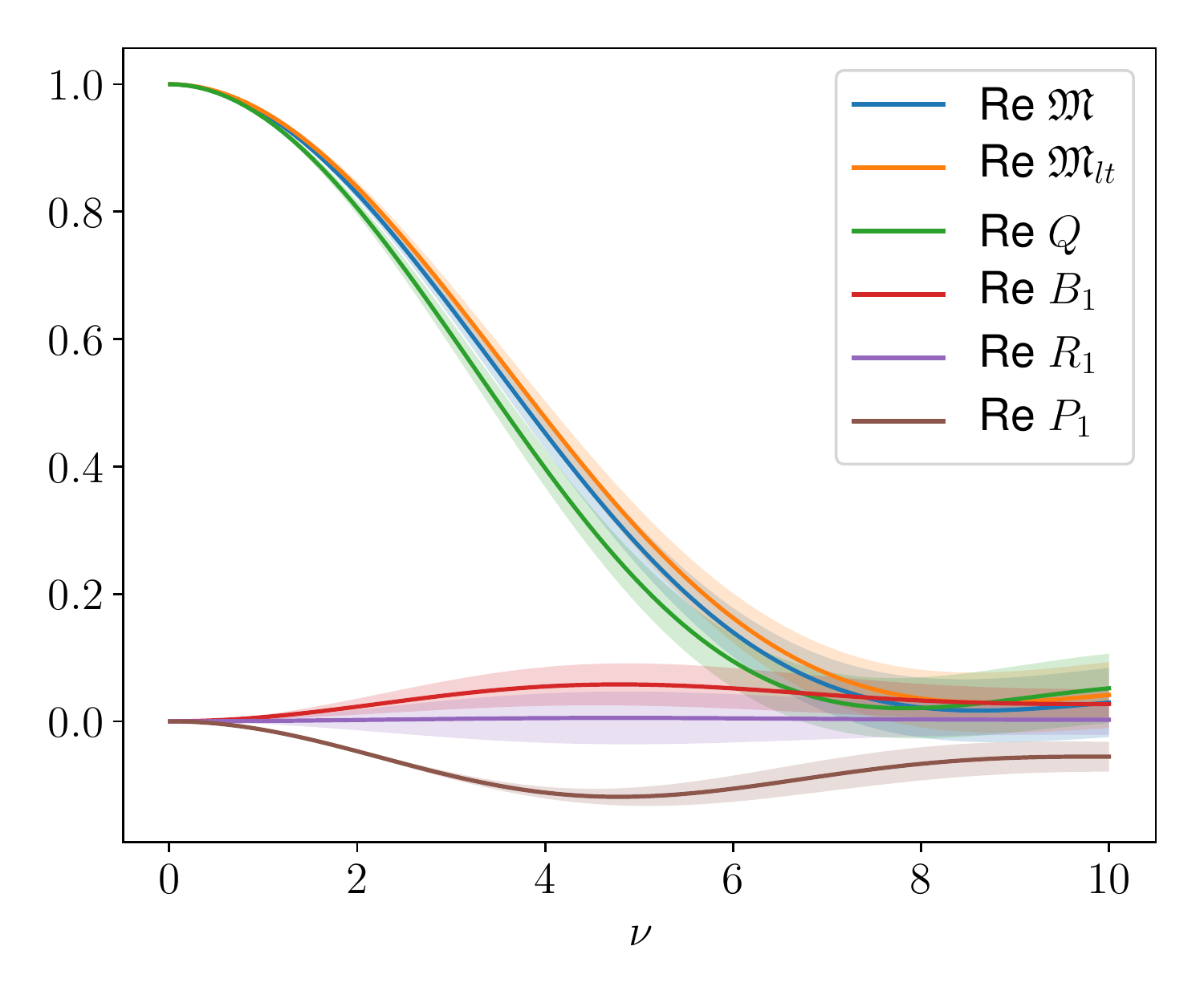}
\includegraphics[width=0.48\textwidth]{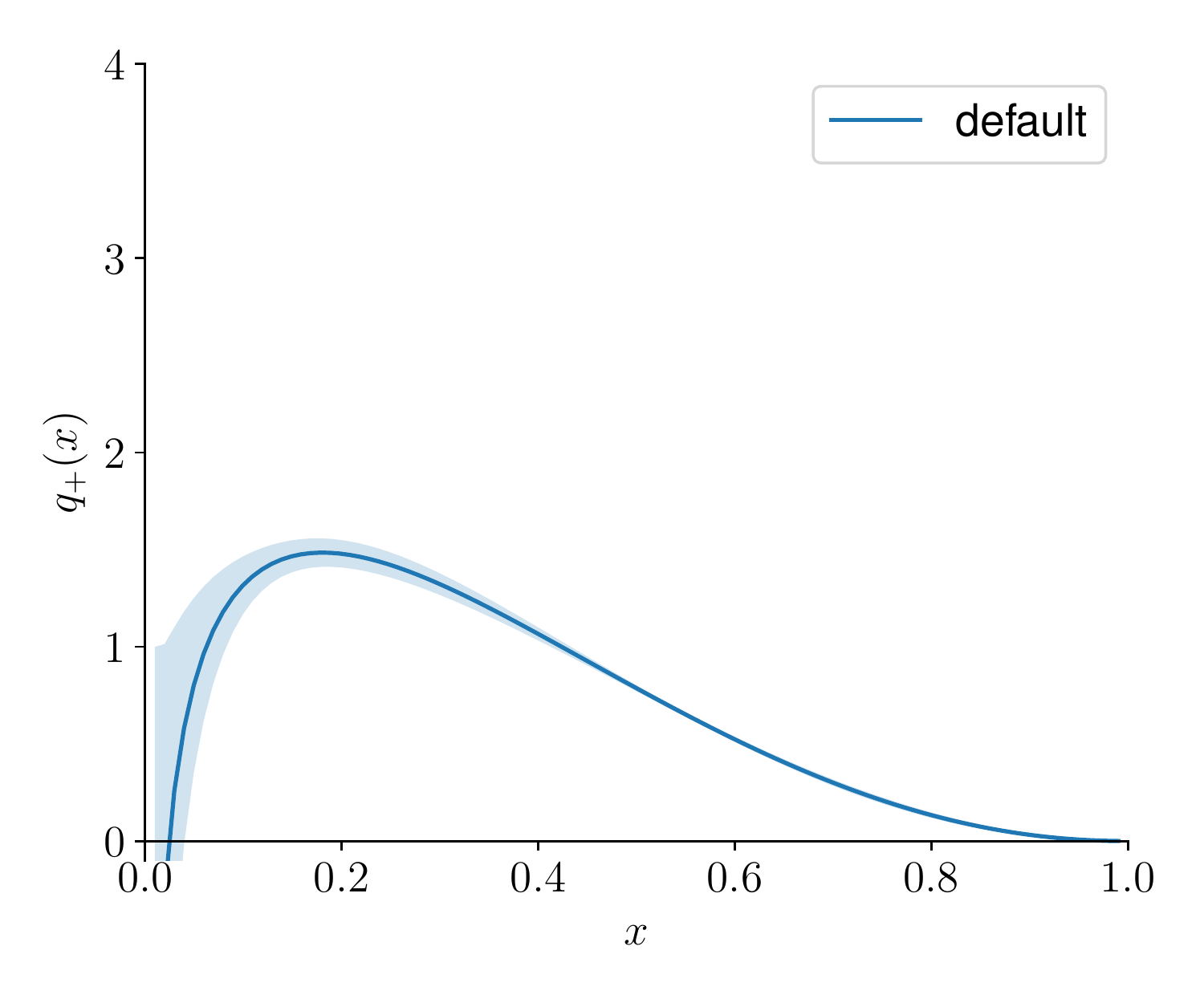}
\includegraphics[width=0.48\textwidth]{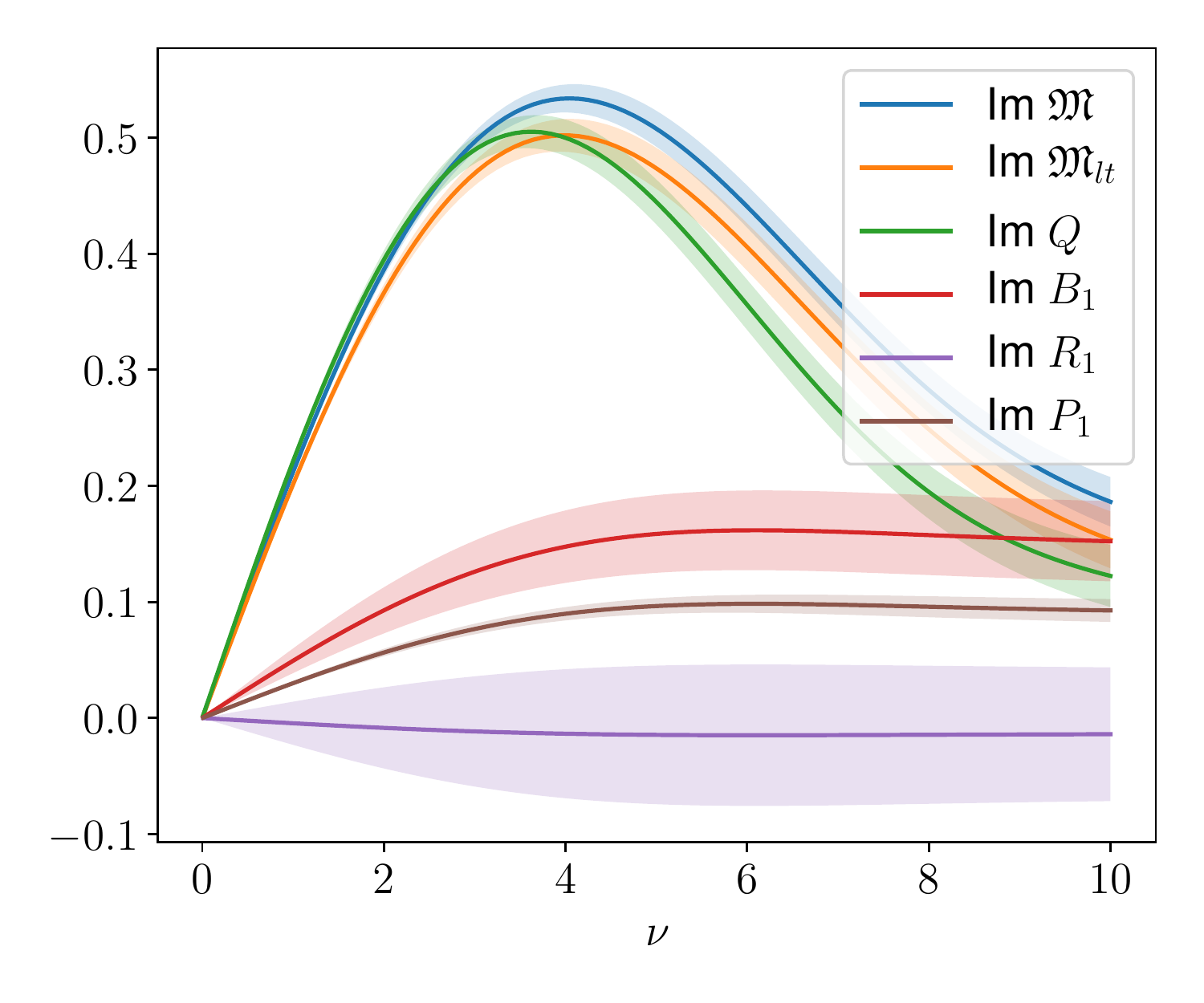}
\caption{\label{fig:single_model}The results of fitting to the model ``default''. The upper and lower plots come from the fits to the real and imaginary components respectively. The PDFs are given on the left and on the right, the nuisance terms are compared to the ITD, reduced pseudo-ITD, and the leading twist part of the reduced pseudo-ITD. For the reduced pseudo-ITD, the value $z = 4 * 0.0652$ fm was used as a typical example.}
\end{figure}

The model designated as ``default'' serves as a baseline for this study. This model is the same as the one considered in the previous section where all nuisance terms were included. The models dubbed ``wide'' and ``thin'' are to study the effect of the widths of the prior distributions. The model named ``limited'' is to study the case where only the nuisance term $P_1$, which decreased the $L^2/$d.o.f. most significantly in Sec.~\ref{sec:which_terms}, is included. As can be seen in Fig.~\ref{fig:single_model}, the model ``default'' qualitatively reproduces many of the known features of the PDF. The $L^2$ and $\chi^2$ of this fit are given in Tab.~\ref{tab:single_model_chi2}. Due to the limited extent in Ioffe time, the low $x$ behavior is not well resolved, allowing for solutions which converge and diverge as $x\to 0$ for the $q_+$ and $q_-$ distributions respectively. In a previous study of mock data~\cite{Karpie:2019eiq}, we found that significantly larger values for the maximum Ioffe time than what is currently achievable in lattice QCD are required to resolve the region of $x<0.2$ accurately. The size of the nuisance parameter terms are also smaller than the dominant leading twist component until $\nu$ has become large. The contribution from $R_1$ is completely consistent with 0. On the other hand, $B_1$ and $P_1$ are not consistent with zero for a range of Ioffe times. 

\begin{figure}[!htp]
\centering
\includegraphics[width=0.48\textwidth]{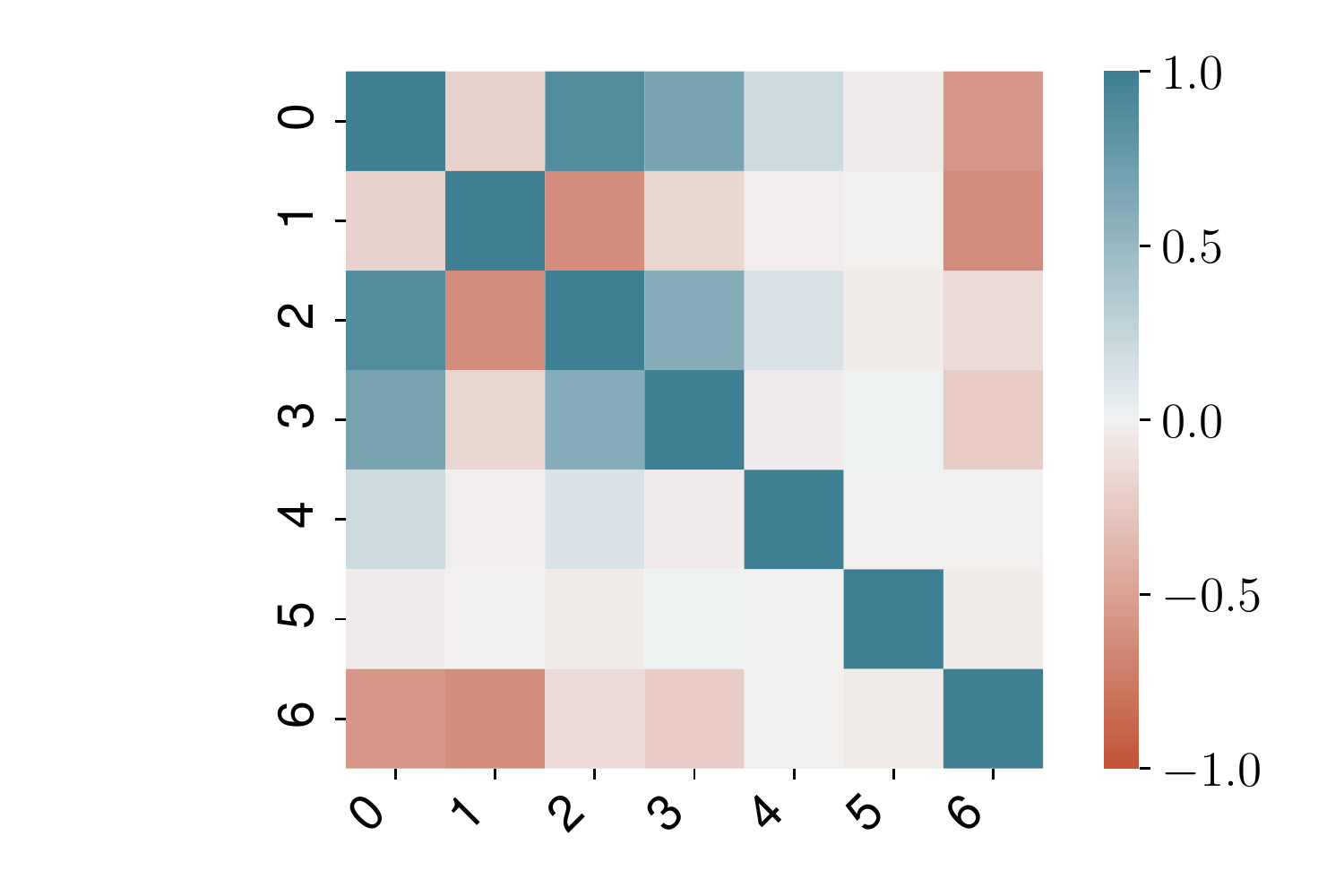}
\includegraphics[width=0.48\textwidth]{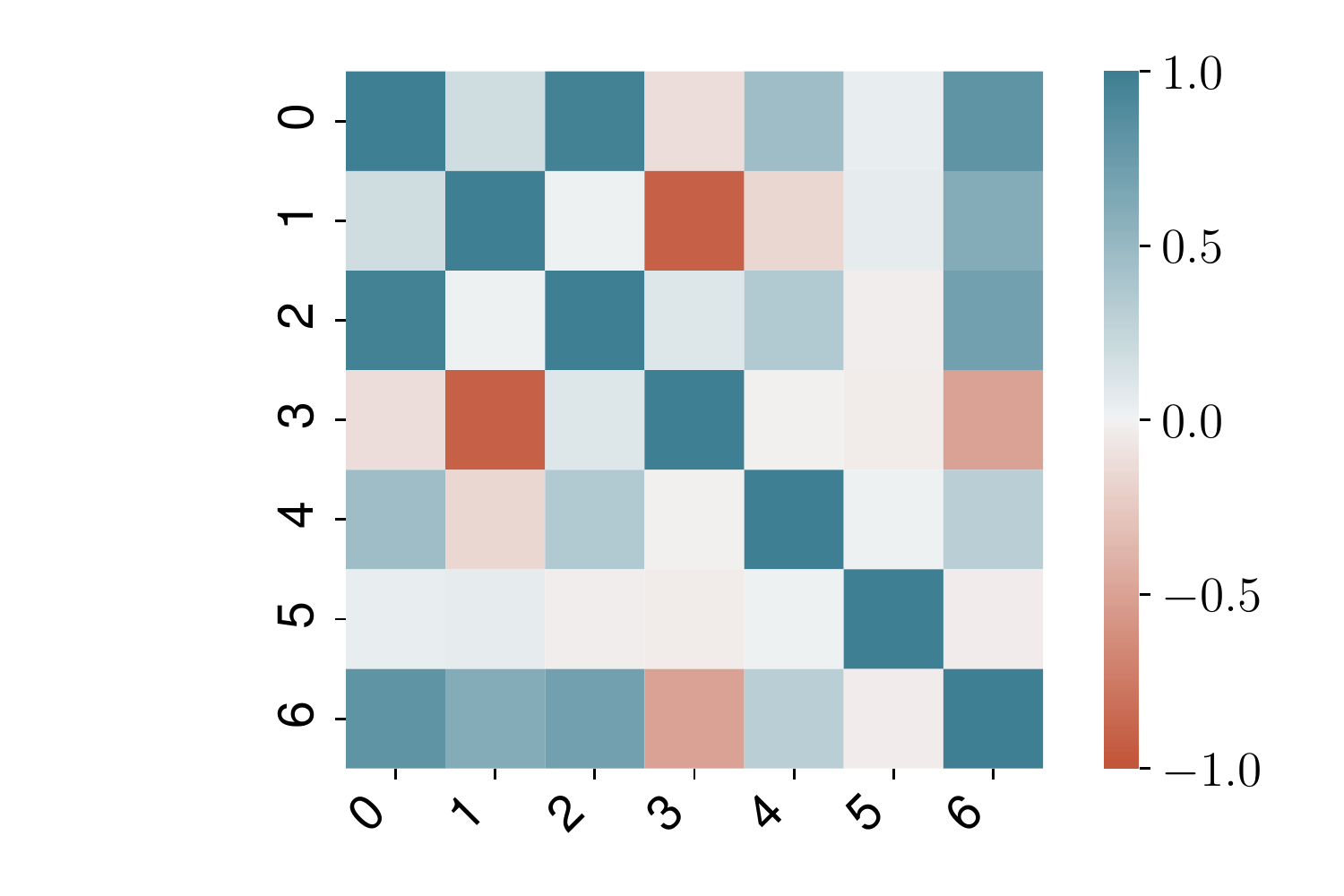}
\caption{\label{fig:single_model_correlation}The normalized covariance matrix between the parameter of the ``default'' model fit. The real component is on the left and the imaginary component on the right. The IDs of the parameters used on the labels are given in Tabs.~\ref{tab:single_model_parameters_real} and~\ref{tab:single_model_parameters_imag}.}
\end{figure}

 The values of the model parameters are given in  Tabs.~\ref{tab:single_model_parameters_real} and~\ref{tab:single_model_parameters_imag}. Fig.~\ref{fig:single_model_correlation} shows a normalized correlation matrix between the various model parameters, whose labels correspond to the numbers given in the tables. As can be seen the coefficient for $R_1$ is largely uncorrelated with the rest of the model parameters. Its central value and error are also very similar to those of the prior distribution. These features imply that it is not being constrained by the data, but by the prior distribution only. The higher twist and $O(\frac{a}{z})$ lattice spacing parameters, $b_{R/I,1}^{(\alpha,\beta)}$ and $p_{R/I,1}^{(\alpha,\beta)}$ on the other hand do seem to be controlled by the data to a greater extent, especially $p_{R/I,1}^{(\alpha,\beta)}$ which we found to be the most important for lowering $L^2/$d.o.f.

\begin{table*}[t] 
\centering
\begin{tabular}{ l c | c  }
parameter & ID & value \\ \hline\hline
$\alpha$ & 0 &  -0.45(14) \\\hline
$\beta$ & 1 & 0.93(20) \\\hline\hline
$_-d_1^{(\alpha,\beta)}$ & 2 & -0.29(31) \\\hline
$_-d_2^{(\alpha,\beta)}$ & 3 & -0.77(6) \\\hline
$b_{R,1}^{(\alpha,\beta)}$ & 4 & 0.13(6) \\\hline
$r_{R,1}^{(\alpha,\beta)}$ & 5 & 0.01(10) \\\hline
$p_{R,1}^{(\alpha,\beta)}$ & 6 & -0.27(5) \\\hline
\hline\end{tabular}
\caption{\label{tab:single_model_parameters_real}\footnotesize The values of the parameters from fitting the real component to the model ``default''. The ID numbers correspond to the labels in Fig.~\ref{fig:single_model_correlation}. }
\end{table*}

\begin{table*}[t] 
\centering
\begin{tabular}{ l c | c  }
parameter & ID & value \\ \hline\hline
$\alpha$ & 0 & -0.69(7) \\\hline
$\beta$ & 1 & 2.11(13) \\\hline\hline
$_+d_0^{(\alpha,\beta)}$ & 2 & 0.29(15) \\\hline
$_+d_1^{(\alpha,\beta)}$ & 3 &  -1.29(12) \\\hline
$b_{I,1}^{(\alpha,\beta)}$ & 4 & 0.26(5) \\\hline
$r_{I,1}^{(\alpha,\beta)}$ & 5 & -0.02(10) \\\hline
$p_{I,1}^{(\alpha,\beta)}$ & 6 & 0.16(2) \\\hline
\hline\end{tabular}
\caption{\label{tab:single_model_parameters_imag}\footnotesize The values of the parameters from fitting the imaginary component to the model ``default''. The ID numbers correspond to the labels in Fig.~\ref{fig:single_model_correlation}. }
\end{table*}

\begin{figure}[!htp]
\centering
\includegraphics[width=0.48\textwidth]{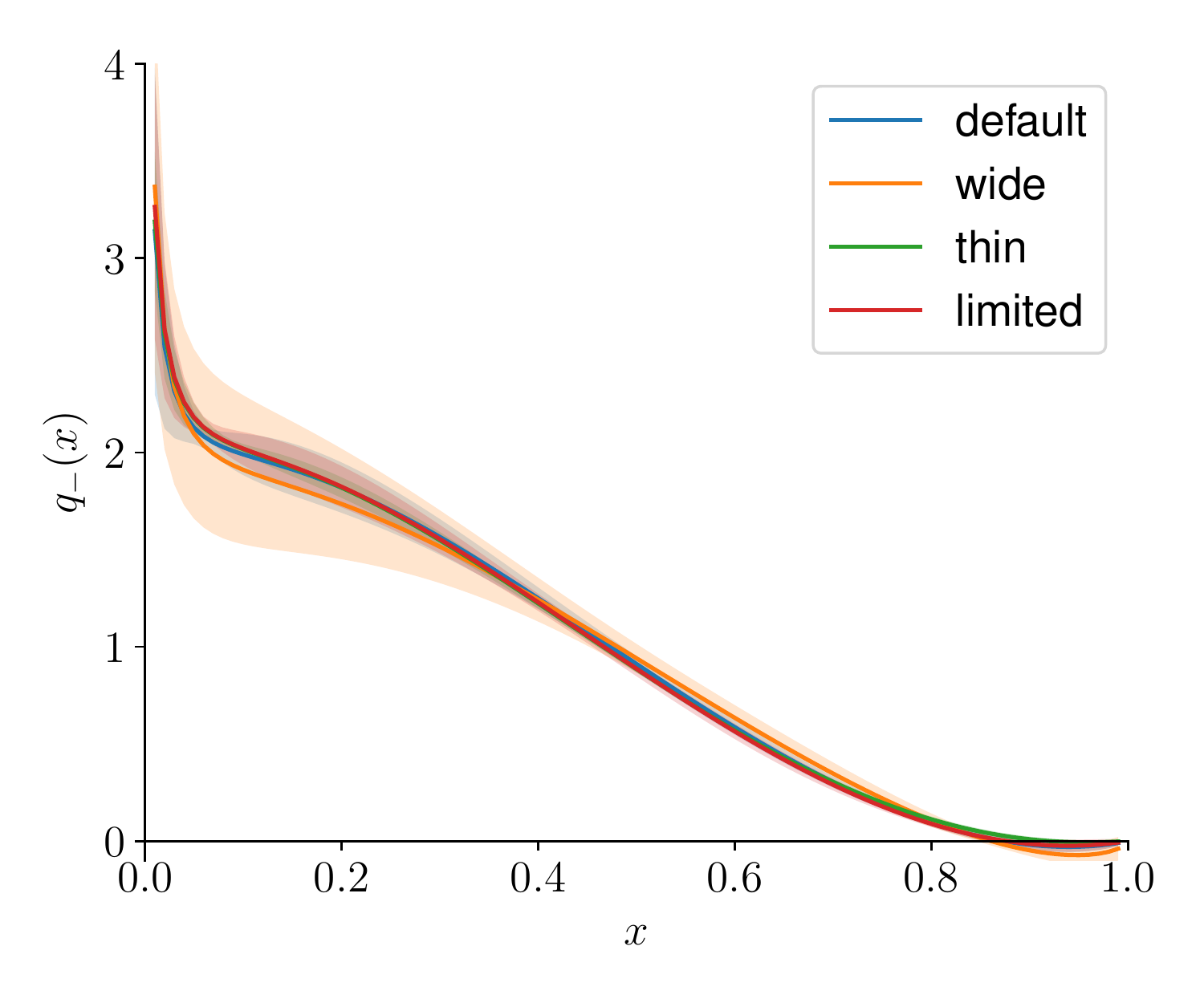}
\includegraphics[width=0.48\textwidth]{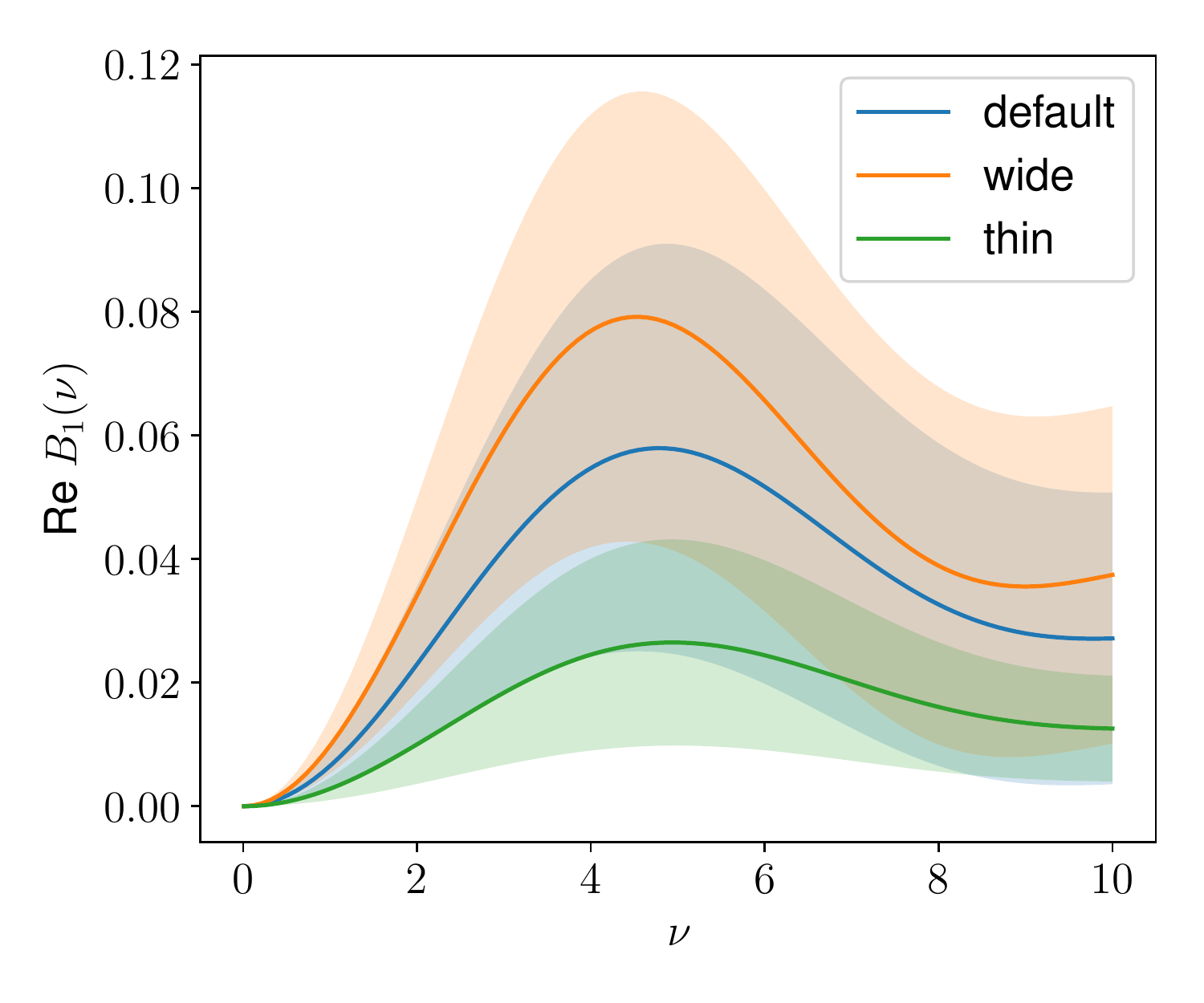}
\includegraphics[width=0.48\textwidth]{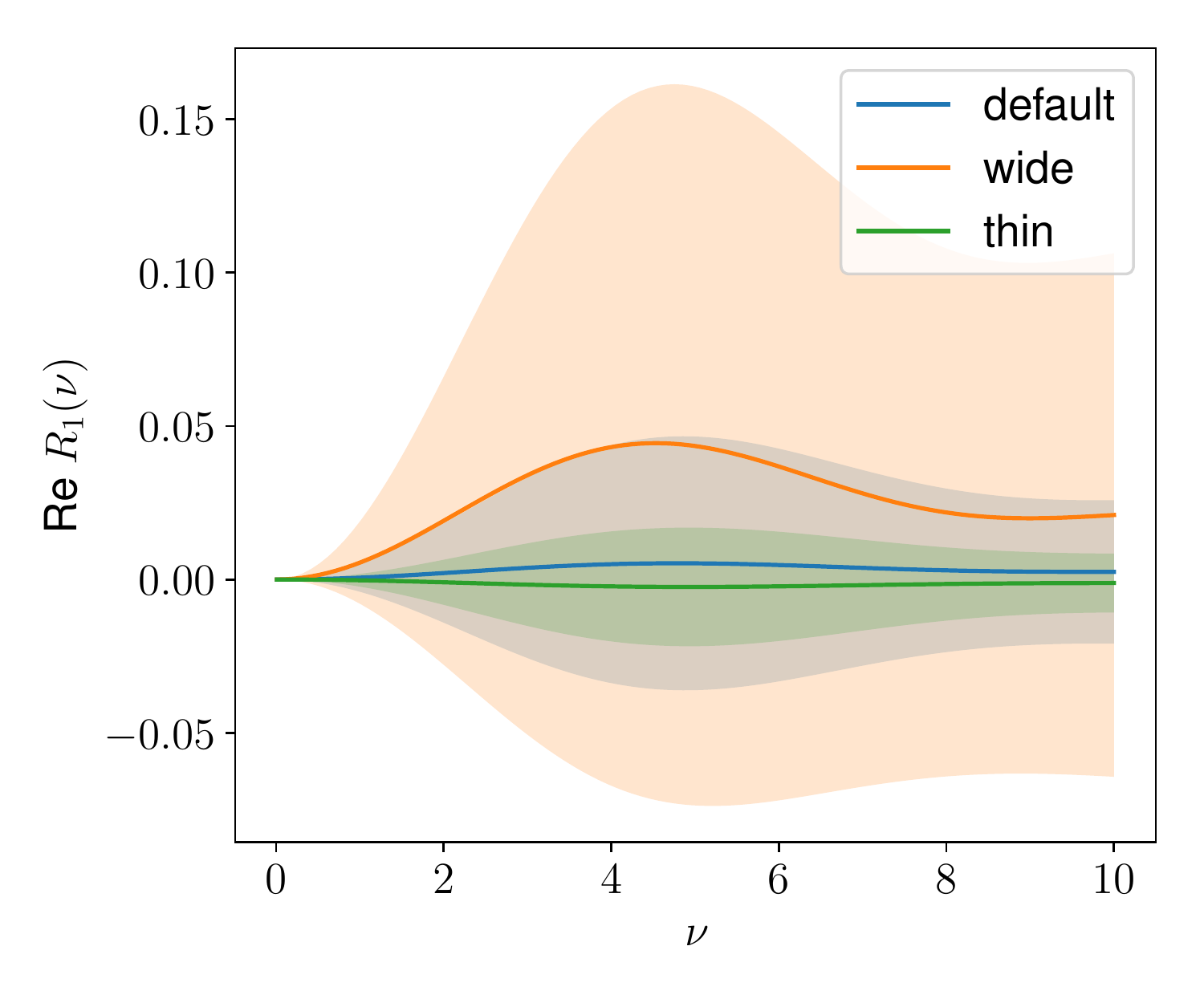}
\includegraphics[width=0.48\textwidth]{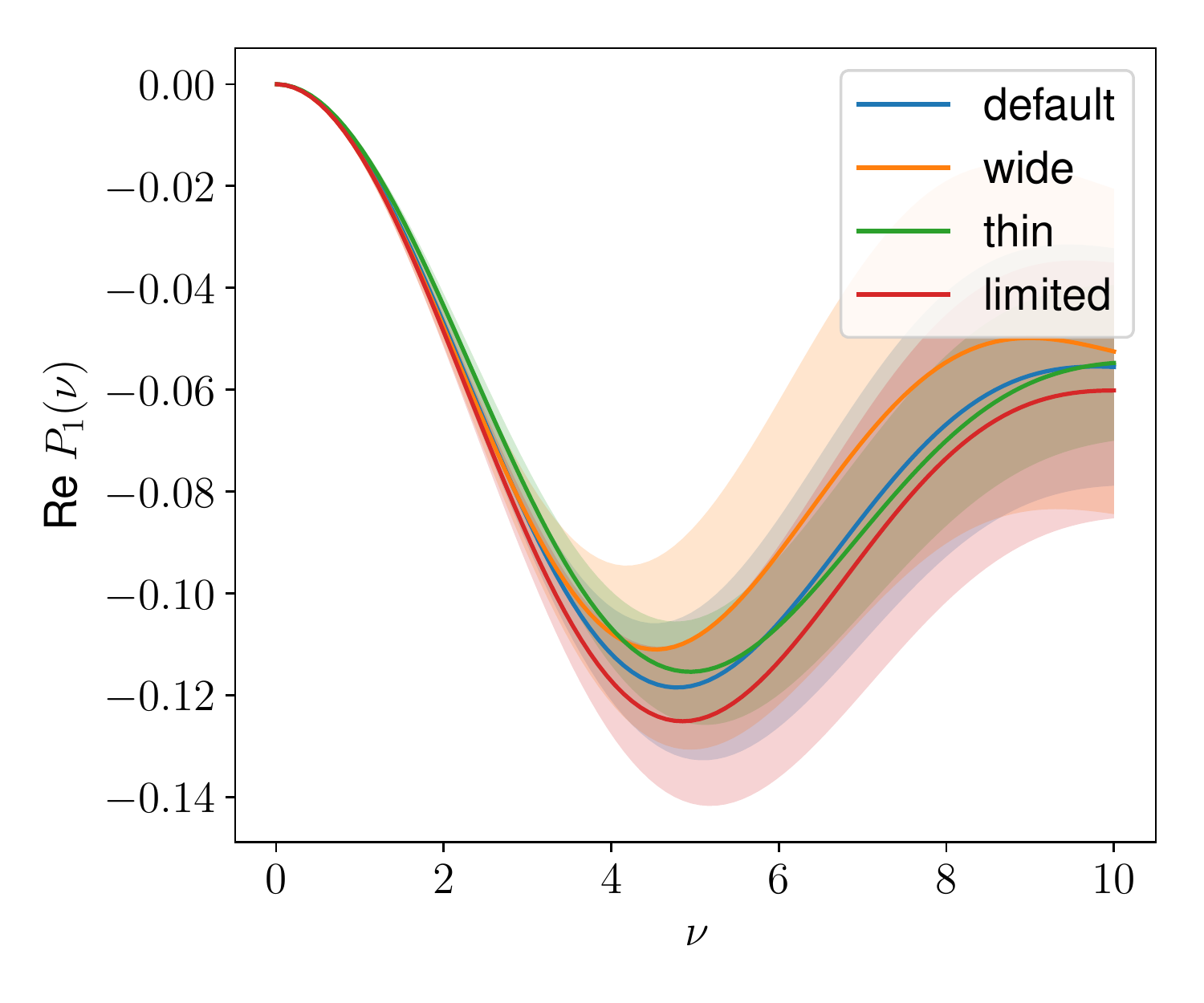}
\caption{\label{fig:single_model_comparison_real}The results of fitting to the models of Tab~\ref{tab:single_model}. The PDF is given by the upper left, $B_1(\nu)$ is given by the upper right, $R_1(\nu)$ is given by the lower left, and $P_1(\nu)$ is given by the lower right.}
\end{figure}

\begin{figure}[!htp]
\centering
\includegraphics[width=0.48\textwidth]{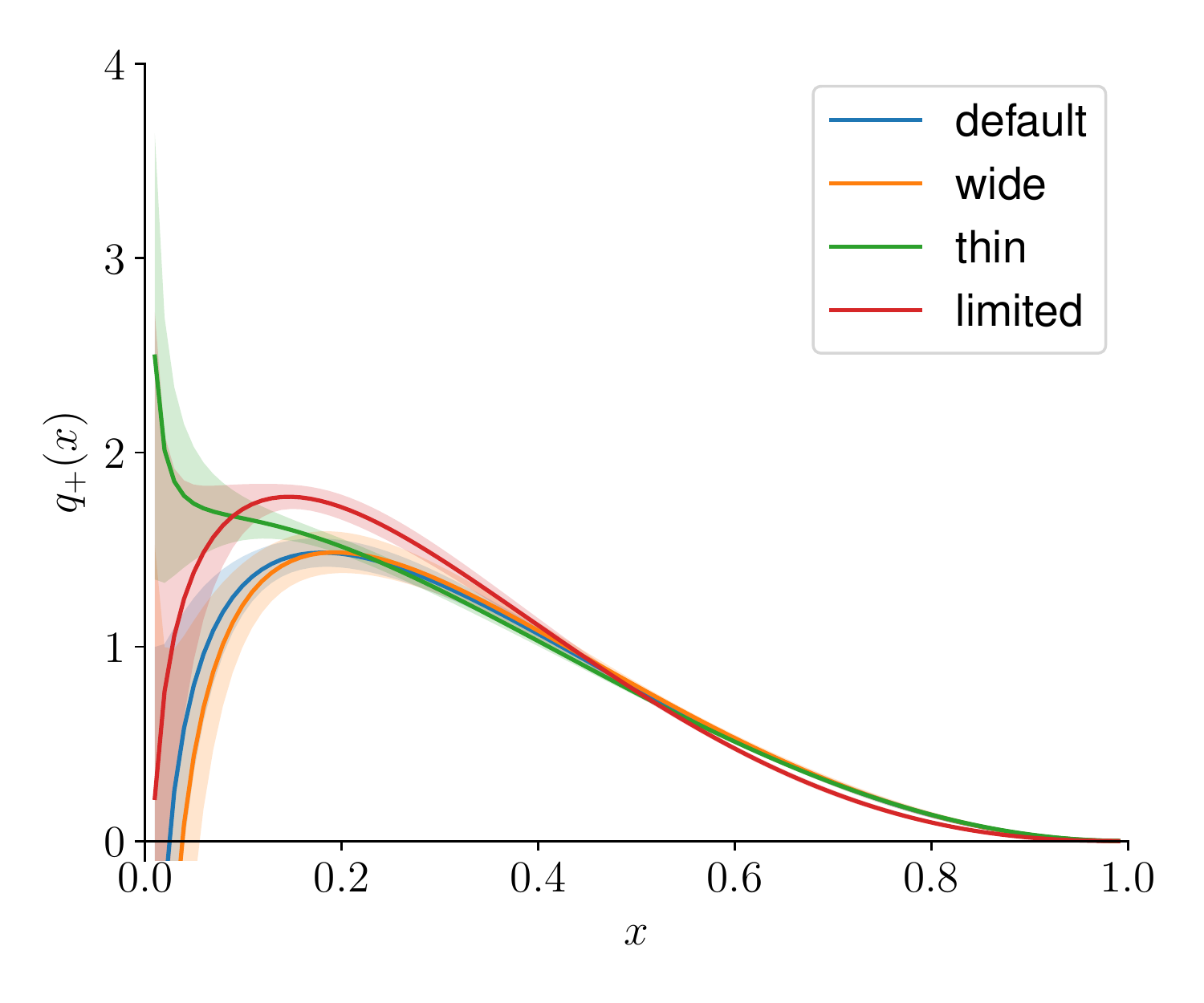}
\includegraphics[width=0.48\textwidth]{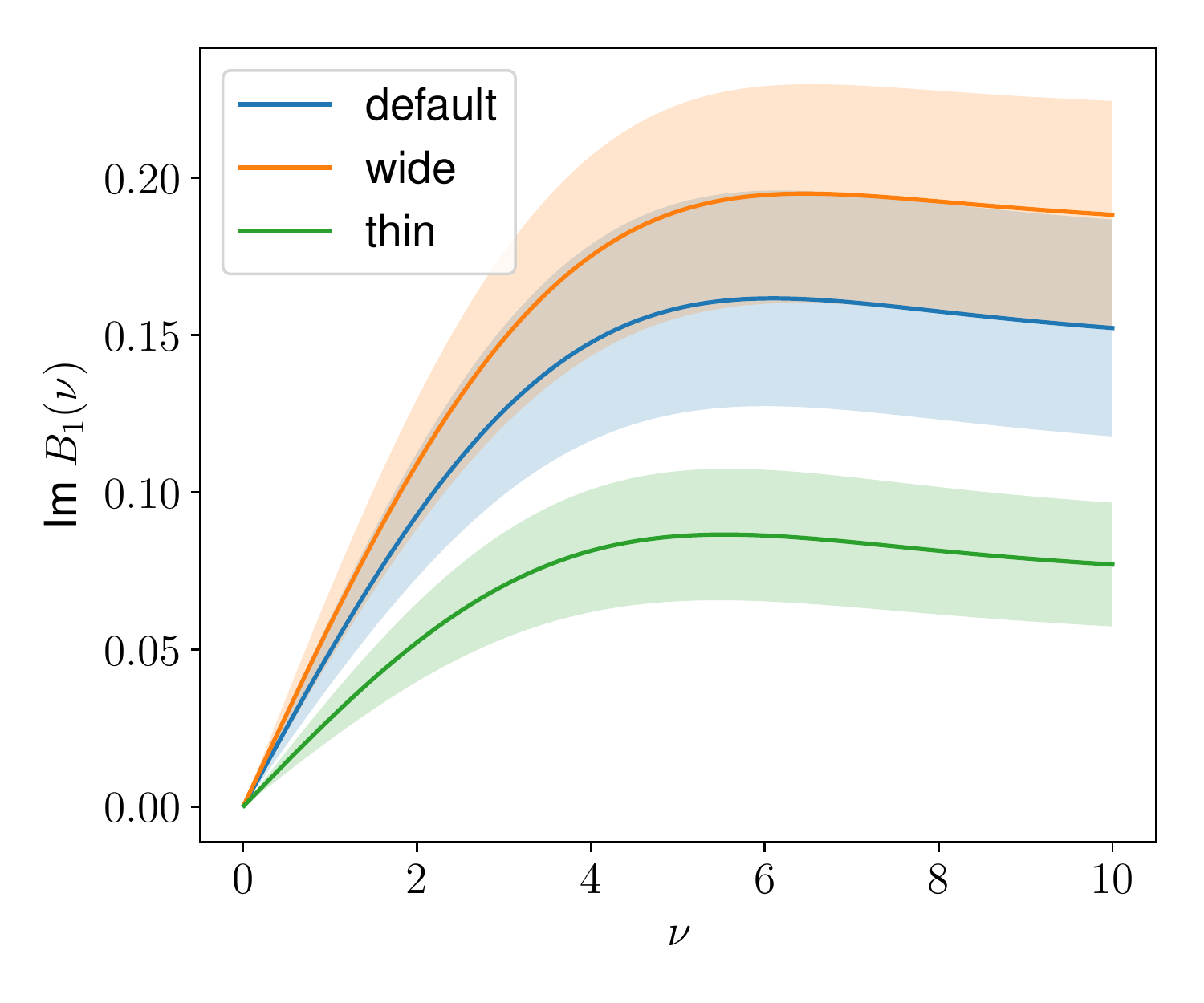}
\includegraphics[width=0.48\textwidth]{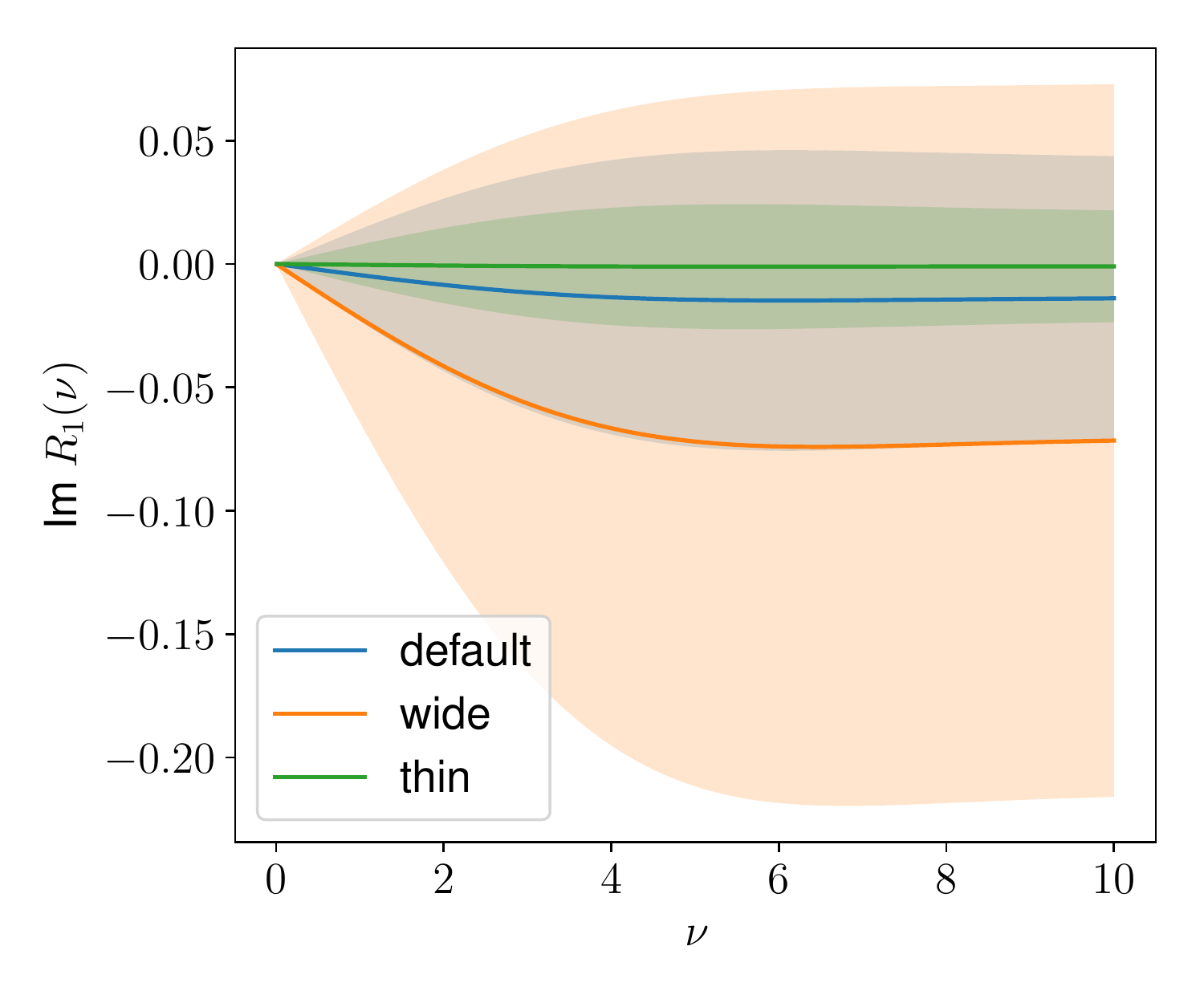}
\includegraphics[width=0.48\textwidth]{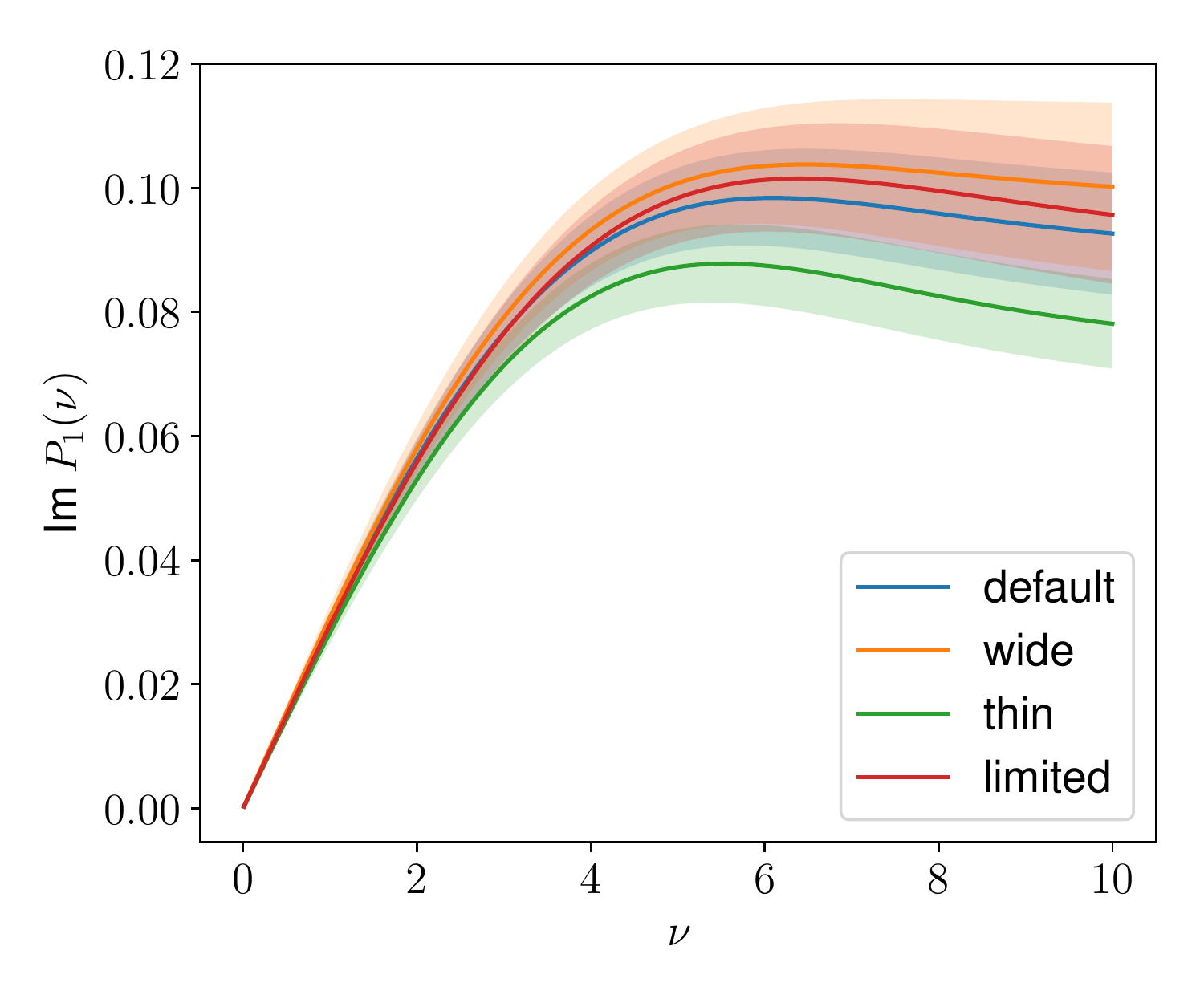}
\caption{\label{fig:single_model_comparison_imag}The results of fitting to the models of Tab~\ref{tab:single_model}. The PDF is given by the upper left, $B_1(\nu)$ is given by the upper right, $R_1(\nu)$ is given by the lower left, and $P_1(\nu)$ is given by the lower right.}
\end{figure}

In Figs.~\ref{fig:single_model_comparison_real} and~\ref{fig:single_model_comparison_imag}, the results of all the models described in Tab.~\ref{tab:single_model} are compared. The $L^2$ and $\chi^2$ of these fits are given in Tab.~\ref{tab:single_model_chi2}. The models all give largely consistent results to each other. As expected ``thin'' and ``wide'' gave results with smaller and wider statistical errors than ``default'' respectively. For the higher twist terms, the ``wide'' and ``thin'' results actually seem to deviate slightly from the ``default'' model, compared to the other terms, but with little effect on the resulting PDF.

\subsection{Varying the number of parameters}\label{sec:vary_numbers}

In order to study the model dependence, the number of parameters for the PDF and for each of the nuisance terms are all be varied. The prior distributions are set to those of the model ``default'' in Sec.~\ref{sec:study_priors}. Changing the number of parameters varies the flexibility of the model. Based upon the size of the $\sigma_n(\nu)$ and $\eta_n(\nu)$ functions, the terms with the lowest $n$ will dominate the result. It appears that terms with $n>5$ will have an entirely negligible effect in the given range of Ioffe time. For this study, the maximum $n$ is 3 for any given term. The number of polynomials in each of the nuisance terms is allowed to vary from 0 to the number of polynomials in the ITD term. The models are labeled with 4 numbers corresponding to $(N_\pm,  N_{R/I,b}, N_{R/I,r}, N_{R/I,p})$.  

\begin{figure}[!htp]
\centering
\includegraphics[width=0.48\textwidth]{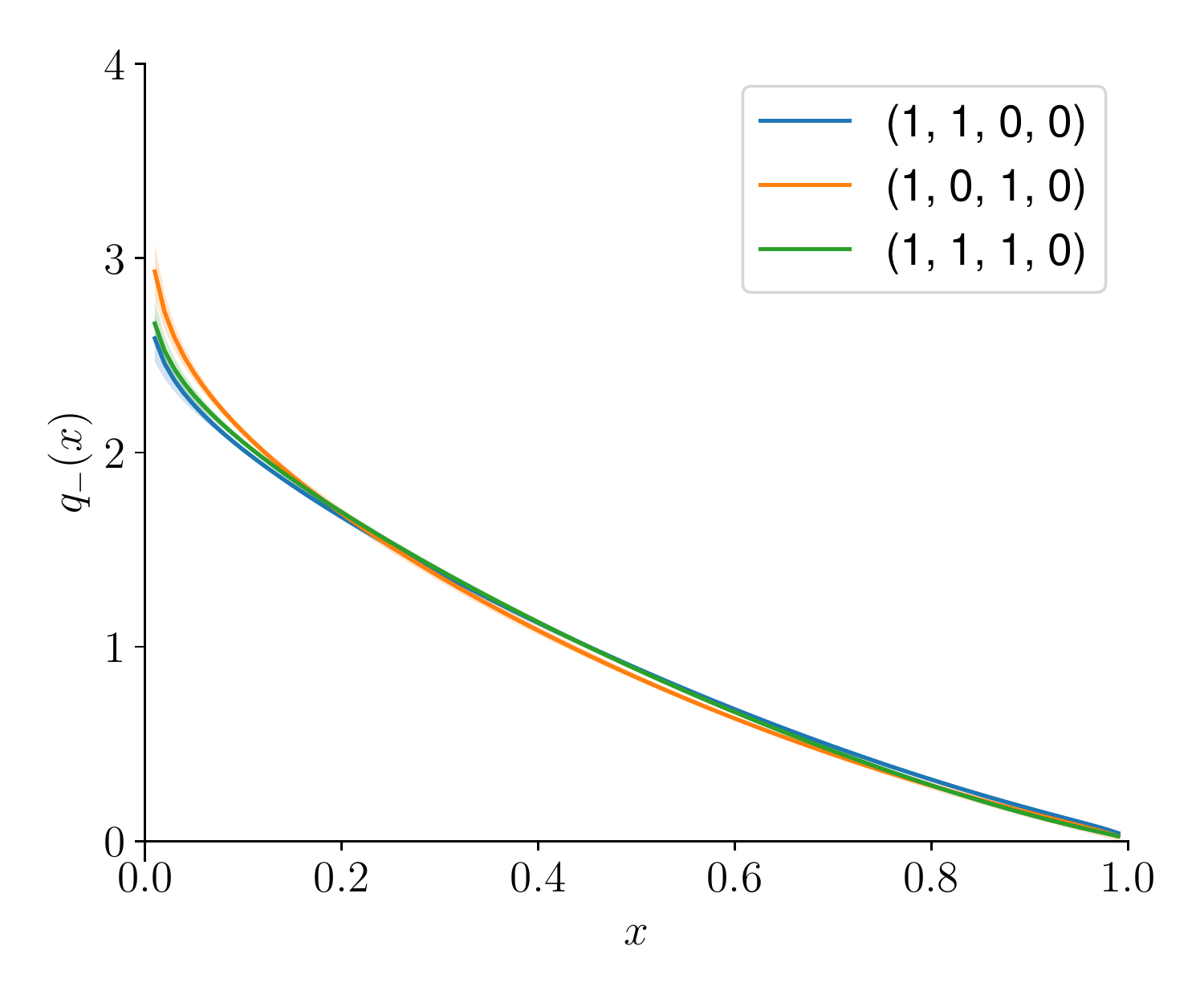}
\includegraphics[width=0.48\textwidth]{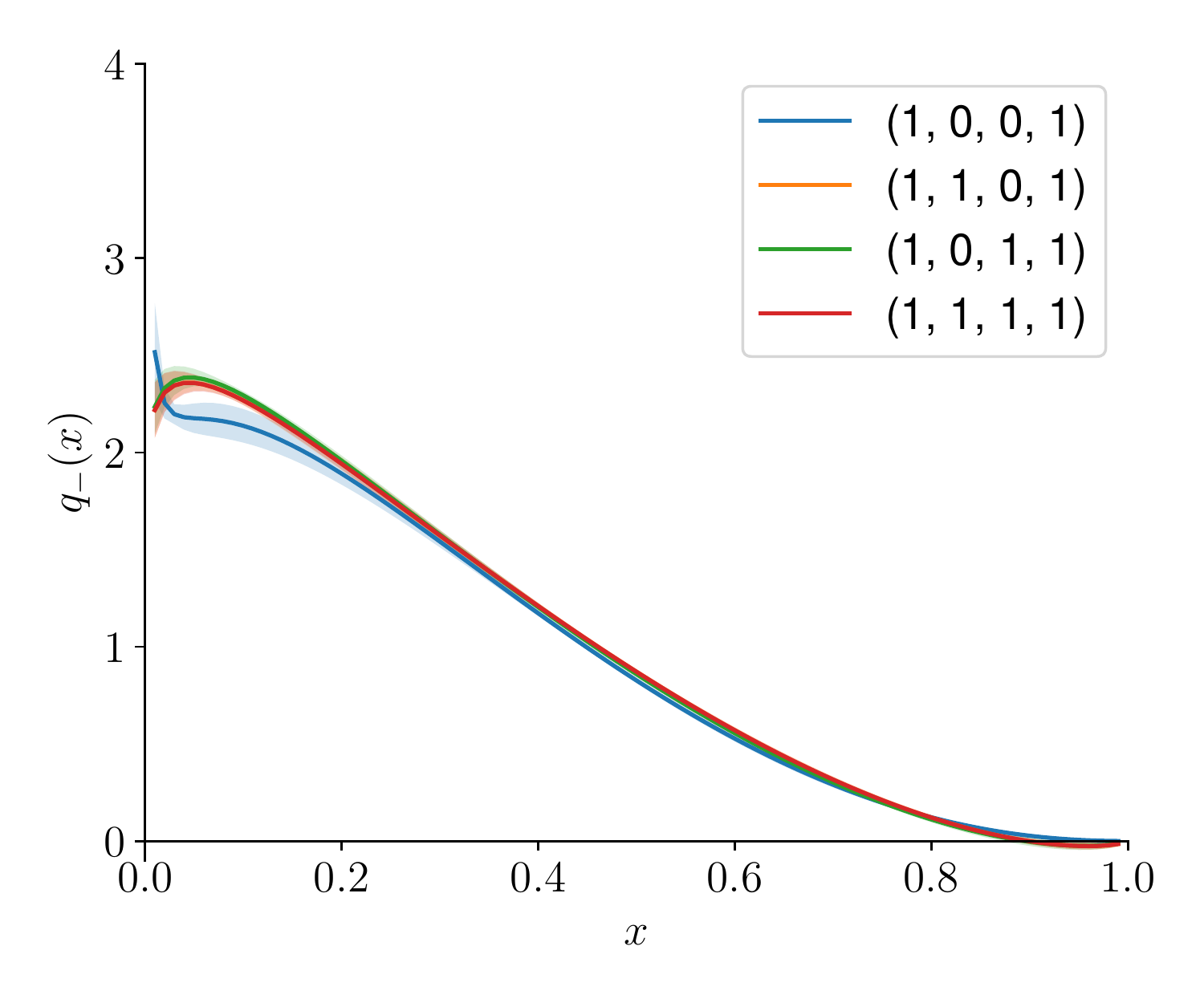}
\includegraphics[width=0.48\textwidth]{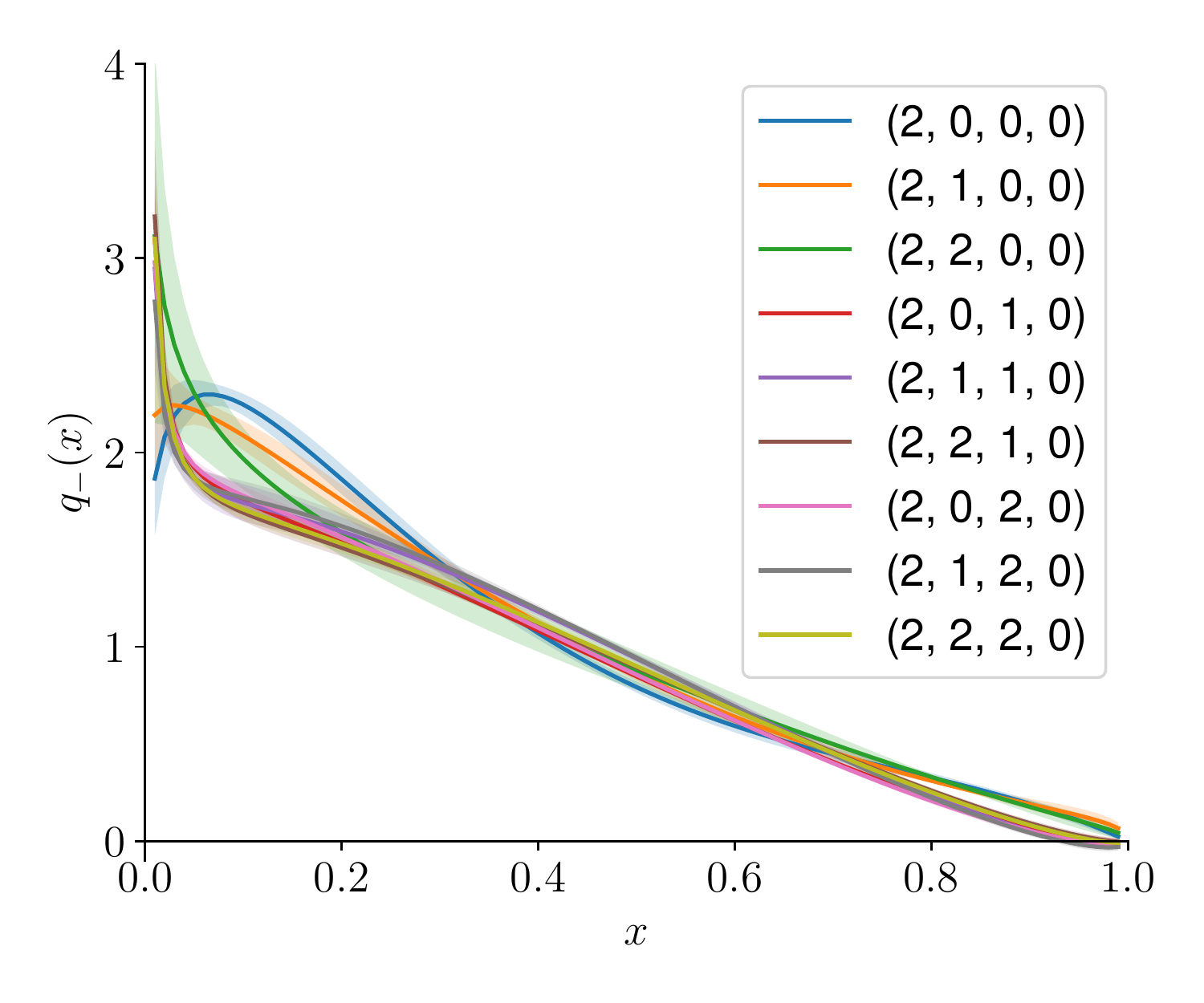}
\includegraphics[width=0.48\textwidth]{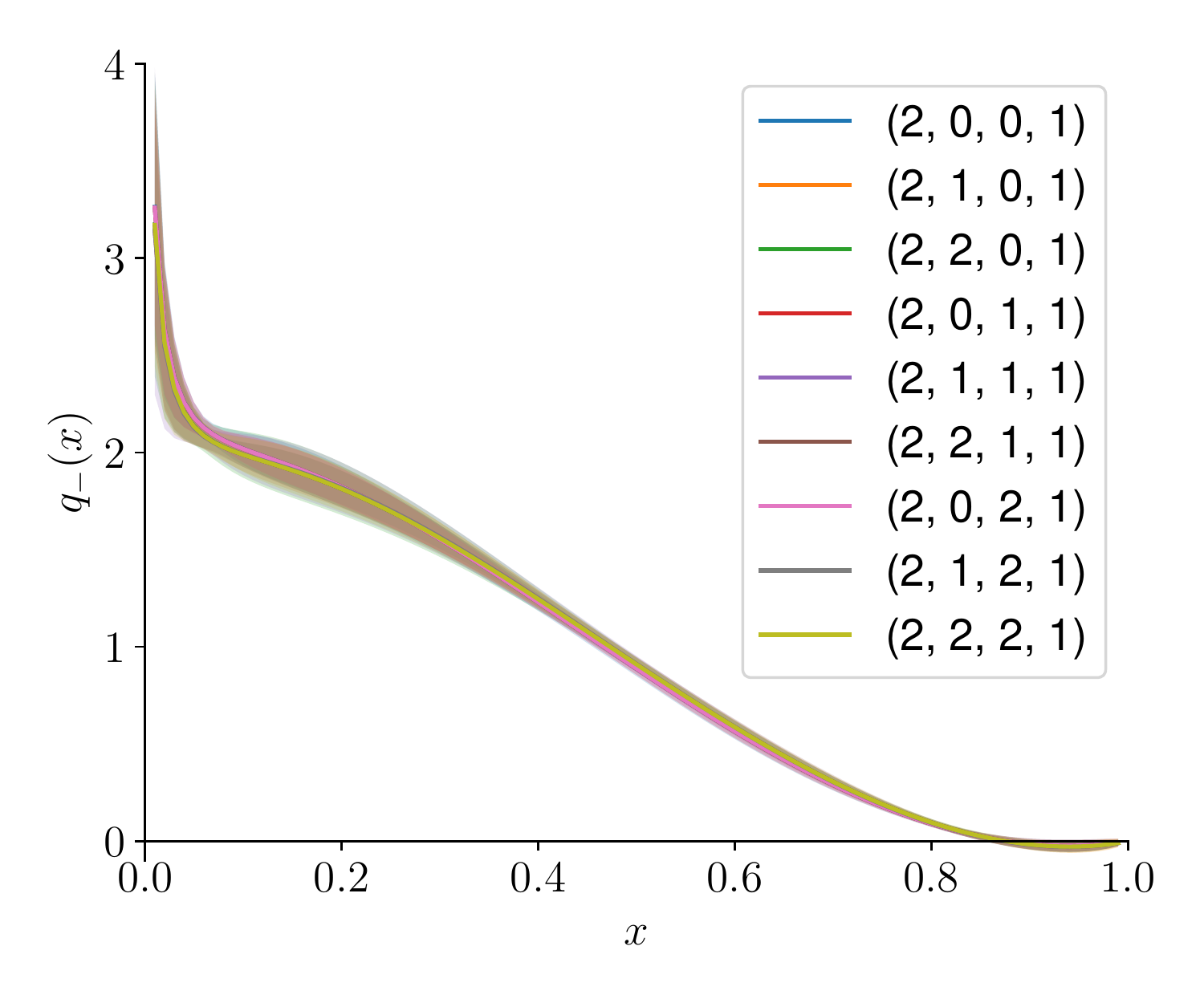}
\includegraphics[width=0.48\textwidth]{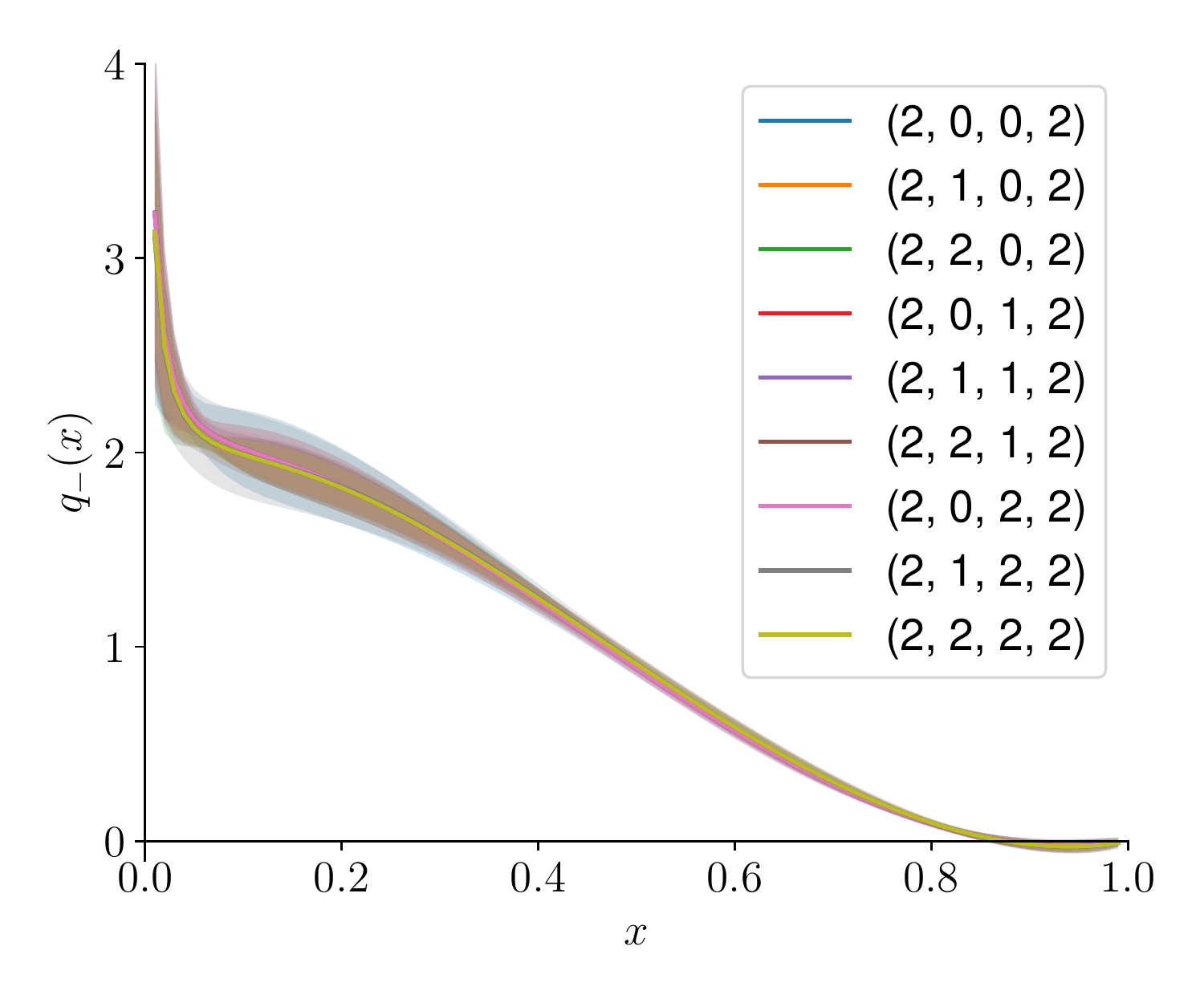}
\caption{\label{fig:many_model_pdfs_real}The PDF results from fitting the real component to the models. The numbers in the legend correspond to $(N_\pm,  N_{R/I,b}, N_{R/I,r}, N_{R/I,p})$.}
\end{figure}

\begin{figure}[!htp]
\centering
\includegraphics[width=0.48\textwidth]{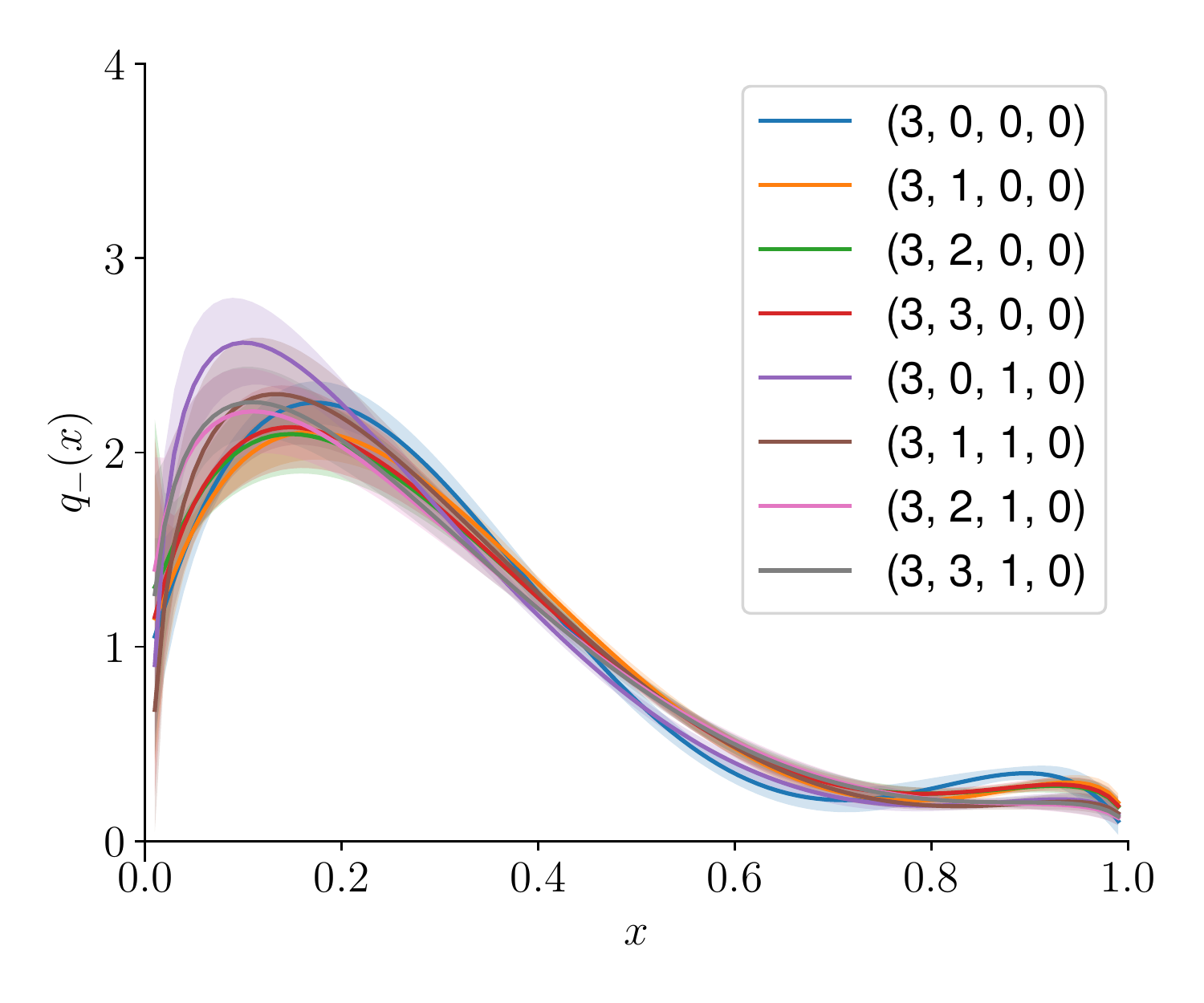}
\includegraphics[width=0.48\textwidth]{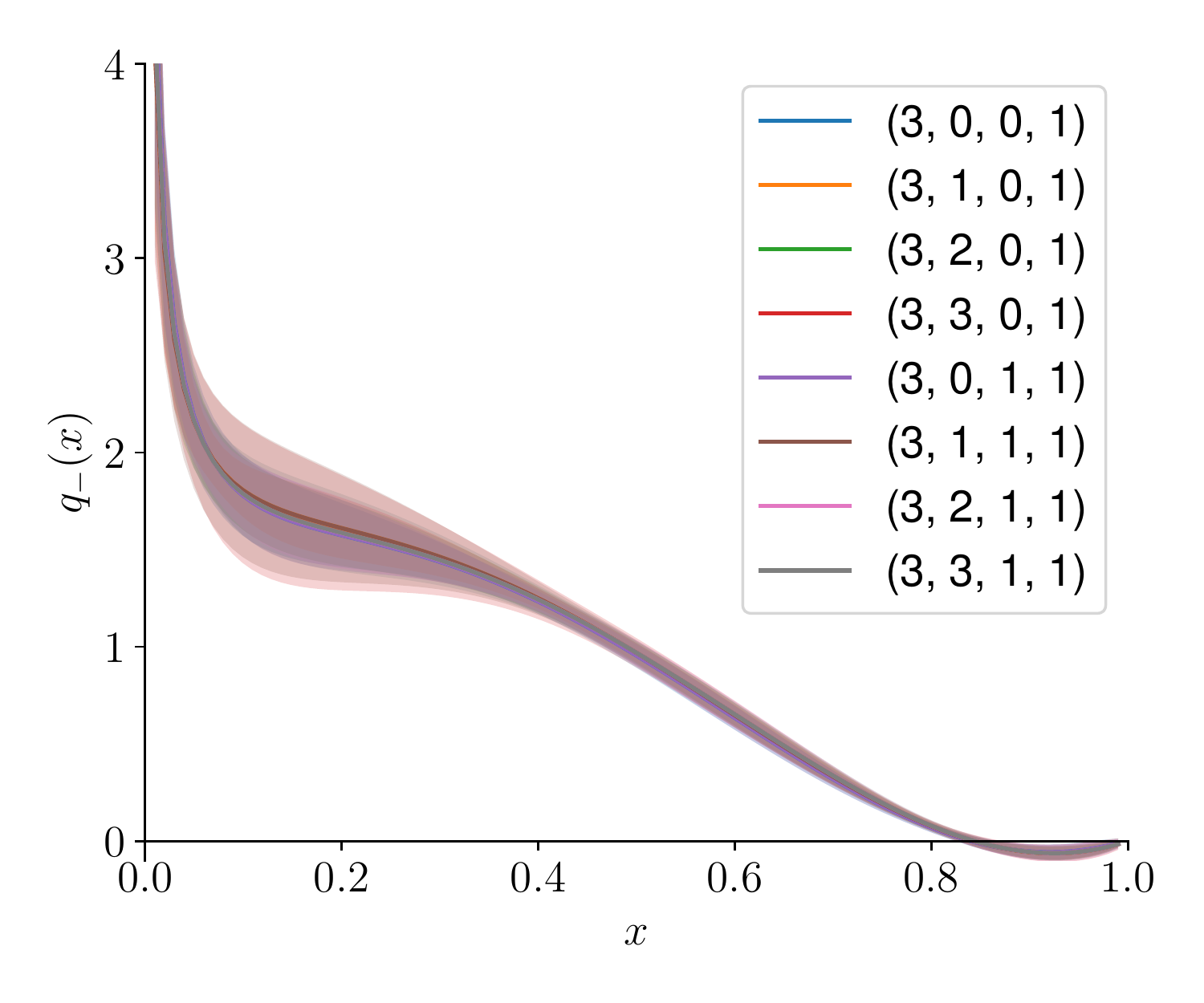}
\includegraphics[width=0.48\textwidth]{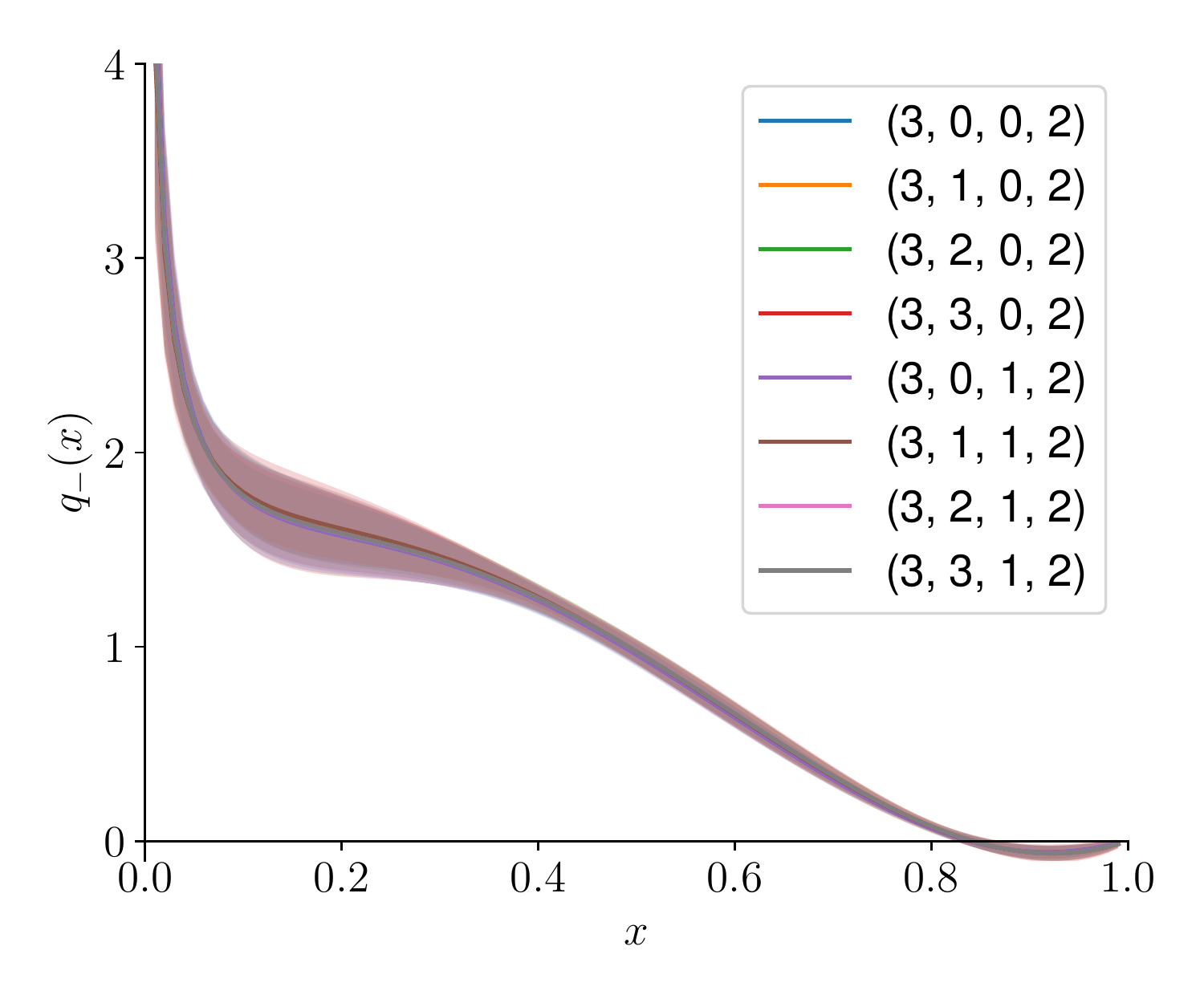}
\includegraphics[width=0.48\textwidth]{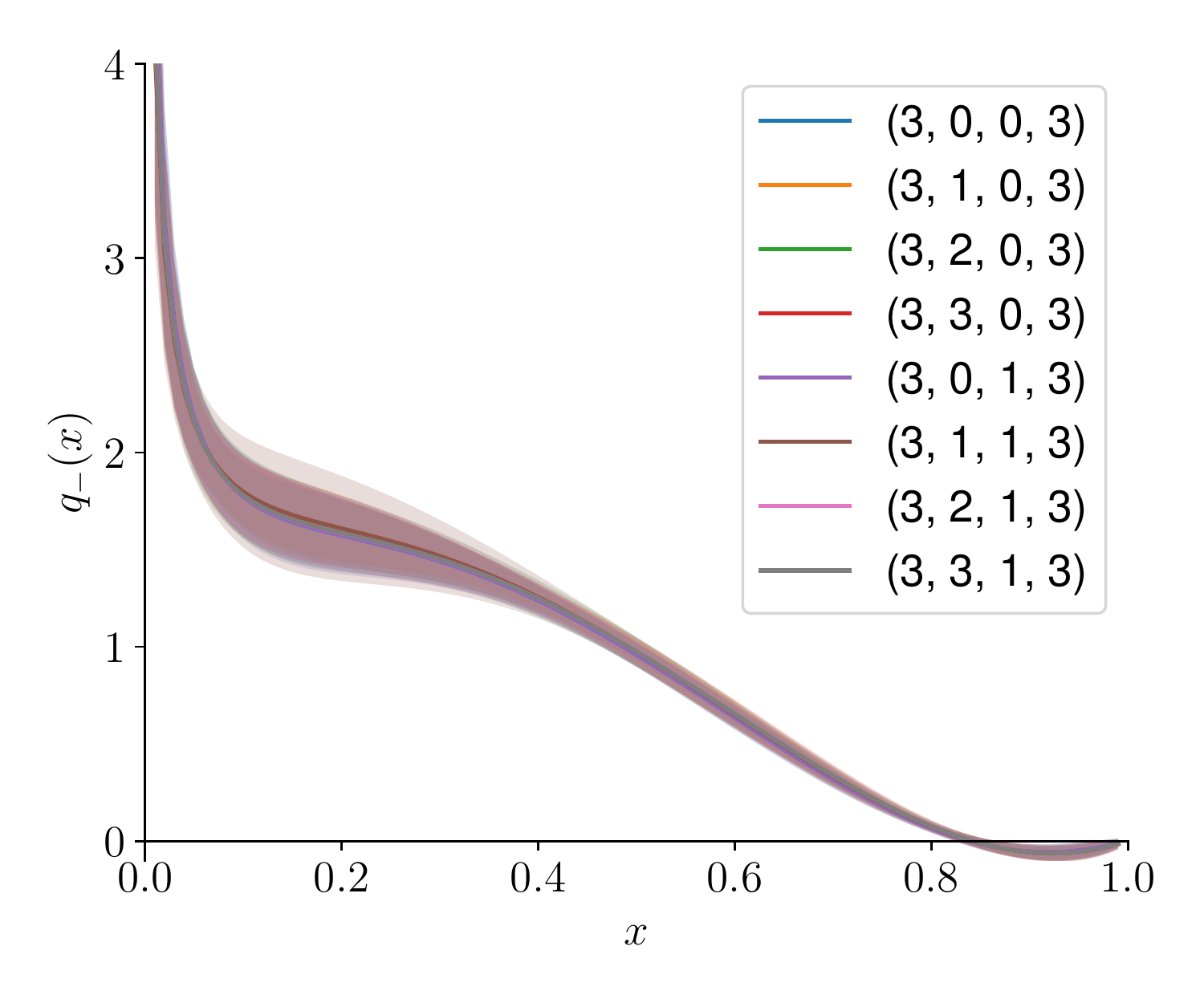}
\caption{\label{fig:many_model_pdfs_real2}The PDF results from fitting the real component to the models. The numbers in the legend correspond to $(N_\pm,  N_{R/I,b}, N_{R/I,r}, N_{R/I,p})$.}
\end{figure}

\begin{figure}[!htp]
\centering
\includegraphics[width=0.48\textwidth]{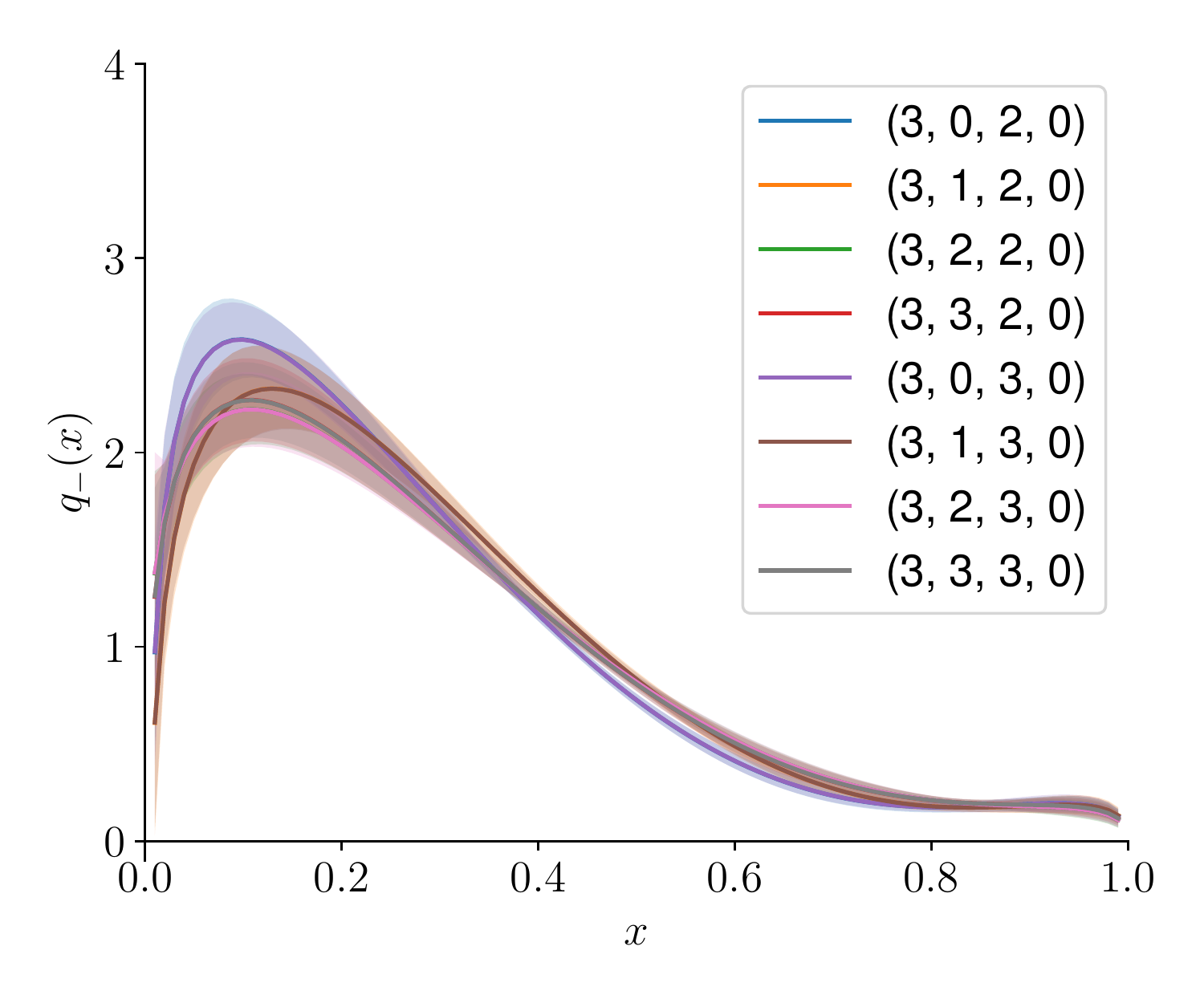}
\includegraphics[width=0.48\textwidth]{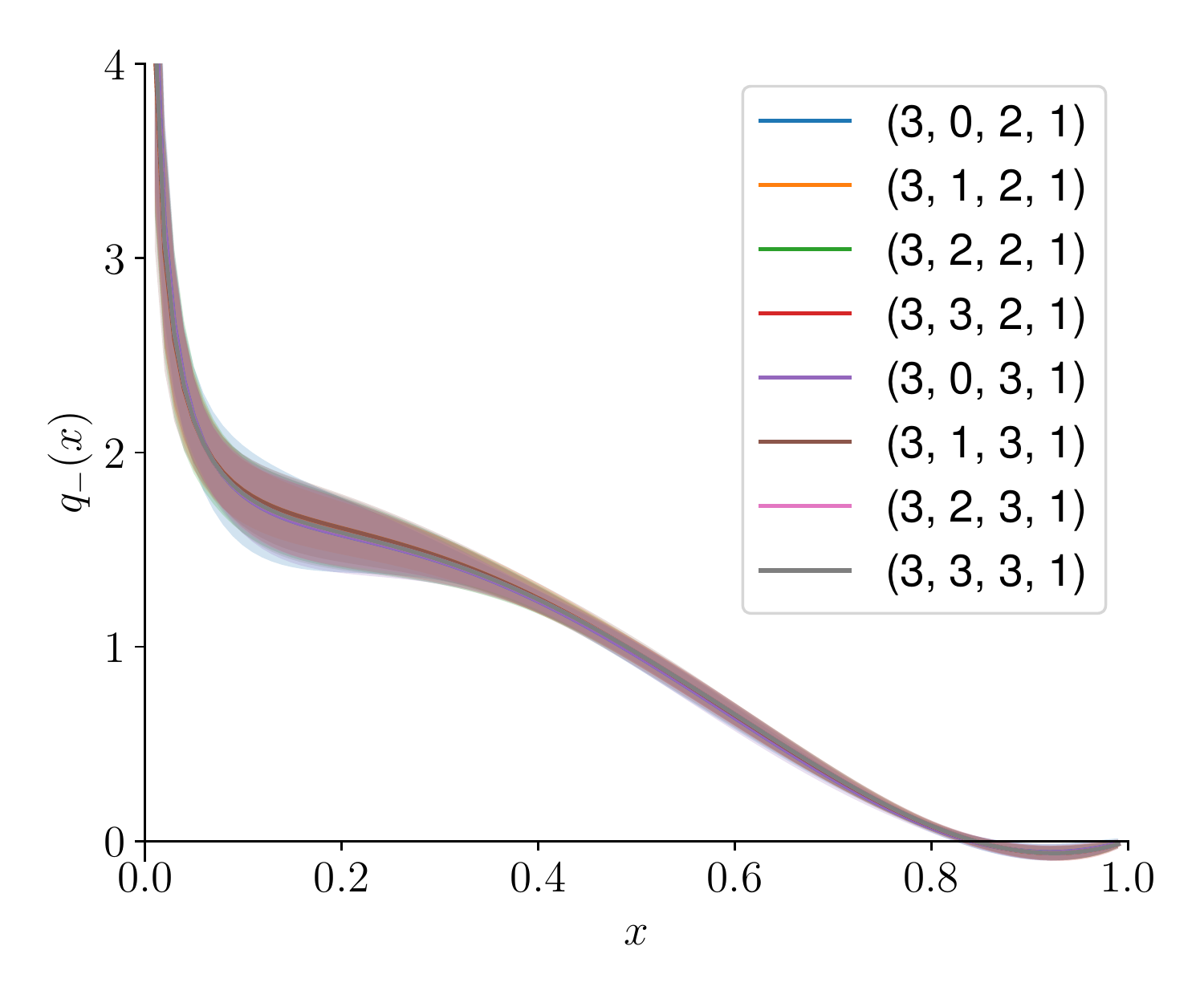}
\includegraphics[width=0.48\textwidth]{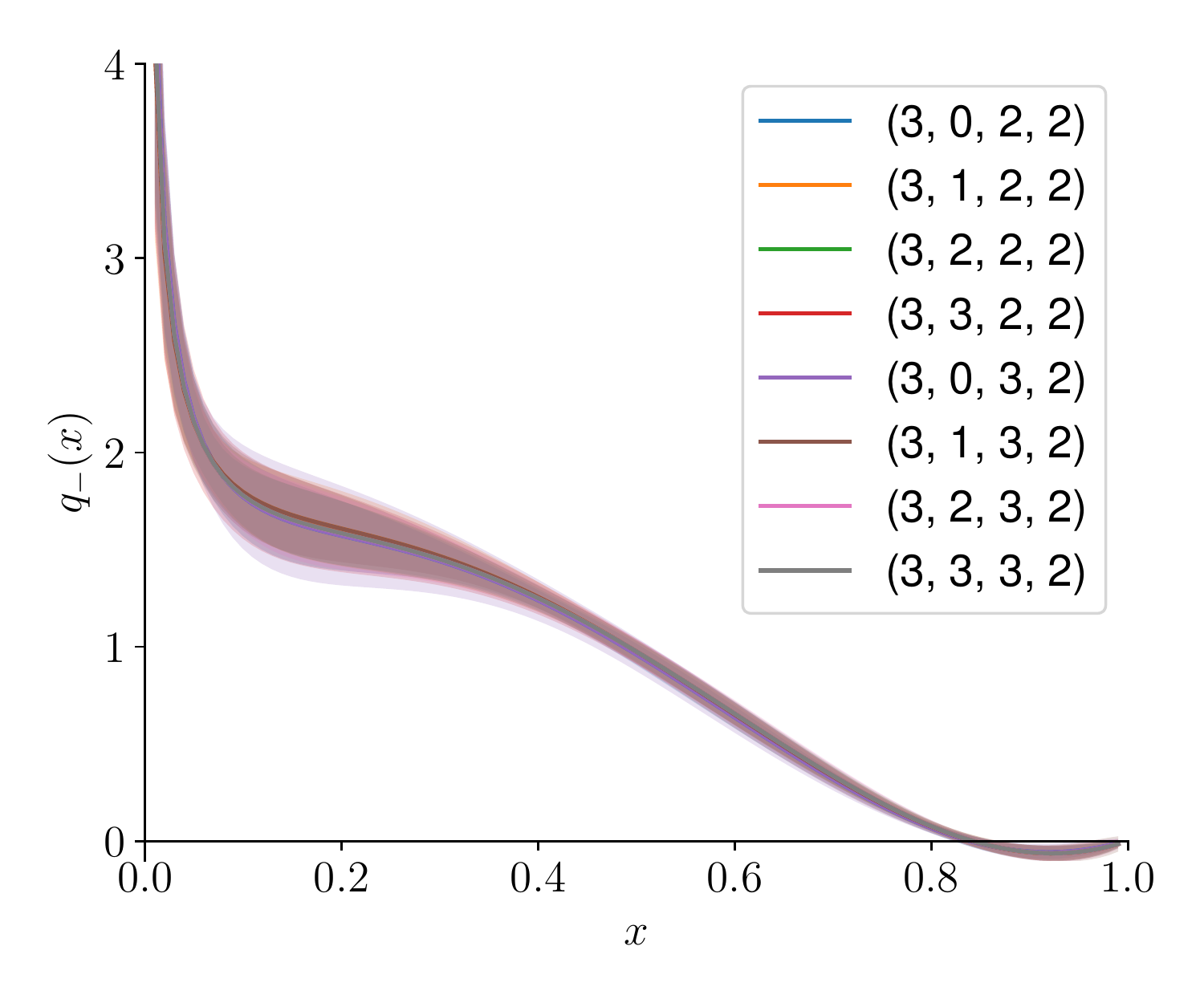}
\includegraphics[width=0.48\textwidth]{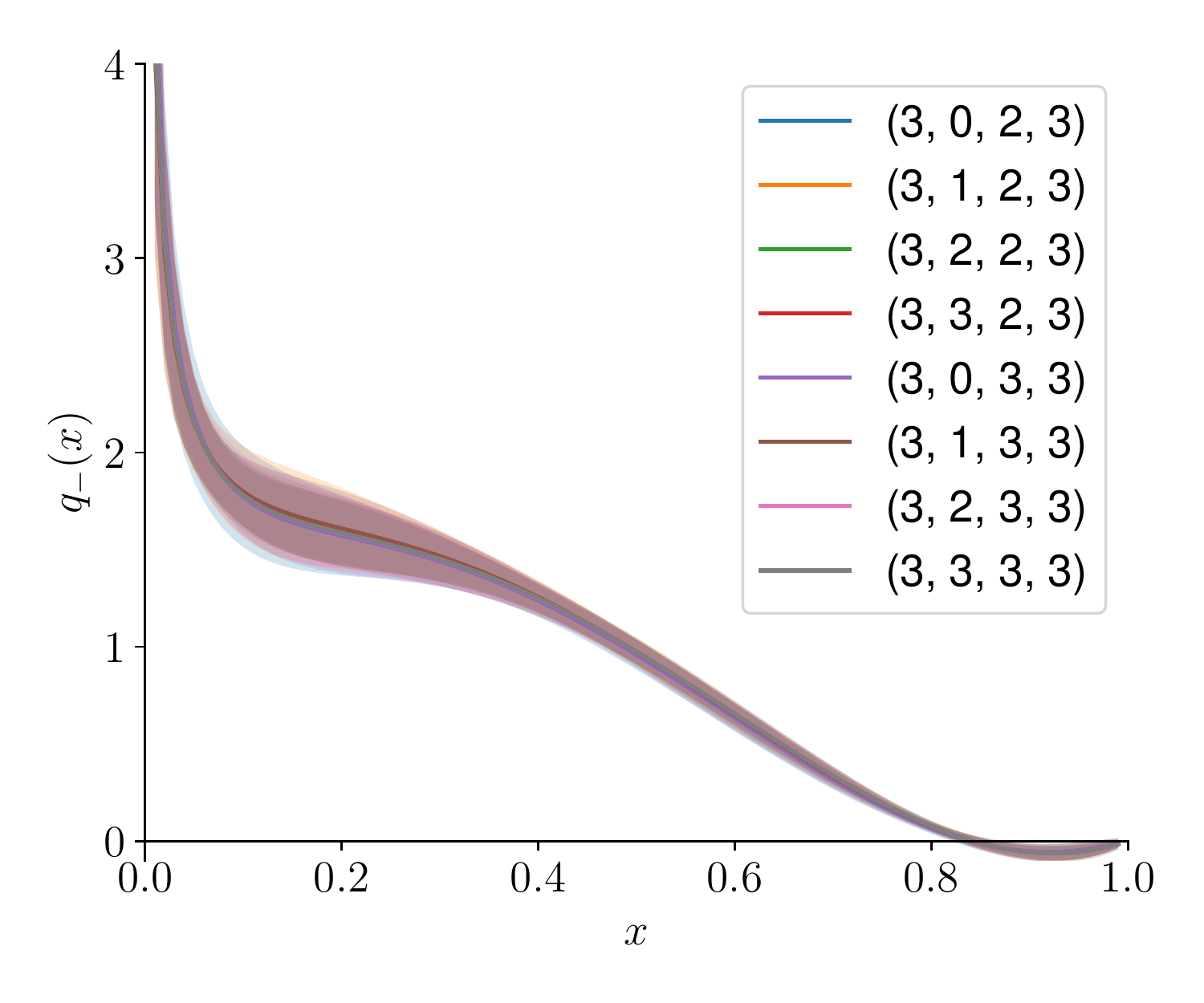}
\caption{\label{fig:many_model_pdfs_real3}The PDF results from fitting the real component to the models. The numbers in the legend correspond to $(N_\pm,  N_{R/I,b}, N_{R/I,r}, N_{R/I,p})$.}
\end{figure}

\begin{figure}[!htp]
\centering
\includegraphics[width=0.48\textwidth]{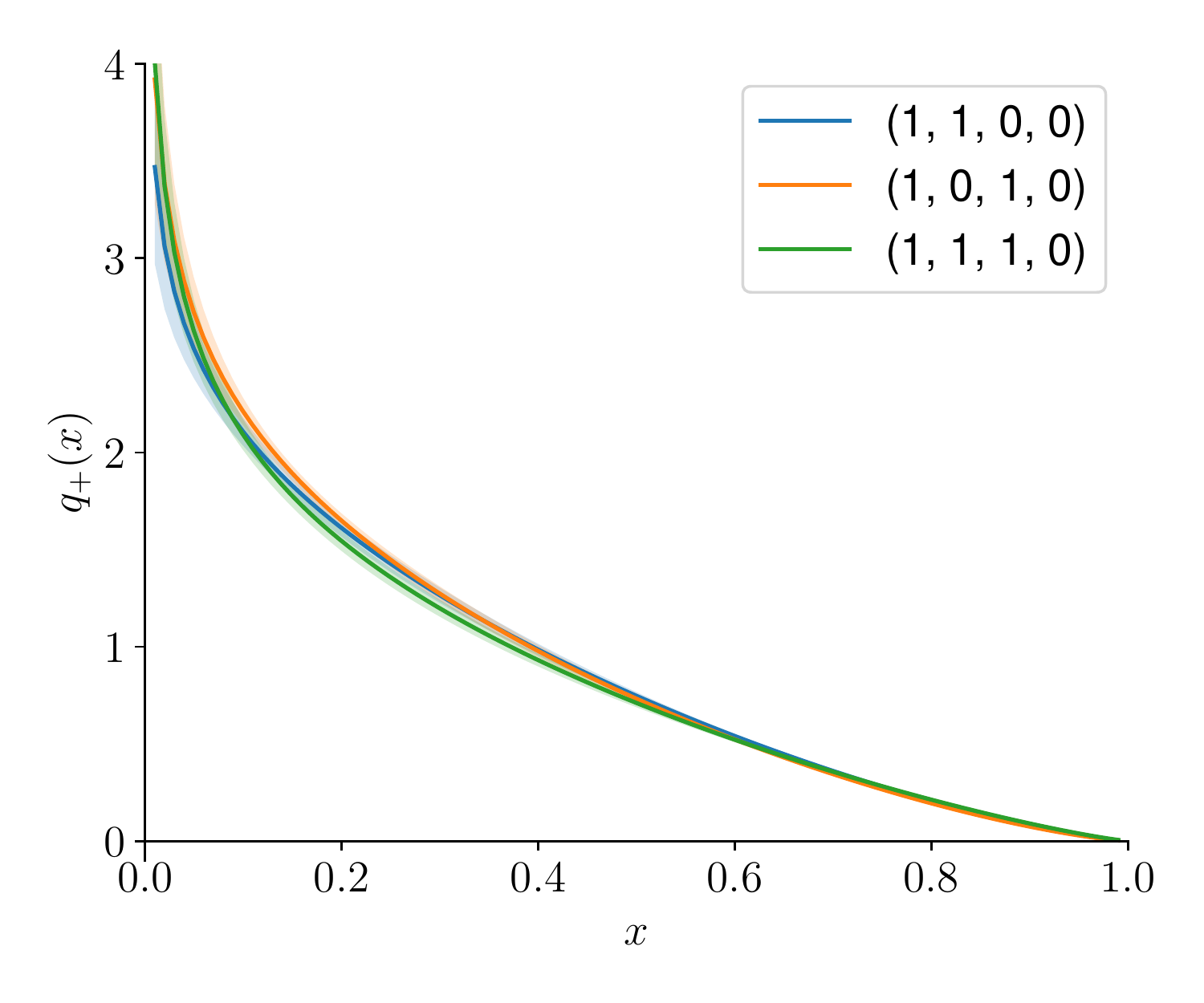}
\includegraphics[width=0.48\textwidth]{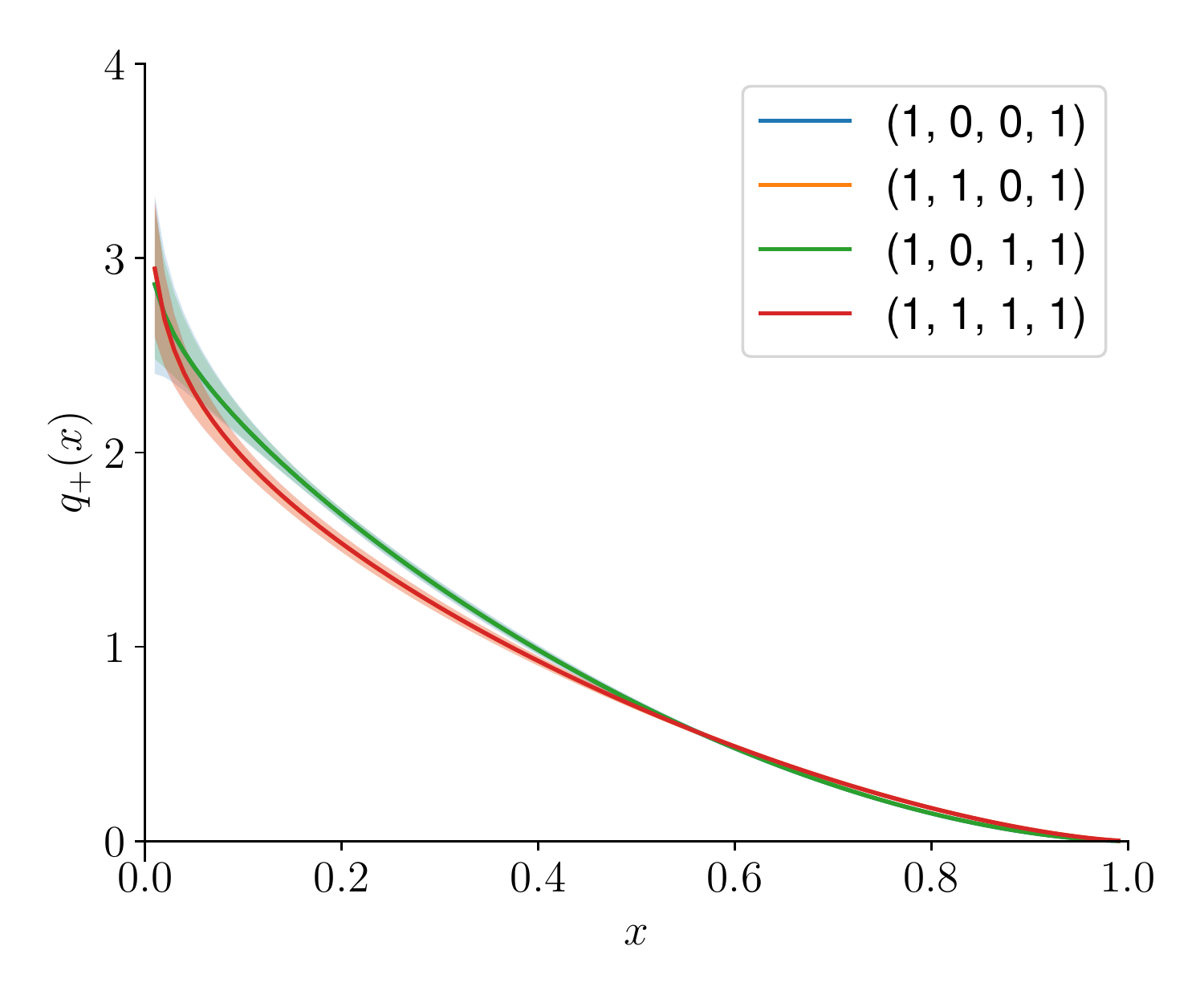}
\includegraphics[width=0.48\textwidth]{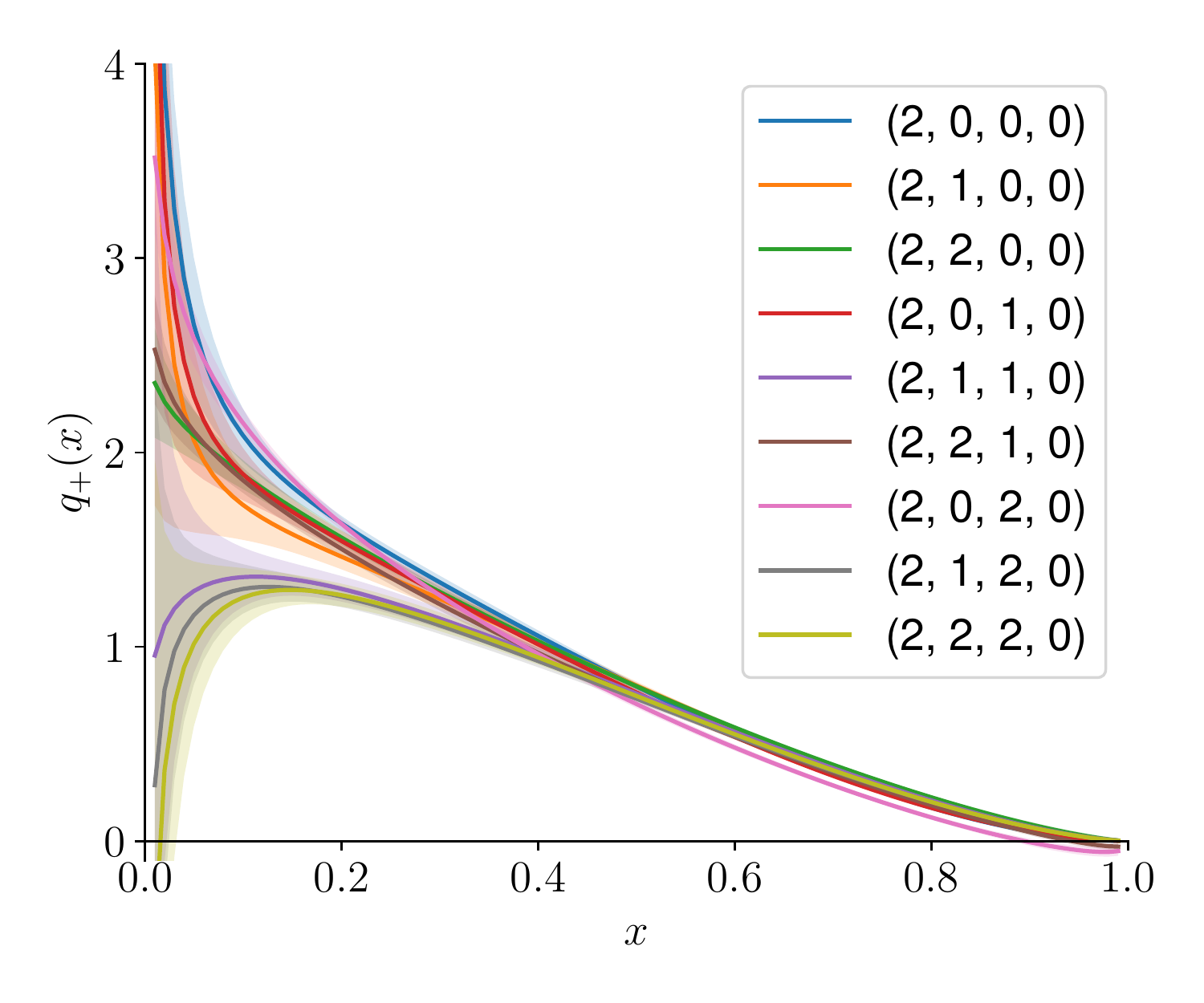}
\includegraphics[width=0.48\textwidth]{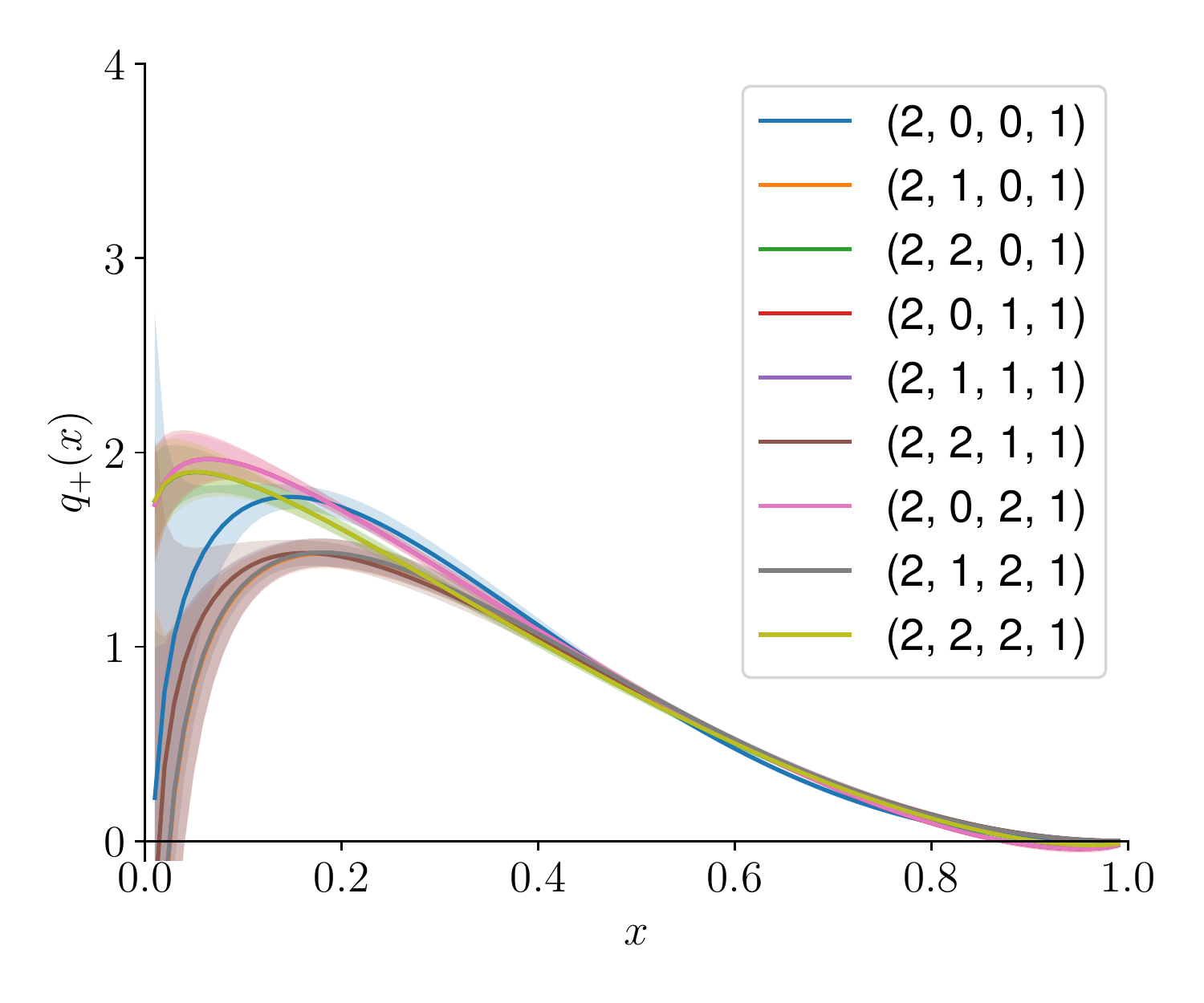}
\includegraphics[width=0.48\textwidth]{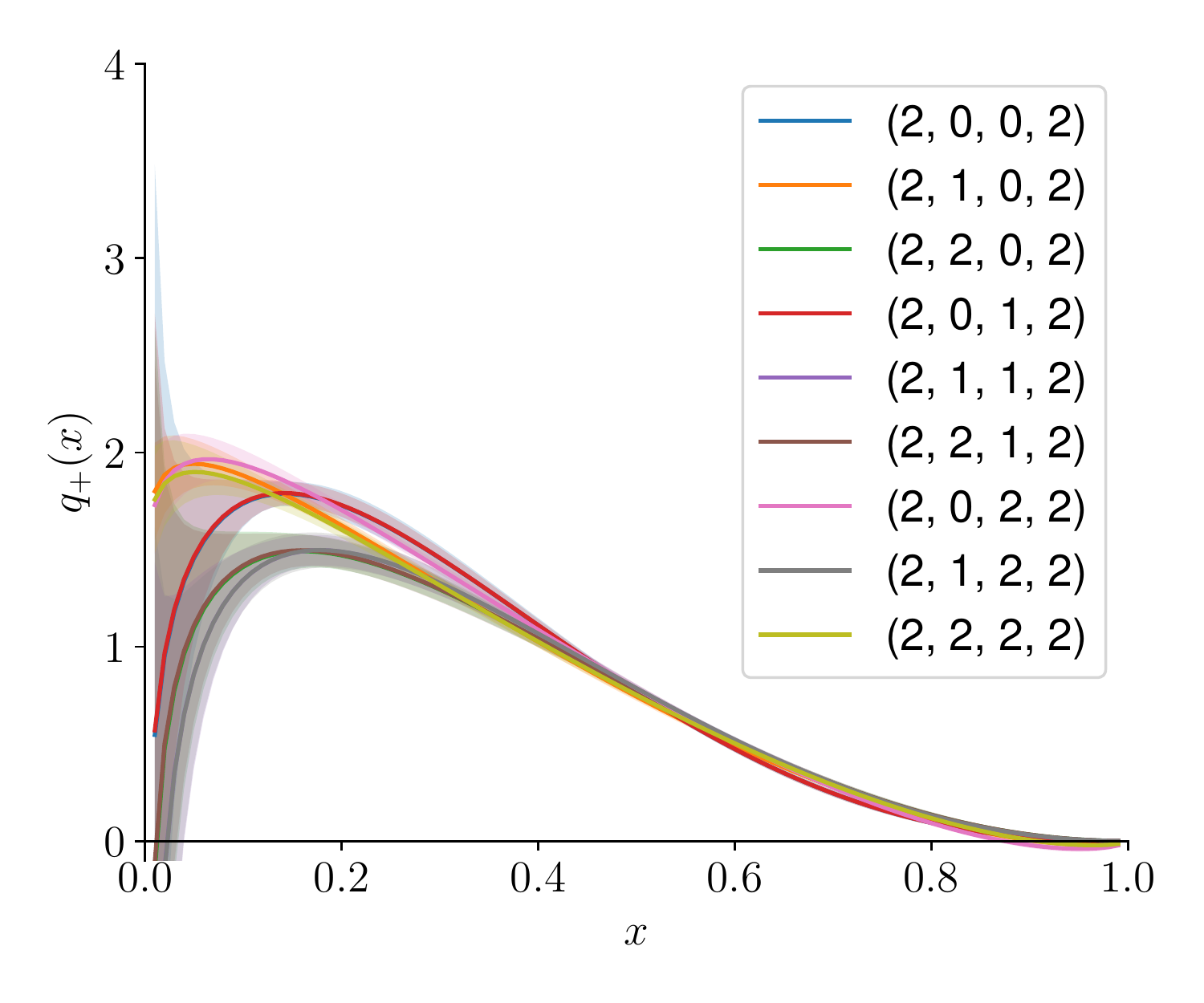}
\caption{\label{fig:many_model_pdfs_imag}The PDF results from fitting the imaginary component to the models. The numbers in the legend correspond to $(N_\pm,  N_{R/I,b}, N_{R/I,r}, N_{R/I,p})$.}
\end{figure}

\begin{figure}[!htp]
\centering
\includegraphics[width=0.48\textwidth]{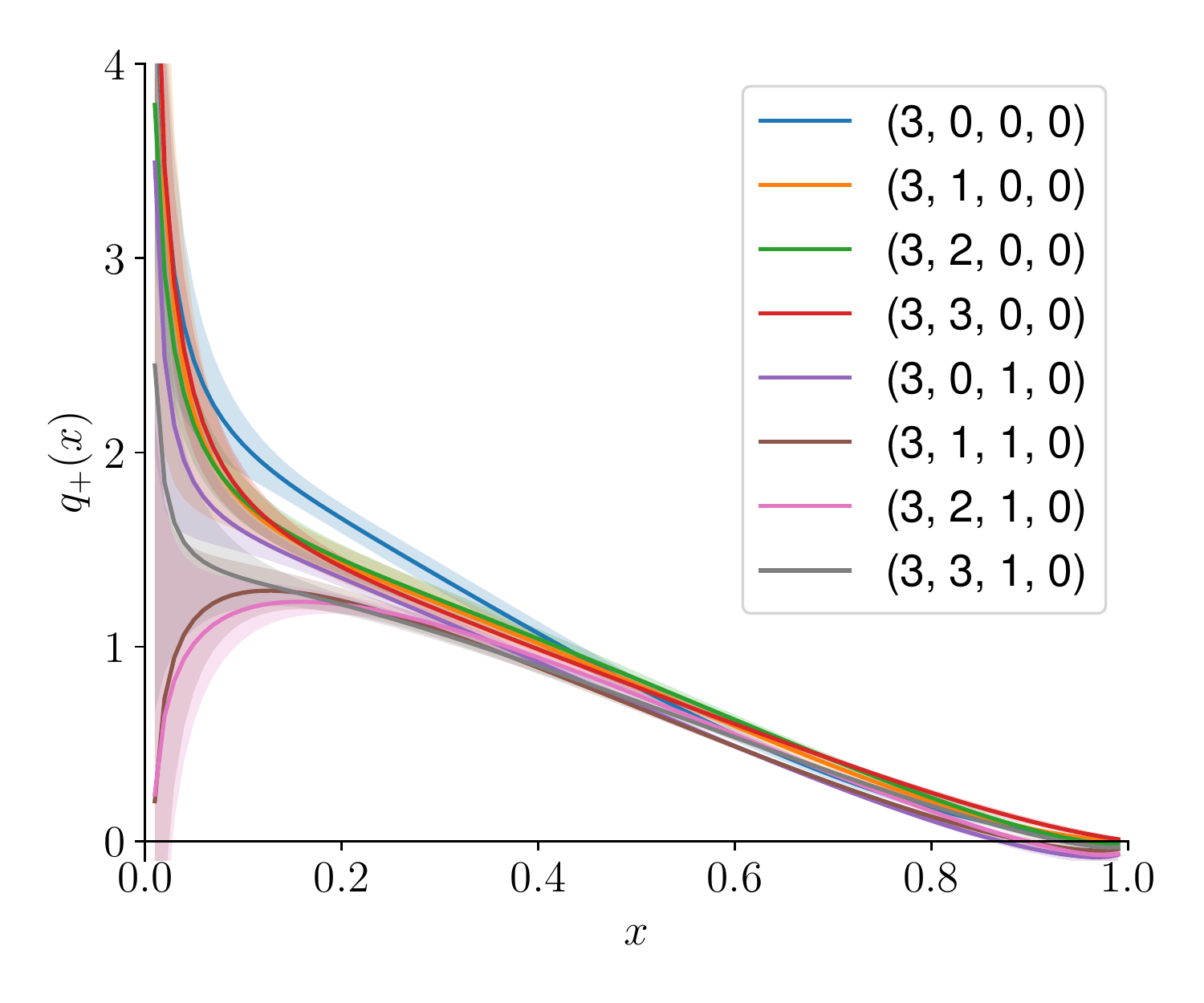}
\includegraphics[width=0.48\textwidth]{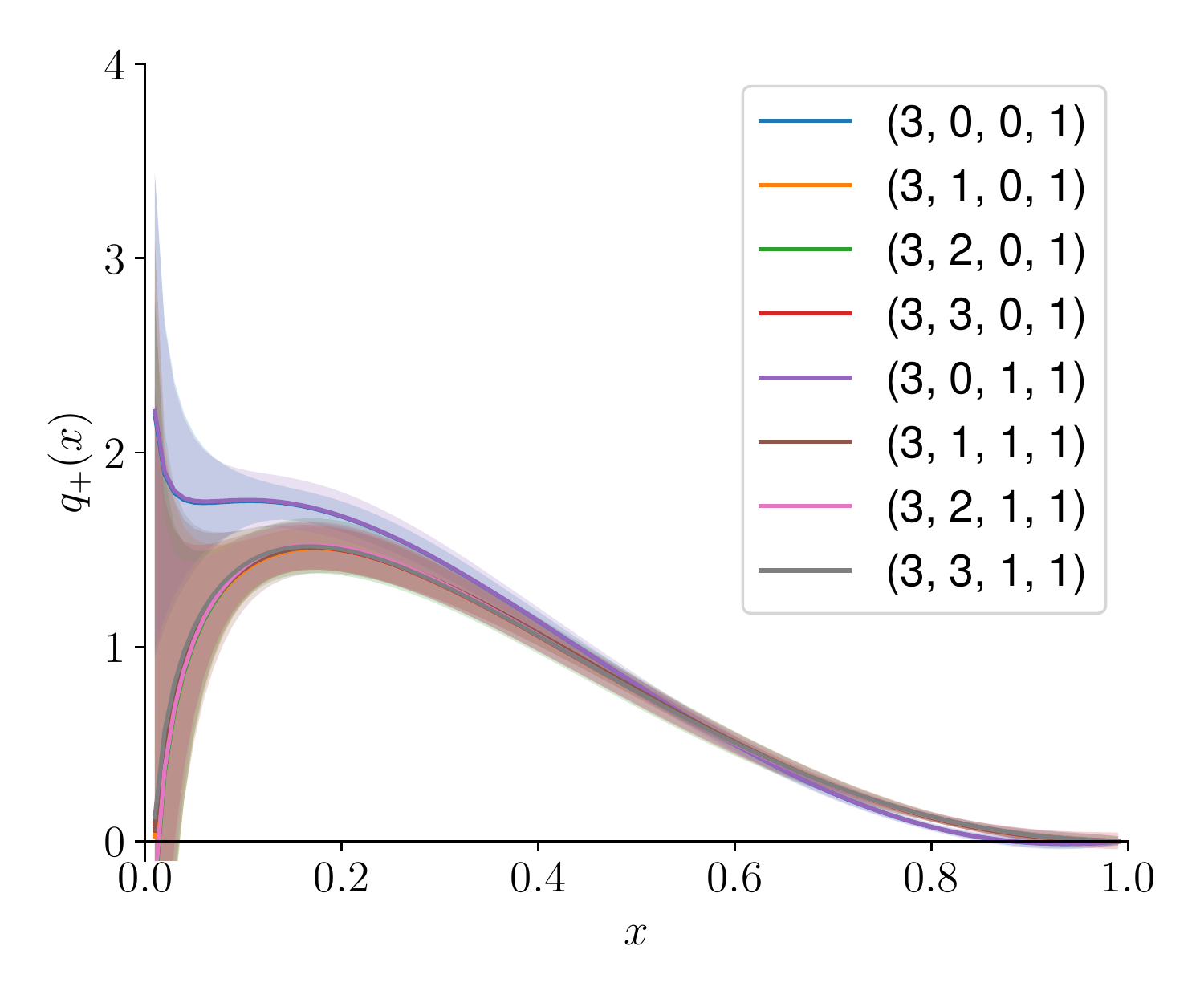}
\includegraphics[width=0.48\textwidth]{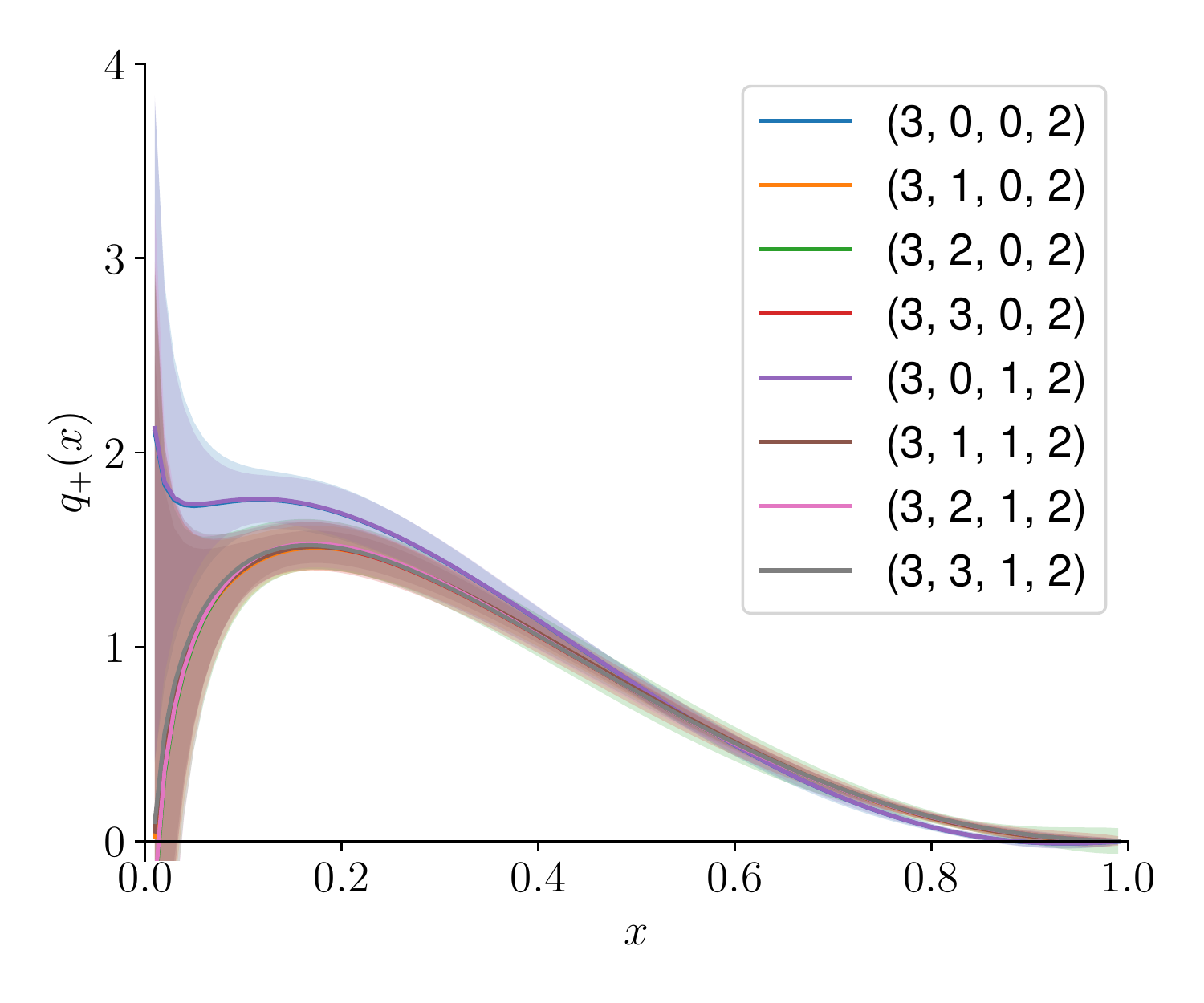}
\includegraphics[width=0.48\textwidth]{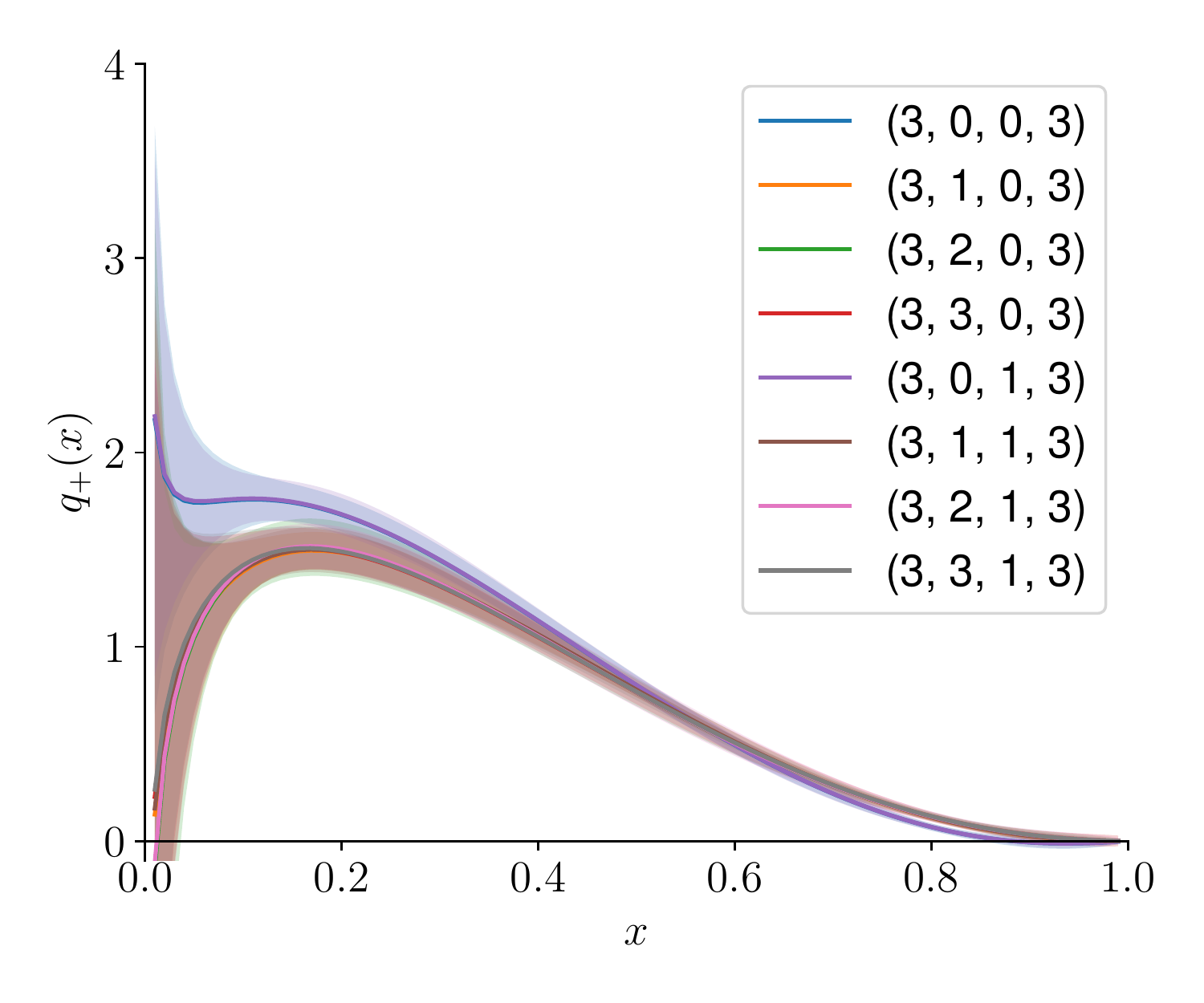}
\caption{\label{fig:many_model_pdfs_imag2}The PDF results from fitting the imaginary component to the models. The numbers in the legend correspond to $(N_\pm,  N_{R/I,b}, N_{R/I,r}, N_{R/I,p})$.}
\end{figure}

\begin{figure}[!htp]
\centering
\includegraphics[width=0.48\textwidth]{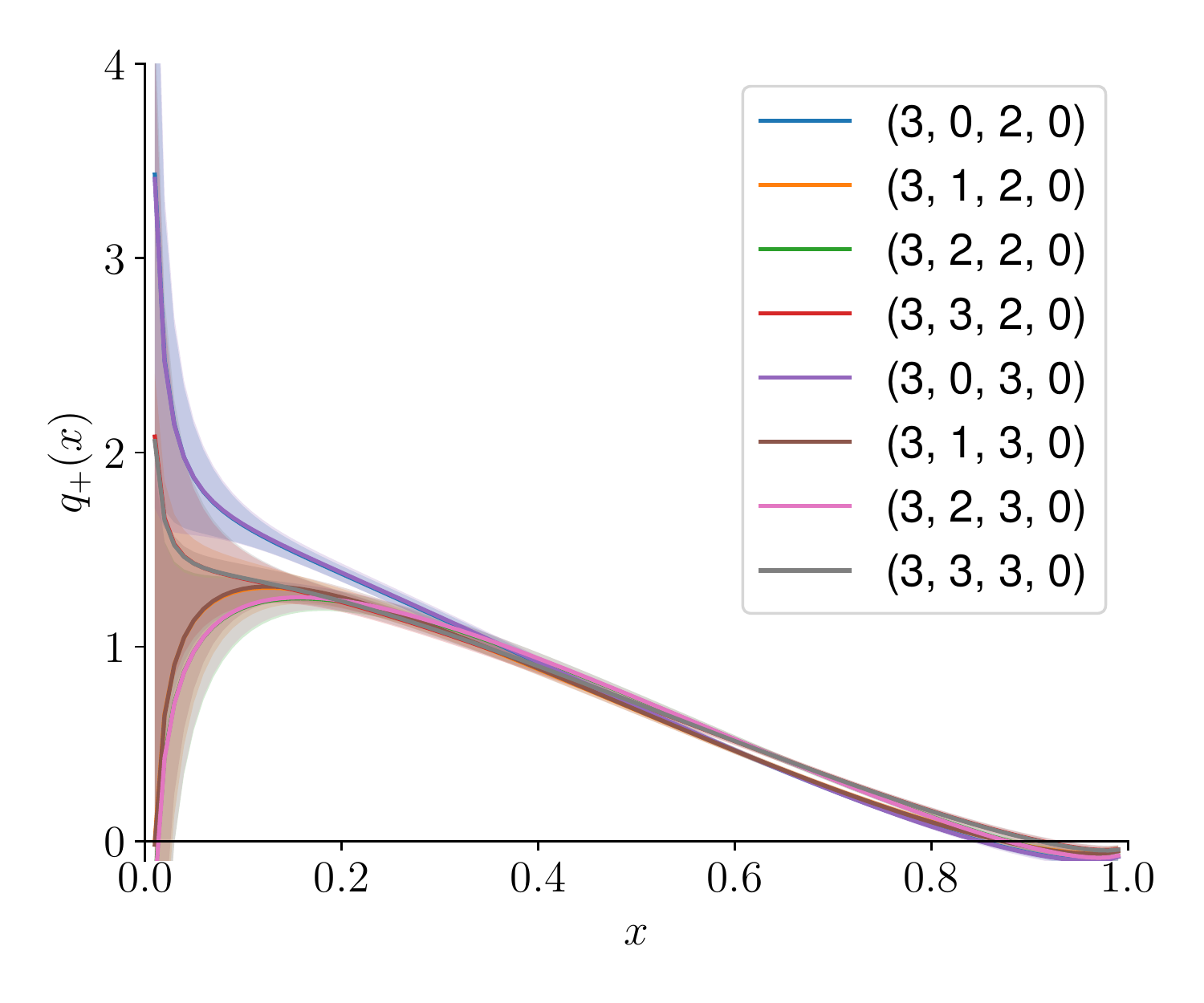}
\includegraphics[width=0.48\textwidth]{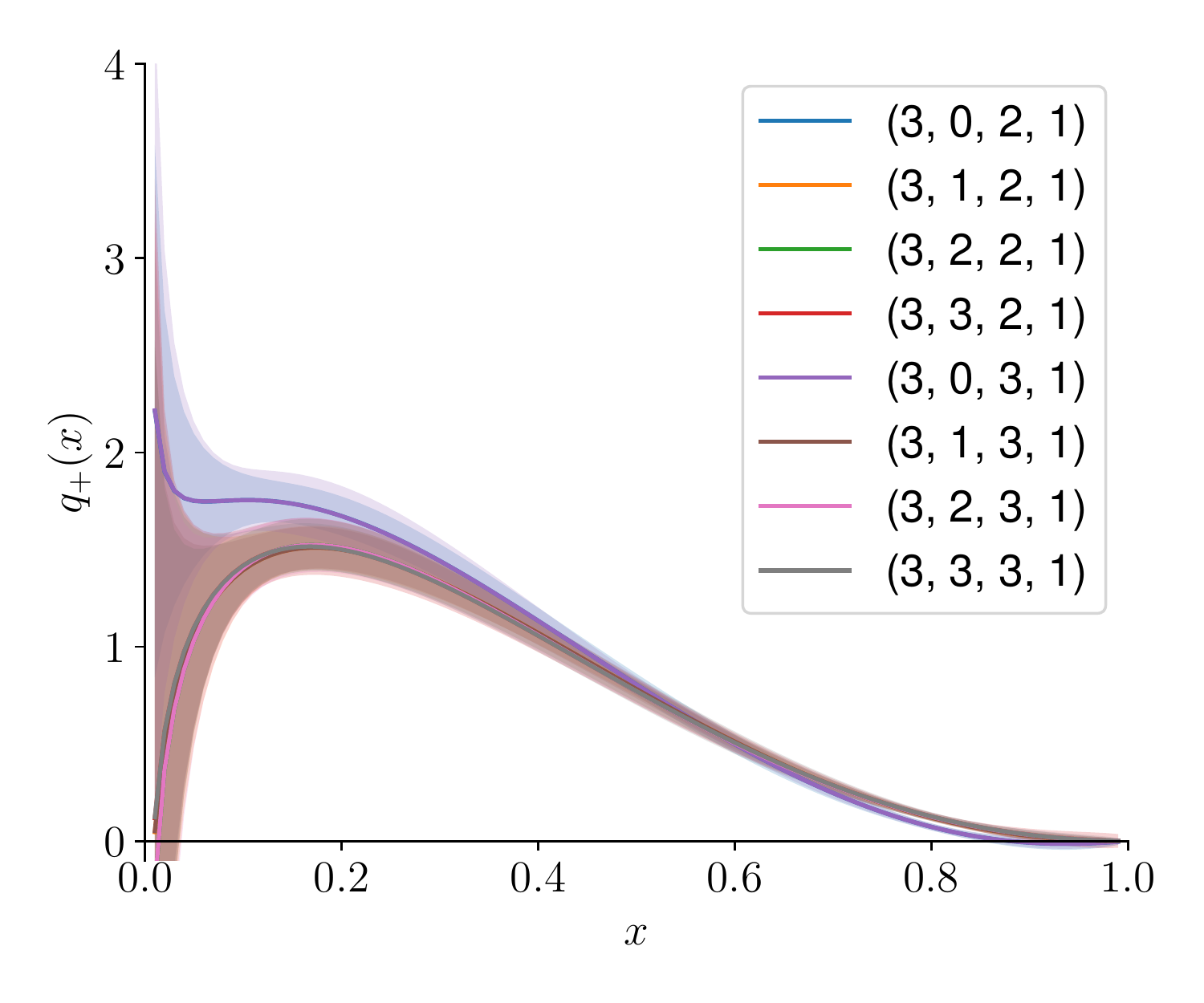}
\includegraphics[width=0.48\textwidth]{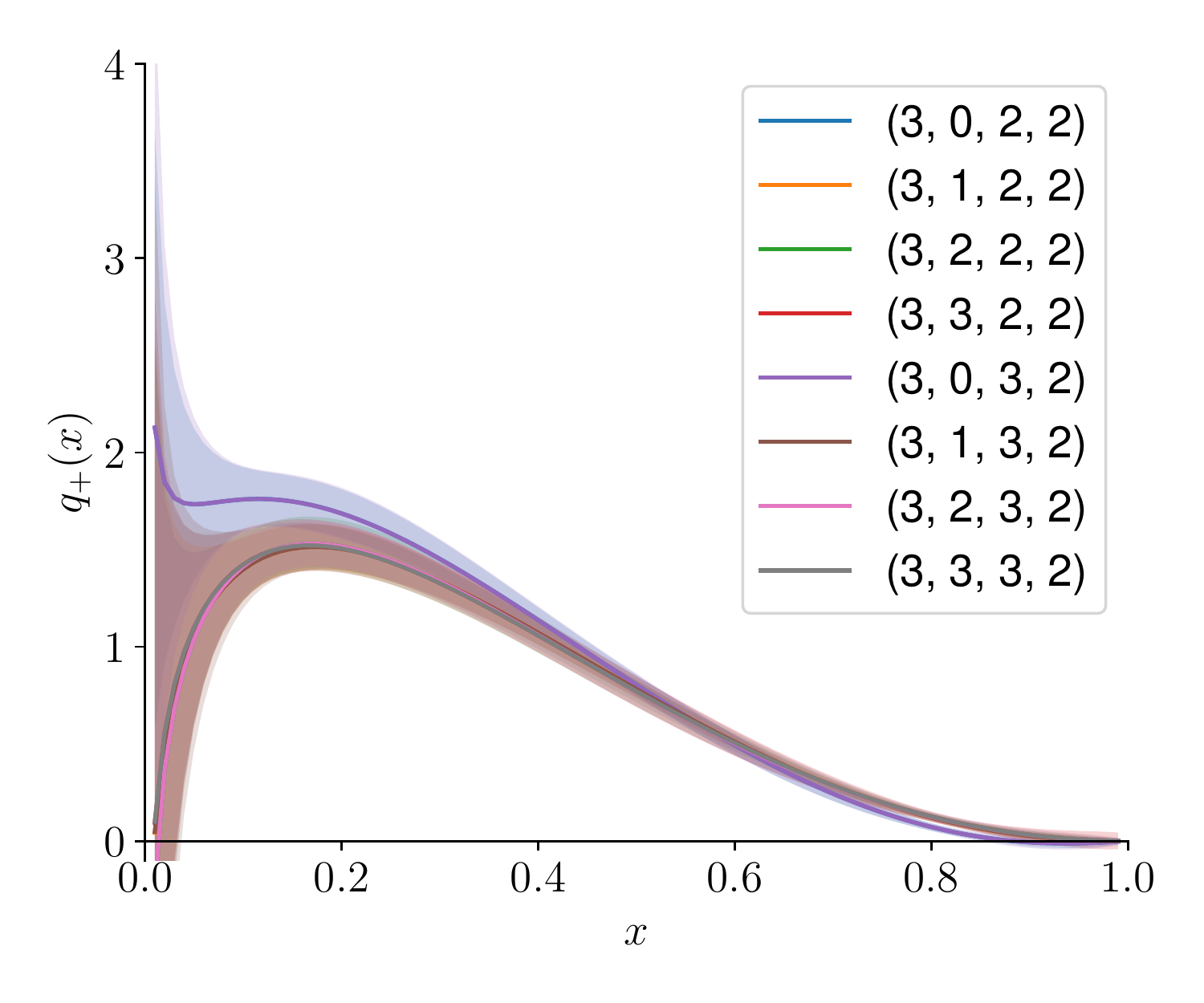}
\includegraphics[width=0.48\textwidth]{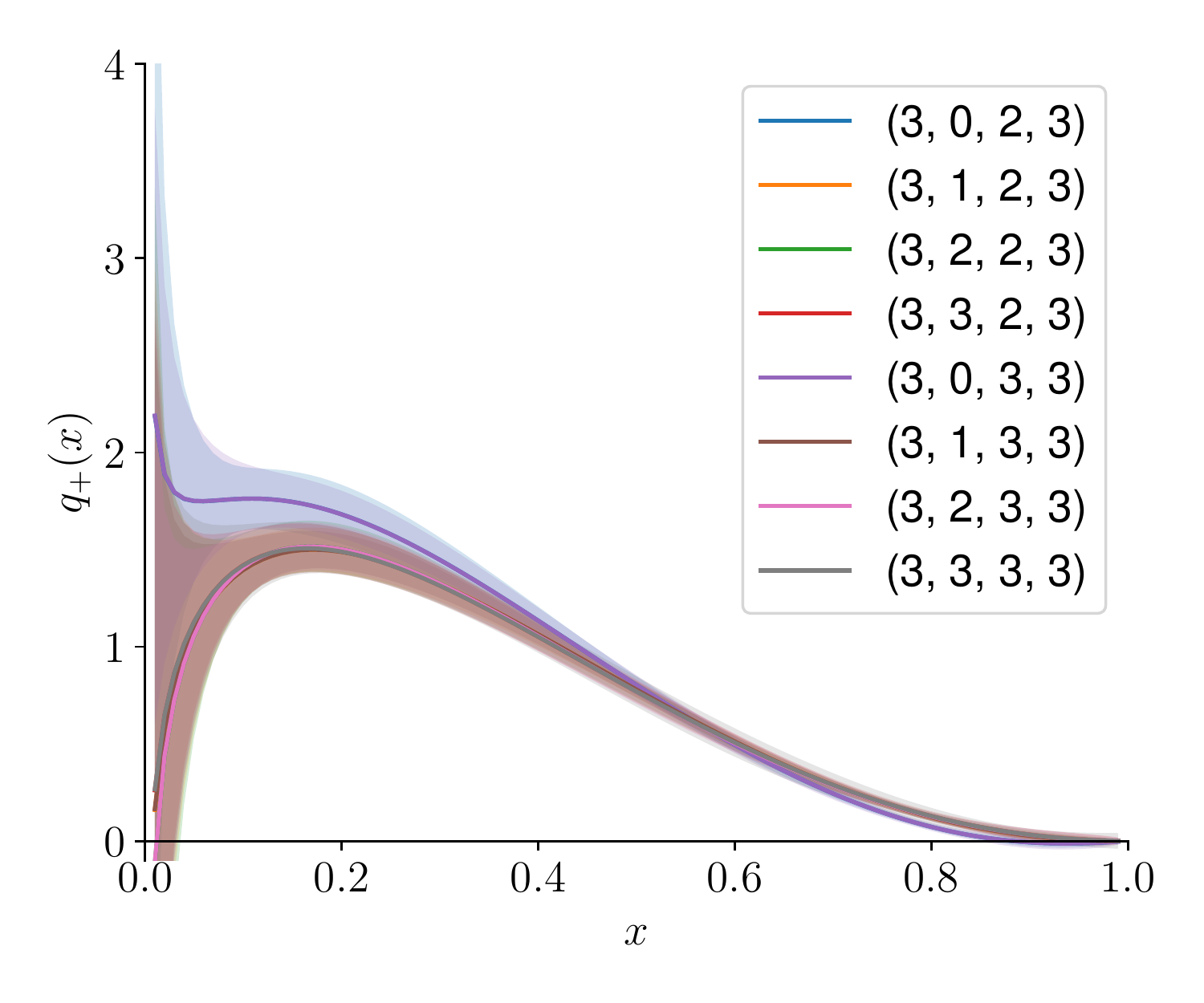}
\caption{\label{fig:many_model_pdfs_imag3}The PDF results from fitting the imaginary component to the models. The numbers in the legend correspond to $(N_\pm,  N_{R/I,b}, N_{R/I,r}, N_{R/I,p})$.}
\end{figure}

The results of the PDFs are shown in Figs.~\ref{fig:many_model_pdfs_real}-\ref{fig:many_model_pdfs_imag3}. When the $P_1$ nuisance terms are included, the PDFs are largely consistent. Some of the $q_+$ distributions begin to have significant differences for the region of $x \leq 0.4$. These discrepancies are to be expected from a fit with a limited range in Ioffe time, as the study in~\cite{Karpie:2019eiq} showed. When $P_1$ is included, the $q_-$ distribution consistently diverges at $x=0$, but $q_+$ may converge or diverge.  Without that $P_1$ term, the PDFs differ not only with the PDFs from fits with $P_1$ but also amongst themselves.

\begin{figure}[!htp]
\centering
\includegraphics[width=0.48\textwidth]{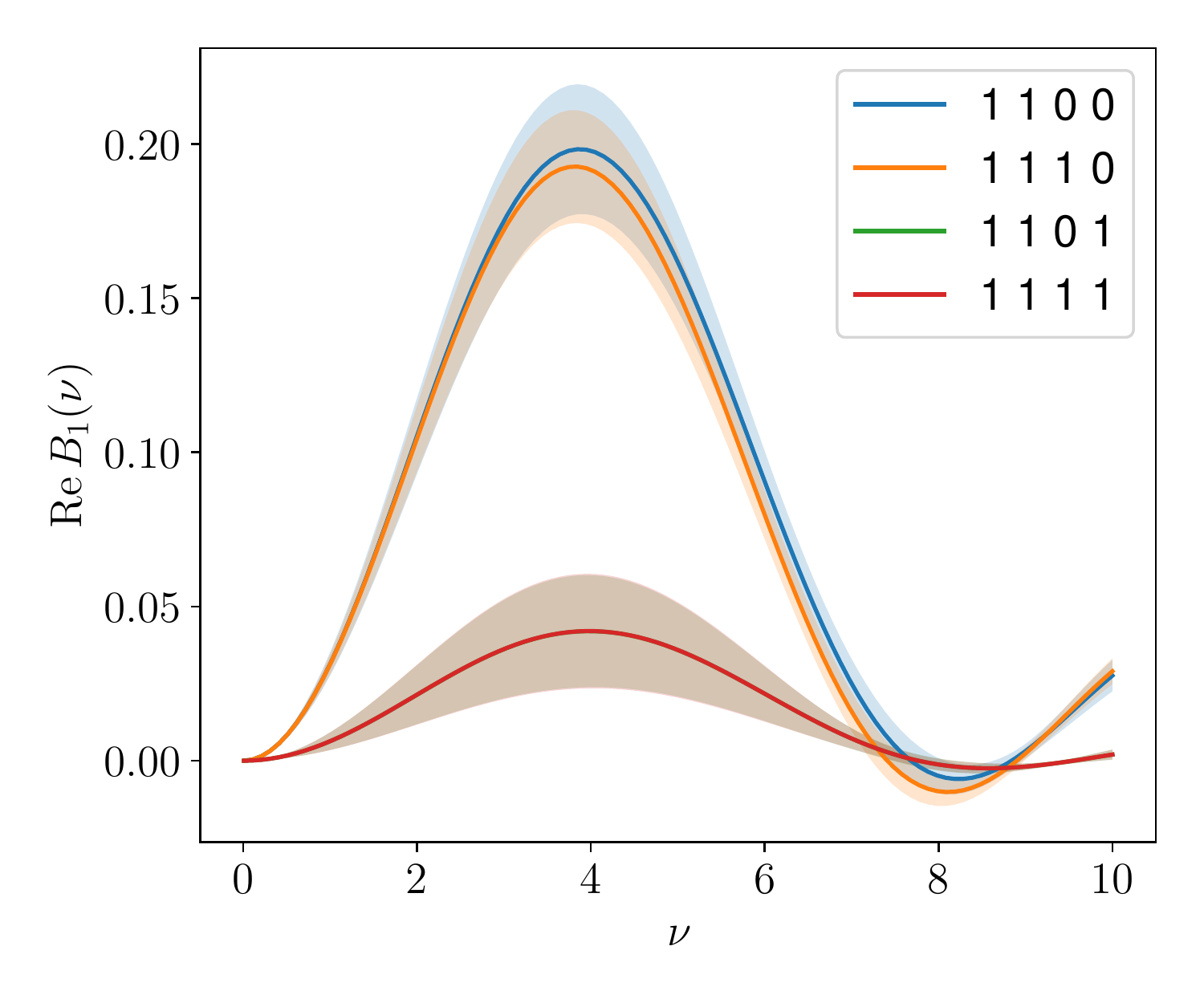}
\includegraphics[width=0.48\textwidth]{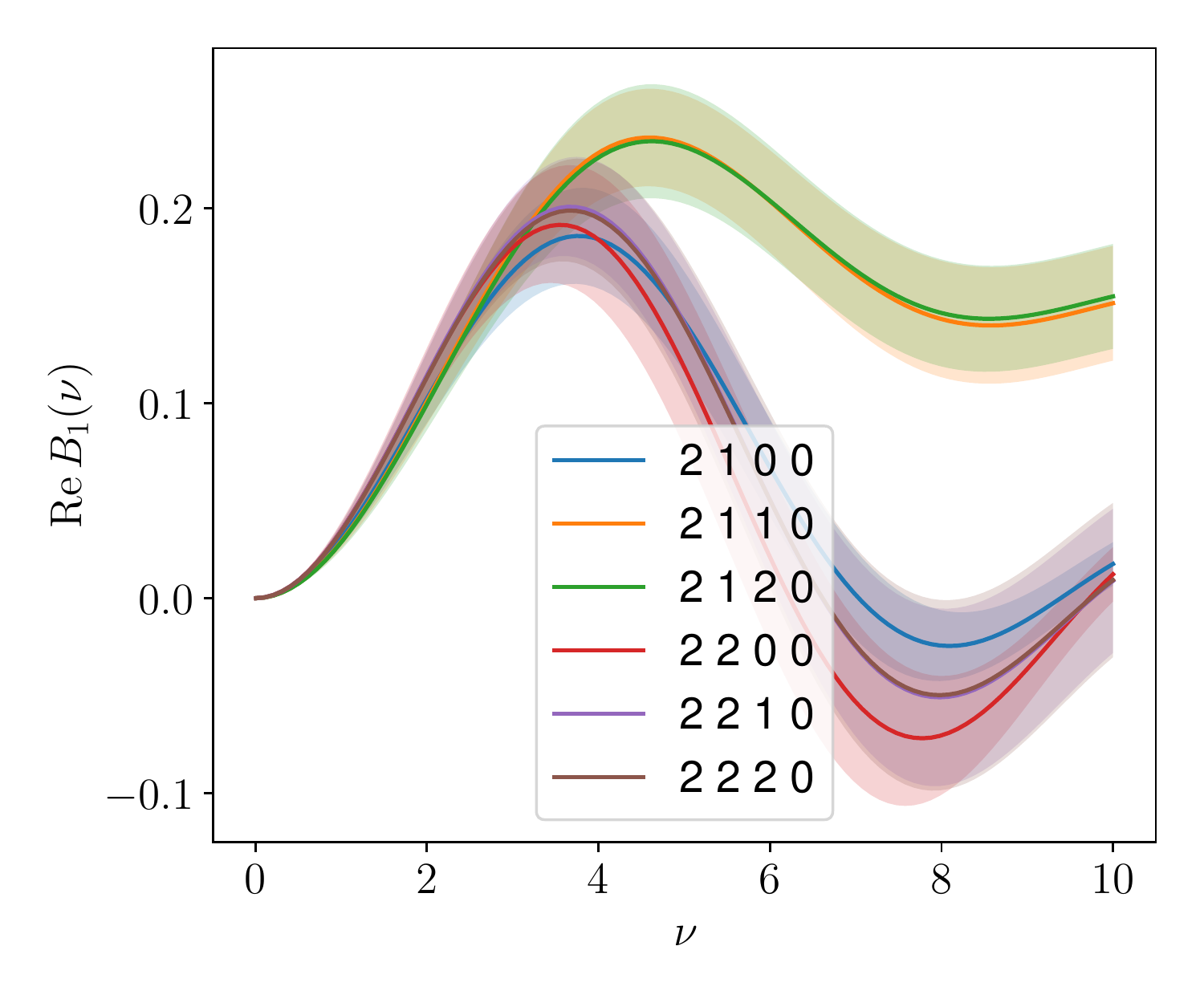}
\includegraphics[width=0.48\textwidth]{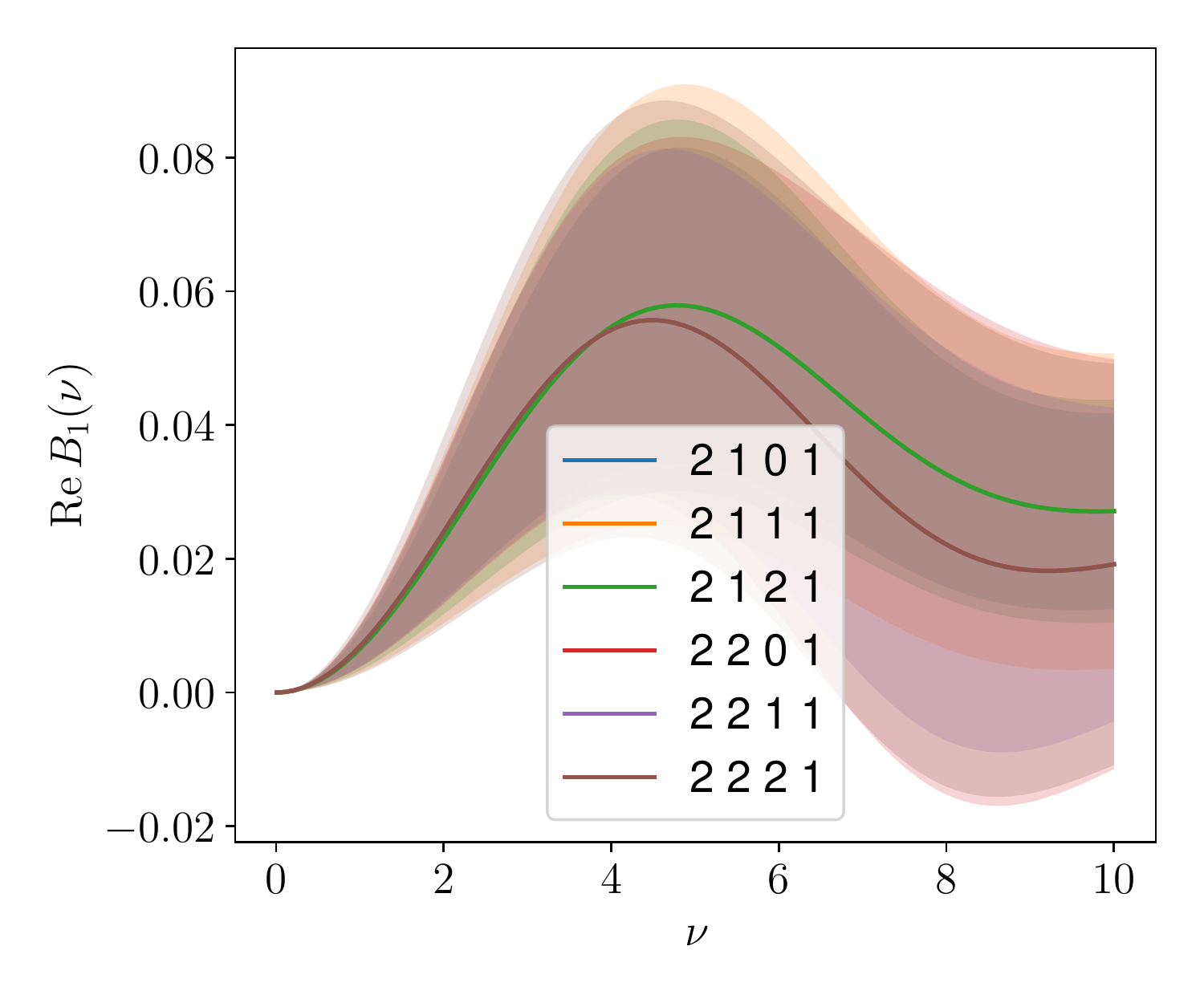}
\includegraphics[width=0.48\textwidth]{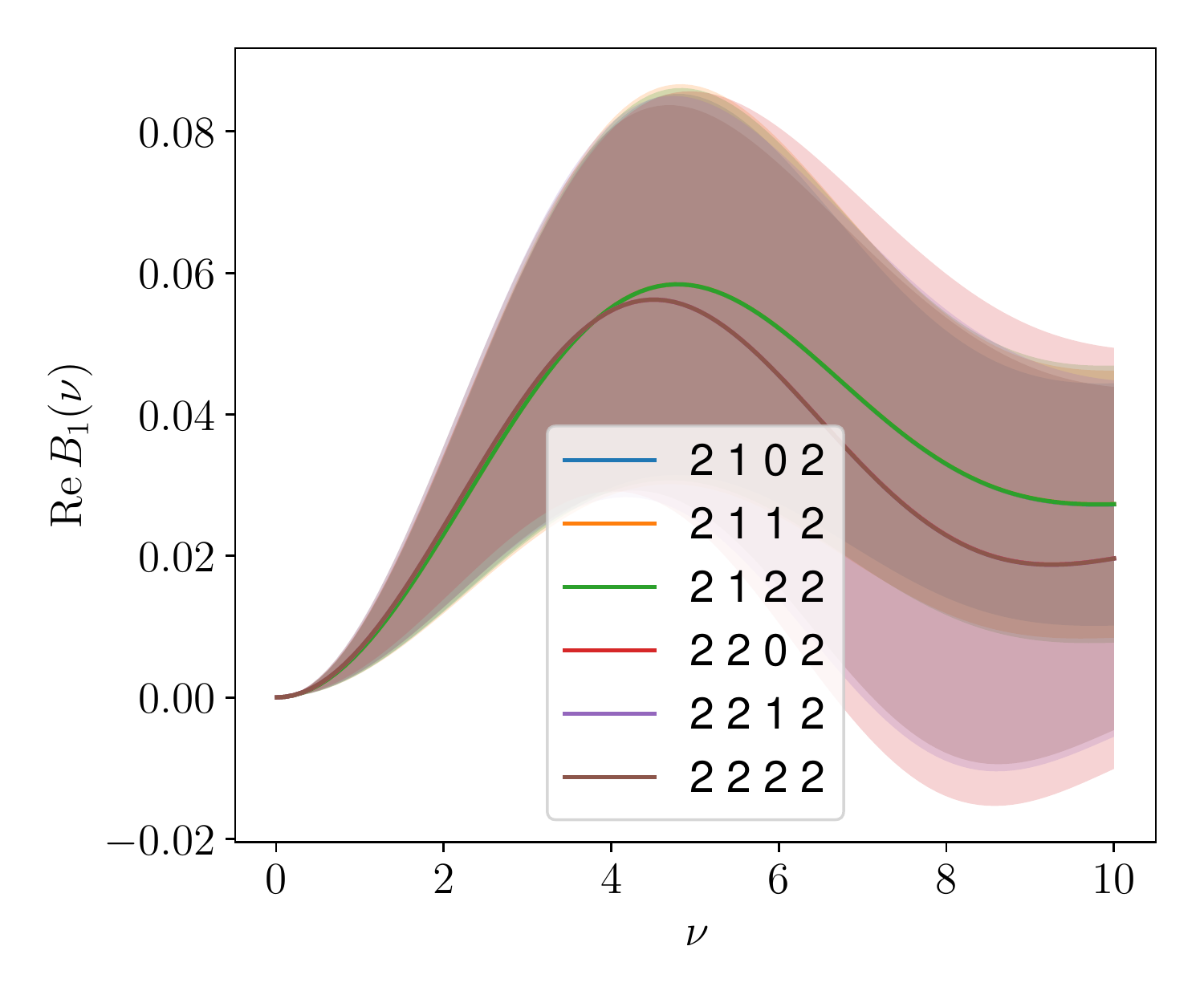}
\caption{\label{fig:many_model_ht_real1}The higher twist term, $B_1$, results from fitting the real component to the models. The numbers in the legend correspond to $(N_\pm,  N_{R/I,b}, N_{R/I,r}, N_{R/I,p})$.}
\end{figure}

\begin{figure}[!htp]
\centering
\includegraphics[width=0.48\textwidth]{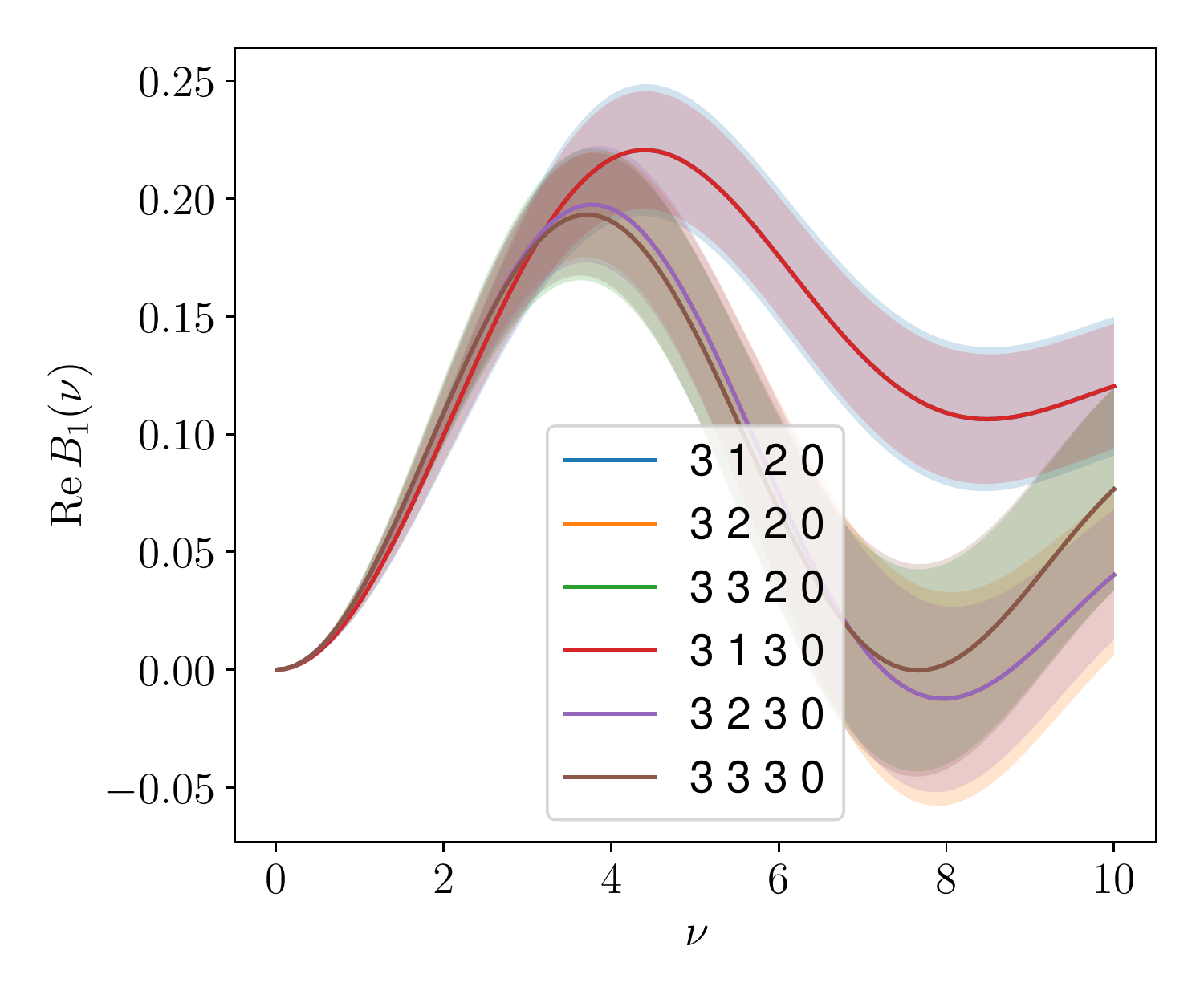}
\includegraphics[width=0.48\textwidth]{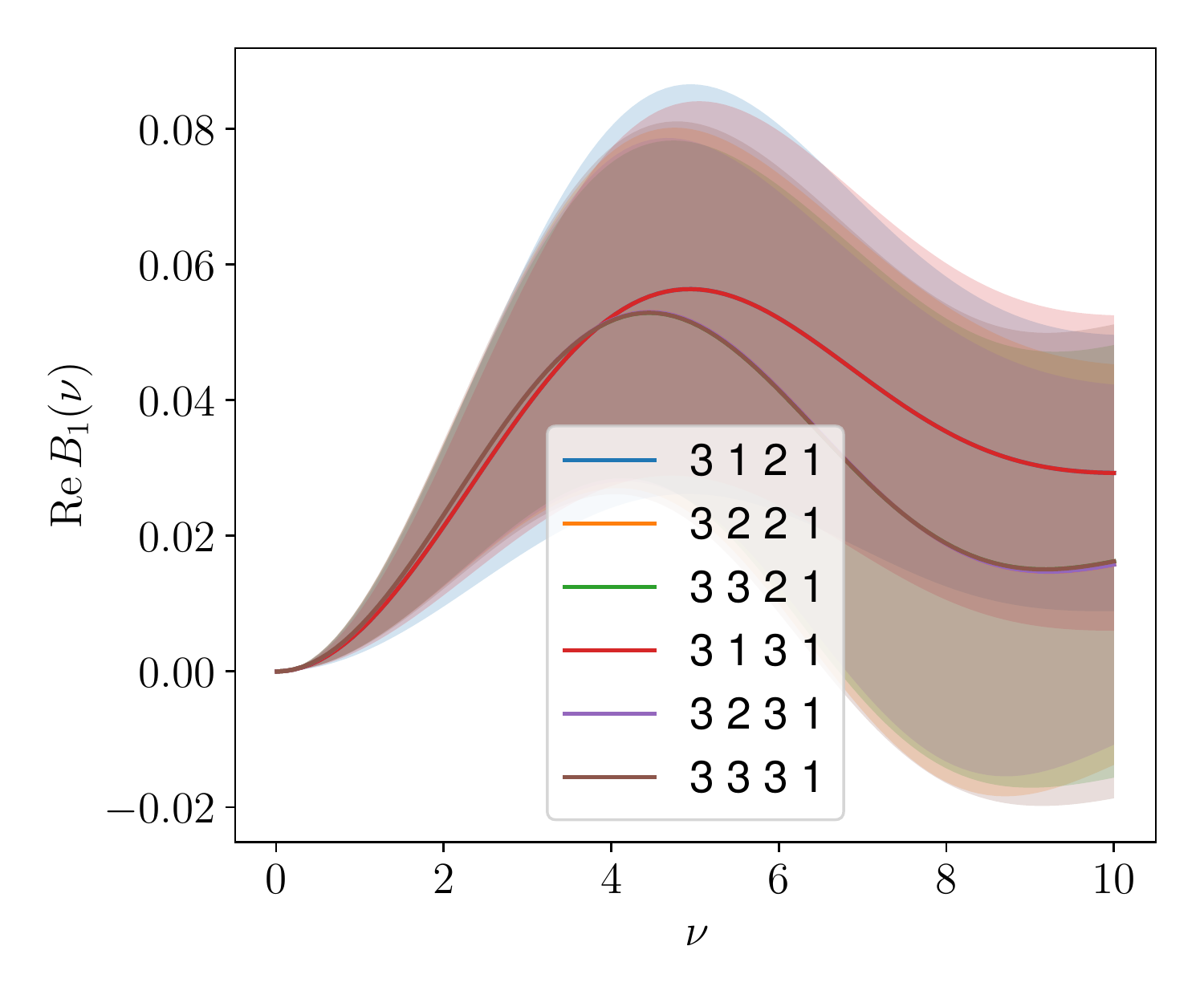}
\includegraphics[width=0.48\textwidth]{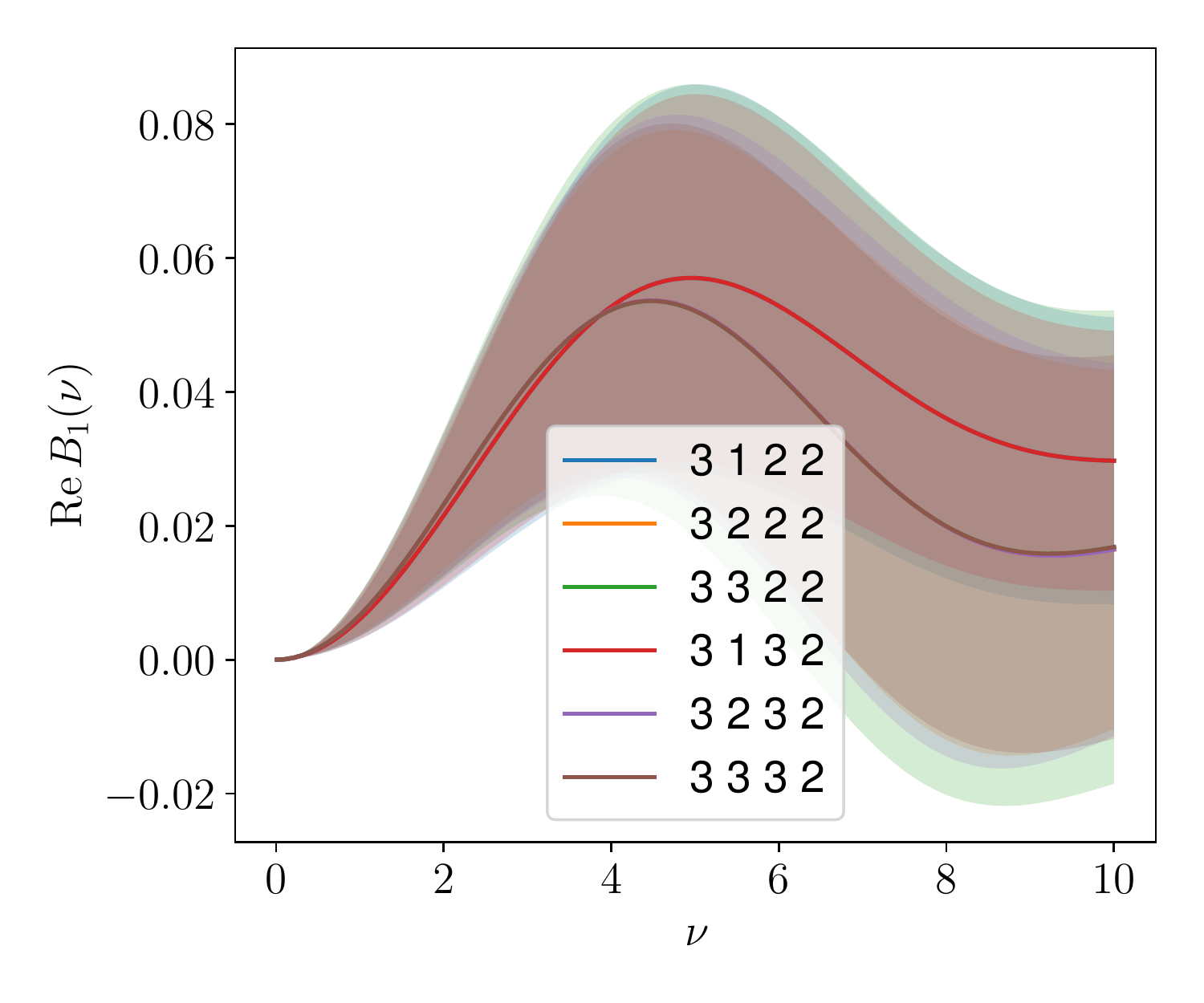}
\includegraphics[width=0.48\textwidth]{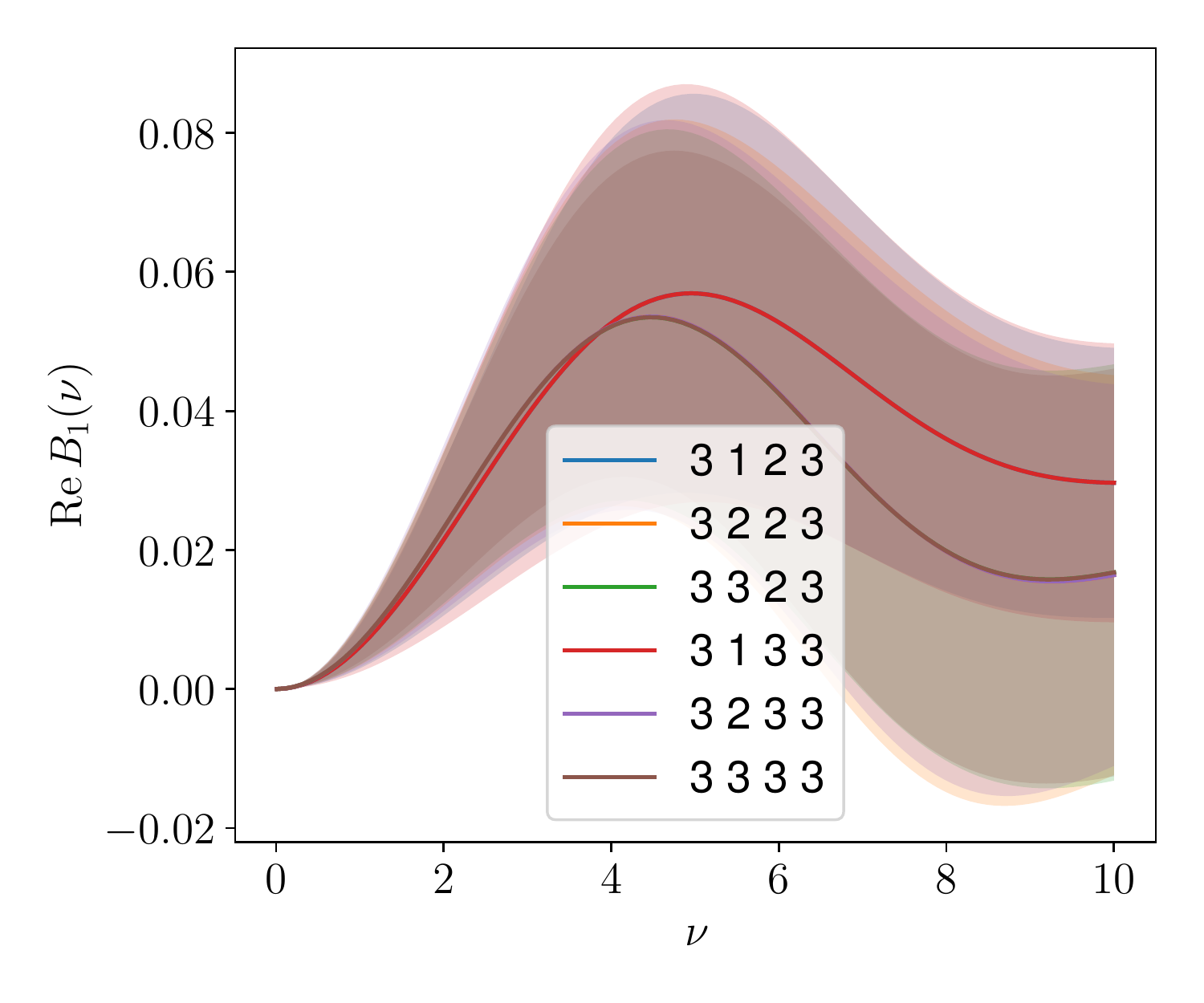}
\caption{\label{fig:many_model_ht_real2}The higher twist term, $B_1$, results from fitting the real component to the models. The numbers in the legend correspond to $(N_\pm,  N_{R/I,b}, N_{R/I,r}, N_{R/I,p})$.}
\end{figure}

\begin{figure}[!htp]
\centering
\includegraphics[width=0.48\textwidth]{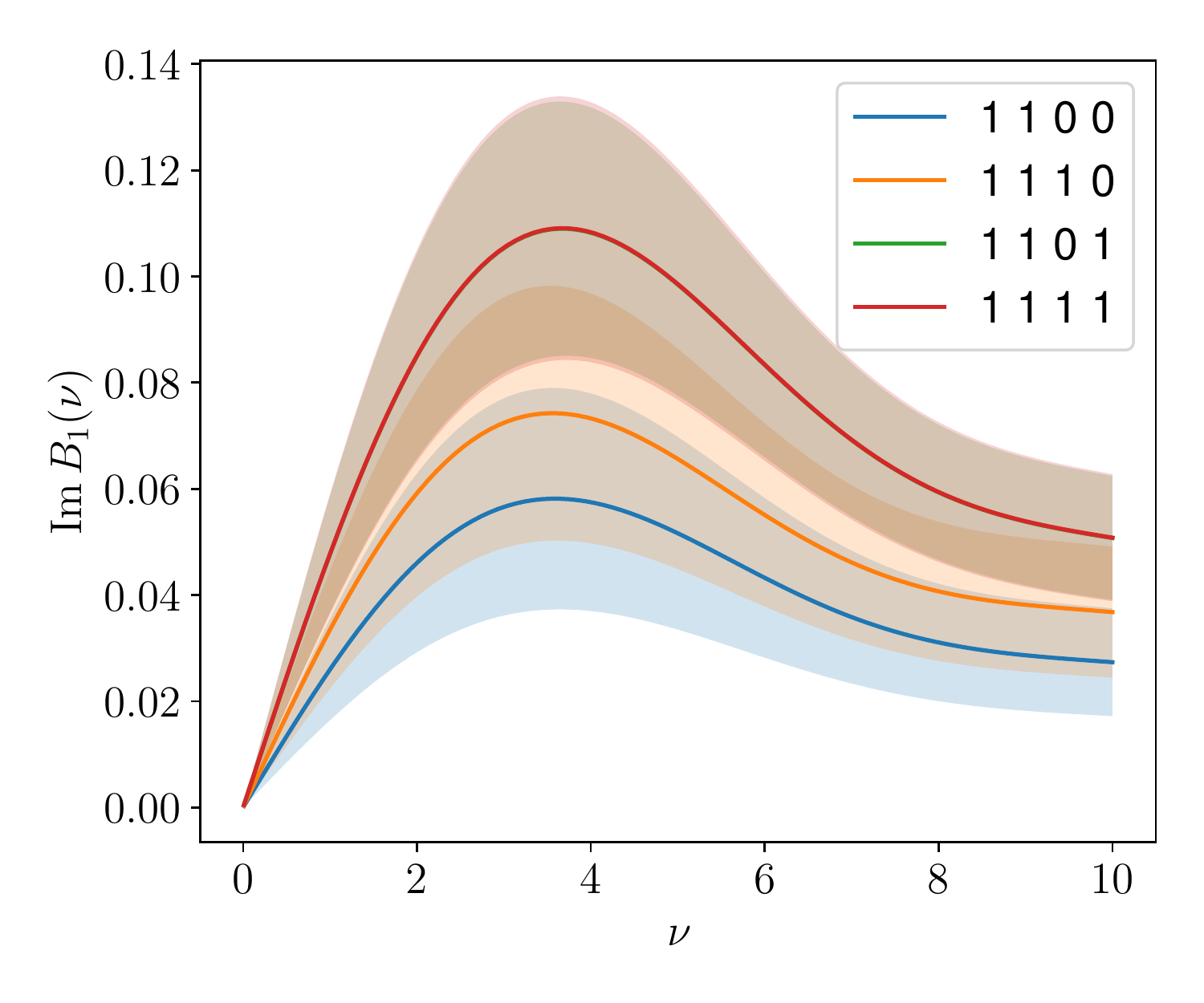}
\includegraphics[width=0.48\textwidth]{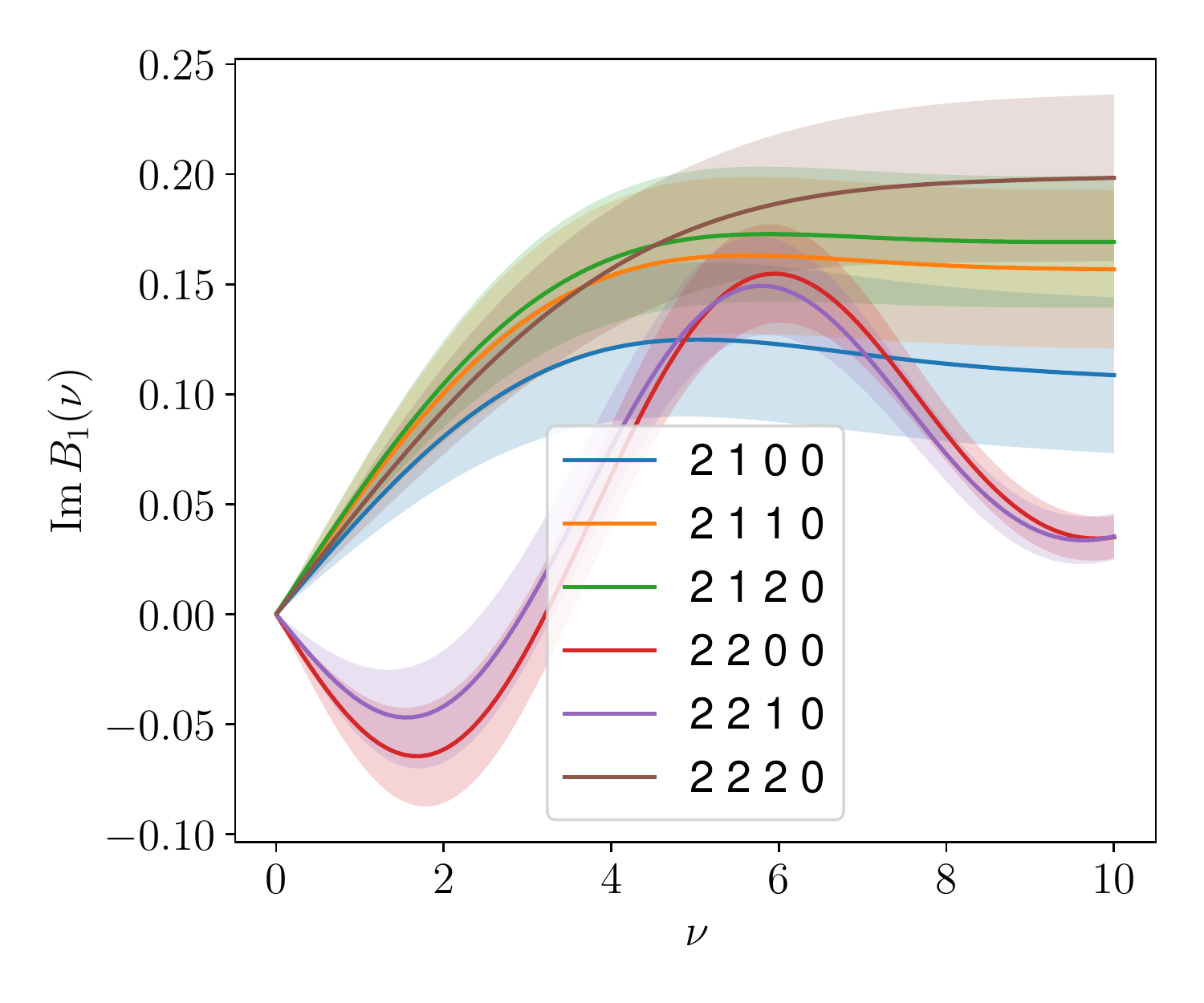}
\includegraphics[width=0.48\textwidth]{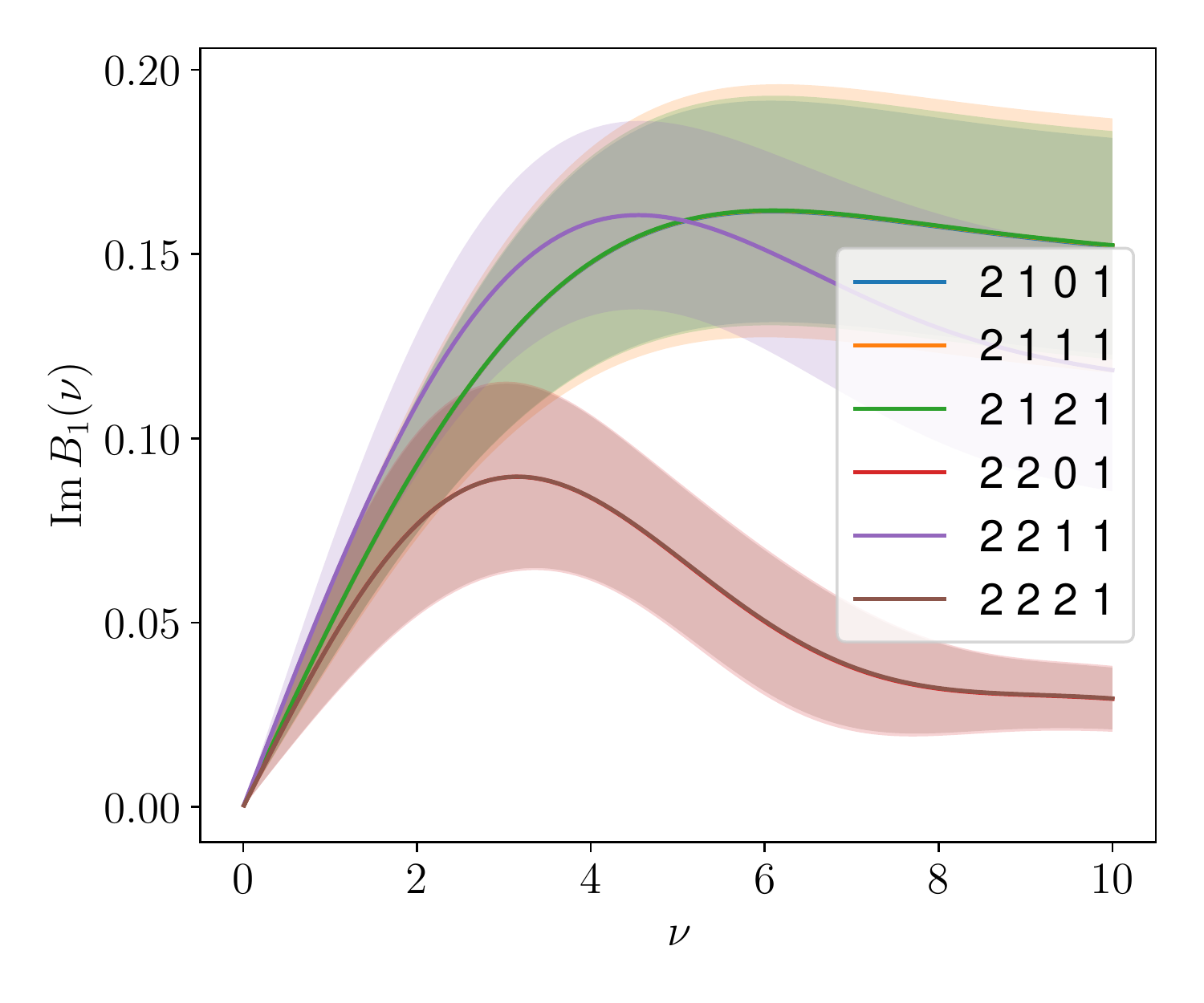}
\includegraphics[width=0.48\textwidth]{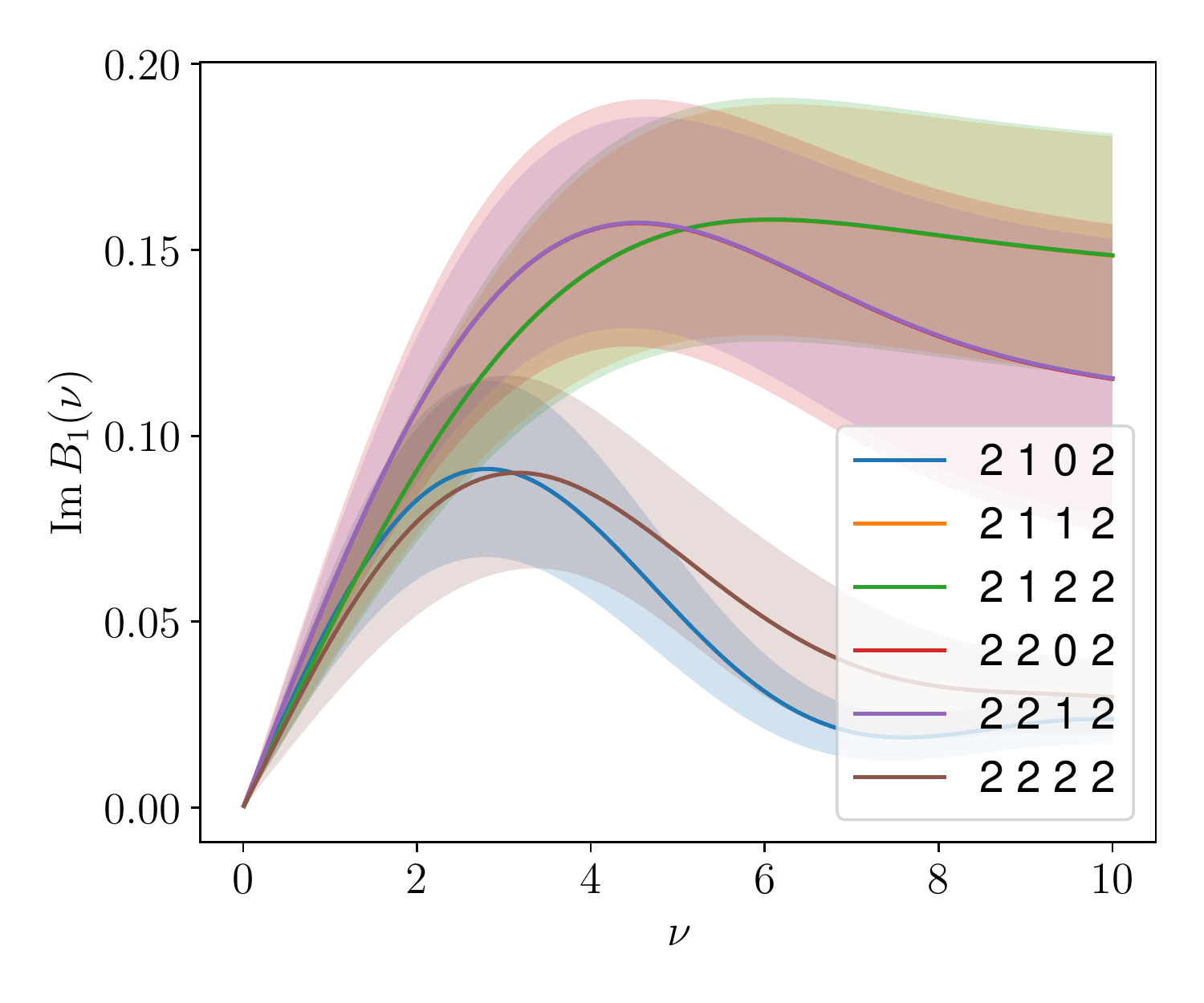}
\caption{\label{fig:many_model_ht_imag1}The higher twist term, $B_1$, results from fitting the imaginary component to the models. The numbers in the legend correspond to $(N_\pm,  N_{R/I,b}, N_{R/I,r}, N_{R/I,p})$.}
\end{figure}

\begin{figure}[!htp]
\centering
\includegraphics[width=0.48\textwidth]{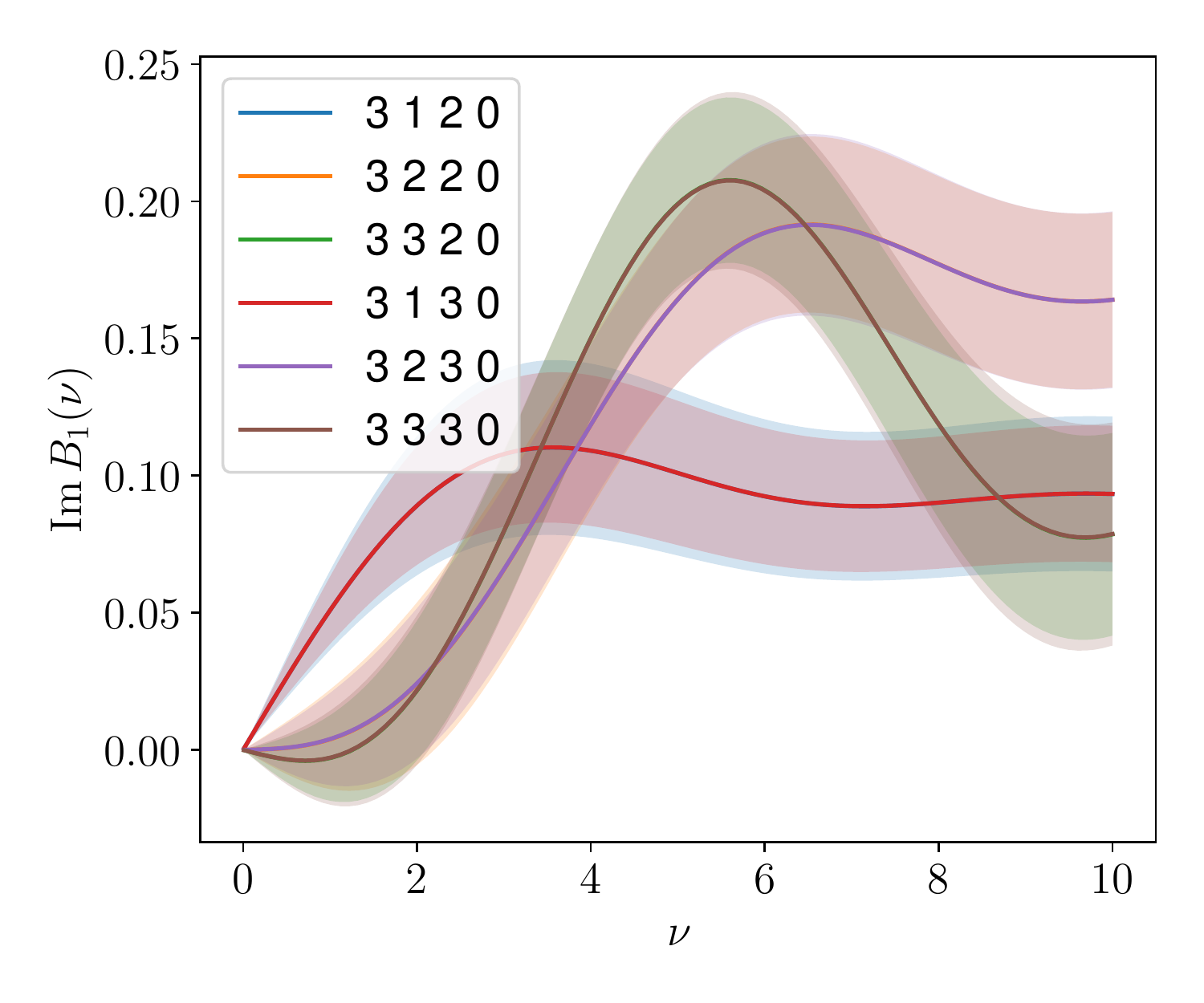}
\includegraphics[width=0.48\textwidth]{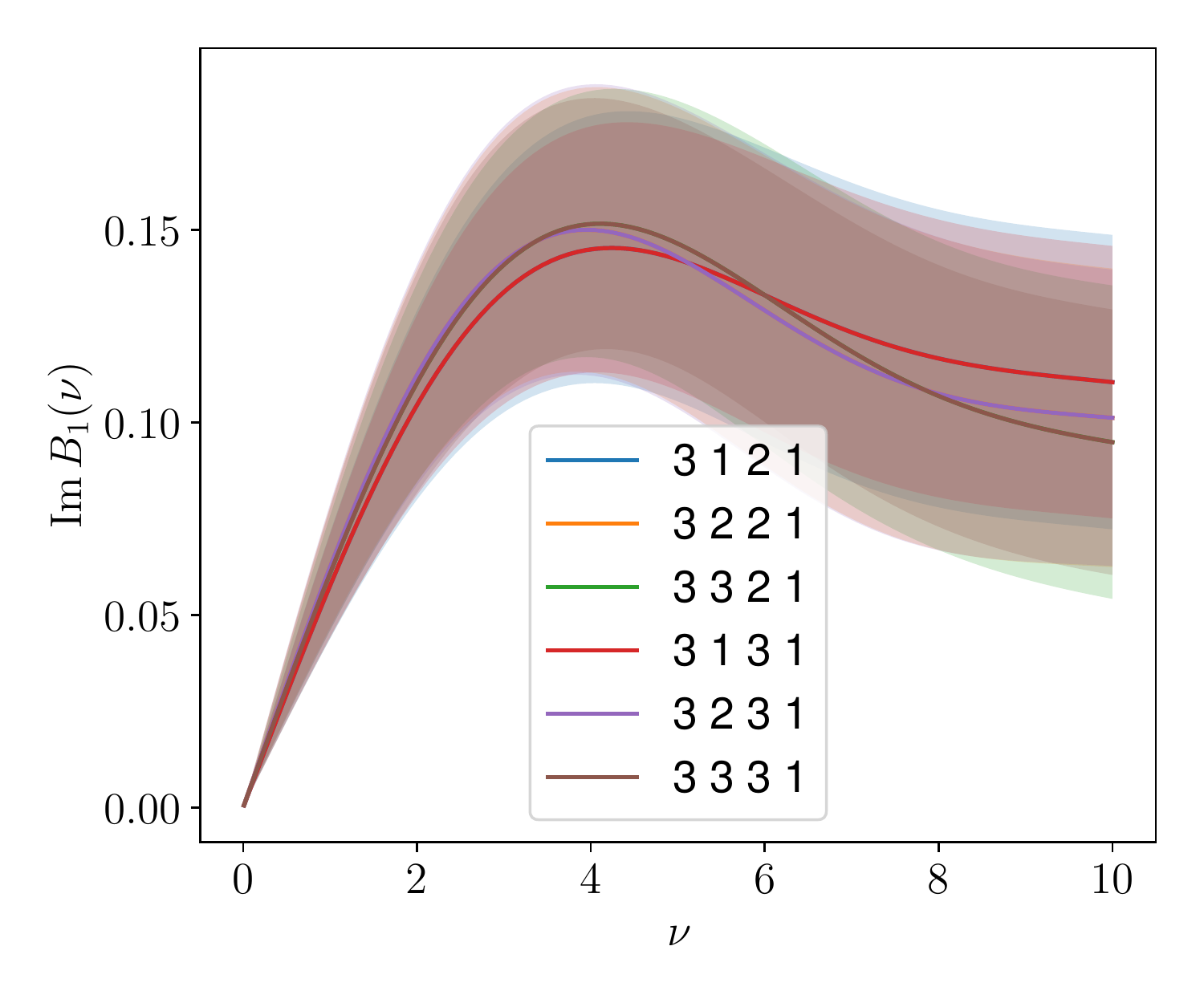}
\includegraphics[width=0.48\textwidth]{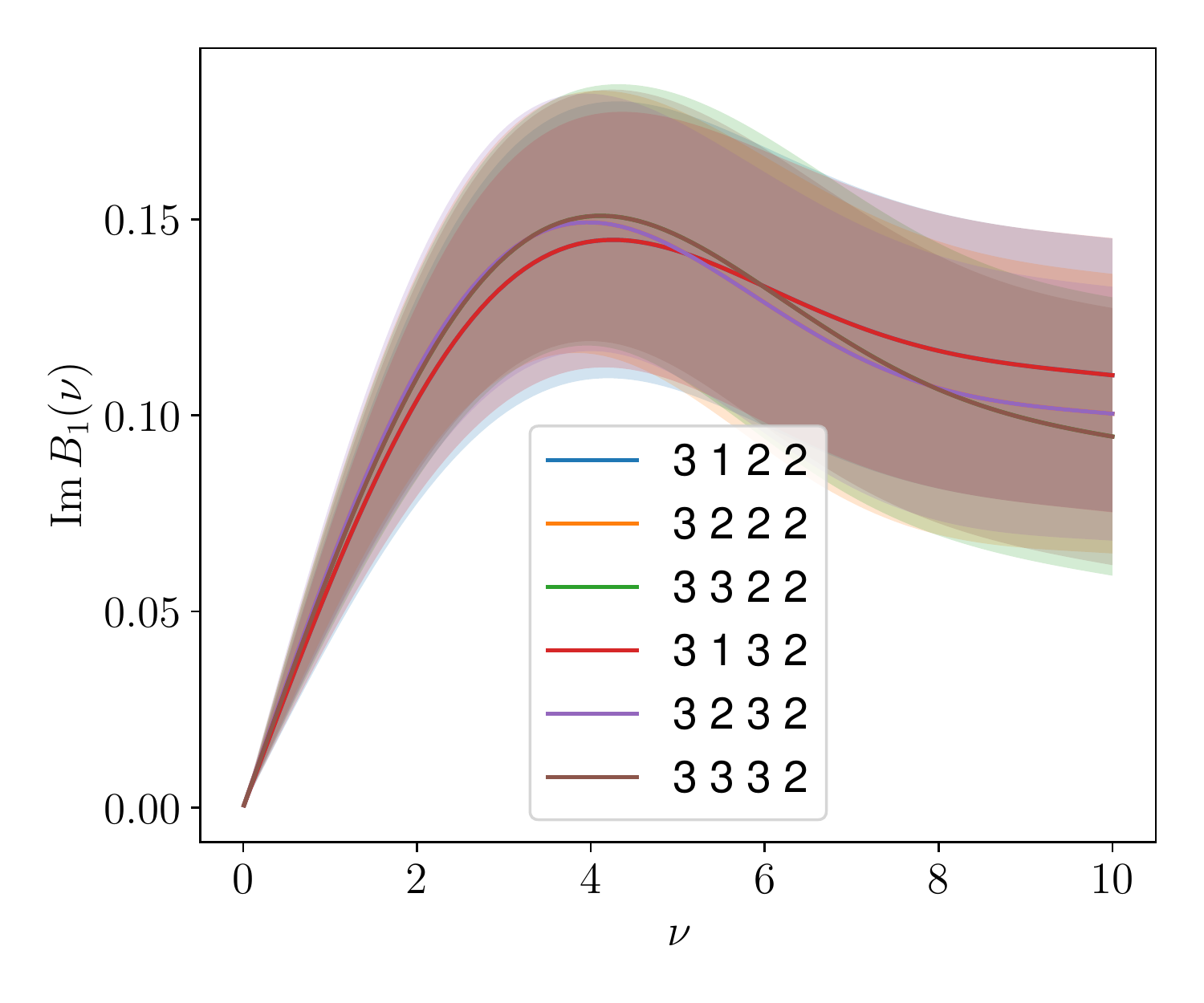}
\includegraphics[width=0.48\textwidth]{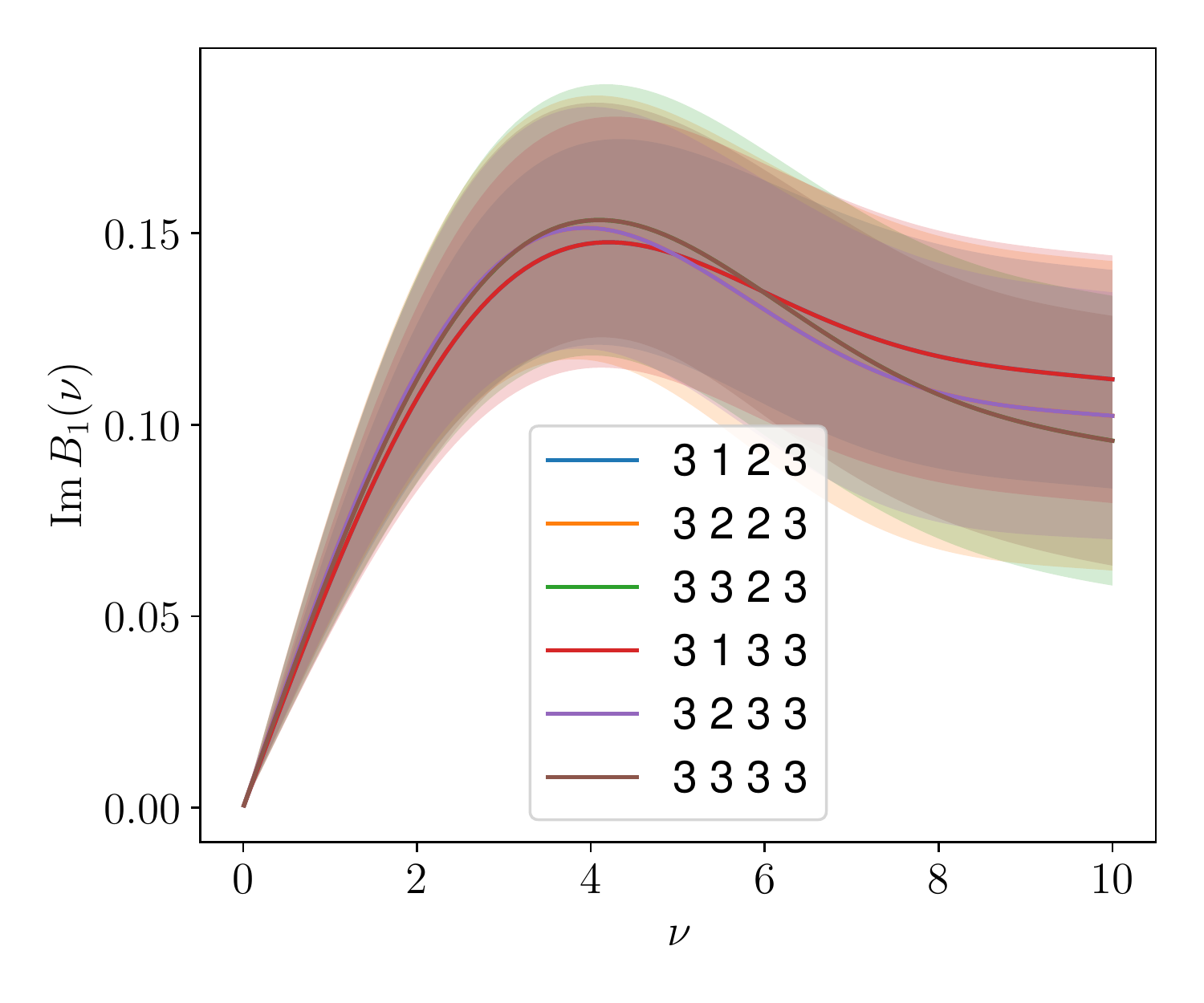}
\caption{\label{fig:many_model_ht_imag2}The higher twist term, $B_1$, results from fitting the imaginary component to the models. The numbers in the legend correspond to $(N_\pm,  N_{R/I,b}, N_{R/I,r}, N_{R/I,p})$.}
\end{figure}

\begin{figure}[!htp]
\centering
\includegraphics[width=0.48\textwidth]{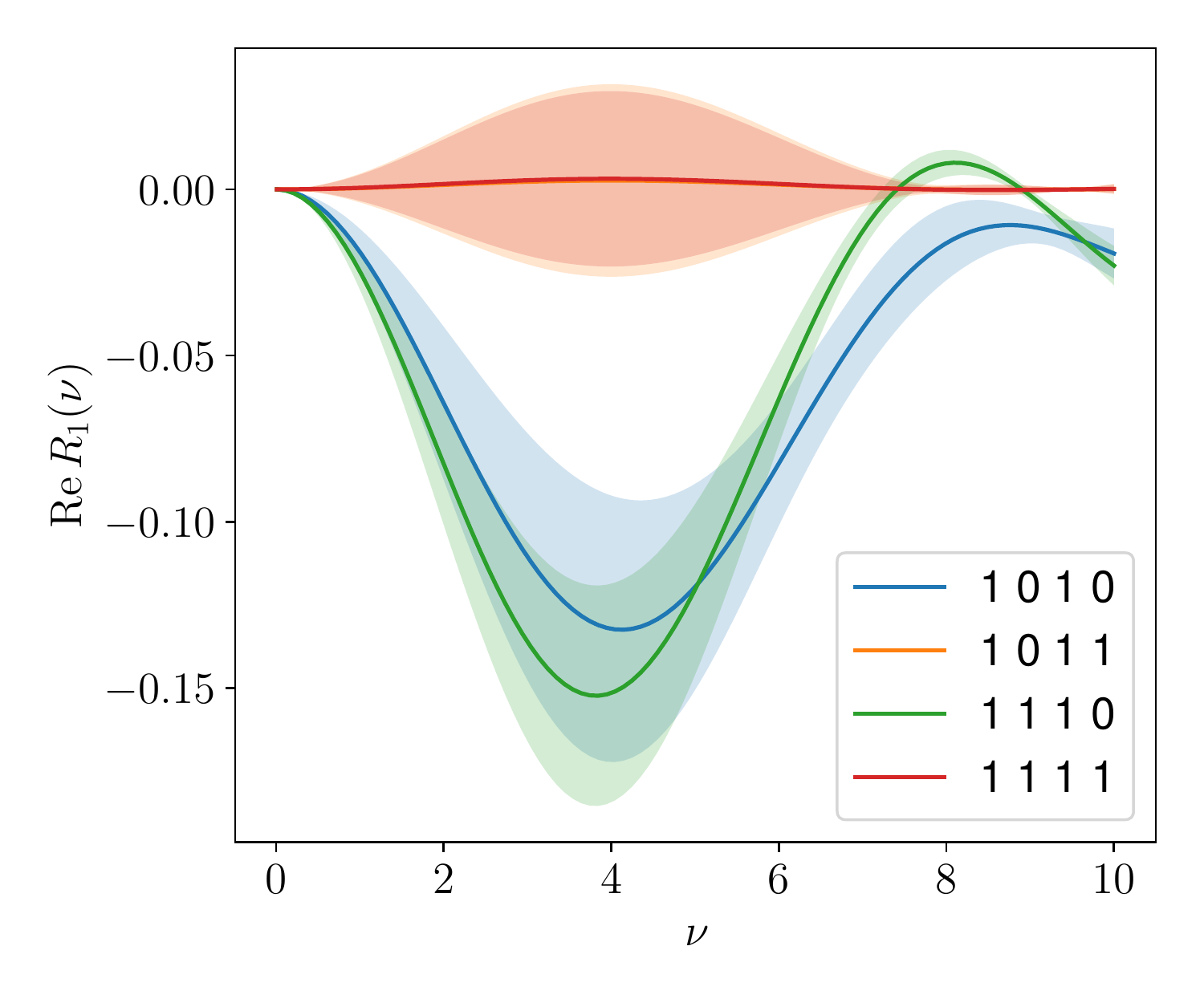}
\includegraphics[width=0.48\textwidth]{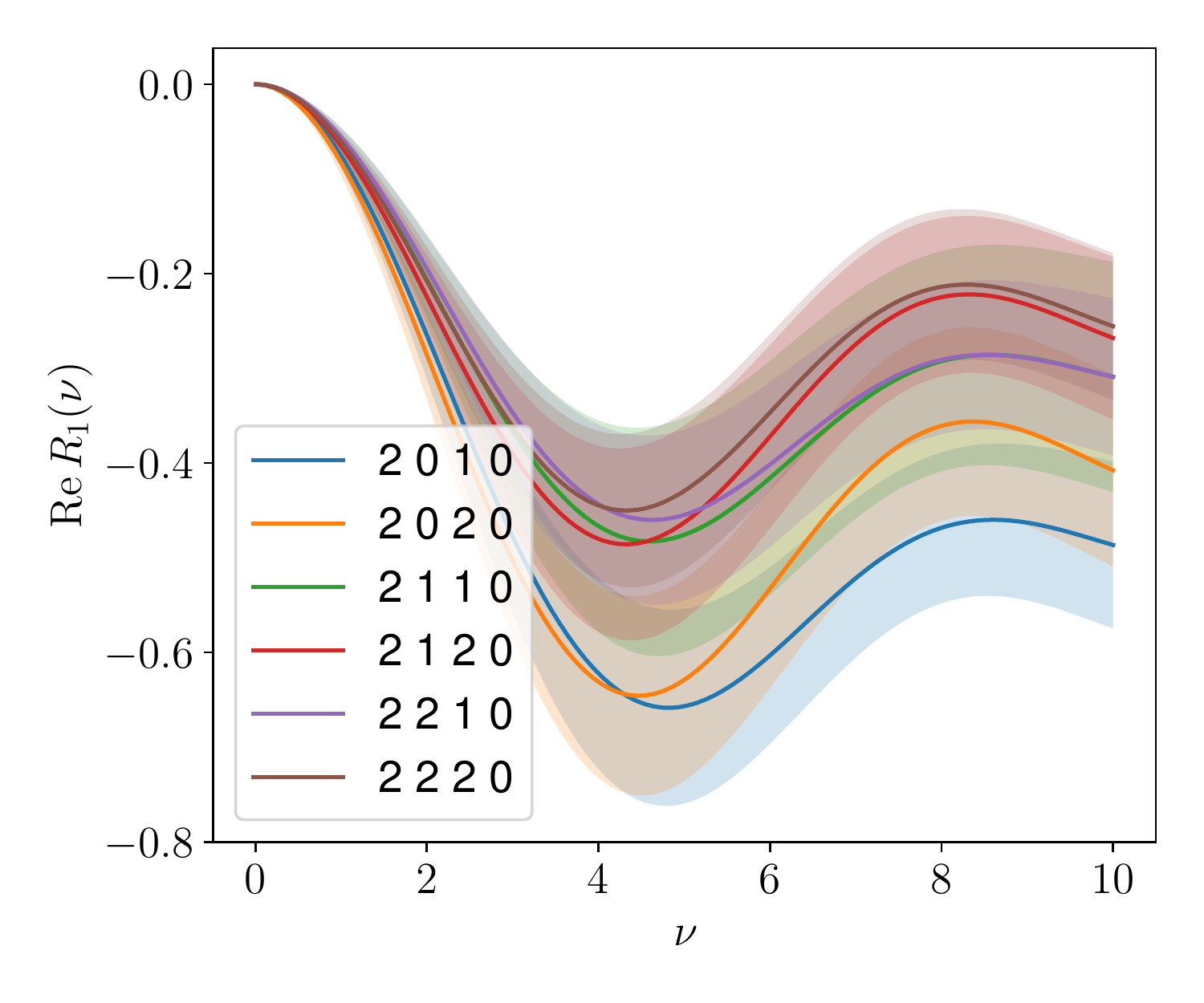}
\includegraphics[width=0.48\textwidth]{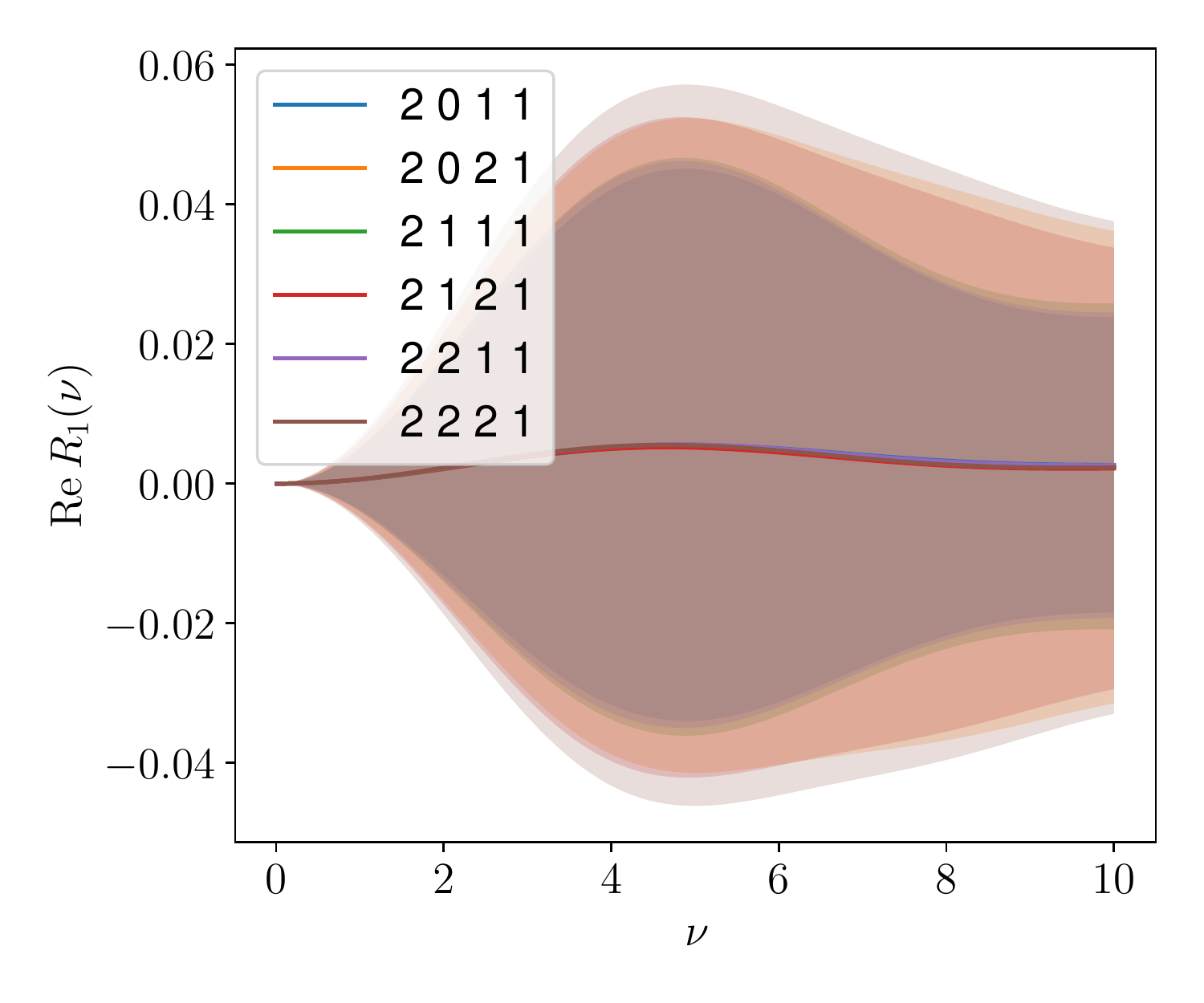}
\includegraphics[width=0.48\textwidth]{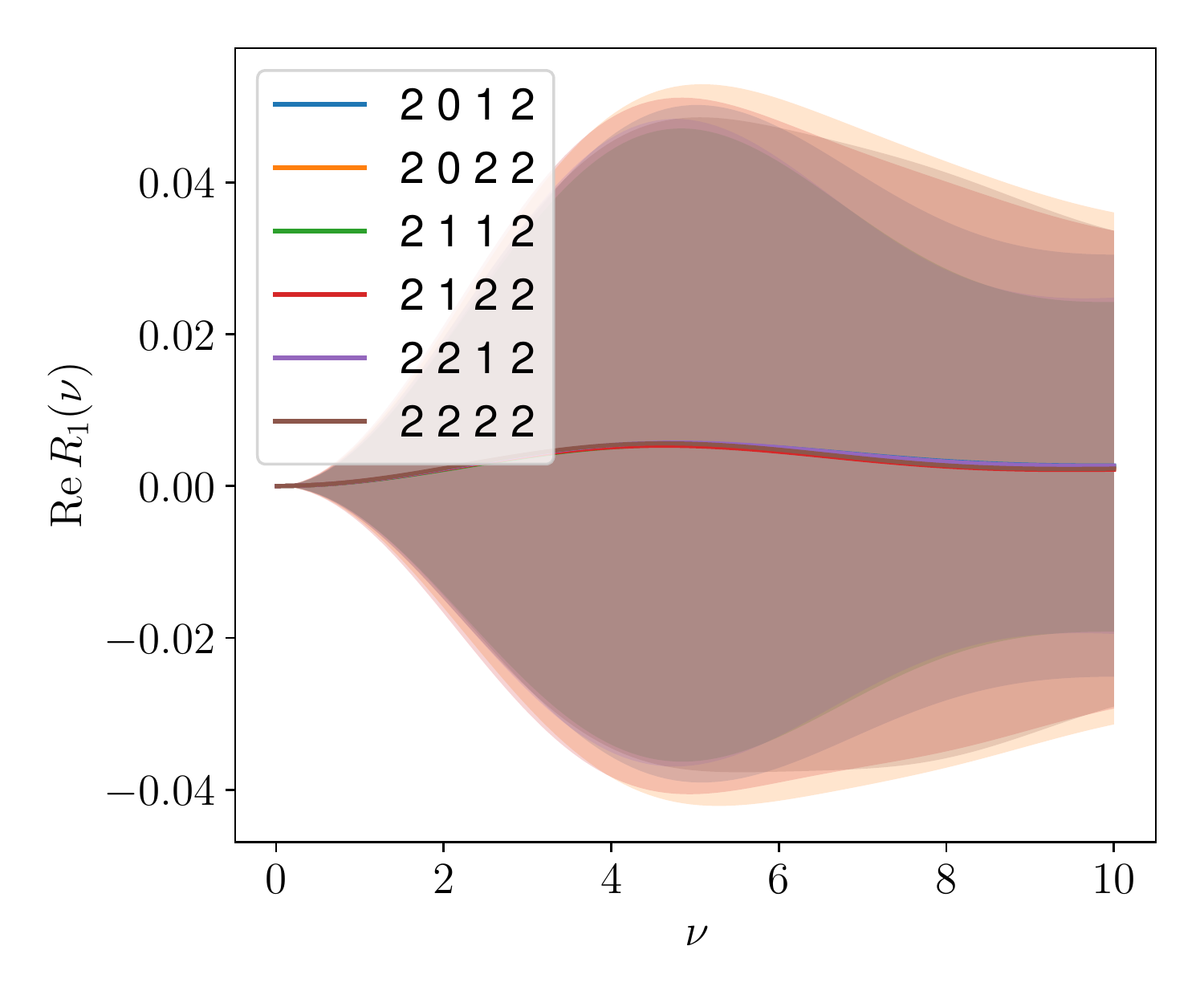}
\caption{\label{fig:many_model_oa_real1}The higher twist term, $B_1$, results from fitting the real component to the models. The numbers in the legend correspond to $(N_\pm,  N_{R/I,b}, N_{R/I,r}, N_{R/I,p})$.}
\end{figure}

\begin{figure}[!htp]
\centering
\includegraphics[width=0.48\textwidth]{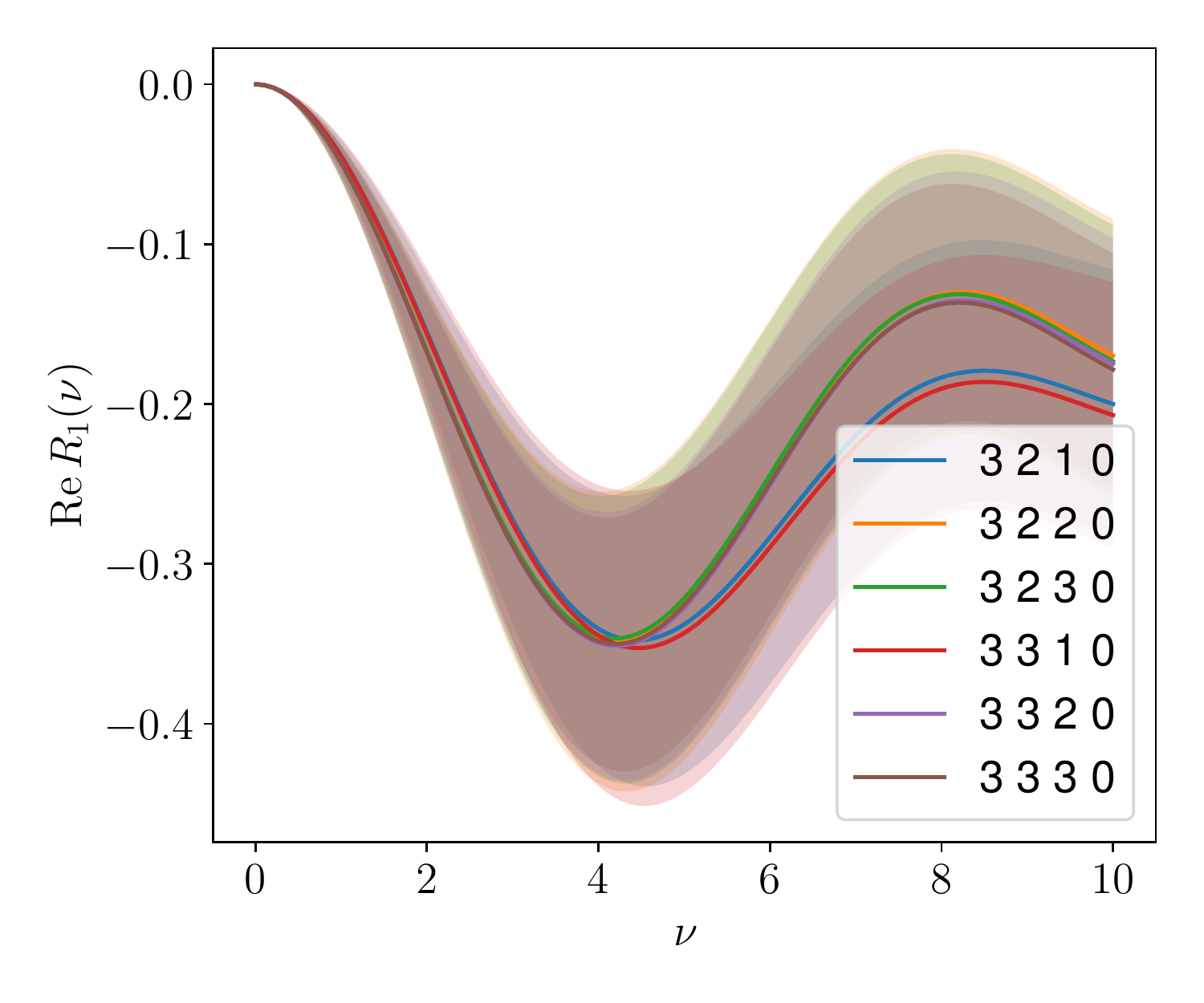}
\includegraphics[width=0.48\textwidth]{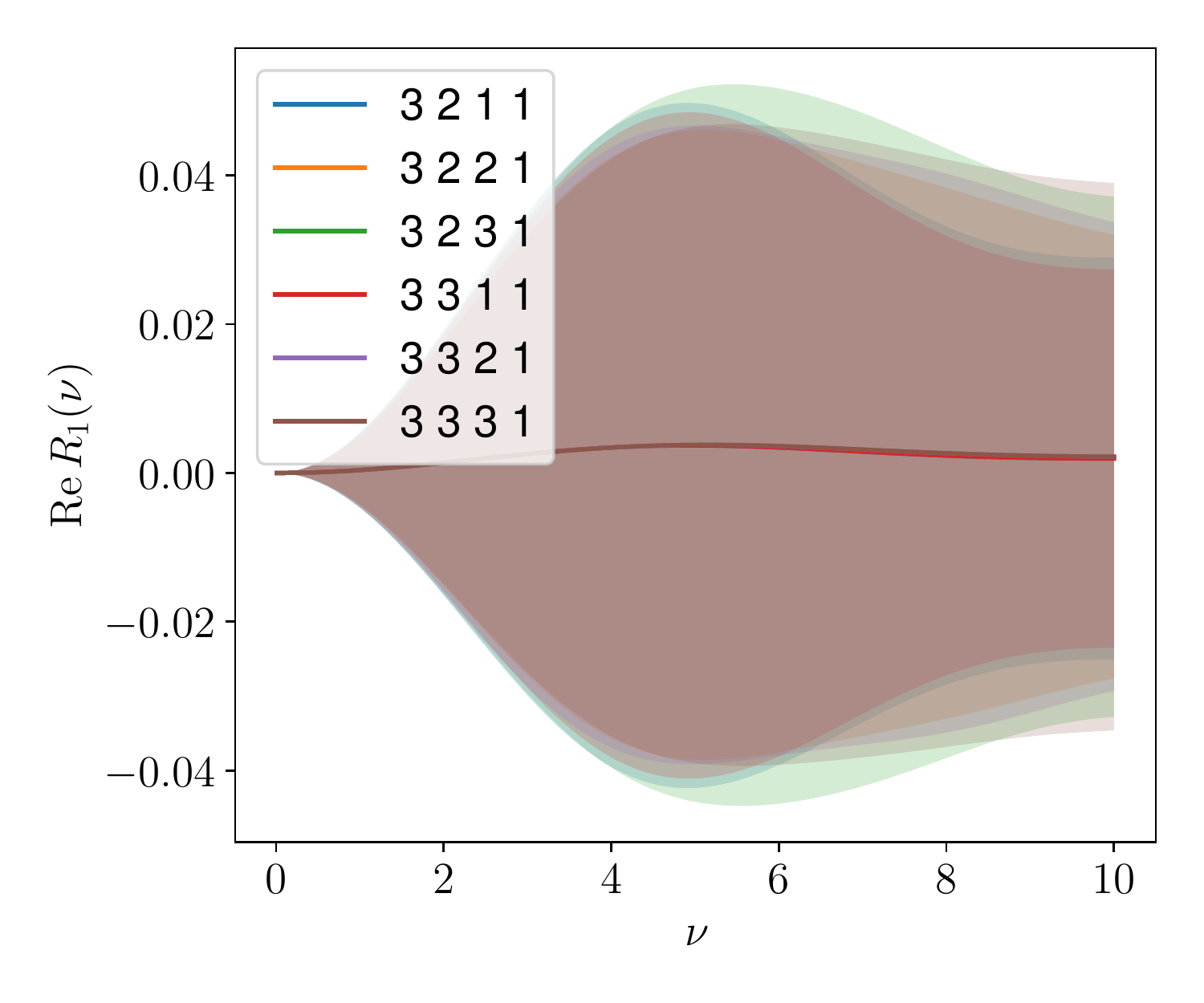}
\includegraphics[width=0.48\textwidth]{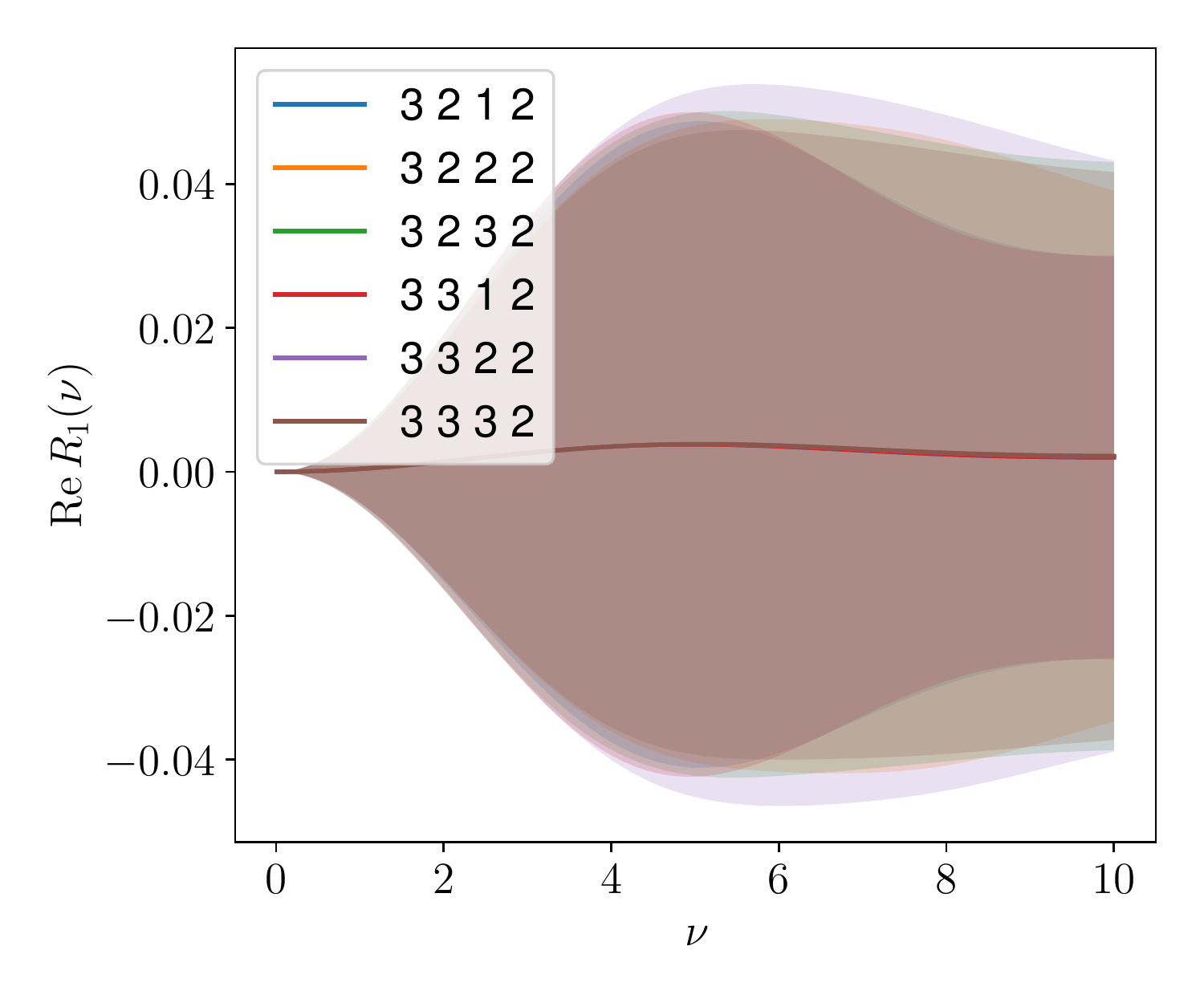}
\includegraphics[width=0.48\textwidth]{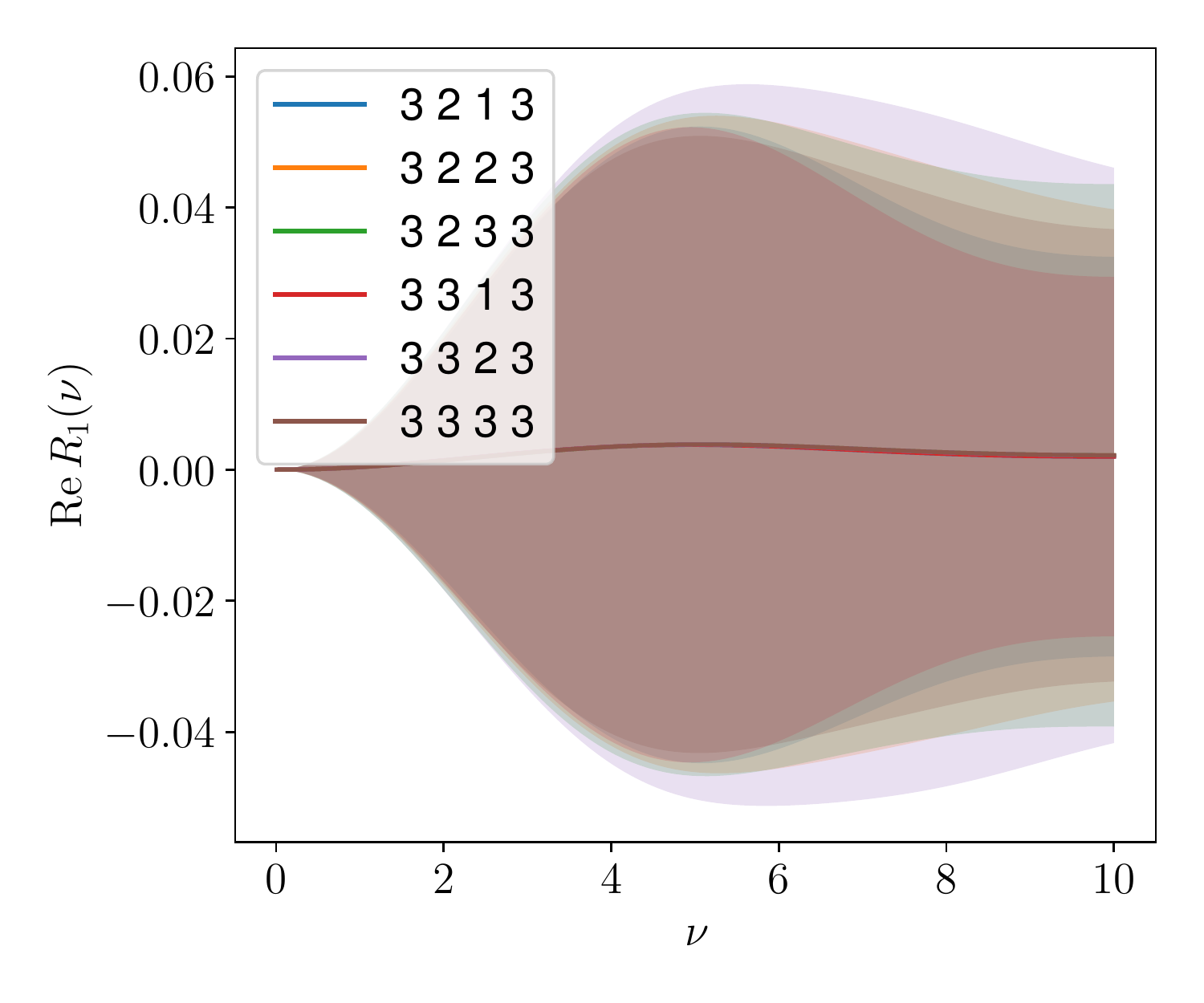}
\caption{\label{fig:many_model_oa_real2}The higher twist term, $B_1$, results from fitting the real component to the models. The numbers in the legend correspond to $(N_\pm,  N_{R/I,b}, N_{R/I,r}, N_{R/I,p})$.}
\end{figure}

\begin{figure}[!htp]
\centering
\includegraphics[width=0.48\textwidth]{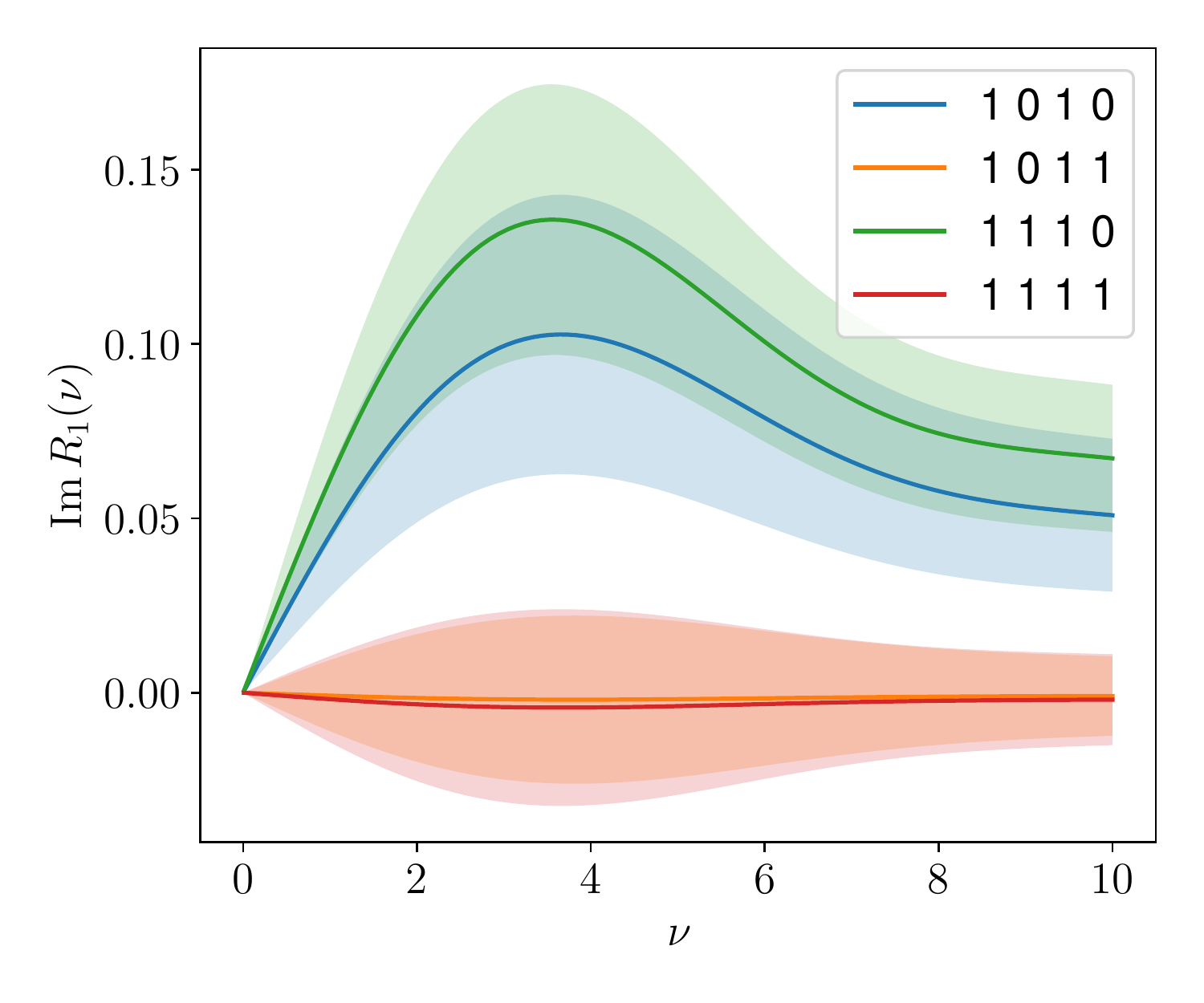}
\includegraphics[width=0.48\textwidth]{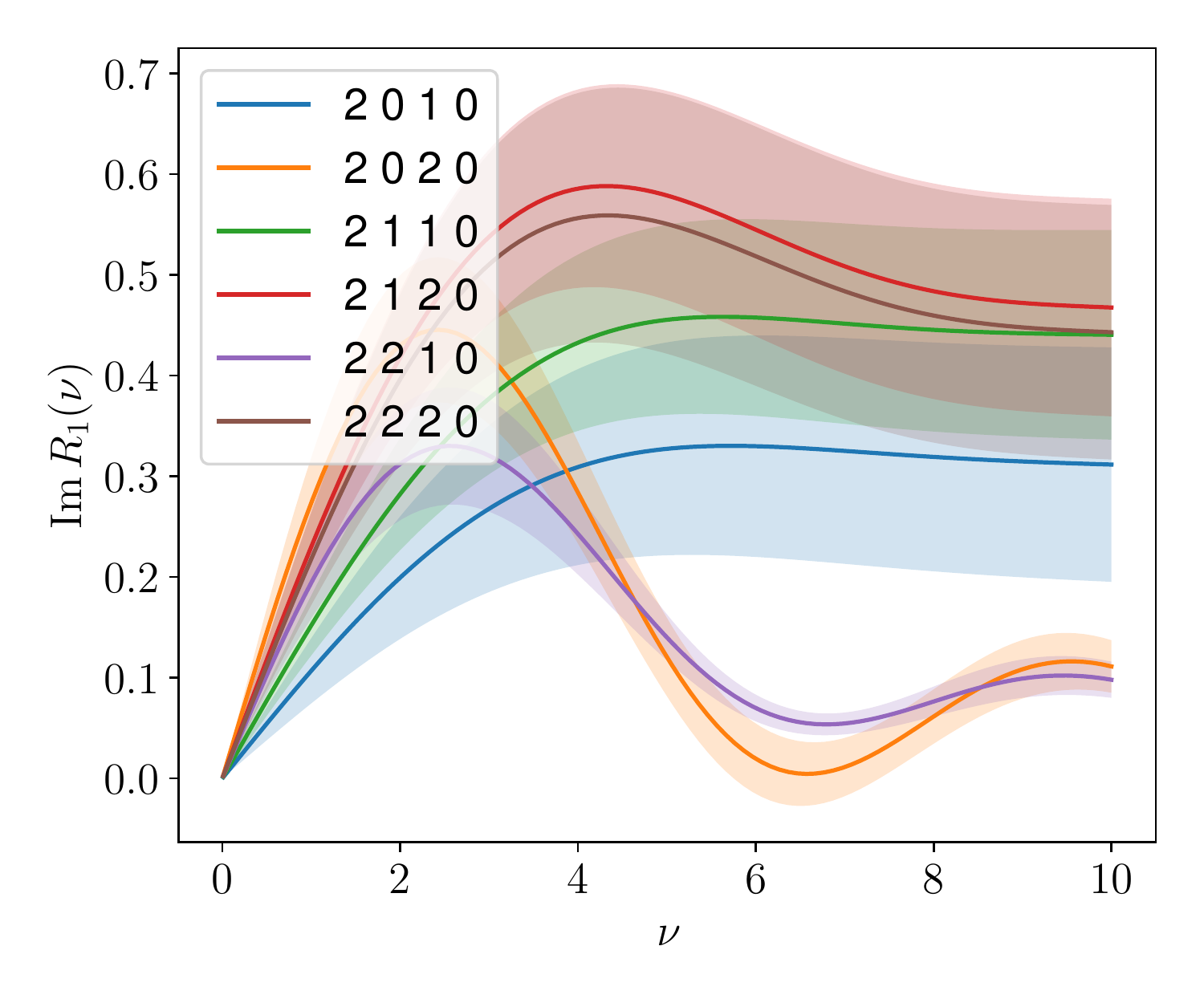}
\includegraphics[width=0.48\textwidth]{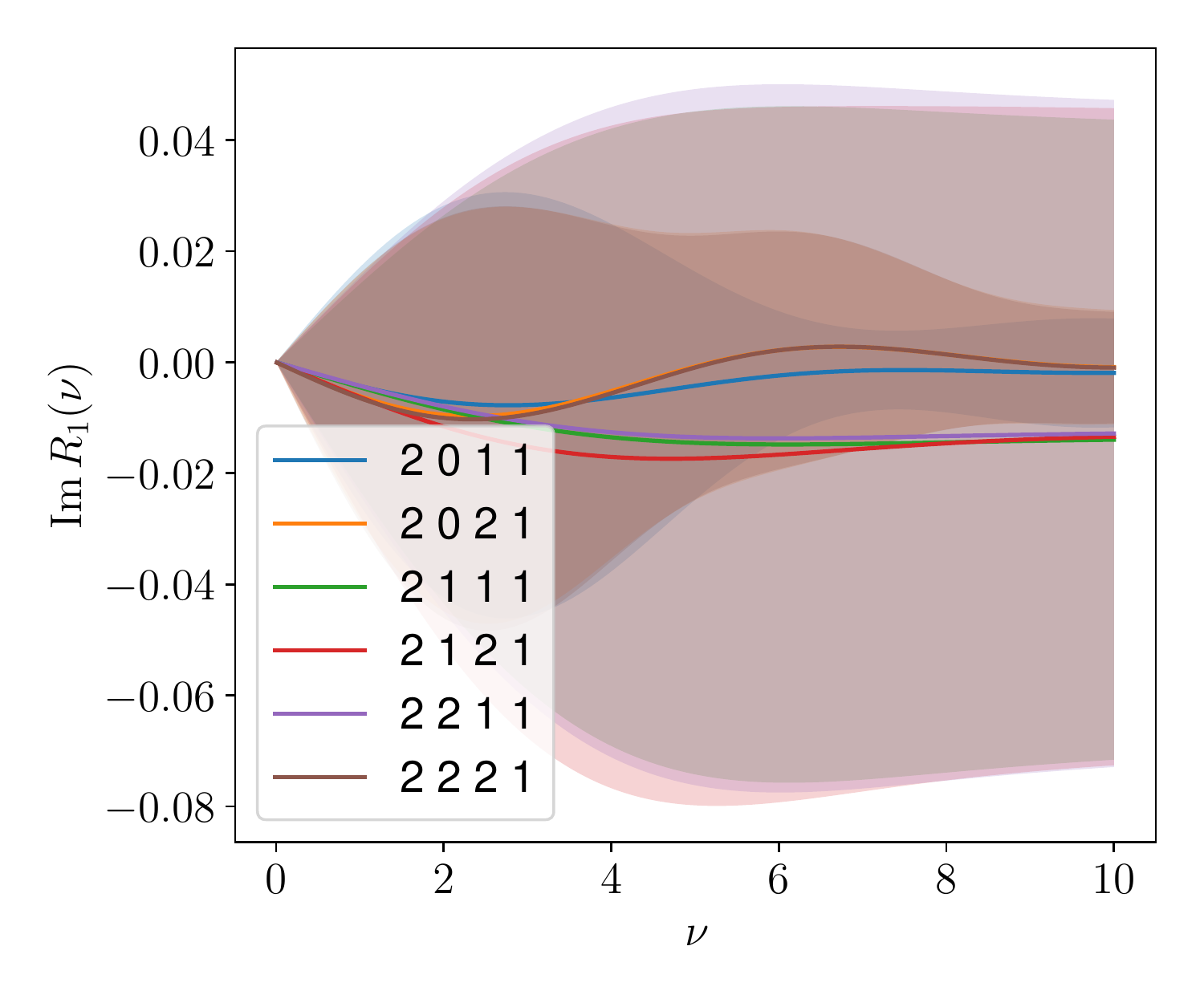}
\includegraphics[width=0.48\textwidth]{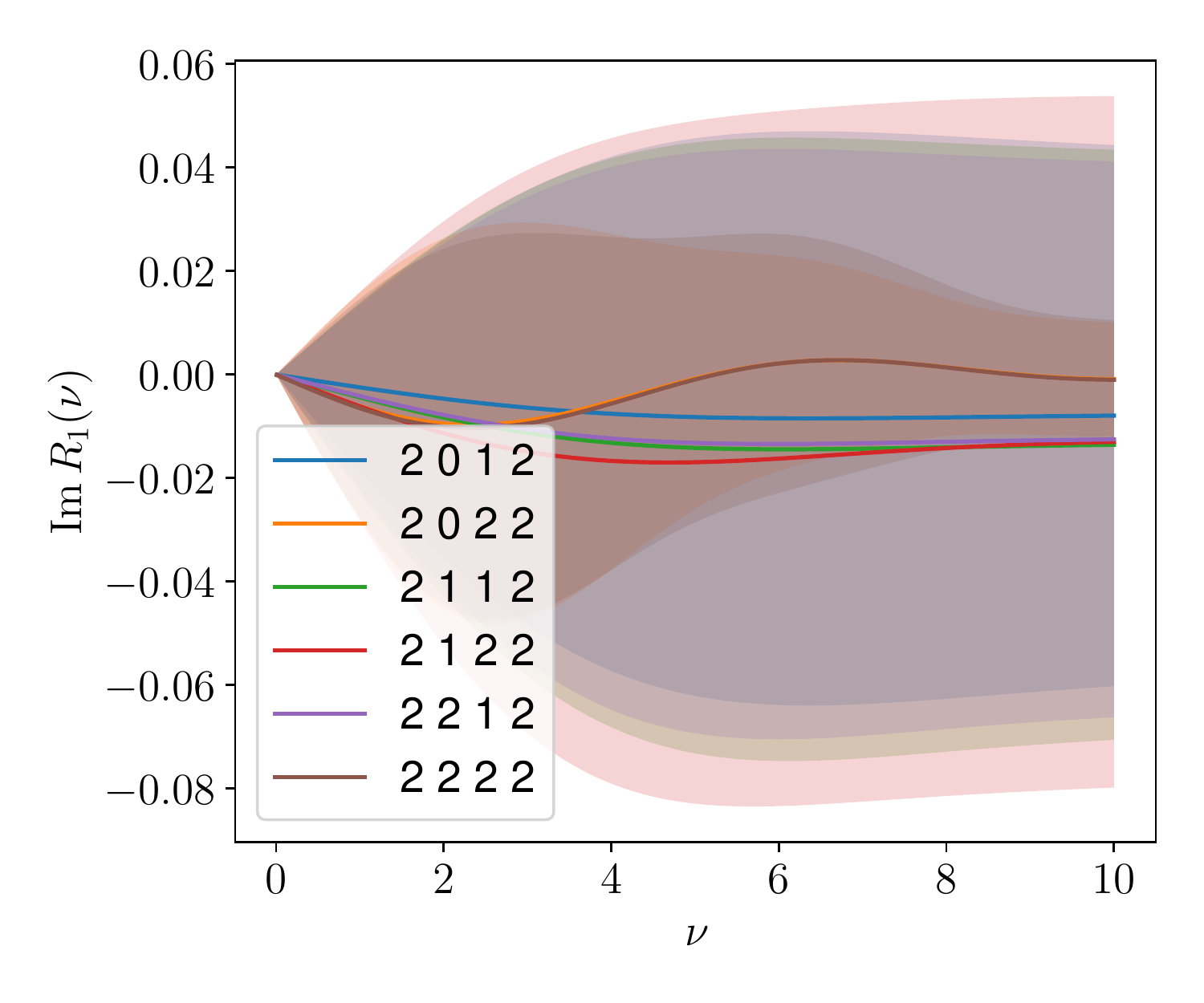}
\caption{\label{fig:many_model_oa_imag1}The higher twist term, $B_1$, results from fitting the imaginary component to the models. The numbers in the legend correspond to $(N_\pm,  N_{R/I,b}, N_{R/I,r}, N_{R/I,p})$.}
\end{figure}

\begin{figure}[!htp]
\centering
\includegraphics[width=0.48\textwidth]{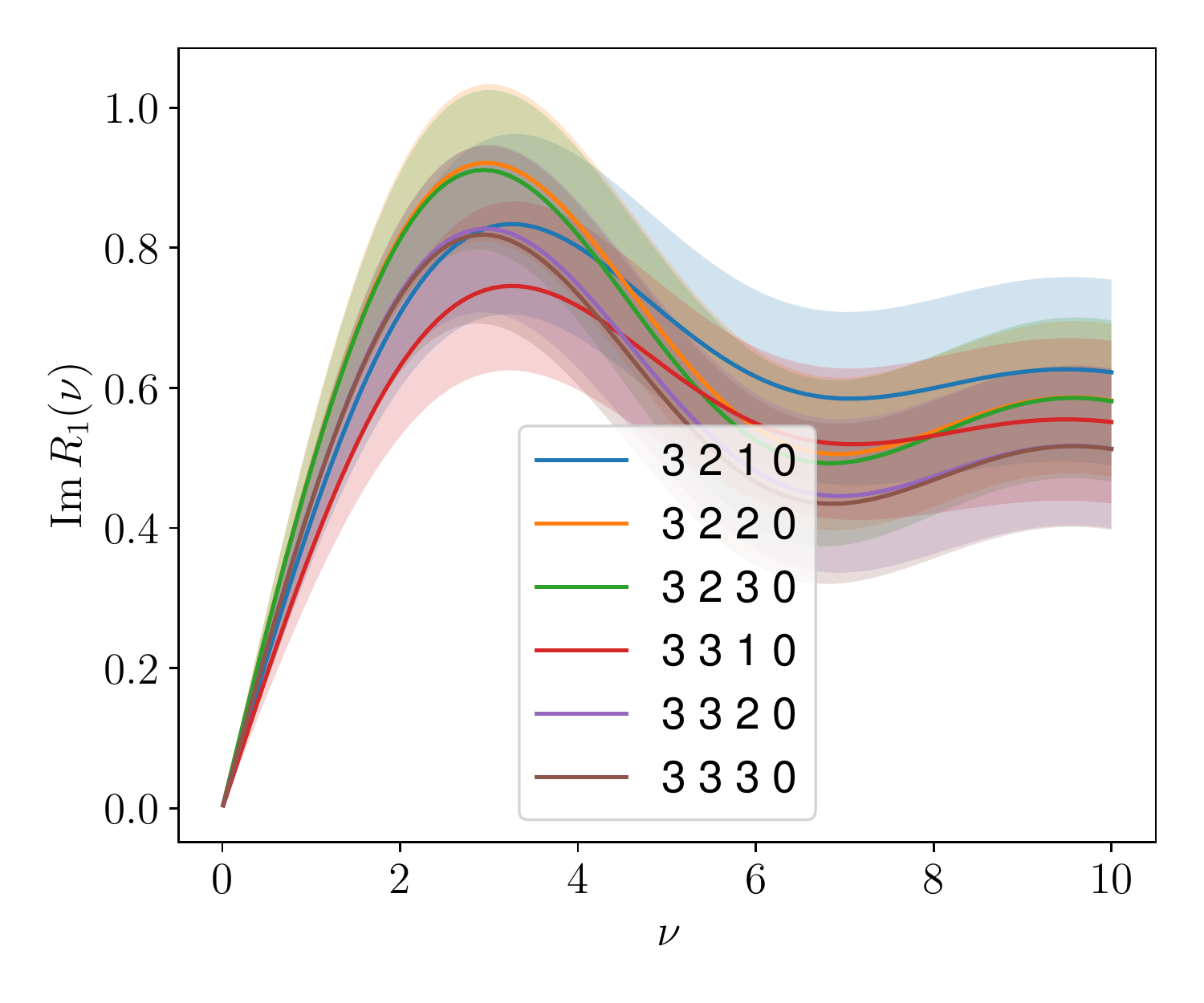}
\includegraphics[width=0.48\textwidth]{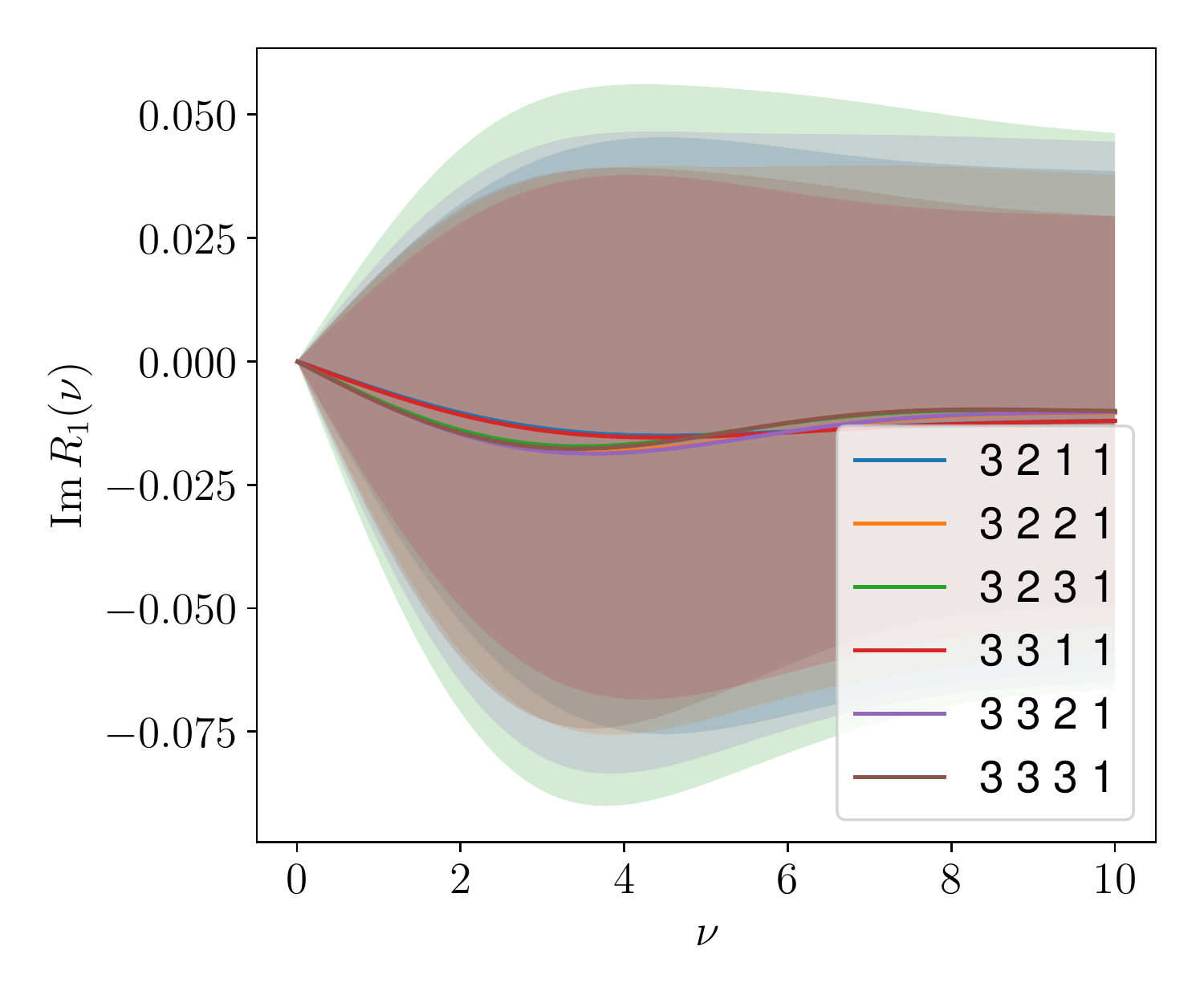}
\includegraphics[width=0.48\textwidth]{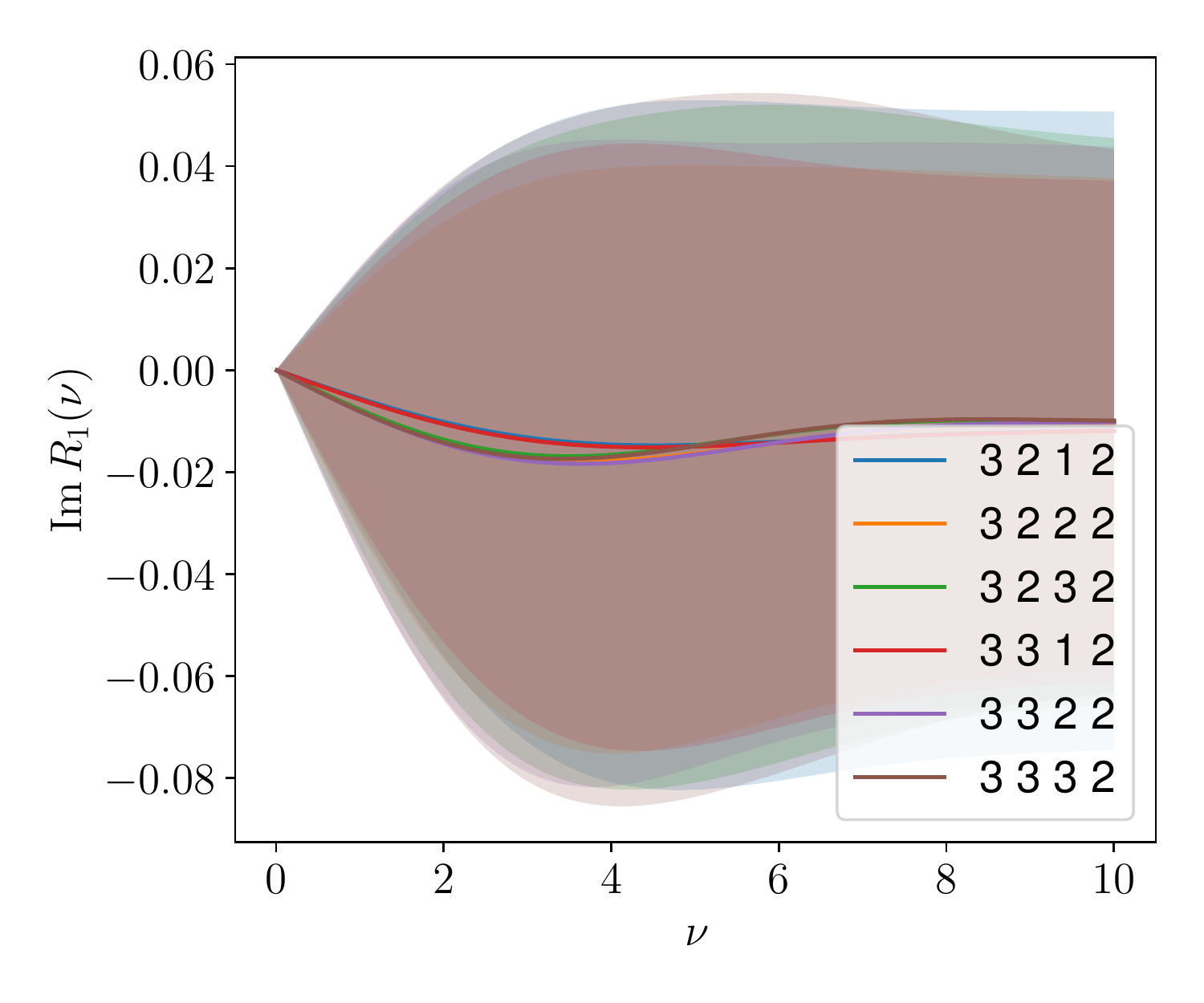}
\includegraphics[width=0.48\textwidth]{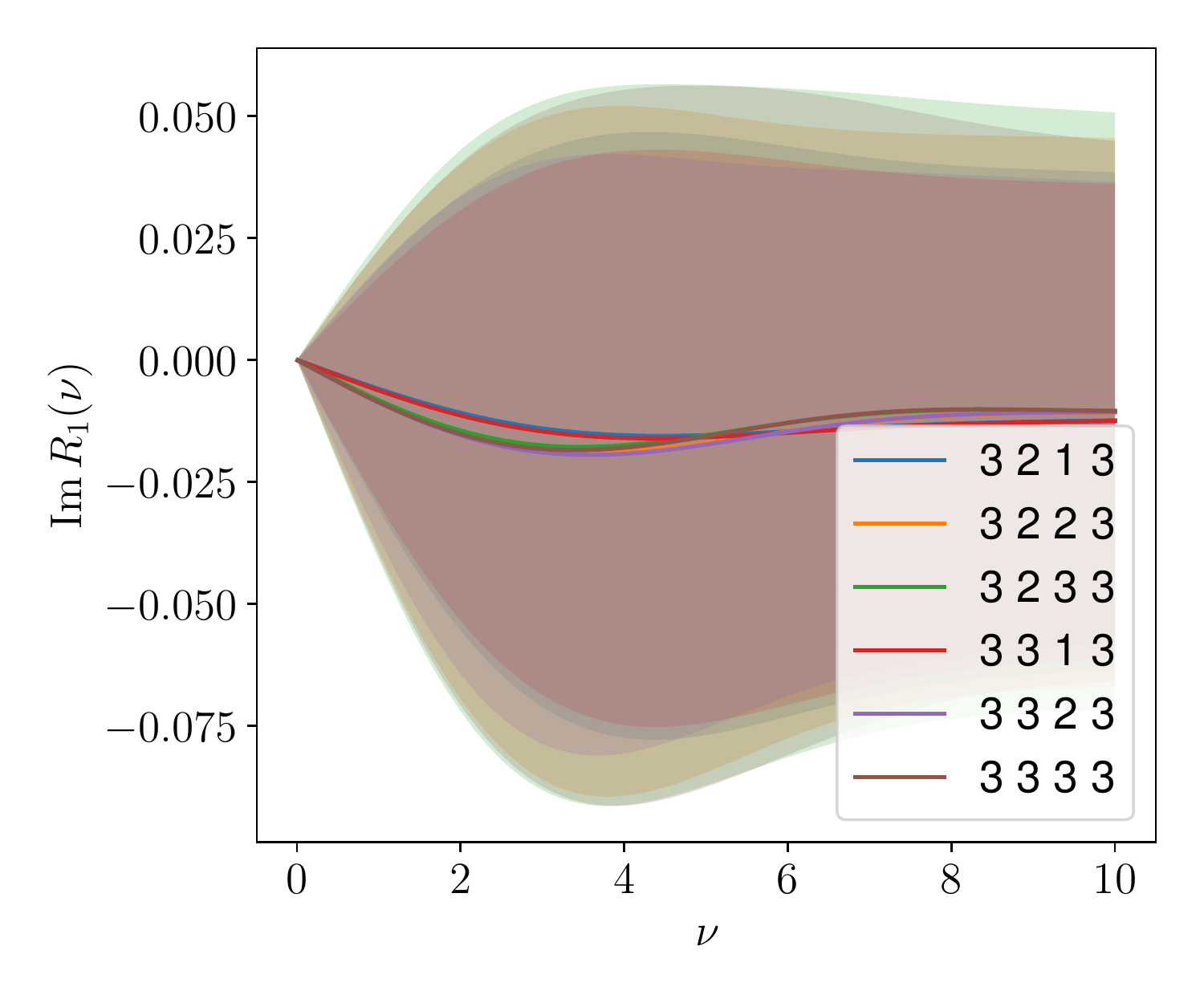}
\caption{\label{fig:many_model_oa_imag2}The higher twist term, $B_1$, results from fitting the imaginary component to the models. The numbers in the legend correspond to $(N_\pm,  N_{R/I,b}, N_{R/I,r}, N_{R/I,p})$.}
\end{figure}

\begin{figure}[!htp]
\centering
\includegraphics[width=0.48\textwidth]{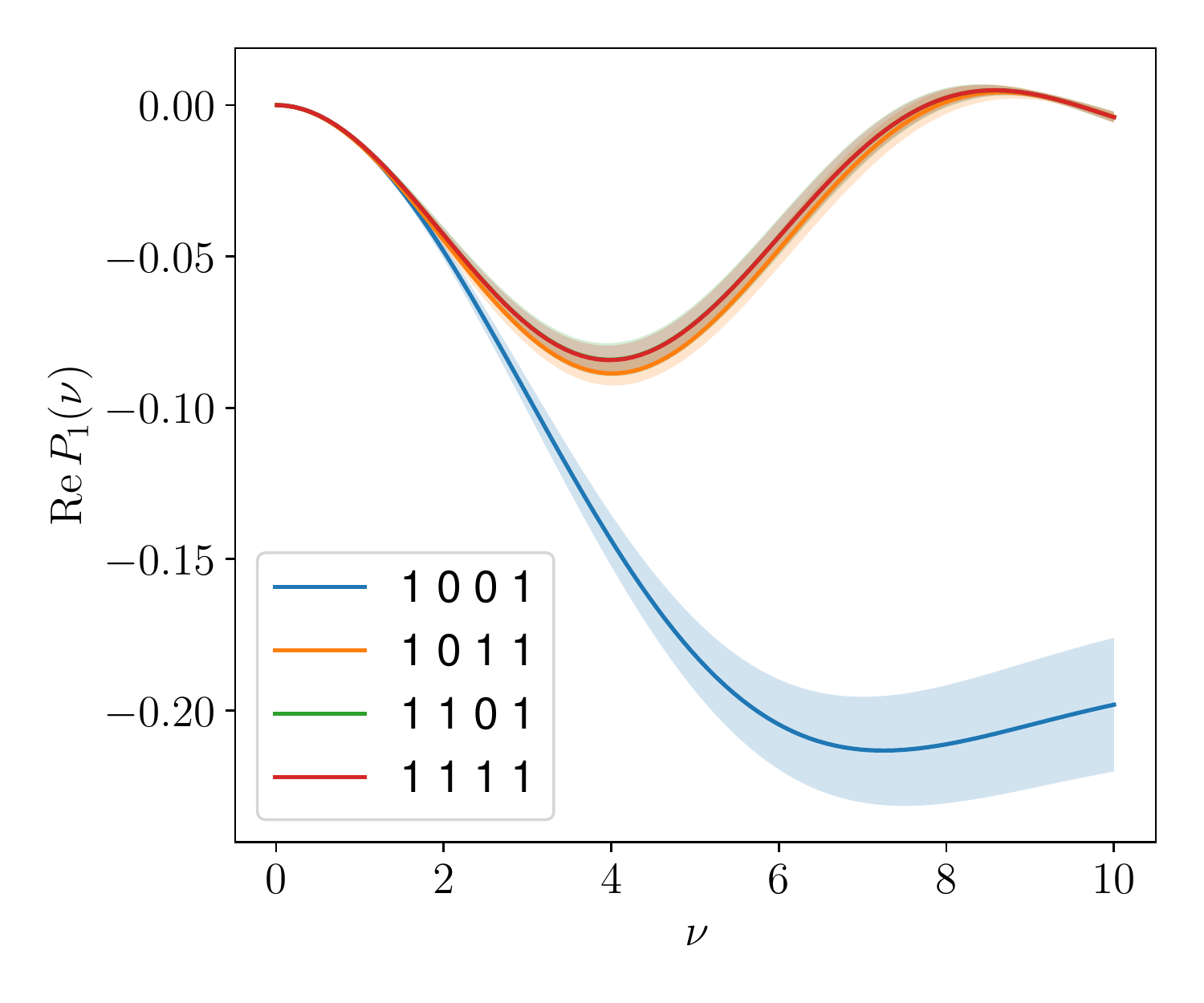}
\includegraphics[width=0.48\textwidth]{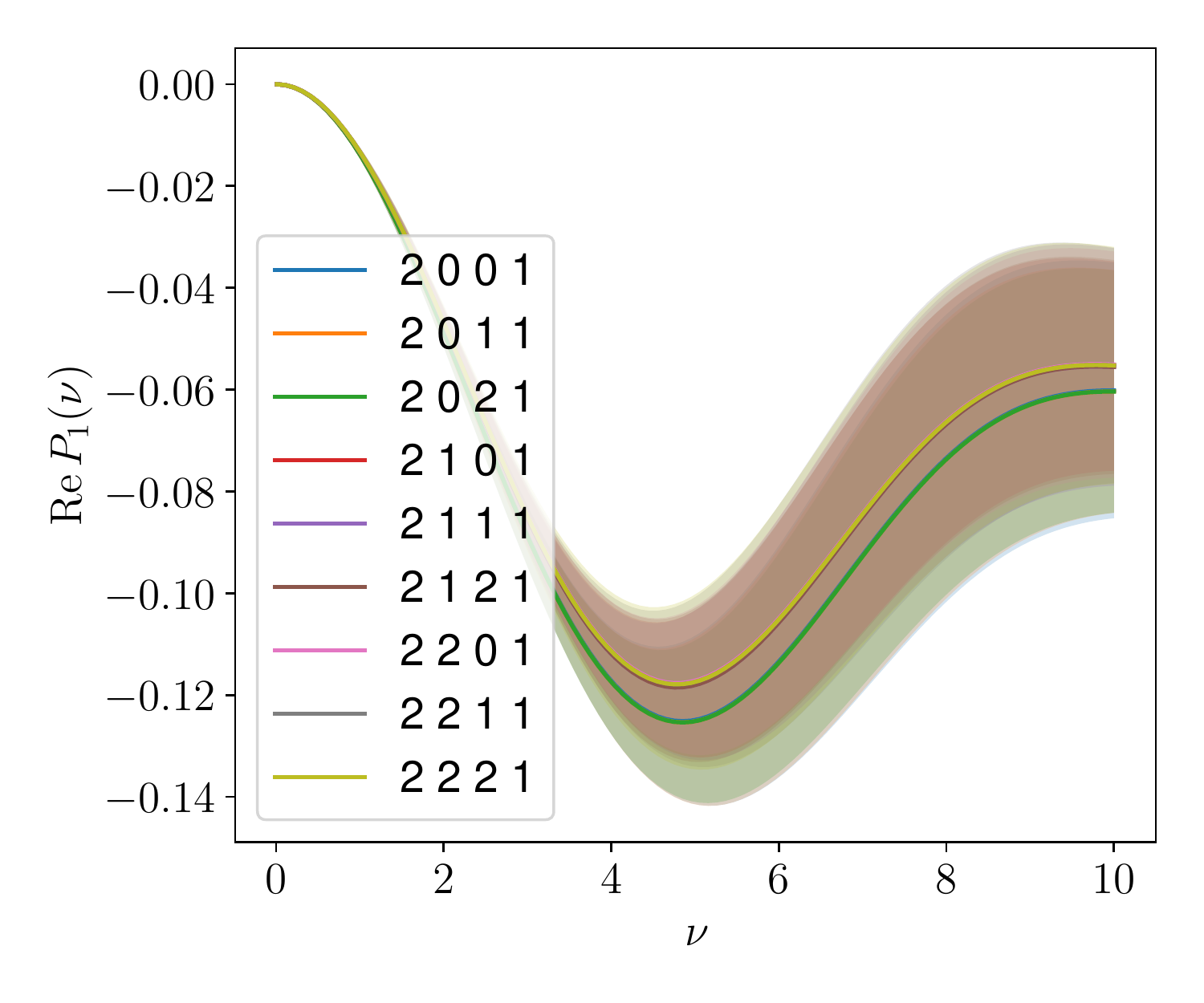}
\includegraphics[width=0.48\textwidth]{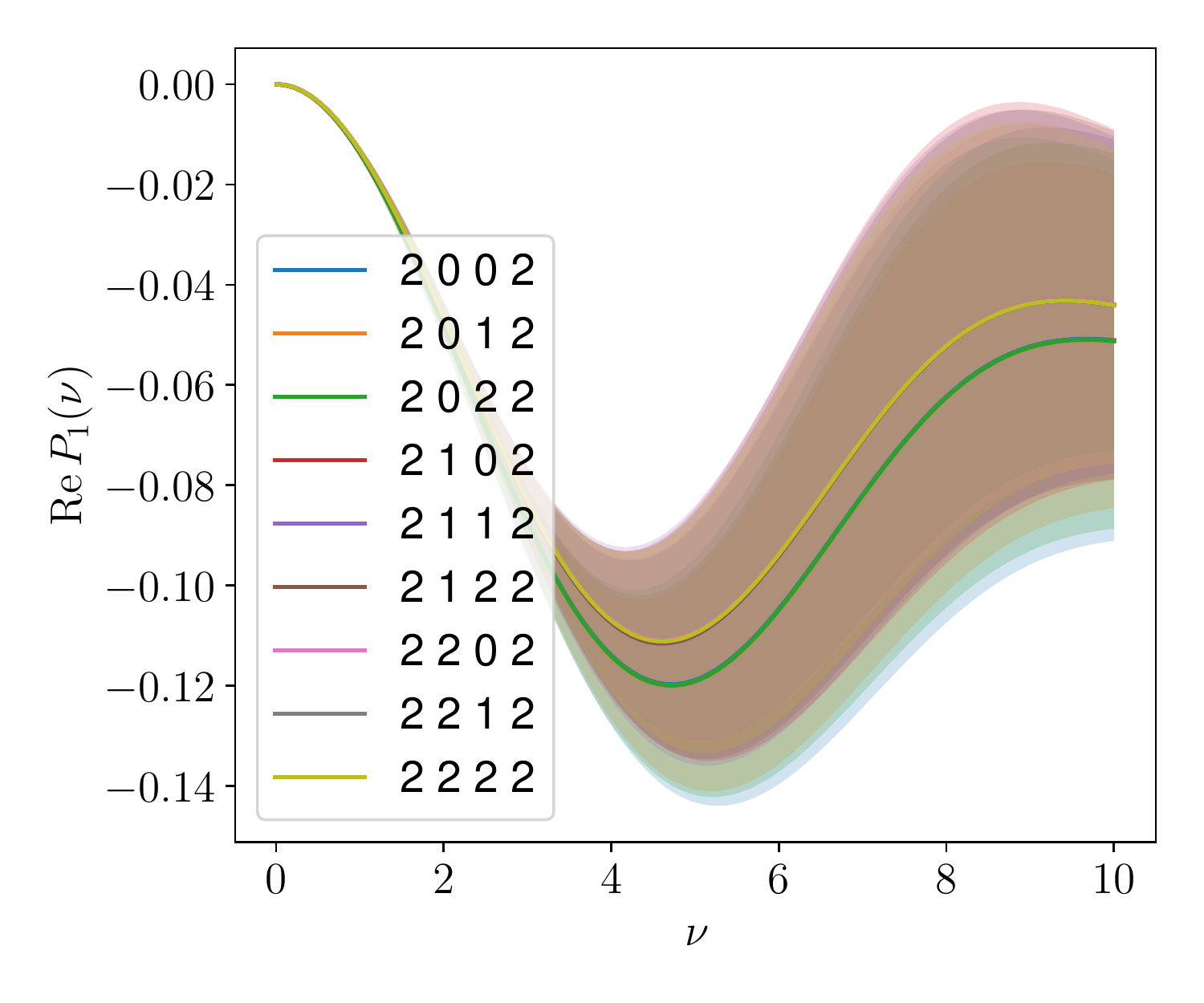}
\includegraphics[width=0.48\textwidth]{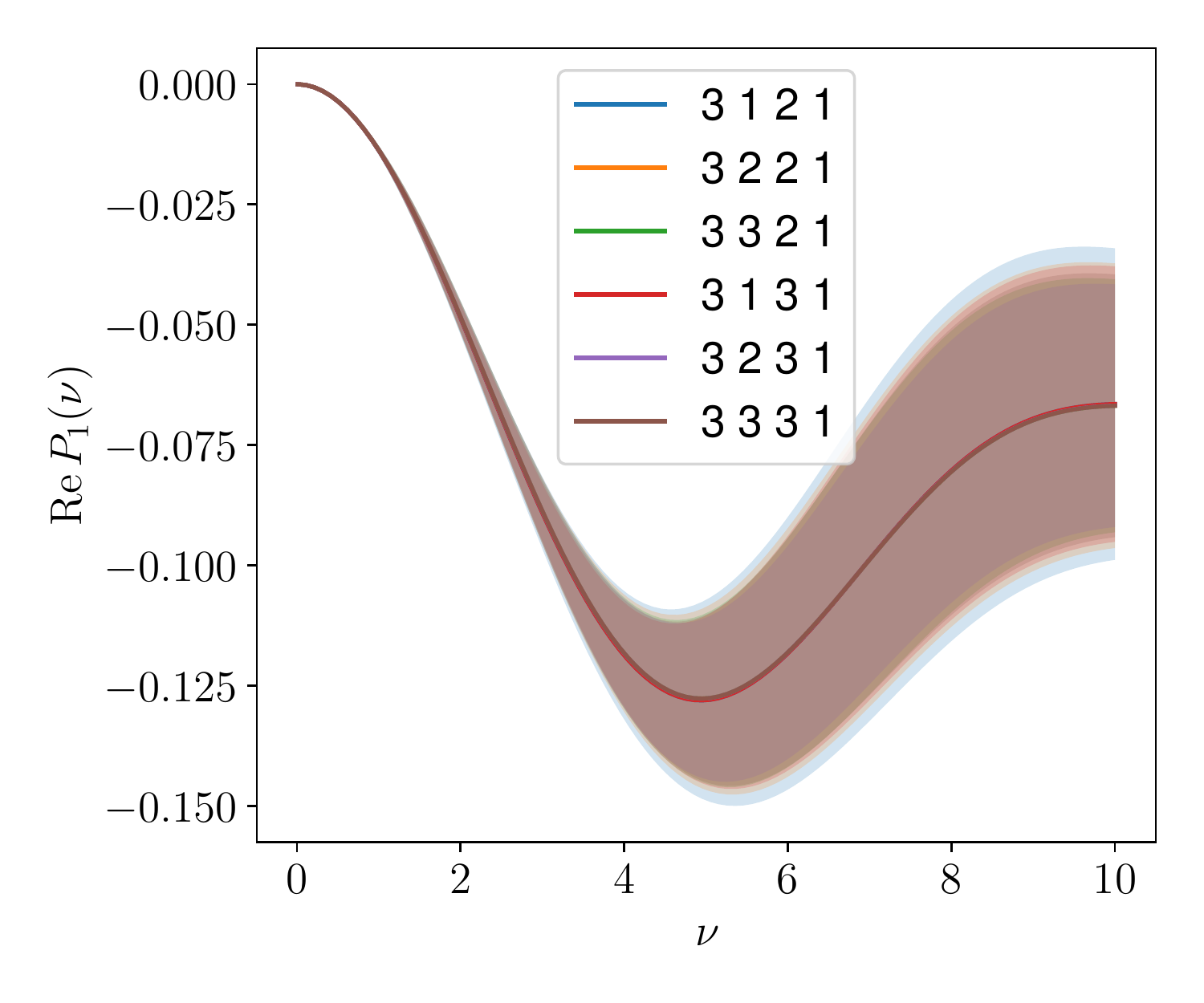}
\includegraphics[width=0.48\textwidth]{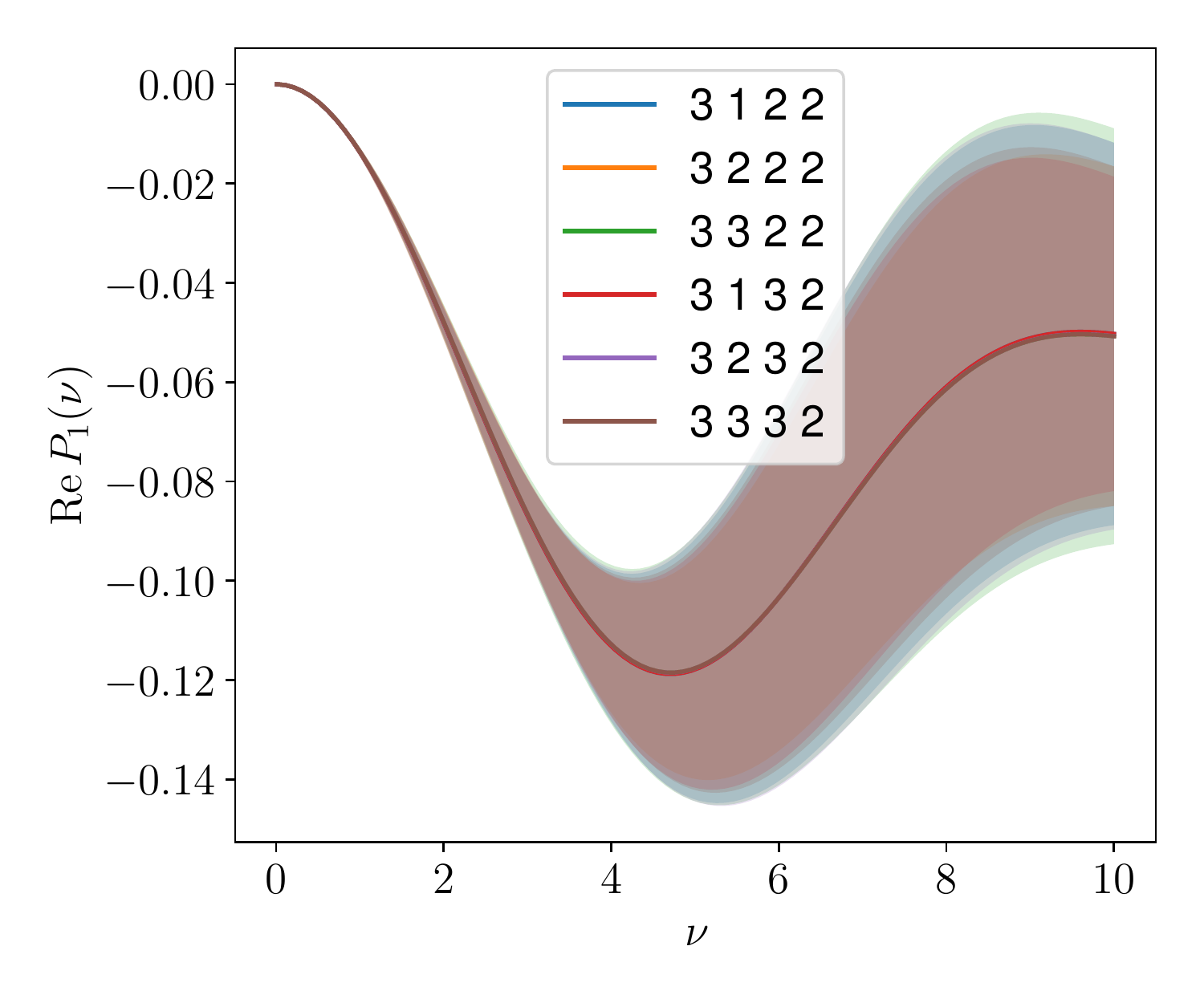}
\includegraphics[width=0.48\textwidth]{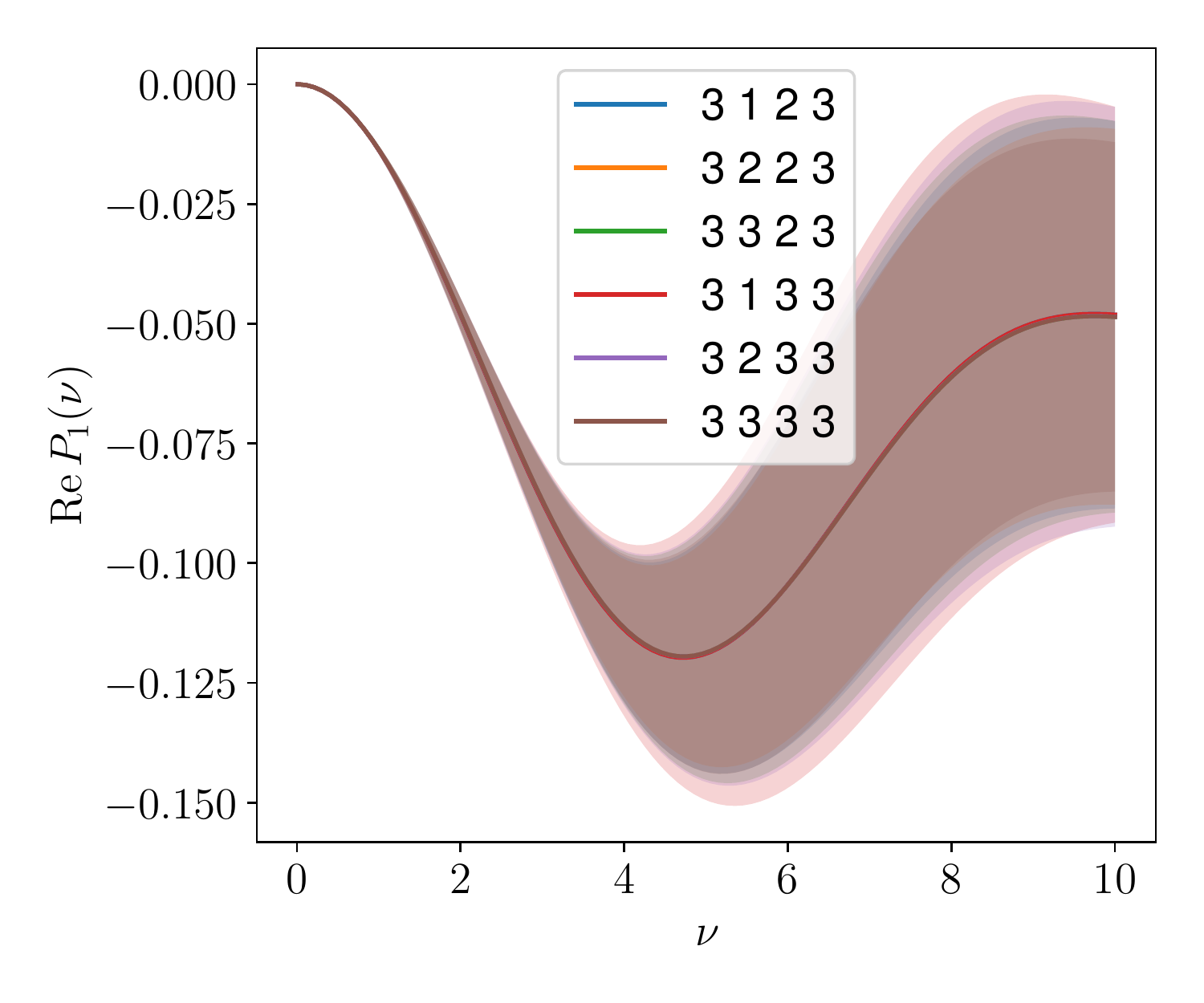}
\caption{\label{fig:many_model_az_real2}The higher twist term, $B_1$, results from fitting the real component to the models. The numbers in the legend correspond to $(N_\pm,  N_{R/I,b}, N_{R/I,r}, N_{R/I,p})$.}
\end{figure}

\begin{figure}[!htp]
\centering
\includegraphics[width=0.48\textwidth]{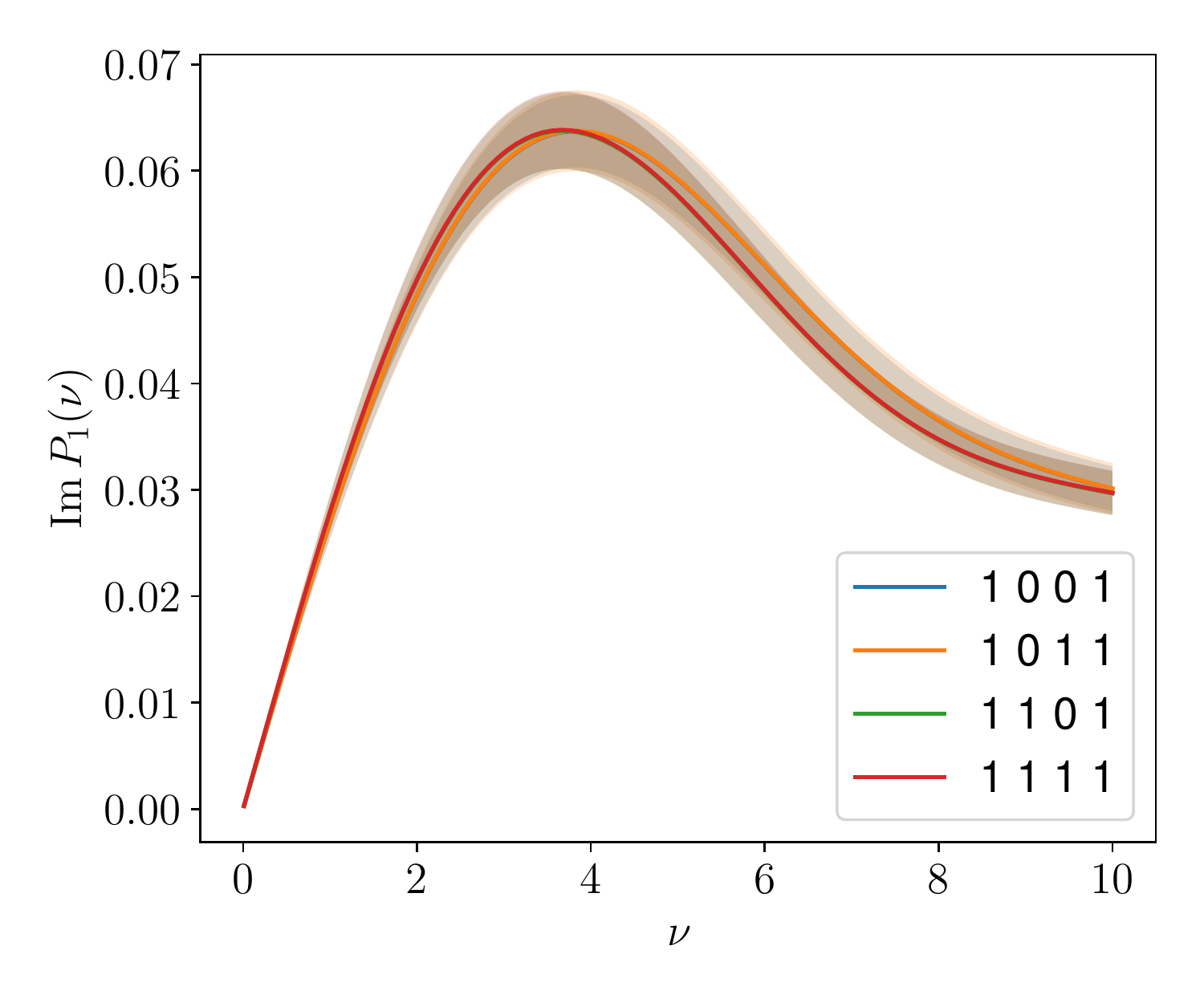}
\includegraphics[width=0.48\textwidth]{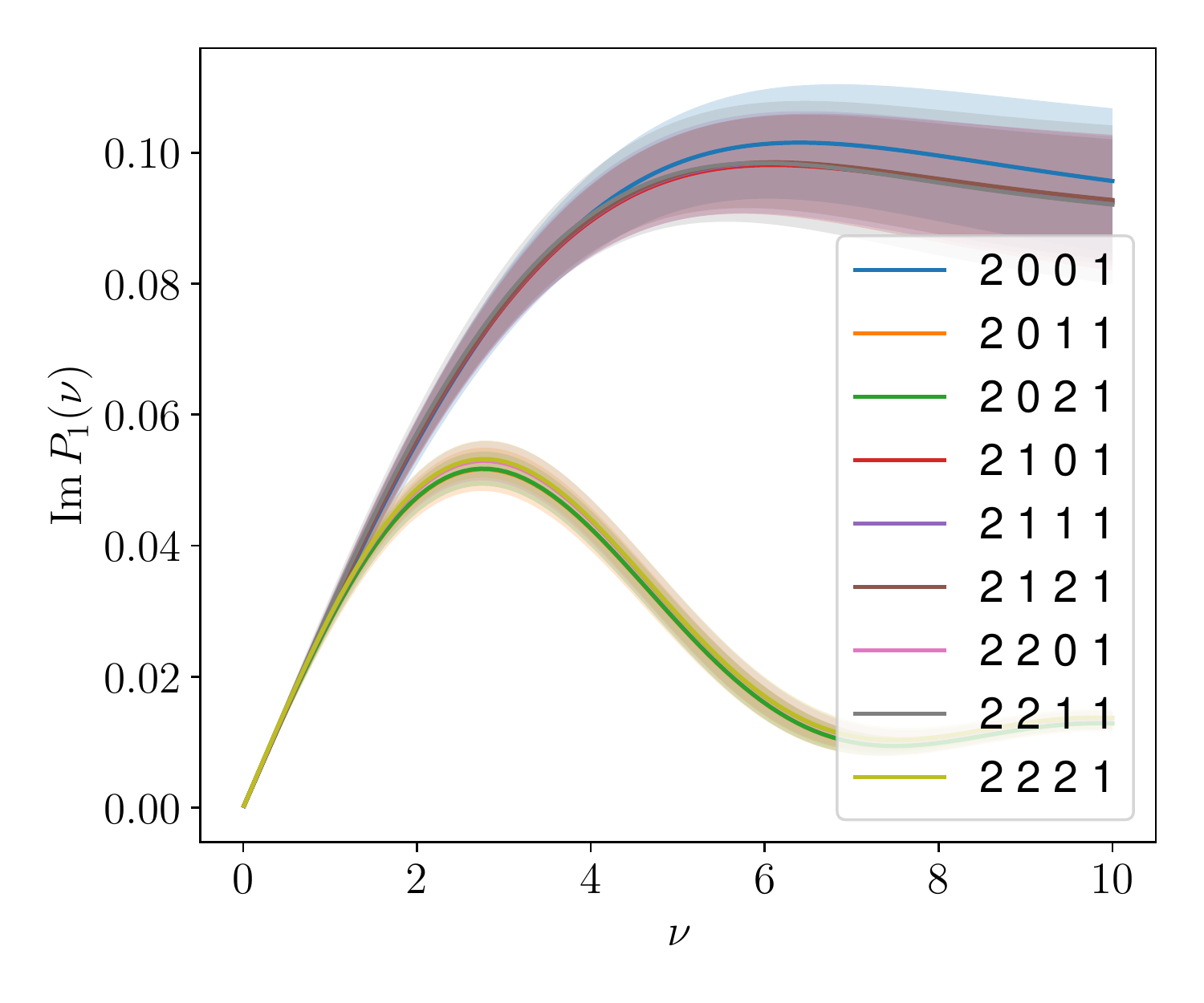}
\includegraphics[width=0.48\textwidth]{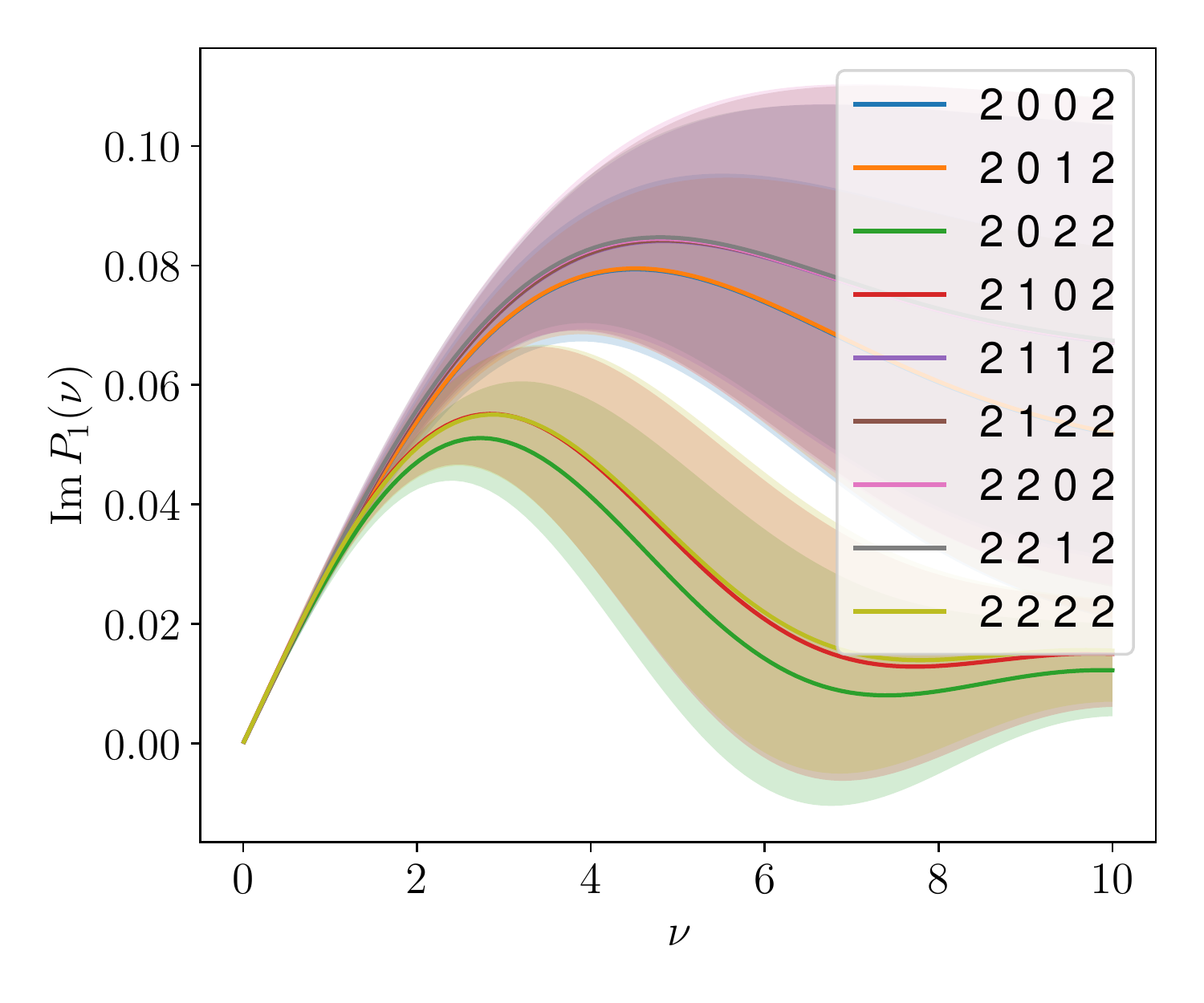}
\includegraphics[width=0.48\textwidth]{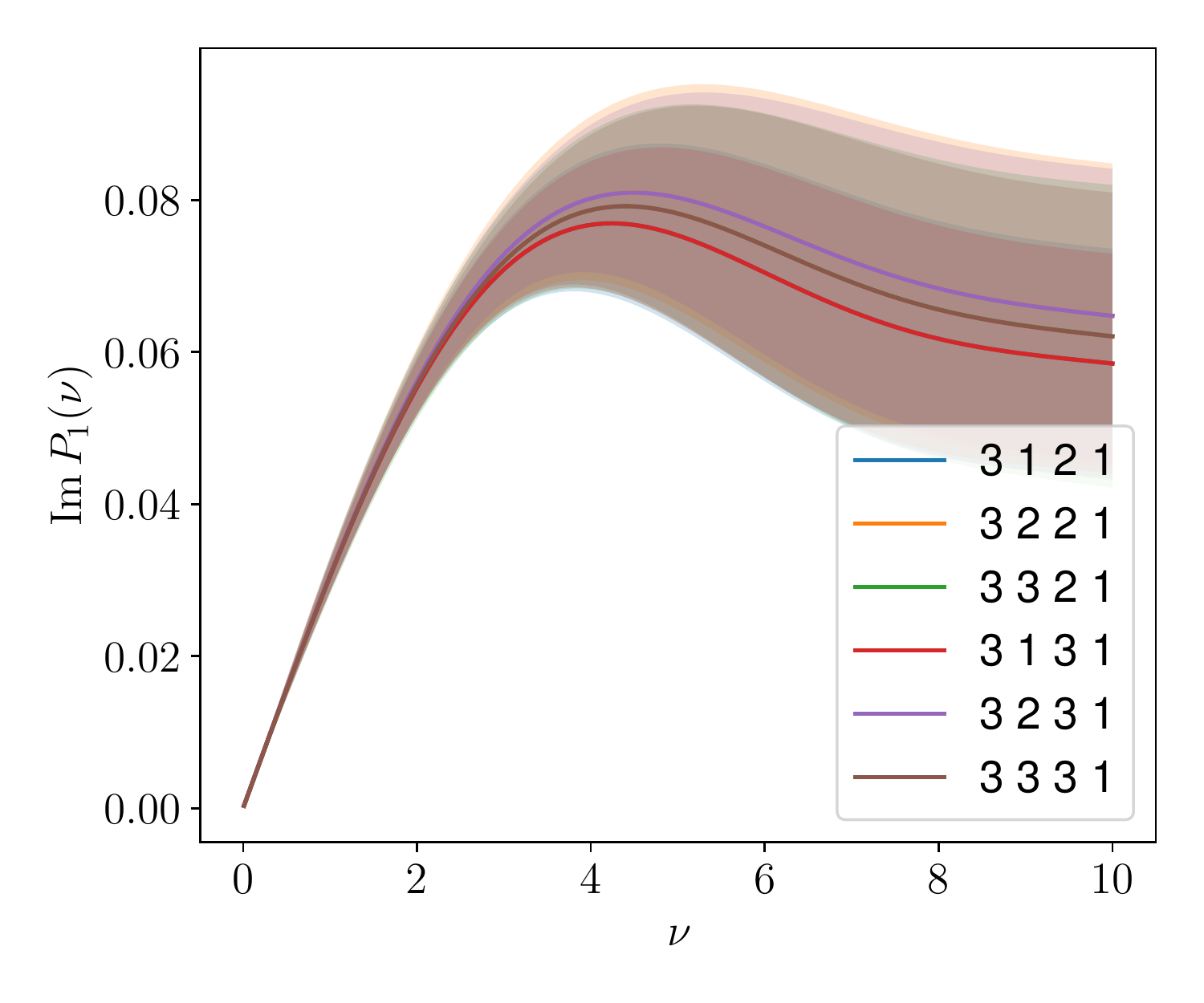}
\includegraphics[width=0.48\textwidth]{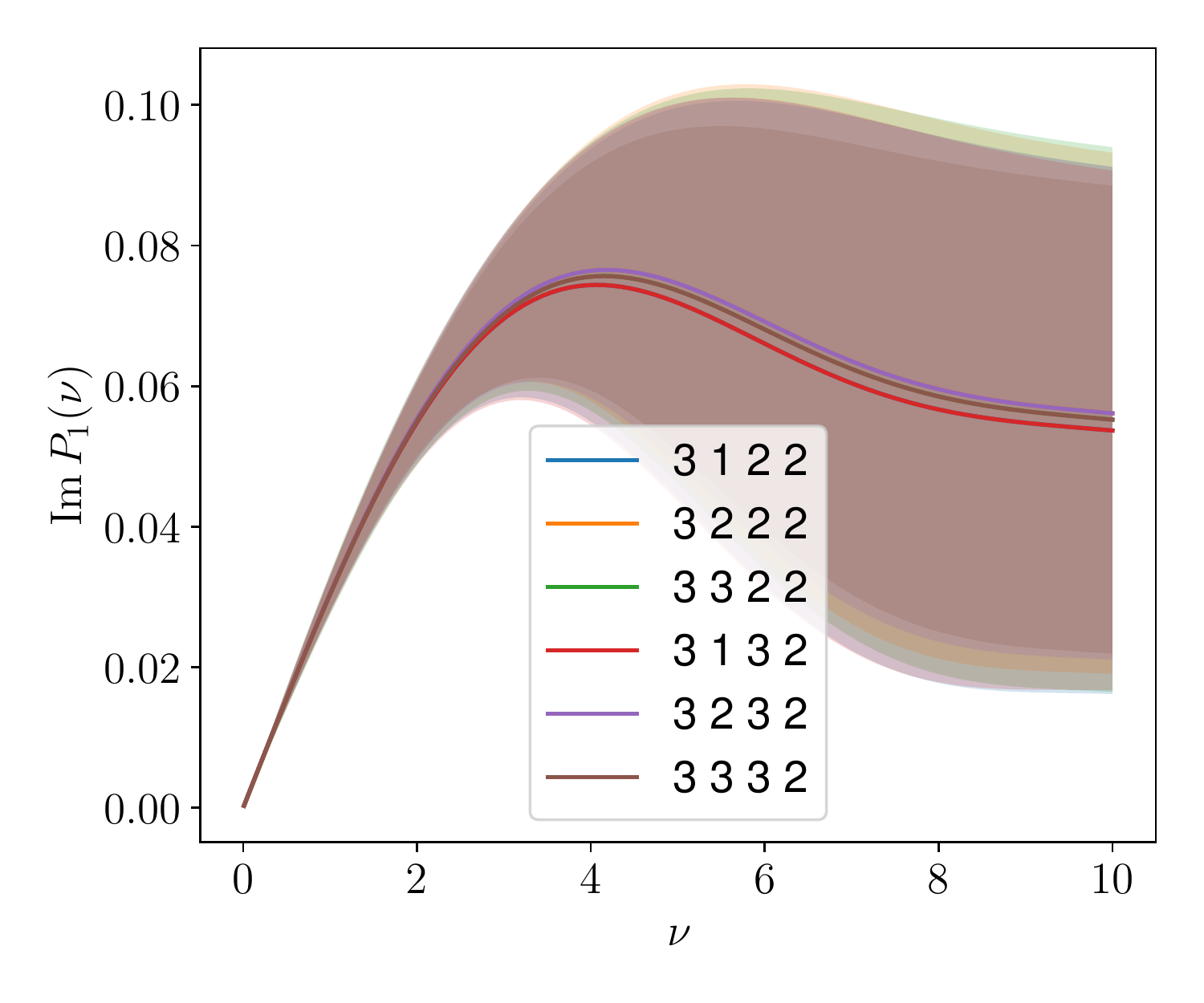}
\includegraphics[width=0.48\textwidth]{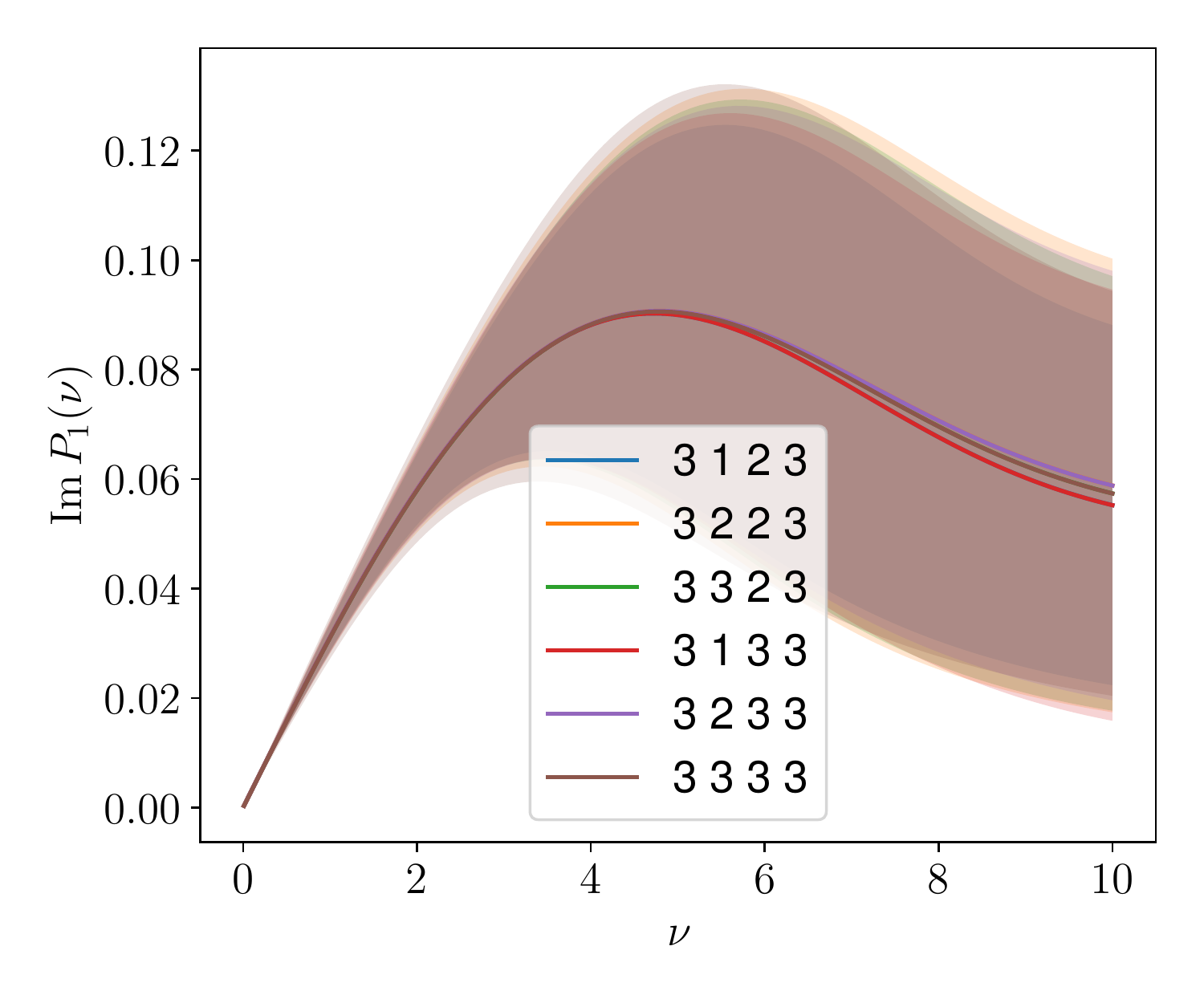}
\caption{\label{fig:many_model_az_imag2}The higher twist term, $B_1$, results from fitting the imaginary component to the models. The numbers in the legend correspond to $(N_\pm,  N_{R/I,b}, N_{R/I,r}, N_{R/I,p})$.}
\end{figure}

The nuisance terms are shown in Figs.~\ref{fig:many_model_ht_real1}-\ref{fig:many_model_az_imag2}. The $O({a}/{z})$ term $P_1$ also appears to grow to a peak around $\nu\sim 4$ and either plateaus or goes to zero. The location of this peak is expected from $\sigma^{(\alpha,\beta)}_1$ and $\eta^{(\alpha,\beta)}_1$ when $N_{R/I,p}=1$ but is robust even for $N_{R/I,p}=2,3$. This peak or plateau is negative for the real component and positive for the imaginary component. The $R_1$ term is only ever non-zero when there is no $P_1$ term. This implies that in those models it is attempting to compensate for the absence of $P_1$.

For the majority of the models, the higher twist term $B_1$ appears to grow to a peak around $\nu\sim4$ and either plateaus or shrinks slightly. Similarly to $P_1$ the location of this peak is expected from the shape of $\sigma^{(\alpha,\beta)}_1$ and $\eta^{(\alpha,\beta)}_1$ when $N_{R/I,p}=1$. For a wide range of models, the size of the $B_1$ term is smaller than 0.15, sometimes even less than 0.1. The naive expectation of the higher twist contribution was that the coefficient of $\Lambda_{\rm QCD}^2 z^2$ would be order 1. This result implies that even larger $z^2$ could be safely used. One may worry about the convergence of the higher twist sum and the size of the neglected twist 6 and higher terms. Since those are not included in the fit, the effects of those terms are being accumulated, if imperfectly, into $B_1$. For the size of $z^2$ used here, and given the fact that $B_1$ is small the twist 6 contributions the reduced matrix element must be also small.



\subsection{Model weighted averages}\label{sec:aic}
There exist a number of methods for combining results from different models in order to create an average. In this study, we utilize the Akaike Information Criterion (AIC)~\cite{aic} for this goal. The AIC is given by $a_i = 2k_i + 2L_i^2$, where $k_i$ is the number of parameters in the $i^{\rm th}$ model and $L_i^2$ is the negative log of the posterior probability distribution for that model. It should be noted that $L_i^2$ differs from the $L^2$ used previously by the proper normalization of the likelihood probability and the prior distributions. If there were no prior distributions and the same data were used in the fit, then this would only be an irrelevant constant. Since our models use different numbers of parameters, each with their own prior, the total normalization factor differs and must be taken into account. When using a relatively few number of data points (n), it is common to use the corrected AIC, called AICc~\cite{aicc}, $A_i = a_i + \frac{2k(k+1)}{n-k-1}$ which approaches the AIC when $n$ becomes large. We adopt the AICc for our analysis. 

The AICc can be used to create weights for averaging results from different models. The weights can be interpreted as the relative likelihood of that given model compared to the rest. The average value from $N$ models is given by
\beq
x =\sum_{i=1}^N w_i x_i \quad ; \qquad w_i =\frac{ e^{-\frac{A_i}{2} }}{\sum_{i=1}^N e^{-\frac{A_i}{2}}}\,,
\eeq
where $x_i$ is the part of the $i^{\rm th}$ model which describes the observable $x$, e.g. the PDF or various model parameters. This weighted average combines knowledge of the likelihood of a given model alongside a factor to avoid overfitting. Ultimately, this procedure can be improved by adding models which are less related to each other than simply varying the number of terms. If sufficiently many distinct models are used, the possible biases from choices of model can be averaged away through this AICc weighted average. For example, including fits which use neural networks, which were performed for lattice PDFs in~\cite{Cichy:2019ebf,DelDebbio:2020rgv}, which likely would have distinct model dependent biases from the Jacobi polynomials fits. Unfortunately in this preliminary study which only uses Jacobi polynomial based models, the systematic errors may not be sufficiently distinct.

\begin{figure}[!htp]
\centering
\includegraphics[width=0.48\textwidth]{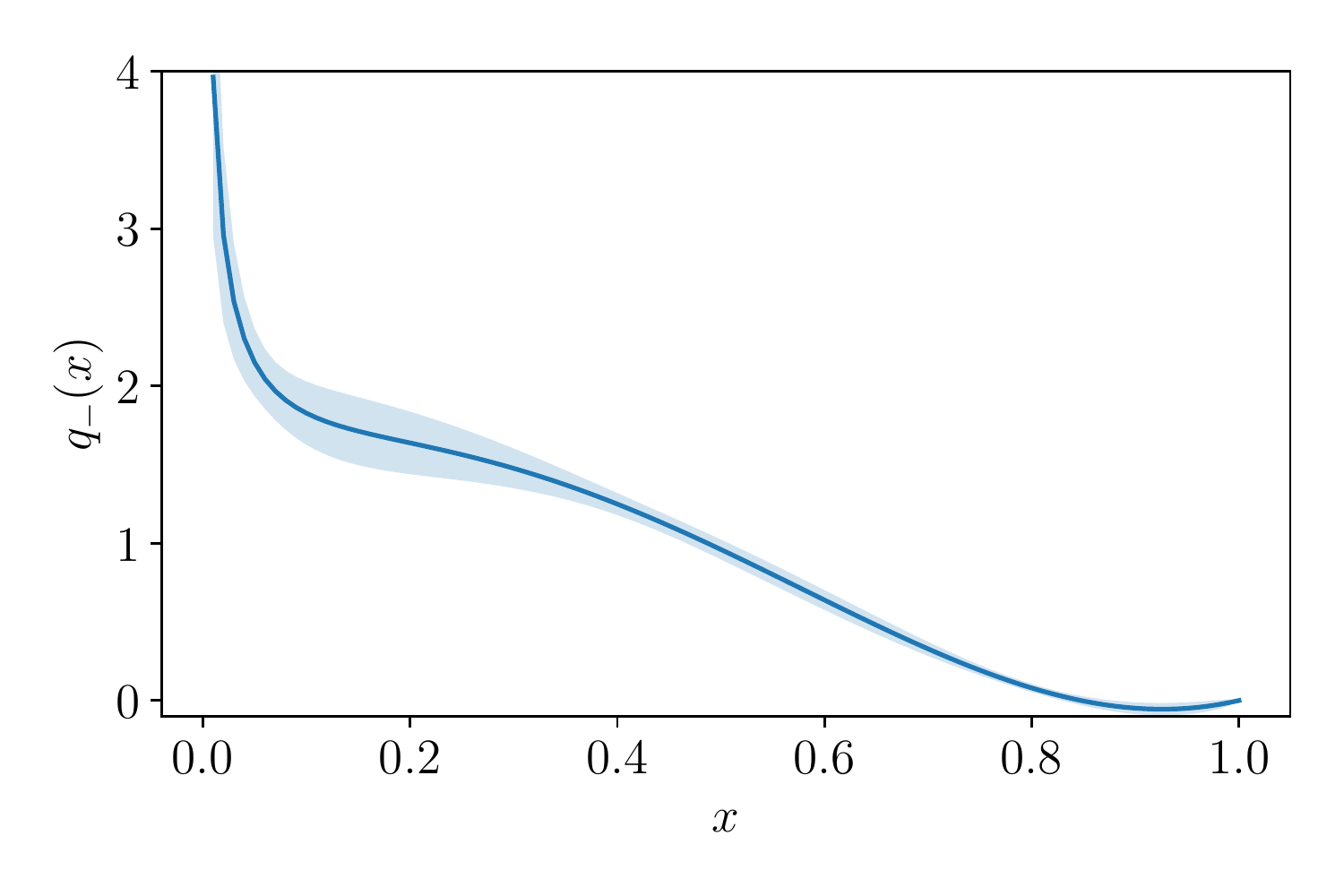}
\includegraphics[width=0.48\textwidth]{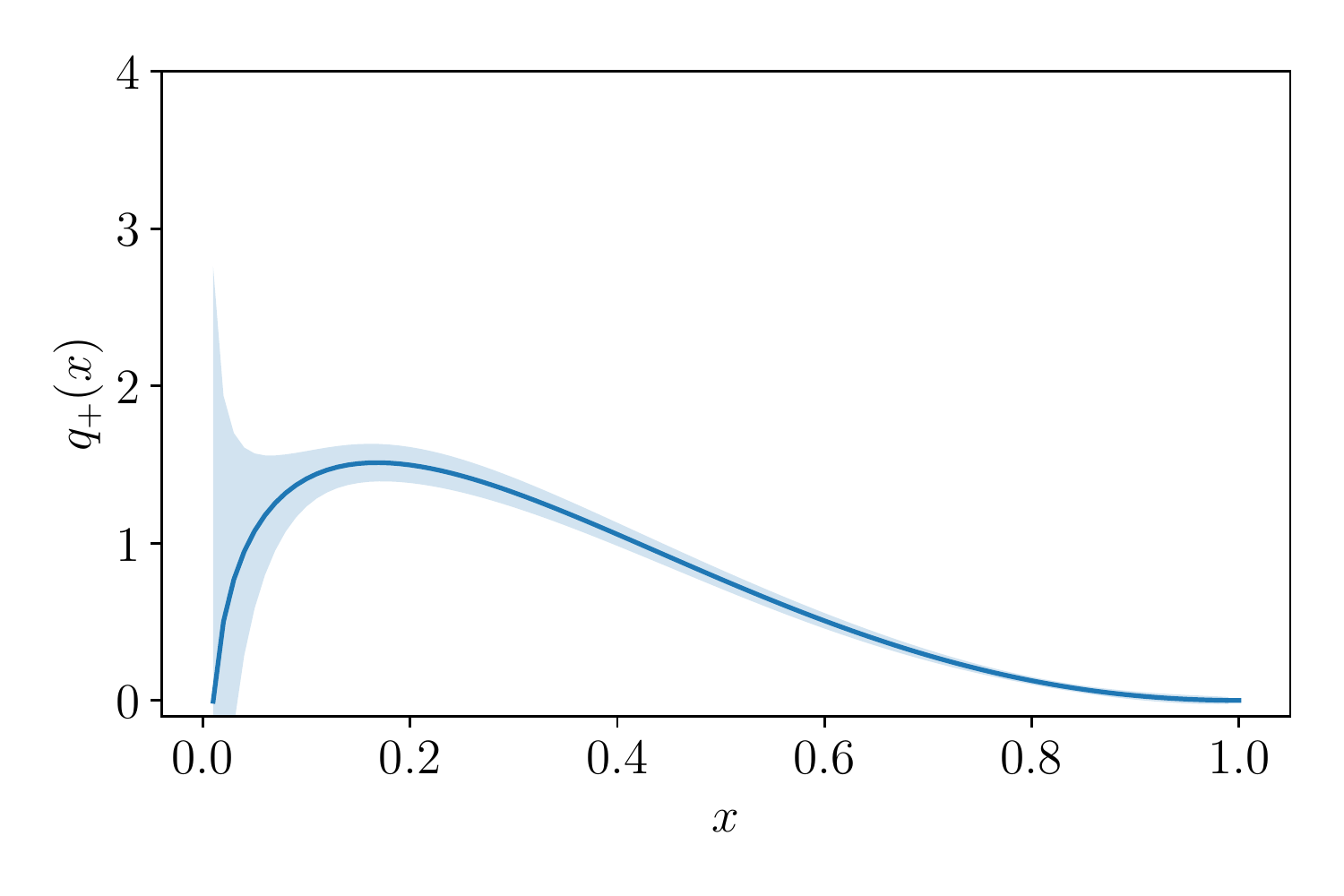}
\caption{\label{fig:many_model_aic} The results of the AICc weighted average of the models of Sec.~\ref{sec:vary_numbers}.}
\end{figure}

Based upon our studies of the models, we should select which ones to include into the model averaging. The models without a $P_1$ term do differ from the rest of the models, which may give a reason to exclude them. Since their $L^2$ was so much higher than the rest, they are exponentially suppressed in the AICc average. We include them in the average anyway in order not to bias ourselves at all due to their discrepancy. The model average of the PDFs is shown in Fig.~\ref{fig:many_model_aic}. These PDFs share many of the features shown in Figs.~\ref{fig:many_model_pdfs_real}-~\ref{fig:many_model_pdfs_imag3}. 

\subsection{Comparison with global fits}
\begin{figure}[!htp]
\centering
\includegraphics[width=0.48\textwidth]{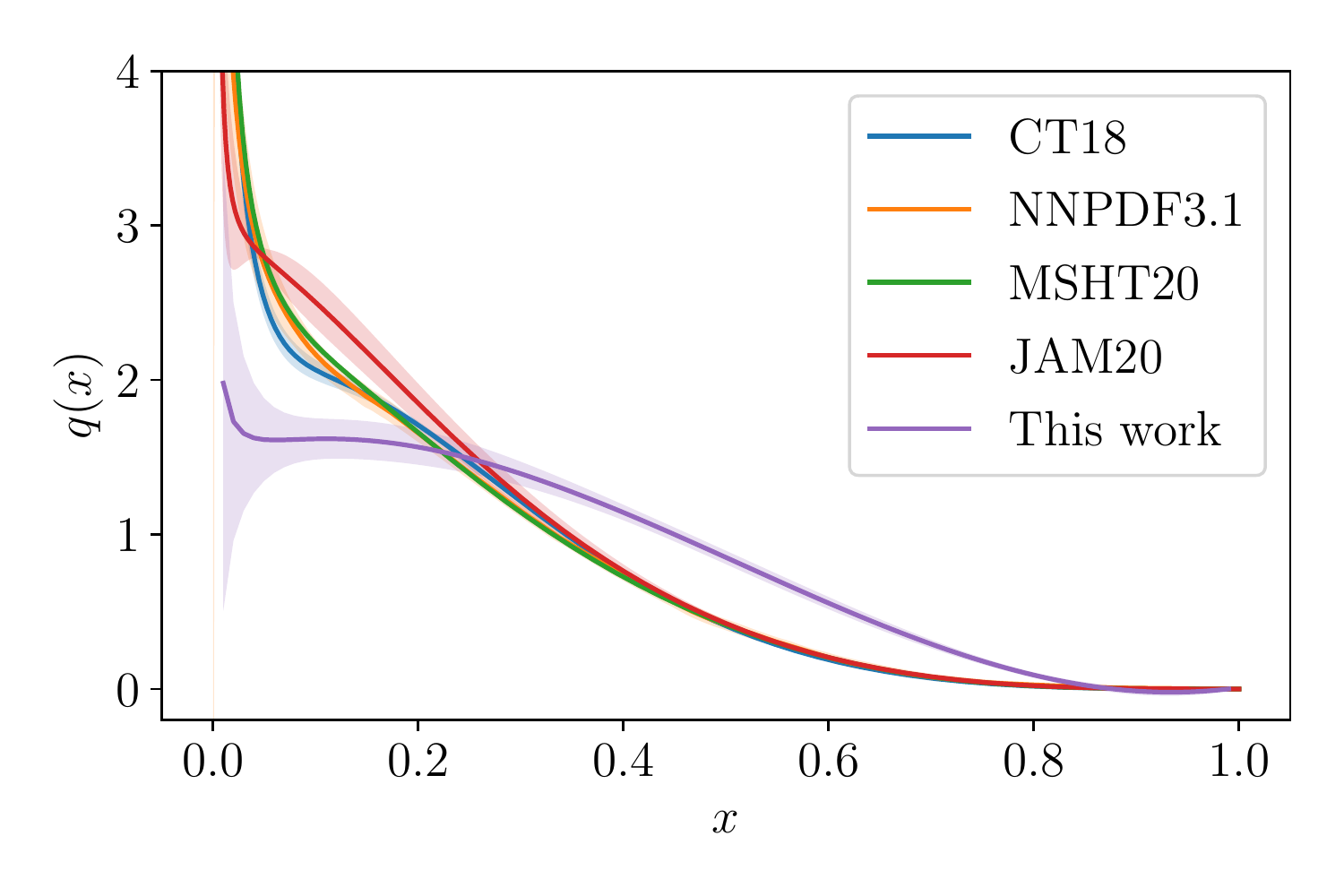}
\includegraphics[width=0.48\textwidth]{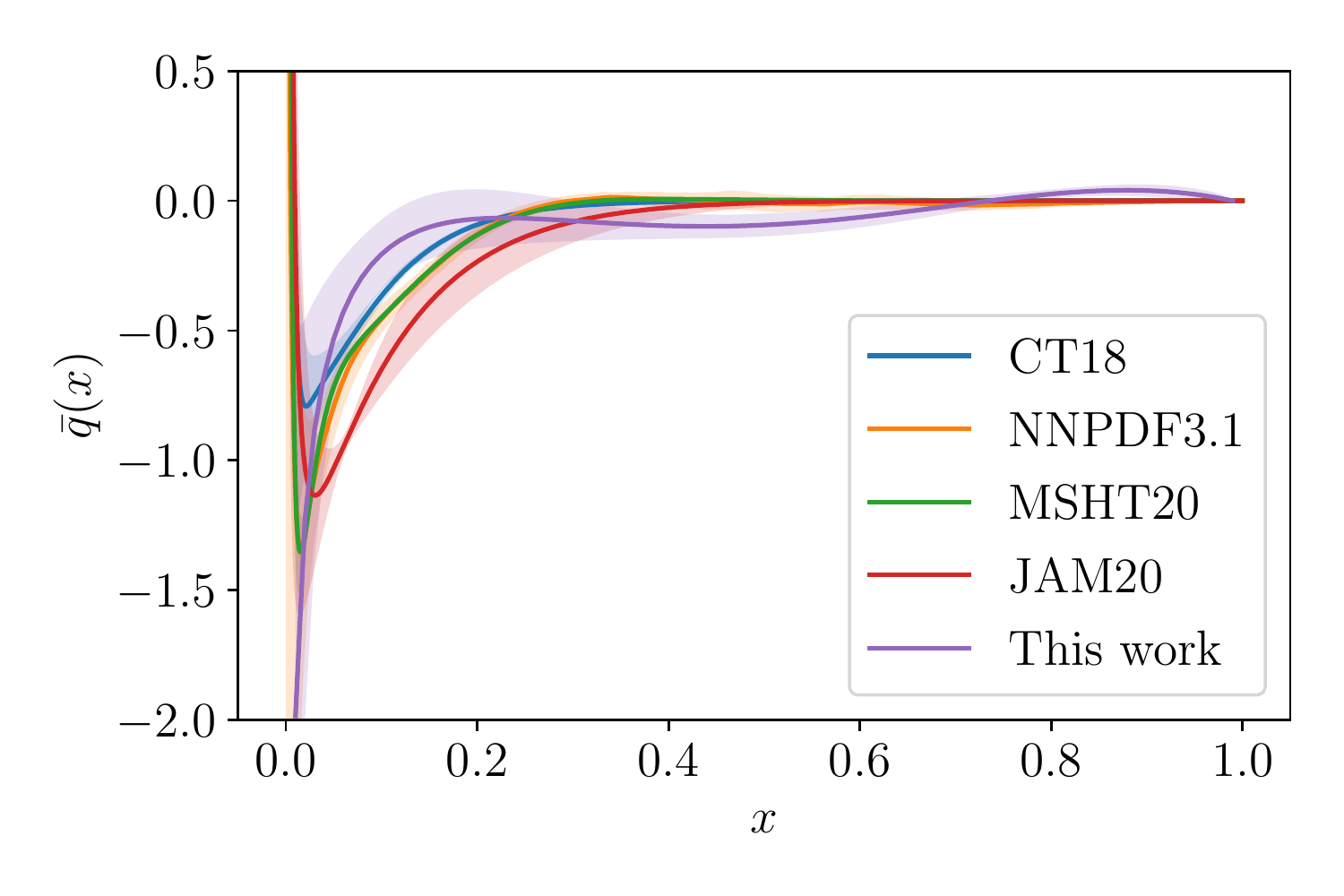}
\includegraphics[width=0.48\textwidth]{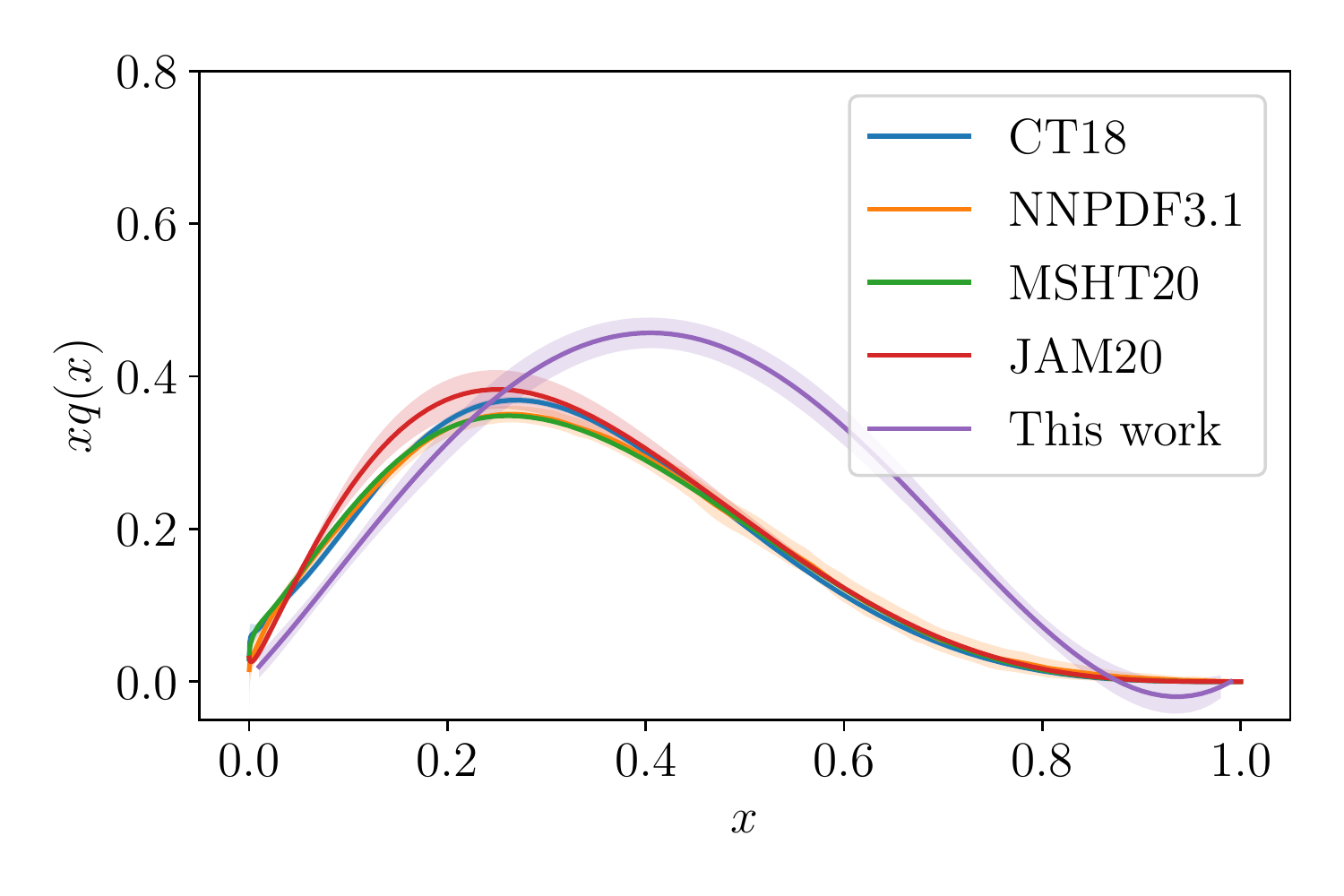}
\includegraphics[width=0.48\textwidth]{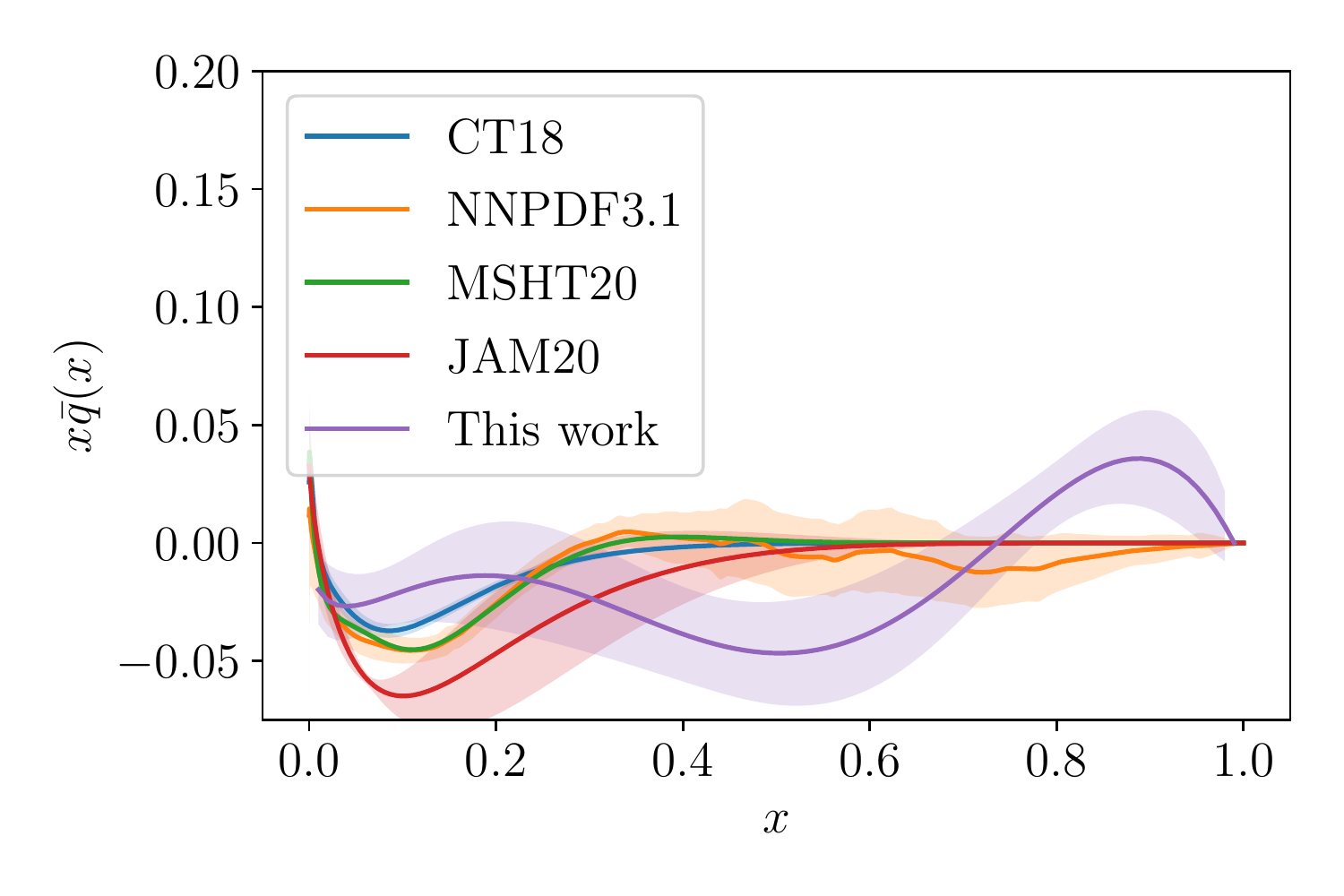}
\caption{\label{fig:global} A comparison of the AICc averaged results to the global fit PDFs, CT18~\cite{Hou:2019qau}, NNPDF 3.1~\cite{Ball:2017nwa}, MSHT'20~\cite{Bailey:2020ooq}, and JAM20~\cite{Moffat:2021dji}. The upper plots are the parton distributions and the lower plots are the distributions weighted by $x$ to emphasize the large $x$ region. }
\end{figure}

A comparison with the phenomenological fits to global collider experiment results can be made, though it is possibly premature given the heavy pion mass of these ensembles. Previous lattice calculations of local matrix elements, obtained moments of the PDF~\cite{Constantinou:2014tga} at these heavy quark masses, with values higher than those from global fits, so we may not expect agreement at the larger $x$ region which dominates those moments. Fig.~\ref{fig:global} shows the AICc averaged results for $q$ and $\bar{q}$ alongside these global fits~\cite{Hou:2019qau,Bailey:2020ooq,Ball:2017nwa,Moffat:2021dji} which were obtained using LHAPDF~\cite{Buckley:2014ana}. The AICc averaged result for $q$ is indeed larger than the phenomenological result for the majority of $x$.

\FloatBarrier

\section{Conclusions}\label{sec:conc}
In this work we have studied the continuum limit extrapolation of the nucleon PDF computed on the lattice via the method of Short Distance Factorization and Ioffe time pseudo-distributions. In this method, we have employed three lattice ensembles all of which have a lattice spacing less than 0.08fm and this work constitutes a significant improvement with respect to our first study~\cite{Joo:2019jct} of discretization errors in lattice computations of Ioffe time distributions. In the first work, we had worked with rather coarse lattices, 0.127fm and 0.091fm while the lattice spacings of the three ensembles employed in the current study, are at the limit of what the current lattice methodologies can achieve without encountering issues related to the freezing of topology and critical slowing down. Currently, taking the continuum limit on the lattice at the physical pion mass with Wilson type quarks is not possible since the generation of at least three ensembles at the physical point with Wilson type fermions is numerically a formidable task for the time being. We therefore consider ensembles at a heavier pion mass all of them at 440 MeV. A continuum extrapolation at the physical pion mass will be performed in future studies once the computational resources render such an endeavor feasible. Additionally, it is well known that the contamination from excited states can plague severely the results of a lattice computation especially as one is approaching the physical pion mass. In this work, we employ the sGEVP method in order to have a better control of the excited states. We also stressed the necessity of dealing explicitly with the various systematic errors that are unavoidable in any lattice calculation. We adopted a Bayesian approach where we build explicit models of the PDF and of the associated systematic errors in order to describe our data. These models contain unknown parameters whose most likely values are determined given some prior information and using Bayes' theorem. Another novelty of this work is that in this article we are using the Jacobi polynomials as a way of tackling the unavoidable inverse problem that one encounters when trying to obtain the Bjorken-$x$ PDF from Ioffe time lattice data. Up to now, we had mainly been using parameterizations of polynomials of $\sqrt{x}$  or neural networks. We advocate that the Jacobi polynomials constitute a very versatile and flexible approach that eliminates the introduction of model dependence.       
\FloatBarrier
\section{ Acknowledgements}
We would like to thank all the members of the Hadstruc collaboration for fruitful and stimulating exchanges. 
We are grateful to the ALPHA collaboration and the CLS effort for sharing some of their ensembles with us. In particular, we would like to thank Mattia Dalla Brida, Hubert Simma and Rainer Sommer. SZ thanks Balint Toth for enlightening discussions. 
The authors also thank C. Egerer and R. Sufian for helpful conversations. This work is supported by Jefferson
Science Associates, LLC under U.S. DOE Contract \#DE-AC05-06OR23177.
KO was supported in part by U.S.  DOE grant \mbox{
  \#DE-FG02-04ER41302} and in part by the Center for Nuclear Femtography grants \#C2-2020-FEMT-006, \#C2019-FEMT-002-05.
    AR was supported in part by U.S. DOE Grant
\mbox{\#DE-FG02-97ER41028. }  J.K. was supported
in part by the U.S. Department of Energy under contract
DE-FG02-04ER41302, Department of Energy Office of Science Graduate
Student Research fellowships, through the U.S. Department of Energy,
Office of Science, Office of Workforce Development for Teachers and
Scientists, Office of Science Graduate Student Research (SCGSR)
program and is supported by U.S. Department of Energy grant DE-SC0011941.  
The authors gratefully acknowledge the computing time
granted by the John von Neumann Institute for Computing (NIC) and
provided on the supercomputer JURECA at J\"ulich Supercomputing Centre
(JSC)~\cite{jureca}. This work benefited also from access to the Joliot-Curie supercomputer of the TGCC (CEA) in France as part of a "grand challenge" project (project id: gch413) awarded by GENCI (Grand Equipement National de Calcul Intensif). We would also like to thank the Texas Advanced Computing Center (TACC) at the University of Texas at Austin for providing HPC resources
on Frontera~\cite{frontera} that have contributed to the results in this paper. 
 We acknowledge the facilities of the USQCD Collaboration used for this research in part, which are funded by the Office of Science of the U.S. Department of Energy. This work was performed in part using computing
facilities at the College of William and Mary which were provided by
contributions from the National Science Foundation (MRI grant
PHY-1626177), and the Commonwealth of Virginia Equipment Trust Fund.
The authors acknowledge William \& Mary Research Computing for providing computational resources and/or technical support that have contributed to the results reported within this paper. This work used the Extreme Science and Engineering Discovery Environment (XSEDE), which is supported by National Science Foundation grant number ACI-1548562~\cite{xsede}.
 In addition, this work used resources at
NERSC, a DOE Office of Science User Facility supported by the Office
of Science of the U.S. Department of Energy under Contract
\#DE-AC02-05CH11231, as well as resources of the Oak Ridge Leadership Computing Facility at the Oak Ridge National Laboratory, which is supported by the Office of Science of the U.S. Department of Energy under Contract No. \mbox{\#DE-AC05-00OR22725}. In addition, this work was made possible using results obtained  at NERSC, a DOE Office of Science User Facility supported by the Office of Science of the U.S. Department of Energy under Contract \mbox{\#DE-AC02-05CH11231}, as well as resources of the Oak Ridge Leadership Computing Facility (ALCC and INCITE) at the Oak Ridge National Laboratory, which is supported by the Office of Science of the U.S. Department of Energy under Contract No. \mbox{\#DE-AC05-00OR22725}.  The software libraries used on these machines were Chroma~\cite{Edwards:2004sx}, QUDA ~\cite{Clark:2009wm,Babich:2010mu}, QDP-JIT~\cite{Winter:2014dka} and QPhiX~\cite{Joo:2013lwm,optimising} developed with  
 support from the U.S. Department of Energy, Office of Science, Office of Advanced Scientific Computing Research and Office of Nuclear Physics, Scientific Discovery through Advanced Computing (SciDAC) program, and of the U.S. Department of Energy Exascale Computing Project.
\\\\\\\\\\\\\

\newpage

%%%%%%%%%%%%%%%%%%%%%%%%%%%%%%%%%%%%%%%%%%%%%%%%%%%%%%%%%
%%                              BIBLIO
%%%%%%%%%%%%%%%%%%%%%%%%%%%%%%%%%%%%%%%%%%%%%%%%%%%%%%%%%

%\bibliographystyle{unsrt} % not to be used with revtex
\bibliography{dynppdfs.bib}
\bibliographystyle{jhep}

\end{document}